\documentclass[rmp,aps,twocolumn,10pt]{revtex4-1}
\usepackage[utf8]{inputenc} 
\usepackage[T1]{fontenc}
\usepackage{textcomp} 
\usepackage{lmodern}
\usepackage{graphicx}
\graphicspath{{figures/}}
\usepackage{amsmath,amssymb,xcolor,bm}
\usepackage{dsfont}
\usepackage{siunitx}
\usepackage{physics}
\usepackage[colorlinks=true,allcolors=blue]{hyperref}
\hypersetup{bookmarksdepth=2}

\renewcommand{\d}{\mathrm{d}}          % differential d
\renewcommand{\O}{\mathcal{O}}         % observable / order symbol
\newcommand{\calS}{\mathcal{S}}         % action

\renewcommand{\r}{\mathbf{r}}          % position (NB: with [angstrom], use $\mathring{\A}$ for the angstrom symbol)
\renewcommand{\v}{\mathbf{v}}          % velocity
\renewcommand{\k}{\mathbf{k}}          % wavevector
\newcommand{\q}{\mathbf{q}}            % wavevector (parallel)
\newcommand{\p}{\mathbf{p}}            % momentum
\newcommand{\n}{\mathbf{n}}            % density (Keldysh vector)
\renewcommand{\P}{\mathbf{P}}          % polarization
\newcommand{\E}{\mathbf{E}}            % electric field
\newcommand{\F}{\mathbf{F}}            % force
\newcommand{\D}{\mathbf{D}}            % displacement field
\newcommand{\G}{\mathbf{G}}            % Green's function matrix
\newcommand{\N}{\mathbf{N}}            % density matrix in Keldysh space

\newcommand{\psibar}{\bar{\psi}}
\newcommand{\bpsi}{\boldsymbol{\psi}}
\newcommand{\bpsibar}{\bar{\boldsymbol{\psi}}}
\newcommand{\bphi}{\boldsymbol{\phi}}

\renewcommand{\l}{\ell}                % liquid subscript (friction sections)
\newcommand{\w}{\ell}            % liquid subscript (Keldysh sections)
\newcommand{\e}{\mathrm{e}}            % electron subscript 
\newcommand{\h}{\mathrm{h}}            % hydron subscript
\newcommand{\s}{\mathrm{s}}            % solid subscript
\newcommand{\R}{\mathrm{R}}			   % Keldysh superscripts
\newcommand{\A}{\mathrm{A}}
\newcommand{\K}{\mathrm{K}}
\renewcommand{\S}{\mathrm{S}}
\newcommand{\ext}{\mathrm{ext}}        % external
\newcommand{\ind}{\mathrm{ind}}        % induced
\newcommand{\kB}{k_{\rm B}}            % Boltzmann constant
\newcommand{\kF}{k_{\rm F}}            % Fermi wavevector

\newcommand{\nB}{n_\mathrm{B}}         % Bose-Einstein
\newcommand{\nF}{n_\mathrm{F}}         % Fermi-Dirac

\newcommand\re[1]{\textnormal{Re}\left[#1\right]}
\newcommand\im[1]{\textnormal{Im}\left[#1\right]}

\begin{document}

\title{Fluctuation-induced and quantum effects in nanofluidic transport}

\author{Adrien Sutter$^{\dagger}$}
\affiliation{Laboratoire de Physique de l'\'Ecole Normale Sup\'erieure, Paris, France \\ and The Quantum Plumbing Lab, \'Ecole Polytechnique F\'ed\'erale de Lausanne (EPFL), Lausanne, Switzerland}
\author{Peter Gispert$^{\dagger}$}
\affiliation{The Quantum Plumbing Lab, \'Ecole Polytechnique F\'ed\'erale de Lausanne (EPFL), Lausanne, Switzerland}
\author{Baptiste Coquinot}
\affiliation{Institute of Science and Technology Austria (ISTA), Klosterneuburg, Austria}
\author{Lyd\'eric Bocquet}
\affiliation{Laboratoire de Physique de l'\'Ecole Normale Sup\'erieure, Paris, France}
\author{Nikita Kavokine}
\email{nikita.kavokine@epfl.ch}
\affiliation{The Quantum Plumbing Lab, \'Ecole Polytechnique F\'ed\'erale de Lausanne (EPFL), Lausanne, Switzerland~}
\author{\small $\dagger$: these authors contributed equally. }

\date{\today}
\begin{abstract}
The hydrodynamic wall has traditionally been considered a featureless object, whose only role is to provide a boundary for fluid flow. Yet, there is now ample evidence that at nanometer scales, liquid flows are sensitive to the wall's internal -- in particular, electronic -- degrees of freedom. Here, after reviewing the experimental evidence for nanoscale liquid-electron couplings, we present the theoretical advances that have allowed for their quantitative understanding. We discuss how a quantum description of the liquid-solid interface reveals the influence of electron dynamics on classical fluid transport, in the form of the fluctuation-induced quantum friction effect. 
Quantum friction is at the root of liquid-electron coupled transport phenomena, that may be combined into a hydro-electronic transport matrix. We present analytical formulas for the hydro-electronic transport coefficients, that allow for their quantitative estimation in practical cases; we further outline the potential consequences of coupled liquid-electron transport for the water-energy nexus. Fluctuation-induced and quantum effects at liquid-solid interfaces represent an emerging interface between fluid dynamics and condensed-matter physics, and a largely uncharted territory for both theory and experiment. 
\end{abstract} 
\maketitle
\tableofcontents

\section{Introduction and Outline}

Nanofluidics is the study of fluid transport through channels with nanoscale dimensions -- that is, on the order of a few molecular sizes. Nanoscale fluid transport represents the frontier where the continuum of hydrodynamics meets the atomic nature of matter. This is what sets nanofluidics apart from its parent field, microfluidics. While microfluidic transport is readily described within continuum hydrodynamics, one may expect the emergence of new physical laws when pushing channel dimensions down to the nanoscale \cite{schochtransportphenomena2008}. Beyond its fundamental importance, the understanding of these laws is at the heart of several important technologies \cite{bocquetphysicstechnological2014,alurufluidselectrolytesconfinement2023}. A prime example is electrochemical energy storage: electric double layer capacitors, or \emph{supercapacitors}, employ nanoporous carbon electrodes and liquid electrolytes to store energy at much higher density than conventional capacitors \cite{chmiolaanomalousincreasecarbon2006}. Furthermore, several of the water splitting technologies that are used for hydrogen gas production rely on nanoporous membranes and their gas exclusion or ion selectivity properties \cite{parkmembranestrategies2022}. The same is true for membrane-based water desalination, and its reverse process, osmotic energy conversion -- the generation of electricity through the mixing of fresh water and seawater -- also known as \emph{blue energy} \cite{phillipfutureseawater2011,sirianewavenues2017}. Last but not least, many key biological processes, such as neurotransmission or blood filtration by the kidney, rely on ion and water transport through protein-based channels with nanoscale dimensions. The possibility of mimicking these processes in artificial channels has recently positioned brain-inspired ionic computing as a compelling technological perspective -- deeply rooted in the physics of nanoscale fluid transport \cite{robinnanofluidicscrossroads2023,emmerichnanofluidics2024}. 

In the early days of the field, there was an expectation that stronger and stronger nanoconfinement would eventually lead to a breakdown of hydrodynamics -- in the sense of a failure of the Navier-Stokes equation. The main assumption in the Navier-Stokes treatment lies in describing the internal energy dissipation in the fluid -- due to intermolecular collisions -- with a single viscosity coefficient. It turns out, however, that the coarse-graining procedure underlying the notion of viscosity is extremely robust, with scaling arguments predicting that it should hold down to $\sim 1~\rm nm$ confinement \cite{bocquetnanofluidicsbulkinterfaces2010}. Today, experiments have reached well-controlled fluid confinement down to the nanometer and even sub-nanometer scale, and indeed they have not significantly challenged the notion of viscosity \cite{kavokinefluidsnanoscalecontinuum2021}. However, many of those experiments do point to a failure of hydrodynamics, not as a failure of viscosity {\it per se}, but in a more subtle way, as discussed below. Overall, fluid transport at the ultimate scales deviates in many respects from the predictions of the continuum framework.

The crossover from microfluidics to nanofluidics occurs way above molecular scale confinement. Already at the 100 nm scale, interfacial effects introduce a non-negligible coupling between hydrodynamic flow and ionic current. This coupling is captured by a transport matrix that relates the thermodynamic forces that may be applied to a nanochannel and the corresponding fluxes. The forces include pressure, electric field, solute concentration and temperature gradients. The corresponding fluxes are the liquid mass flow, the ionic current in solution, the solute particle current, the heat current \cite{marbachosmosismolecular2019}. Restricting the forces to pressure gradient and electric field, the transport matrix takes the form 
\begin{equation}
\left(
\begin{array}{c} 
Q \\ 
I 
\end{array}
\right) 
=
\left(
\begin{array}{cc}
\mathcal{L} & \mu_{\rm eo} \\
\mu_{\rm eo} & G_{\rm i} 
\end{array}
\right)
\left(
  \begin{array}{c}
    \Delta P/L \\ \Delta V / L 
  \end{array}
\right). 
\label{transport_matrix}
\end{equation}
Here, the pressure gradient $\Delta P/L$ and voltage gradient $\Delta V /L$ ($L$ being the channel length) induce a liquid flow rate $Q$ and an ionic current $I$. The diagonal terms of the matrix are the hydraulic permeance $\cal L$ and the ionic conductance $G_{\rm i}$. The off-diagonal terms, which must be equal due to Onsager symmetry, describe the cross-coupling effects. \emph{Electro-osmosis} refers to a liquid flow induced by a voltage drop, while a \emph{streaming current} is an ionic current induced by a pressure drop~\cite{marbachosmosismolecular2019}. Both effect are controlled by the same transport coefficient: the electro-osmotic mobility $\mu_{\rm eo}$. Coupled transport effects are a hallmark of nanofluidics and have been extensively reviewed over the past years \cite{bocquetnanofluidicsbulkinterfaces2010,kavokinefluidsnanoscalecontinuum2021,emmerichnanofluidics2024}. They are also at the heart of important applications, such as electro-osmotic pumping \cite{abdelghani-idrissiresonantosmotic2025}, or the diffusio-osmotic mechanism for blue energy harvesting \cite{siriagiantosmoticenergy2013}. 

From the theoretical point of view, the Poisson-Nernst-Planck-Stokes (PNPS) transport framework has very often been taken as the starting point for the determination of nanofluidic transport coefficients \cite{kavokinefluidsnanoscalecontinuum2021}. It is a set of continuum and mean-field equations that couple ion transport to solvent flow. For a $1:1$ electrolyte solution, the ion concentrations $\rho_{\pm}$ satisfy Nernst-Planck transport equations:
\begin{equation}
\frac{\partial \rho_{\pm}}{\partial t} 
= \nabla \cdot \left( D_{\pm} \nabla \rho_{\pm} \pm \frac{eD_{\pm}}{k_B T} (\nabla \phi)\rho_{\pm} - v \rho_{\pm} \right), 
\end{equation}
where $D_{\pm}$ are the ion diffusion coefficients. The electrostatic potential $\phi$ solves the Poisson equation: 
\begin{equation}
\Delta \phi = -e \frac{\rho_+ - \rho_-}{\epsilon}, 
\end{equation}
where $\epsilon$ is the dielectric permittivity of the solvent, and the flow field $v$ satisfies the Stokes equation: 
\begin{equation}
\eta \Delta v - e \left( \rho_{+} - \rho_{-} \right) \nabla \phi = \nabla P, 
\end{equation}
with $\eta$ the solvent viscosity. These equations must be supplemented by boundary conditions at the channel wall: traditionally, Navier partial slip for the hydrodynamics and a fixed surface charge for the electrostatics.
Qualitatively, the PNPS equations describe the random diffusion of dissolved ions, biased by hydrodynamic advection and electrostatic forces. These forces combine the effect of external charges (such as the surface charge) and of the ionic mean field; they further enter as a body force in the Stokes equation that governs the hydrodynamic flow. 
Within such a formulation, the theory of nanofluidics retains much of the universality that is the hallmark of hydrodynamics. All transport properties are determined by a few coarse-grained parameters: viscosity $\eta$, diffusion coefficient $D$, slip length $b$, surface charge density $\Sigma$, and the bulk salt concentrations \cite{herreropoissonboltzmann2024}. Let us note that, at equilibrium, the PNPS framework reduces to the well-known Poisson-Boltzmann (PB) equations. 

While remarkably successful at capturing coupled transport, as well as non-linear phenomena -- such as diode effects \cite{vlassiouknanofluidicdiode2007,picallonanofluidicosmotic2013} -- the PNPS framework has encountered serious limitations at scales smaller than $\sim 10~\rm nm$: the so-called \emph{single-digit nanopore} regime \cite{fauchercriticalknowledge2019}. Two main reasons can be identified. First, the power of the PNPS equations is in linking the surface to the bulk, by establishing how the various quantities (velocities, concentrations) evolve along the confined direction. But at sufficiently strong confinement, all quantities become essentially uniform across the channel, trivializing the PNPS equations. This is particularly apparent at the level of the Stokes equation, where the viscosity drops out under plug flow conditions: hydrodynamics does not fail, but it also no longer matters \cite{mouterdemolecularstreamingits2019}. Second, when the transport is all surface, the surface can no longer be described as a boundary condition -- a featureless hydrodynamic wall. Zooming in to the 10 nm scale and below, it is clear that the wall is not simply where the fluid ends. The wall is a condensed-matter system in its own right, with its microscopic structure and internal degrees of freedom: electrons and phonons. One must therefore consider the coupled dynamics of the fluid and wall degrees of freedom. In the case of electrons, these are intrinsically quantum dynamics: thus, nanofluidic channels can be viewed as hybrid quantum-classical systems, where quantum electron dynamics may imprint on fluid transport properties \cite{kavokinefluctuationinducedquantum2022}. 

Obtaining a complete theoretical, or even numerical framework that fully captures the interaction of all the relevant degrees of freedom in nanofluidic transport at the single-digit scale remains a challenge, and a goal of current nanofluidics research. However, recent theoretical developments have made possible the coupled description of microscopic liquid and electron dynamics, through the application of a quantum field theory framework to the liquid-electron system \cite{kavokinefluctuationinducedquantum2022,coquinotquantumfeedback2023}. The main prediction of this framework was a fluctuation-induced quantum friction (QF) effect -- a mechanism of liquid-solid friction that amounts to direct momentum transfer from the flowing liquid to the electrons in the solid. As such, QF introduces an interaction between fluid transport within a channel and electronic transport in the channel wall: the electronic current and corresponding voltage thus enter the transport matrix in Eq.~\eqref{transport_matrix} as a new flux-force pair, opening up qualitatively new functionalities for nanofluidic transport \cite{coquinothydroelectricenergy2024}. 

Here, we review the QF theoretical framework and outline its physical consequences. The Article is meant as a toolbox for theoretical and experimental researchers alike. We start with a toy model description that highlights the main physical principles behind QF and coupled liquid-electron transport; we further provide simplified formulas that can be applied to a given experimental situation for estimating the order of magnitude of these effects. We then delve into the  quantum field theory formalism, and show on rigorous grounds how QF and associated phenomena are obtained from perturbation theory in the liquid-electron interaction. 

In more detail, the outline of this Article is as follows. In Sec. II, we provide a survey of the main experimental results that have challenged the notion of hydrodynamic wall in nanofluidic systems, thus motivating the QF framework. This Section is not meant to be exhaustive, and we refer the reader to some excellent recent reviews \cite{fauchercriticalknowledge2019,alurufluidselectrolytesconfinement2023,emmerichnanofluidics2024,licarbonnanotubenanofluidics2025} for a complete picture of the experimental state-of-the-art. In Sec. III, we establish the theoretical context and review the key precursors of QF theory. These are mainly the theories of fluctuation-induced forces in the liquid and gas phases, starting with the famous Lifshitz theory of the van der Waals force. At this occasion we also introduce \emph{surface response functions} -- compact descriptors of a semi-infinite medium's response to an applied potential, that will be important ingredients for the theory presented in the next sections. Sec. IV delves into the microscopic origin of liquid-solid friction, within a classical Langevin equation framework. This simplified framework allows us to contrast the usual hydrodynamic friction that originates from surface roughness with the fluctuation-induced contribution -- the classical limit of QF. We provide quantitative estimates of QF for a few model liquid-solid systems and discuss its quantum character. In Sec V, we examine the practical consequences of QF in the form of coupled fluid-electron transport. We introduce the hydro-electronic transport matrix and provide order-of-magnitude estimates for its coefficients. In Sec. VI, we introduce the quantum field theory (Keldysh) formalism that allows for the derivation of the results in the previous sections on rigorous grounds. We establish the Feynman rules for the perturbation series in the liquid-electron interaction, and show how Feynman diagrams illuminate the microscopic mechanisms at play. Sec. VII contrasts analytical theory with molecular simulation approaches to nanofluidic systems, discussing their potential to capture electron dynamics. Finally, we outline in Sec. VIII perspectives, challenges and open questions. 
  
\section{Experimental signatures: beyond the hydrodynamic wall}

Over the last twenty years, nanofluidic experiments have revealed that channel walls influence the structure and dynamics of the confined liquid beyond merely geometrical constraints and static surface charges. Today, there is ample evidence that, under nanoscale confinement, the internal degrees of freedom of the channel wall couple to the confined liquid and give rise to subtle interactions across the liquid-solid interface.

\subsection{Static structure and phase behavior}
Several experiments indicate that the static structure of ionic solutions and ionic liquids under nanoconfinement is sensitive to the electronic properties of the confining walls, in particular to the degree of metallicity. This dependence plays an important role in electrostatic double-layer capacitors, also known as supercapacitors. The simplest model of a conventional capacitor consists of two oppositely charged plates separated by a dielectric medium. Upon application of a potential difference $\Delta V$, the two plates have opposite charges $\pm \Delta Q$, related to the applied potential through the areal capacitance $C = \Delta Q / (\Delta V \, A) = \epsilon / d$, where $A$ is the area of each capacitor plate, $d$ the plate separation and $\epsilon$ the dielectric constant of the medium in between. In electrolytes, a capacitor configuration is naturally provided by the electrical double layer formed at a solid-liquid interface: the surface charge of the solid electrode is screened by an oppositely charged, ion-rich layer in the adjacent liquid. In supercapacitors, the thickness $d$ of the double layer is on the molecular scale, resulting in extremely high capacitance.  

A material often used for supercapacitor electrodes is nanoporous carbide-derived carbon, owing to its high surface-to-volume ratio and well-controlled nanometric pore-size distribution that extends down to effectively 1D pores. Experiments have shown that for both pure ionic liquids and ionic solutions, the maximum areal capacitance is achieved when the pore size matches the ion size \cite{chmiolaanomalousincreasecarbon2006, largeotrelationionsize2008}. This effect has also been observed in 2D slits between graphene sheets \cite{leetunablesubnanoporesgraphene2016}, and is interpreted as resulting from an interplay of geometric confinement and strong screening of ionic interactions by the electrons in the pore walls \cite{kondrattheorysimulationsionic2023}. 

\begin{figure}
	\begin{center}
		\includegraphics[width=\columnwidth]{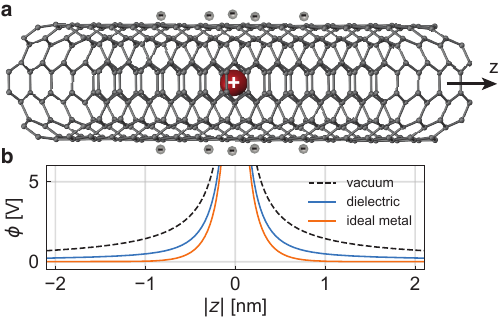}
		\caption{\textbf{Superionic state.} (a) Positive ion in the center of a (6,6) carbon nanotube with radius $R = \SI{0.4}{\nm}$. (b) Resulting Coulomb potential along the tube axis in vacuum (without the nanotube), for a nanopore inside a bulk dielectric material with dielectric constant $\epsilon=3$ and for a nanotube modeled as an ideal metal. The ideal-metal boundary condition leads to an exponentially screened effective interaction along the tube axis, known as the \emph{superionic state} \cite{kondratsuperionicstatedoublelayer2011}.}
		\label{fig:superionic-state}
	\end{center}
\end{figure}

In concentrated ionic liquids, mean-field Poisson-Boltzmann theory does not hold \cite{kornyshevdoublelayerionicliquids2007}. It has to be extended to account for finite size and electrostatic correlation effects, which result in overscreening at charged surfaces: the ion density exhibits an oscillatory behavior because each charged layer overcompensates the previous one \cite{kondratsuperionicstatedoublelayer2011, dudkasuperionicliquidsconducting2019}. Effectively, the surface charge is screened over longer distances due to these oscillations, thereby limiting the resulting capacitance. Molecular dynamics simulations of realistic microporous carbon electrodes held at constant potential \cite{merletmolecularorigin2012} have revealed that geometric confinement limits overscreening, leading to more efficient charge storage. The simulations further reveal that the ions partially dehydrate to pack more densely into the pore. This partial dehydration is enabled by the screening of the ions' Coulomb potential by image charges in the pore wall, which partially compensate for the loss of the solvation shell. Such a dense packing of like-charge ions into a pore, enabled by electronic screening, has also been predicted theoretically and dubbed a \emph{superionic state} \cite{kondratsuperionicstatedoublelayer2011, dudkasuperionicliquidsconducting2019}, see figure~\ref{fig:superionic-state}.

The simulations of \textcite{merletmolecularorigin2012} effectively assumed an ideal metal screening behavior for the carbon walls. A theoretical investigation of ionic screening inside 1D and 2D pores with walls described within the Thomas-Fermi model \cite{rochesterinterionicinteractionsconducting2013} -- which interpolates between insulating and metallic behavior -- has shown that the superionic state can still arise if a non-vanishing Thomas-Fermi length is taken into account, albeit with quantitative differences. The influence of the Thomas-Fermi length on properties of the solid-liquid interface -- such as wettability -- has been further explored theoretically \cite{kaiser2017electrostatic,schlaichelectronicscreeningusing2022}. Recently, the surface response function formalism has extended these studies to account for realistic electronic screening properties beyond the Thomas-Fermi model, in both 2D and 1D geometries \cite{gispertelectrostaticscreeningnanotubes2025, kavokineInteractionConfinement2022}.

On the experimental side, a direct impact of the wall metallicity on ionic liquid structure was particularly apparent in a study of confinement-induced capillary freezing \cite{comtetnanoscalecapillary2017,laine2020nanotribology}. \textcite{comtetnanoscalecapillary2017} used a micrometric tungsten tip supported on a quartz tuning fork to confine an ionic liquid between the tip and a flat substrate made of either mica, graphite, doped silicon or platinum. They detected the freezing of the ionic liquid as a sharp increase in the mechanical impedance of the tuning fork beyond a critical confinement length. Strikingly, the critical confinement depended strongly on the underlying substrate. It increased with substrate conductivity, from \SI{15}{\nm} for mica to \SI{160}{\nm} for platinum. The phase transition was rationalized by a competition between entropy (favoring the liquid state) and surface energy (favoring the frozen state) -- evaluated within a Thomas-Fermi model for the electrostatic screening by the substrate. Stronger screening (shorter Thomas-Fermi length) results in a reduced surface energy and thus earlier onset of the freezing transition. 

\subsection{Dynamics and friction}
The transport properties of nanofluidic channels depend strongly on the channel wall material. In many cases, this dependence lacks a satisfactory theoretical explanation and suggests a subtle role played by the channel wall's internal degrees of freedom.

\subsubsection{Surface charge regulation and surface transport}
Ion transport in nanoscale channels is strongly affected by the surface charge of the channel wall \cite{kavokinefluidsnanoscalecontinuum2021}. The surface charge attracts counterions into the Debye layer, thereby determining the number of charge carriers inside the channel, hence the conductance. The net charge of the Debye layer gives rise to coupled ion-fluid transport phenomena. For instance, an ionic current can be induced not only by an applied voltage, but also by a pressure gradient (streaming current) or a salt concentration gradient (diffusio-osmotic current).

\begin{figure*}
	\centering
	\includegraphics[width=\textwidth]{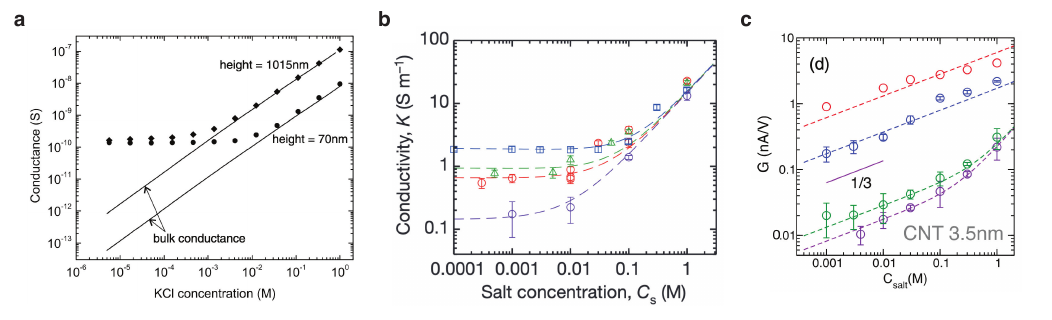}
	\caption{\textbf{Ion conductivity and surface charge.} (a) Electrical conductance of aqueous KCl solution in fused silica channels of \SI{1015}{\nm} and \SI{70}{\nm} height as a function of KCl concentration, reproduced from \textcite{stein2004surface}. Due to the surface charge on the channel walls, the conductance exhibits a plateau at low concentrations. (b) Electrical conductivity of aqueous KCl solution in single transmembrane boron nitride nanotubes at $\rm pH = 5$ as a function of KCl concentration.  The tubes are about \SI{1}{\mu\m}  long and their radii are (from bottom to top) \SI{40}{\nm}, \SI{29}{\nm}, \SI{22}{\nm} and \SI{15}{\nm}, reproduced from \textcite{siriagiantosmoticenergy2013}. Due to the surface charge on the nanotube wall, the conductivity plateaus at low concentrations. (c) Electrical conductance of aqueous KCl solution in a single trans-membrane carbon nanotube with radius $R = \SI{3.5}{\nm}$ and length $L = \SI{3}{\mu\m}$ as a function of KCl concentration. The conductance exhibits a power-law dependence on concentration with exponent 1/3 at low concentrations, and its magnitude depends on pH (from bottom to top pH = 4, 6, 8, 10). This behavior can be rationalized based on a charge regulation model. Figure reproduced from \textcite{secchiscalingbehaviorionic2016}.
	}
	\label{fig:surface-charge}
\end{figure*}

In the PNPS framework, the simplest electrostatic boundary condition is a fixed surface charge \cite{schochtransportphenomena2008} -- that does not depend on salt concentration or external forces. The corresponding conductance versus salt concentration curve exhibits a characteristic plateau at low concentration, which is indeed observed for channels with silica walls \cite{stein2004surface} or in boron-nitride nanotubes \cite{siriagiantosmoticenergy2013}, see figure \ref{fig:surface-charge}(a),(b). However, more recent experiments, performed in carbon nanotubes (CNTs) and 2D channels with atomically flat graphite or boron nitride walls, do not show a conductance plateau, but rather a power law scaling behavior of conductance versus salt concentration at low concentration \cite{liu2010translocation,guo2015giant,secchiscalingbehaviorionic2016, pangorigingiantionic2011, esfandiarSizeEffect2017}, see figure \ref{fig:surface-charge}(c). 
The scaling exponents have been interpreted in terms of charge-regulation models, in which the surface charge depends on the salt concentration \cite{secchiscalingbehaviorionic2016, cuicouplingiontransport2025}. These models hypothesize that the surface charge arises from ion adsorption: then, the concentration dependence of the adsorption equilibrium can account for the power-law behavior of the conductance \cite{secchiscalingbehaviorionic2016}, with a variety of possible scaling exponents (from 0 to 1) depending on the details of the adsorption mechanism \cite{uematsu2018}. 
In some experiments \cite{mouterdemolecularstreamingits2019, emmerichenhancednanofluidictransport2022}, coupled transport coefficients could not be interpreted in terms of a fixed surface charge, even with charge regulation. These studies introduced a mobile surface charge, which was assumed to originate from physisorbed ions that can diffuse along the surface. The surface charge mobility was found to strongly affect the surface-induced transport phenomena \cite{mouterde2018}.

Independently testing the assumptions made by these transport models has been a major challenge. \emph{Ab initio} molecular dynamics simulations \cite{grosjean2019versatile} indicate that hydroxyde ions chemisorb on boron nitride surfaces, yielding a fixed surface charge, and physisorb on graphene surfaces, retaining lateral mobility. More recent simulations suggest that hydronium ions may also physisorb at the water-graphene interface \cite{advinculaProtonsAccumulateGraphene2025a}. Experimentally, a study based on sum frequency generation (SFG) spectroscopy \cite{wangspontaneoussurfacecharging2025} found a negative surface charge on planar hBN with contributions from both physisorbed and chemisorbed hydroxide ions. In the case of graphene, in addition to possible ion adsorption, there is experimental evidence for hole doping (positive charge) at acidic pH arising from electron transfer to dissolved oxygen \cite{parkRedoxgovernedCharge2019}. For CNTs, however, experiments suggest an extrinsic origin of the surface charge. Pang \emph{et al.} fabricated CNT devices with a gate electrode separated from the single-walled CNT by a thin dielectric layer of $\rm SiO_2$ \cite{pangorigingiantionic2011}. By monitoring the ionic conductance as a function of gate voltage, they determined that the current is mostly carried by positive ions, indicating a negative surface charge, which was attributed to charged groups in the $\rm{SiO_2} $ layer. Interestingly, the effect of the electrostatic gate was weaker for metallic tubes than for semiconducting tubes, which was attributed to electronic screening of the gate potential. More recent experiments with CNT-on-$\rm SiO_2$ field effect transistor devices \cite{pashayevelectronicfingerprintsconfined2025} found a systematic shift of the charge neutrality point upon exposure to water, which was explained in terms of a surface charge arising from ion adsorption; the adsorption equilibrium appeared to be independent of the CNT electronic nature. 

Overall, ion transport at the ultimate scales cannot be understood in the conventional terms of a fixed surface charge. A surface charge with mobility and dynamical regulation can no longer be considered a simple boundary condition: it is rather a degree of freedom in its own right that couples to both the liquid and the solid. Going beyond a phenomenological description of surface charge is a challenge for nanofluidics theory.

\subsubsection{Material dependence of hydrodynamic resistance}

The hydrodynamic resistance $\mathcal{R}$ relates the liquid flow rate through a channel $Q$ to the applied pressure $\Delta P$: $Q = \Delta P / \mathcal{R} $. It can be decomposed into an entrance resistance $\mathcal{R}_\e$, which is dominant for shorter channels, and an internal resistance $\mathcal{R}_i$, which scales with the channel length. In a macroscopic channel, the internal resistance depends only on the liquid's viscosity $\eta$ and the channel dimensions. In a nanoscale channel, however, the internal resistance depends also on the liquid-solid friction coefficient $\lambda$, or equivalently the slip length $b = \eta/\lambda$ (see Sec. IV for further details). Several experimental studies have reported that both entrance resistance and slip length are sensitive to the channel wall material, and particularly to its electronic properties. 

In principle, some degree of material dependence for the slip length is expected within the well-established theory of roughness-induced liquid-solid friction. Within this theory (detailed in Sec. IV), friction is determined by the corrugation of the energy landscape on which interfacial water molecules move, i.e., the atomic-scale roughness of the solid surface. Such a description implies a quasi-universal relationship between slippage and contact angle \cite{bocquetnanofluidicsbulkinterfaces2010, huang2008}. However, 2D materials such as hBN and graphene, as well as their tubular counterparts, violate this relationship, implying that microscopic properties beyond wettability determine liquid-solid friction.

\paragraph{Long nanotubes.} Perhaps most strikingly, membrane-based assays of water transport through single-digit CNTs (with a diameter smaller than 10 nm) have revealed slip lengths on the micron scale -- much larger than on any other crystalline surface \cite{holtfastmasstransport2006,majumderenhancedflowcarbon2005,yangfastwatertransport2023}. Such slip lengths would indicate very strong hydrophobicity; yet, CNTs were shown to spontaneously fill with water when exposed to a humid atmosphere \cite{agrawalObservationExtreme2017}. This can at least in part be attributed to a disjoining pressure effect, originating in enhanced liquid structuring under strong confinement \cite{thomasWaterFlow2009, gravelleAnomalousCapillary2016} and/or entropic stabilization \cite{pascal2011entropy}. However, the very large water slip lengths observed in CNTs remain unexplained, with systematic discrepancies between experiments and simulations \cite{thiemannwaterflow2022}. There is also a significant discrepancy among experimental results \cite{alurufluidselectrolytesconfinement2023}, highlighting the difficulty of analyzing membrane transport measurements and calling for unambiguous single-nanotube assays.

\begin{figure}
	\centering
	\includegraphics[width=\columnwidth]{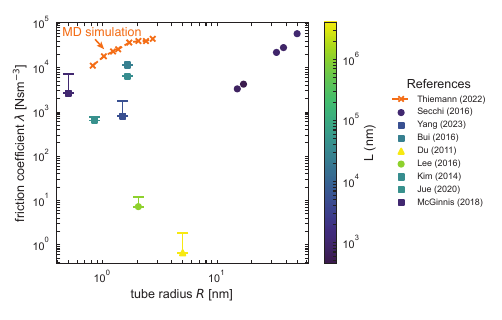}
	\caption{\textbf{Water friction in CNTs.} Water friction coefficient $\lambda$ as a function of CNT radius $R$ from various experiments \cite{secchimassiveradiusdependent2016, yangfastwatertransport2023, buiUltrabreathableProtectiveMembranes2016, duMembranesVerticallyAligned2011, leeMostDensifiedVerticallyaligned2016, kimFabricationFlexibleAligned2014, jueUltraPermeableSingleWalledCarbon2020, mcginnisLargescalePolymericCarbon2018}. All data is from membrane-based assays, except the data in \textcite{secchimassiveradiusdependent2016}, which is from single-tube measurements. For comparison, we also show the friction coefficient predicted by machine-learning-based MD simulations in orange \cite{thiemannwaterflow2022}. The color of the data points represents the CNT length, while the symbol indicates the number of walls (circle: multi-walled, triangle: double-walled, square: single-walled). The friction coefficient values were extracted from the reported flow enhancements over no-slip using the slip-viscosity-corrected Sampson formula for entrance resistance \cite{heiranianRevisitingSampsonsTheory2020}, with error bars indicating how much the slip-viscosity correction affects the result. Several experiments \cite{majumderenhancedflowcarbon2005, majumderMassTransportCarbon2011, holtfastmasstransport2006, trivediEffectVerticallyAligned2016, zhangGasTransportVerticallyaligned2014, baekHighPerformanceAntifouling2014} cannot be represented in this plot because the reported flow rates exceeded the Sampson limit, with or without a slip-viscosity correction. The data reported by \textcite{secchimassiveradiusdependent2016} already account for entrance effects and therefore do not have error bars.
	}
	\label{fig:CNT-friction-coefficients}
\end{figure}

Individual tube permeability measurements have been achieved with multiwall CNTs and BNNTs in the $\rm 30 ~nm$ to $100~\rm nm$ diameter range \cite{secchimassiveradiusdependent2016}. The authors inserted single, micrometer-long nanotubes into the opening of a glass nanopipette. To determine the flow rate, they analyzed the pressure-driven water jet emerging from the CNT outlet using fluorescent particle tracking. The CNTs exhibited a strongly radius-dependent slip length, ranging from $30~\rm nm$ in the largest tubes to $300~\rm nm$ in the smallest. The BNNTs, in contrast, showed no detectable flow slippage, despite a crystal structure very similar to CNTs. This difference between CNTs and BNNTs has been attributed to ion adsorption: the chemisorption of hydroxide ions -- which also accounts for the high surface charge in ion transport measurements -- results in an enhanced surface roughness, hence reduced slippage \cite{xieliquidsolidslip2020}. The radius-dependent slippage in multiwall CNTs has been attributed to the radius dependence of CNT electronic properties and a quantum friction mechanism \cite{kavokinefluctuationinducedquantum2022} -- which is the main subject of this Article.
An overview of the CNT water flow measurements is depicted in figure~\ref{fig:CNT-friction-coefficients}.

In double-wall, single-digit CNTs, indirect evidence from ion transport measurements further suggests a dependence of the slip length on the electronic nature of the CNT (metallic or semiconducting), which so far lacks a theoretical explanation \cite{cuicouplingiontransport2025}. The authors observed a systematic difference in ionic conductance between metallic and semiconducting CNTs. By combining measurements of ionic conductance and streaming current, they disentangled the concentration dependence of surface charge and slip length. At a fixed salt concentration, they deduced a larger slip length in semiconducting CNTs than in metallic CNTs. However, their model neglected surface charge mobility, which cannot be unambiguously assessed using ion transport measurements only. 
The reported slip lengths for tubes with radii from \SI{1.3}{\nm} to \SI{2.7}{\nm} range from about \SI{0.5}{\mu m} to \SI{3}{\mu m} and decrease with increasing radius.

\paragraph{Short nanotubes.} The experiments discussed so far have targeted long nanotubes, whose hydrodynamic resistance is mostly due to liquid-solid friction inside the tube. By contrast, the Noy group has specifically studied short CNTs, whose transport properties are determined entirely by entrance effects. They used $\sim$\SI{10}{\nm} long \emph{carbon nanotube porins}, which spontaneously insert into lipid membranes of large unilamellar vesicles, forming transmembrane channels \cite{tunuguntlaenhancedwaterpermeability2017}. The permeability of these channels can be determined by applying an osmotic shock and monitoring the resulting vesicle shrinkage through light scattering.
Li \emph{et al.} found that, for a diameter of approximately \SI{0.8}{\nm}, metallic CNT porins show higher permeability (that is, lower entrance resistance) than the semiconducting ones \cite{limoleculartransport2024}. Interestingly, in the same study, ion transport showed no dependence on CNT metallicity. 

Additional evidence for the coupling between electrons in short CNTs and the surrounding water has been provided by macroscopic diffusion measurements. Kistwal \emph{et al.} found that the resonant optical excitation of semiconducting CNTs reduces their diffusion coefficient in aqueous solution \cite{kistwallightinducedquantum2025}. The diffusion decrease was linear with the light intensity and was only observed in defect-free CNTs, where excitons are mobile along the CNT. The authors thus proposed that the change in diffusion is due to a coupling between CNT exciton dynamics and the surrounding water, akin to quantum friction. However, the diffusion of a CNT in solution is not directly related to its hydrodynamic permeability, and it remains to be assessed whether the water slip length in a semiconducting CNT can be affected by resonant optical excitation. 

\paragraph{Flat surfaces and 2D channels.} One may expect that the peculiar water transport properties of CNTs and BNNTs can be traced back to their parent 2D materials. However, studies of water slippage on monolayer 2D materials face the challenge of disentangling the material's intrinsic properties from those of the underlying substrate. Atomic force microscopy (AFM) and colloidal probe AFM measurements determined a slip length $\sim 4 - 8~\rm nm$ for water on multilayer graphite \cite{li2022,maalimeasurementsliplength2008a}. The situation is more complex for monolayer graphene. In transport measurements through graphene-covered silica grooves, slip lengths of up to $\sim 200~\rm nm$ were observed, but the results showed a very large spread, between no-slip and 200 nm -- presumably due to a variation in graphene quality and graphene-substrate interaction \cite{xiefastwatertransport2018}. Colloidal probe AFM measurements showed a very small water slip length for graphene on $\rm SiO_2$ ($\sim \SI{1}{\nm}$), but a larger slip length ($\sim \SI{13}{\nm}$) for graphene on silanized silica \cite{li2022}.  In a similar spirit, Cetindag \emph{et al.} performed membrane-based flow measurements in hetero-nanotubes consisting of single-walled CNTs of about $1 - 5~\rm nm$ diameter, coated with a few layers of boron nitride \cite{cetindagIonHydrodynamicTranslucency2024}. The authors measured water slip lengths on the order of \SI{12}{\nm}, substantially lower than the \SI{770}{\nm} in their pristine single-walled CNTs. It remains to be determined whether these results point to a translucency of graphene to the frictional properties of the underlying substrate, or to a modification of the graphene's own properties through interaction with the substrate. 

Slit-like channels assembled from 2D material crystals further show a peculiar material dependence of water transport. However, to our knowledge, no direct pressure-driven water transport measurements could be carried out in those channels so far: all data has been obtained with evaporation measurements, where the driving force is not exactly known \cite{radhamoleculartransport2016}. Nevertheless, approximate slip lengths can be extracted from those measurements, especially in the larger channels. These slip lengths are qualitatively consistent with nanotube measurements: relatively high slippage ($b \sim 60~\rm nm$) with graphite walls, and almost no slippage ($b \sim 1~\rm nm$) with boron nitride (hBN) walls \cite{keerthiwaterfrictionnanofluidic2021}. However, the latter is somewhat surprising, since boron nitride 2D channels do not exhibit a high fixed surface charge in ion transport measurements \cite{esfandiarSizeEffect2017}, in contrast to BNNTs. It is even more surprising that hybrid nanochannels (one graphite wall, one hBN wall) exhibit almost the same evaporation rate as graphite channels. Keerthi \emph{et al.} further observe that roughening one of the graphite walls with oxygen plasma etching has very little effect on the evaporation rate. This is qualitatively consistent with ion transport measurements in etched graphite channels \cite{emmerichenhancednanofluidictransport2022}, which point to significant slippage despite high surface charge.

Recently, Liz\'ee \emph{et al.} exploited the substantial slowdown of molecular relaxation in supercooled glycerol with decreasing temperature \cite{lizeeanomalousfriction2024} to evidence the role of phonon modes in liquid-solid friction. Using a tuning-fork-based AFM, they measured the slip length of glycerol on mica for various temperatures between \SI{0}{\celsius} and \SI{35}{\celsius}. At low temperatures, friction decreases as the temperature increases, which is in accordance with 
a conventional roughness-induced friction mechanism described by an Arrhenius law. 
However, for higher temperatures, the friction increases. This anomalous behavior was rationalized by coupling the molecular motion of glycerol to the phonons of mica within a fluctuation-induced friction mechanism, as discussed in Sec. IV. Increasing the temperature shifts the dominant molecular relaxation mode of glycerol to higher frequencies, resulting in greater spectral overlap with the phonons of mica. This frequency shift opens a vibrational channel of momentum transfer from the liquid to the solid and leads to increased friction.

A conceptually similar mode coupling has been demonstrated between the charge fluctuations of water and the electron motion in graphene \cite{yuelectroncooling2023}, where the Coulomb force mediates the interaction. Using optical pump terahertz probe spectroscopy, hot graphene electrons were created by an ultra-short excitation pulse, and the subsequent electron cooling in the presence of different liquids was monitored. The study reveals direct energy transfer from hot graphene electrons to water ($\rm H_2O$) and heavy water ($\rm D_2O$). Strikingly, heat transfer to water is faster than to heavy water, which is attributed to greater spectral overlap between the charge fluctuations in graphene and those of water than with those of heavy water -- in good agreement with quantum friction theory (Sec. IV-V).

All these results point to subtle microscopic origins of hydrodynamic friction that cannot be captured by coarse-grained descriptors such as hydrophobicity.

\subsubsection{Coupling of nanofluidic flow and electronic transport}
There have been several reports of electronic current generation in a channel wall in response to ion and fluid transport in the channel. The induction of ionic current in the channel by an electronic current in the wall has been observed as well. These effects are clearly beyond the PNPS description, and they challenge the concept of an inert hydrodynamic wall. 

\begin{figure*}
	\centering
	\includegraphics[width=\textwidth]{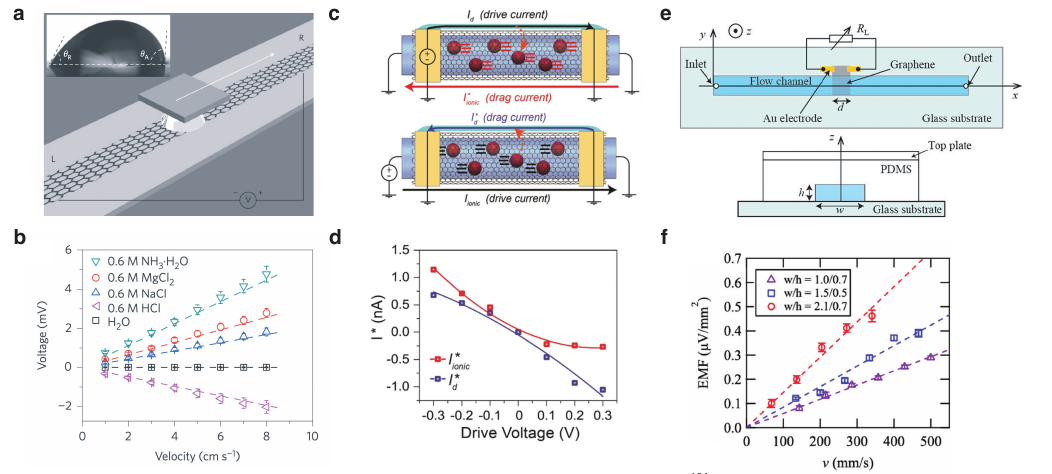}
	\caption{\textbf{Coupling of nanofluidic flows and electronic currents.} (a) A droplet of ionic solution is pulled along a graphene sheet. (b) The induced voltage in the graphene depends linearly on the droplet velocity. The sign of the voltage depends on the ion species. (a,b) Reproduced from \textcite{yingeneratingelectricitymoving2014}. (c) Coupling of ionic and electronic currents in a CNT. An electrical current in the CNT wall induces an ionic current through the CNT. Conversely, an ionic flow through the CNT induces an electronic current in the CNT wall. (d) Experimental data for the electronic-current-induced ion flow (red) and the ion-flow-induced electronic current (blue). (c,d) Reproduced from \textcite{rabinowitzelectricallyactuatedcarbonnanotubebased2020}. (e) Schematic of a fluidic channel used to flow deionized water over electrically-connected CVD graphene, with electrodes shielded from water. (f) The induced voltage at maximum power in the graphene depends linearly on the water velocity. (e,f) Reproduced from \textcite{takedaenhancingelectricitygeneration2025}.
	}
	\label{fig:current-generation}
\end{figure*}

\paragraph{Droplets.} Several studies have demonstrated the generation of an electric current by millimeter-scale sliding drops of ionic solutions over graphene or CNT films. Yin \emph{et al.} showed that the induced voltage in graphene scales linearly with the sliding velocity, and that its magnitude depends on the ion species \cite{yingeneratingelectricitymoving2014}, with no detectable voltage when using deionized water droplets, see figure~\ref{fig:current-generation}(a,b). Sliding droplets have also been shown to generate current in a nearby conductor even across an insulating layer, which suggests a mechanism based on capacitive coupling between ions at the liquid-solid interface and the conductor's electrons \cite{parkIdentificationDropletFlowInduced2017, shinSaltWater2026}. Liz\'ee \emph{et al.} used an AFM tip to displace much smaller, micron-scale liquid droplets on the surface of multilayer graphene \cite{lizee2023}. They observed an electric current in graphene induced by the droplet motion, with a similar order of magnitude regardless of whether the droplet contained an ionic liquid or ion-free silicone oil. Since the current generation was enhanced by the presence of wrinkles on the graphene surface, it was attributed to a phonon drag mechanism, where the droplet motion generates a phonon wind in the solid, which in turn drives an electric current \cite{coquinotquantumfeedback2023,kralnanotubeelectron2001}. 

\paragraph{Ionic currents.} In individual single-wall CNTs, an ionic current through the tube was found to induce an electronic current in the tube wall, and vice versa: an electronic current in the tube wall induced ion transport through the tube \cite{rabinowitzelectricallyactuatedcarbonnanotubebased2020}, see figure~\ref{fig:current-generation}(c,d). The ionic and electronic currents had opposite directions, qualitatively supporting an ionic Coulomb drag mechanism in which image charges in one medium follow real charges in the other. A similar observation was reported for the water-graphene interface \cite{cheninducingelectriccurrent2023}. However, the interpretation in terms of ionic Coulomb drag has been challenged, as it was demonstrated that silver electrodeposition, resulting from the dissolution of the Ag/AgCl electrodes, could account for the current generation \cite{chensilverelectrodepositionag2025}.

\paragraph{Flow of ionic solutions.} Solid-state electric currents have also been observed in response to pressure-driven flow of ionic solutions. 
The early studies by \textcite{ghoshflowinducedvoltagecurrent2004, ghoshcarbonnanotubeflow2003} report a sublinear dependence of the induced open-circuit voltage and short-circuit current on the flow velocity for aqueous KCl solution over bundles of single-walled CNTs. The sign of the flow-induced voltage and current can be tuned by applying a potential bias to the CNT bundles relative to the KCl solution, interpreted as a change between positive and negative charge carriers in the CNTs. Furthermore, the studies report a much weaker signal for multi-walled CNTs and no measurable electrical signal on graphite.
A similar experiment \cite{dhimanharvestingenergywater2011} using an aqueous HCl solution on samples with a few layers of graphene reported flow-induced voltages and currents that were sublinearly dependent on flow velocity but of much larger magnitude.
These studies suggest that the electronic current could originate from image charges in the solid, which are dragged along with the flowing charges in the liquid. However, this interpretation has been debated due to the sublinear (nearly logarithmic) dependence on flow velocity \cite{perssonElectronicFriction2004}.
\textcite{yinharvestingenergywater2012} pointed out that the exposure of metallic electrodes to the ionic solution could lead to artifacts in the induced current.

\paragraph{Flow of ion-free liquids.} 
A few studies report the induction of a solid-state electronic current by the flow of an ion-free liquid. Lee \emph{et al.} demonstrated current generation in graphene by the flow of cyclohexane, pyridine and deionized water  \cite{leeflowinducedvoltagegeneration2013, holeeflowinducedvoltagegeneration2013a}. The effect appeared to depend on the liquid's polarity, with a significantly larger current generated by the water flow. The dependence on flow velocity was non-linear, with an apparent saturation at high velocity. More recently, Takeda \emph{et al.} observed current generation in graphene due to the flow of pure water, with a linear velocity-current relation up to very high flow velocities \cite{kuriyaOutputDensity2020,takedaInvestigatingCorrelation2024, takedaenhancingelectricitygeneration2025}, see figure~\ref{fig:current-generation}(e,f). These results are consistent with a quantitative interpretation in terms of water-graphene quantum friction (see Sec. V).

\section{Interfacial fluctuations in the solid and gas phases}

The experimental results compiled in the previous section collectively illustrate how subtle, microscopic properties of the liquid-solid interface play a key role in nanofluidic transport. Those properties go beyond surface chemistry or coarse-grained descriptors like wettability, and in some cases include explicitly the electronic properties of the channel wall material. 
Traditionally, electronic degrees of freedom have not been explicitly taken into account in the theoretical description of liquid-solid systems; the recent developments that have incorporated electron dynamics into nanofluidic transport theory are the main subject of this Article (sections IV through VI). However, electronic and other charge fluctuation effects have been studied for many years in the context of non-contact interactions between molecules and solids and in the context of the interaction of gas-phase adsorbates with metal surfaces. Dielectric fluctuations have also been invoked as a determinant of the dynamics of ionic and dipolar species in solution. In this section, we review the main theoretical results from those areas, as they are the foundation for the recent developments of nanofluidic transport theory discussed below. 

\subsection{Attractive van der Waals forces: Lifshitz theory}\label{section Lifschitz theory}

The spontaneous charge fluctuations of two neutral bodies -- that can be of thermal or quantum nature -- give rise to an attractive force between them, known either as the van der Waals force or the Casimir force. When two bodies are separated by a distance that is large compared to their own dimensions (consider, for instance, two atoms or molecules in the gas phase), they can be pictured as fluctuating dipoles. Coulomb interactions correlate the dipoles' fluctuations, leading to a net attractive force between them upon averaging.  
Imagine, for instance, a spontaneous positive charge arising on the first body: it will induce a negative image charge in the second body, and the two opposite charges attract. The resulting interaction is long-ranged, showing a power-law decay in the distance $r$. When $r$ is small compared to the characteristic wavelengths of the fluctuating electromagnetic field, the interaction energy scales as $1/r^6$, as originally shown in \cite{london1930}: this scaling is characteristic of the van der Waals force. At larger separations, the Coulomb interactions can no longer be considered instantaneous, and a relativistic treatment predicts a scaling as $1/r^7$ \cite{casimir1948a}, which is the hallmark of the Casimir effect.

\begin{figure}
	\begin{center}
		\includegraphics[width=0.8\columnwidth]{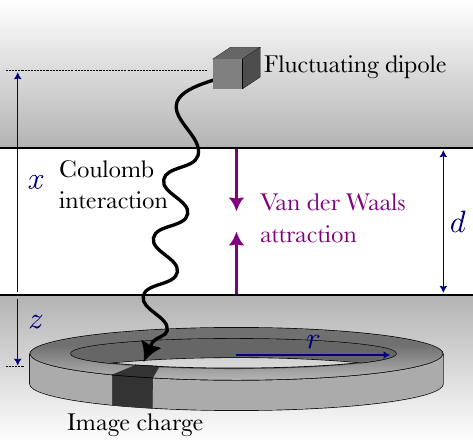}
		\caption{\textbf{Van der Waals force between two extended bodies.} The scaling of the van der Waals interaction between two surfaces with their separation distance $d$ is obtained by summing up interaction energies of the individual atoms -- by convenience, in a cylindrical geometry.}\label{fig Van der Waals}
	\end{center}
\end{figure}

When the separation distance between the two fluctuating bodies is not small compared to their dimensions, a dipolar approximation can no longer be made. However, the scaling of their fluctuation-induced attractive force can be determined by summing up the contributions of atomic pairs, whose interaction is treated within the dipolar approximation.  
Consider, for instance, two semi-infinite bodies, with planar surfaces separated by a distance $d$, and a density of atoms $n$ in each body. In the van der Waals description, the interaction energy between two individual atoms separated by a distance $r$ is $w(r) = -a/r^6$. The interaction energy of an atom of the first body with all the atoms in the second body then reads
\begin{equation}
\begin{split}
	U(x) &= -\int_0^{+\infty} \d z \int_0^{2\pi} r \d \theta \int_0^{+ \infty} \d r \, n \, \frac{a}{[(x+z)^2+r^2]^3} \\ &=-\frac{\pi n a }{6 x^3}.	
\end{split}
\end{equation}
where $x$ is the distance between the atom and the surface of the second body. Summing this expression over all atoms over the first body, as schematized in figure \ref{fig Van der Waals}, the interaction energy per unit area between the two extended bodies is found to be:
\begin{equation}
	u_{\rm vdW} = -\frac{\pi n^2a}{12 d^2}.
\end{equation}
Similarly, the Casimir (retarded) interaction energy scales as $1/d^3$. Thus, the van der Waals and Casimir attractive forces between two parallel plates scale as $1/d^3$ and $1/d^4$, respectively. Although this very simple argument captures the correct scalings of the van der Waals and Casimir forces, it cannot capture quantitatively the multiplicative constant. Indeed, in condensed bodies, the correlations between charge fluctuations on neighboring atomic centers limit computations of van der Waals or Casimir forces from two-atom interaction potentials to scaling arguments.  

Casimir first overcame this difficulty in the particular case of metallic bodies \cite{casimir:1948dh}. 
He noticed that the interaction energy between two metallic plates corresponds to the change in zero-point energy of the electromagnetic field between the two plates when they are separated by a finite distance $d$ compared to when $d \rightarrow \infty$. This now-common way of presenting the Casimir effect is what allowed Casimir to obtain (through a straightforward computation) the attractive force between the two plates of surface $\mathcal{A}$ as:
\begin{equation}\label{casimir force}
	F(d) = \mathcal{A} \, \frac{\pi^2}{240} \frac{\hbar c }{d^4},
\end{equation}
consistent with the scaling discussed above.

Lifschitz unified the van der Waals and Casimir pictures by extending Casimir's point of view beyond perfectly reflective metallic bodies \cite{dzyaloshinksiigeneraltheory1961}. Lifschitz and co-authors focused, like Casimir, on the fluctuating electromagnetic field rather than its sources. But, for the interaction of the two bodies with the electromagnetic field -- instead of a perfect reflection -- they adopted a coarse-grained description in terms of a local dielectric function $\epsilon_{1,2} (\omega)$. Their general result for the fluctuation-induced interaction force is cumbersome, and we state here only the van der Waals limit $c \to \infty$, which is valid if the separation $d$ is smaller than the  wavelengths of the main features in the absorption spectra $\Im \epsilon_{1,2}^{-1}(\omega)$: 
\begin{equation}
	\begin{split}
	F(d) = \mathcal{A} \frac{\hbar}{8 \pi^2 d^3} &\int_0^{+\infty} \frac{\d \omega}{\tanh (\hbar \omega / 2 \kB T)}\dots \\
	&\dots \Im \left[ \frac{\epsilon_1(\omega) - 1}{\epsilon_1(\omega) + 1}\cdot \frac{\epsilon_2(\omega) - 1}{\epsilon_2(\omega) + 1}\right].
	\end{split}
	\label{eq:vdw_lifshitz}
\end{equation}
The Lifshitz theory recovers the expected scaling as $1/d^3$, but now with an expression of the prefactor in terms of dielectric functions -- the analogues of the atomic polarizabilities for two extended bodies. In the van der Waals limit, the fluctuating electromagnetic field between the two bodies consists only of evanescent waves. An evanescent mode is defined by its wavevector $\mathbf{q}$ parallel to the surfaces; its amplitude decays with distance $z$ away from the surfaces as $e^{-qz}$. Thus, for two bodies separated by $d$, only modes with $q \lesssim 1/d$ can significantly contribute to the van der Waals interaction. This sets the validity condition for the local approximation in Lifschitz's theory: the true non-local dielectric functions $\epsilon_{1,2} (\mathbf{q}, \omega)$ must be essentially independent on $\q$ for $q \lesssim 1/d$.

The van der Waals force does not vanish at zero temperature: in the limit $T \to 0$, Eq.~\eqref{eq:vdw_lifshitz} becomes
\begin{equation}\label{reduced vdw}
	F(d) \simeq \mathcal{A} \frac{\hbar \bar{\omega}}{8 \pi^2 d^3}, 
\end{equation}
with
\begin{equation}
\bar{\omega} = \int_0^{+\infty} \d \omega \, \frac{\epsilon_1(\omega) - 1}{\epsilon_1(\omega) + 1}\cdot \frac{\epsilon_2(\omega) - 1}{\epsilon_2(\omega) + 1},
\end{equation}
where $\bar{\omega}$ appears as a characteristic absorption frequency of the two media. Given that fluctuations are purely quantum at zero temperature, the amplitude of the attractive force is controlled by Planck's constant.

In the opposite limit of purely thermal fluctuations ($\hbar \to 0$), the integral in Eq.~\eqref{eq:vdw_lifshitz} simplifies thanks to the Kramers-Kronig relation for the quantity in brackets, and the van der Waals force can then be expressed in terms of the static dielectric permittivities: 
\begin{equation}
	F(d) \simeq \mathcal{A}\frac{\kB T}{8 \pi d^3} \cdot \frac{\epsilon_{1}(0) - 1}{\epsilon_{1}(0) + 1}\cdot \frac{\epsilon_{2}(0) - 1}{\epsilon_{2}(0) + 1},
\end{equation}
with the amplitude controlled by the temperature. In practice, most materials have features in their absorption spectra beyond the thermal frequency $\omega_T \approx 26~\mathrm{meV} \approx 6~\rm{THz}$ at room temperature, so that either Eq.~\eqref{eq:vdw_lifshitz} or Eq.~\eqref{reduced vdw} needs to be used for evaluating the van de Waals force from first principles.

\subsection{Fluctuation-induced friction on single particles}

Except for the assumption of a local dielectric response
-- the lifting of which presents no conceptual difficulty -- the theory of van der Waals forces was essentially complete with the work of Lifshitz. However, it is rather a dynamic analogue of the van der Waals force that is the closest parent of the nanofluidic phenomena discussed in this Article. This dynamical van der Waals force, or van der Waals friction, was predicted only much later, in the 1990s. But the notion of friction induced by fluctuations started being discussed already in the 1960s, in the case of single particles coupled to a fluctuating continuum. 

As a first simple but illuminating example of such a phenomenon, Zwanzig computed the friction force experienced by a spherical particle with finite radius $r_0$ and fixed charge $e$ that is moving in a straight line with fixed velocity $\v$ inside a dielectric medium \cite{zwanzigdielectricfrictionmoving1963}. We reproduce here the main ideas of Zwanzig's original computation, that describes for example the motion of an ion inside a dielectric solvent. The moving ion generates in the surrounding medium an electric displacement field $\D(\r,t)$ given by the gradient of the Coulomb potential $\D(\r, t) = - (e/(4\pi)) \nabla_\r (1/\vert \r - \v t \vert)$. The dielectric medium responds with a retarded polarization field $\P(\r, t)$ determined by the medium's dielectric function $\epsilon$. In terms of the temporal Fourier components of the fields,
\begin{equation}
	\P(\r,\omega) = \left( 1 - \frac{1}{\epsilon(\omega)} \right) \D(\r,\omega).
\end{equation}
This induced polarization generates an electric field $\E_{\textrm{ind}}(\r',t)$ that acts back on the moving ion. Note that the contribution to $\E_{\textrm{ind}}(\r',t)$  from the polarization at another position $\r$ is given by 
\begin{equation}
\E_{\textrm{ind}}(\r',\r, t) = - T(\r'-\r)\cdot \P(\r,t) / \epsilon_0
\end{equation}
where $T(\r'-\r) = - \nabla_{\r'} \nabla_{\r'} (1/\vert \r' - \r \vert)$. The dielectric medium thus reacts to the ion motion with a total force
\begin{equation}
	\F = e \int_{\vert \r - \v t \vert > r_0}\d \r \, \E_{\rm ind}(\v t, \r, t).
\end{equation}
In the limit of small velocity $\v$, this simplifies to: 
\begin{equation}
	\F = - \frac{e^2}{6 \pi \epsilon_0 r_0^3} \left[ \frac{\d\,  \im{\epsilon(\omega)}}{\d\omega } \right]_{\omega=0} \times \v.
	\label{eq:zwanzig}
\end{equation}
This is indeed a friction force that points in the opposite direction to the ion's velocity -- that has been termed \emph{dielectric friction}. Zwanzig's result already contains the important qualitative features of fluctuation-induced friction forces. First, it shows that the friction arises because of retardation in the dielectric response, which is quantified by the imaginary part of the dielectric function. The ion experiences friction because the polarization cloud it creates in the solvent cannot adapt instantaneously to its motion: The ion is constantly trying to escape its own polarization cloud. Second, the imaginary part of the dielectric function represents the charge excitation spectrum of the solvent: one can view dielectric friction as a consequence of the ion generating polarization excitations in the solvent through its Coulomb potential.

Such an excitation picture is more intuitive in the case of an ion moving near a metal surface, since a metal's conduction electrons are more commonly represented in terms of discrete excitations. One may expect that the ion loses energy through the excitation of electron-hole pairs or collective surface plasmon modes in the metal. This particular flavor of dielectric friction is called \emph{electronic friction} and has been widely studied \cite{nunez1980,dedkovelectromagneticfluctuationelectromagneticforces2002, douPerspectiveHow2018}. The idea of electronic friction has further been generalized to particles interacting with a generic classical field, such as a magnetic dipole moving through an Ising ferromagnet \cite{demeryDragForces2010}.

\begin{figure}
	\begin{center}
		\includegraphics[width=\columnwidth]{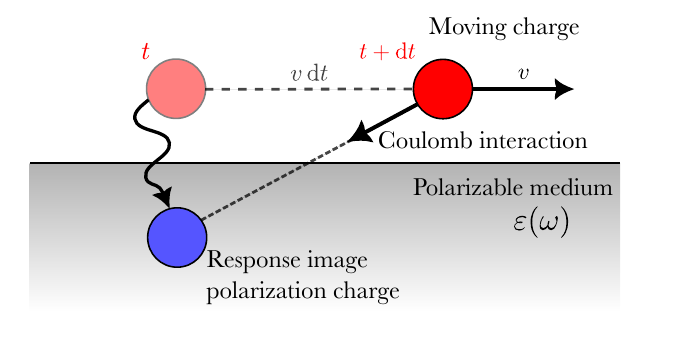}
		\caption{\textbf{Mechanism of electronic friction.} A point charge is sliding above a dielectric material. The solid responds to the potential created by the sliding charge by an image polarization charge, which appears with some delay. This results in a net force on the sliding charge parallel to the direction of motion, which corresponds to \emph{electronic friction}.}\label{fig electronic friction}
	\end{center}
\end{figure}

We do not attempt here to review the full spectrum of theoretical approcahes. Instead, we provide in the following a derivation in the spirit of the original work by Zwanzig, but tackling explicitly the non-local character of the metal's dielectric response \cite{kavokinefluctuationinducedquantum2022}, which will allow us to introduce the \emph{surface response function} -- a key tool for all that follows. 
Let us use cylindrical coordinates $\r=(\mathbf{\rho}, z)$ (with $z$ the direction perpendicular to the solid surface) and consider a point charge $e$, initially at a position $\r_0 = (0,h)$ above the solid, say a metal, that fills the half-space $z<0$. When the charge moves with a velocity $\v$ parallel to the surface, it generates at each point $\r$ in space a Coulomb potential $V(\r,t) = V(\r - (\r_0 +\v t))$ where $V(\r) = e^2/(4 \pi \epsilon_0 \vert \r \vert)$. In the case where the solid's dielectric response is non-local, it is more convenient to work with potentials rather than electric fields. Thus, we describe the solid's response to the applied potential as an induced charge density $\delta n(\r,t)$, as pictured in Figure \ref{fig electronic friction}. Within linear response theory, it is assumed to take the form:
\begin{equation}
	\delta n(\r,t) = \int_{-\infty}^{+\infty} \d t' \int \d \r' \, \chi_\s (\r, \r', t-t') V(\r', t'),
\end{equation}
where the hereby defined $\chi_\s$ is the density response function of the solid. For a homogeneous solid, it can be directly related to the dynamical dielectric permittivity $\epsilon$ according to $\epsilon(\q, \omega)^{-1} = 1 + e^2/(\epsilon_0 q^2) \chi_\s(\q, \omega)$. The induced charge density creates a Coulomb potential that acts back on the external charge, which thus experiences a force
\begin{equation}
\begin{split}
	\F(t) &= - \int_{-\infty}^{+\infty} \d t' \int \d \r \, \d \r' \, \nabla_{\r_0} V(\r_0 + \v t - \r ) \dots \\ & \qquad \qquad \dots \chi_\s(\r, \r', t-t')V(\r' - \r_0 - \v t).
\end{split}
\end{equation}
Assuming the solid is translationally invariant parallel to the interface, we may carry out spatial 2D Fourier transforms, so that the in-plane component of the force $\F(t)$ reads:
\begin{equation}
\begin{split}
	&\F(t) = \frac{e^2}{8 \pi^2 \epsilon_0} \int \d \q \, \frac{i \q}{q} e^{-2qh} \dots \\ & \left[- \frac{e^2}{2 \epsilon_0 q}\int_{-\infty}^0 \d z \, \d z' \, e^{q(z+z')} \chi_\s(\q, z, z', \omega=\q \cdot \v) \right]	
\end{split}
\end{equation}
where we used that the 2D Fourier transform of the 3D Coulomb potential $V(\r)$ is given by $V_\q(z) = \frac{e^2}{2\epsilon_0 q}e^{-q \vert z \vert}$. We define the surface response function of the solid $g_\s(\q, \omega)$ as
\begin{equation}\label{surface response function}
	g_\s(\q, \omega) = - \frac{e^2}{2 \epsilon_0 q}\int_{-\infty}^0 \d z \, \d z' \, e^{q(z+z')} \chi_\s(\q, z, z', \omega),
\end{equation}
which satisfies $g_\s(\q, -\omega) = g_\s(\q, \omega)^*$ for a solid at equilibrium. Therefore, as $\re{g_\s(\q, \q\cdot\v)}$ is even under $\q \rightarrow -\q$, the force experienced by the moving charge becomes:
\begin{equation} \label{electronic friction}
	\F(t) = - \frac{e^2}{8 \pi^2 \epsilon_0} \int \d \q \, \frac{\q}{q} \, e^{-2qh} \, \im{g_\s(\q, \q\cdot \v)}.
\end{equation}
In the limit of small velocities, the friction force scales as $\F = - \gamma_{\rm ion} \v$, so that one may define a friction coefficient 
\begin{equation}
	\gamma_{\rm ion} = \frac{e^2}{8 \pi^2 \epsilon_0 }\int \d \q \, \frac{(\q \cdot \v)^2}{qv^2} e^{-2qh} \left[ \frac{\d \, \im{g_\s(\q, \omega)}}{\d \omega} \right]_{\omega=0}.
\end{equation}
This result is formally analogous to the ones obtained in \cite{ferrell1979419,nunez1980}, based on the computation of the stopping power of a moving charge above a metal surface. The non-local response description clarifies the short-distance behavior of the electronic friction force. In Eq.~\eqref{eq:zwanzig}, the dielectric friction diverges as $1/r_0^3$ in the limit of a point charge ($r_0 \to 0$). Here, we consider a point charge from the onset, but at a finite distance $h$ from the metal surface. The friction force scales as $1/h^3$ at sufficiently large $h$, but it does not diverge as $h \to 0$: it is in fact the electronic response of the metal itself that implements a short-distance cutoff. Electronic friction is thus sensitive to subtle short-wavelength electronic properties, beyond the macroscopic dielectric response.

Further works \cite{perssonelectronicfriction1995, tomassone1997} have generalized these results to electrically neutral molecules moving near a metal surface. Tomassone and Widom considered a polarizable molecule, with frequency-dependent polarizability $\alpha(\omega)$ (defined as the response function of the molecule dipole moment to the electric field, $\mathbf{p}(\omega)= \alpha(\omega) \E(\omega)$) moving above a dielectric medium of dielectric permittivity $\epsilon(\omega)$. They obtained a friction force that is purely induced by fluctuations, $\F = - \gamma_{\rm pol} \v$, with:
\begin{equation}\label{fif polarizable}
\begin{split}
 	\gamma_{\rm pol} = \frac{3}{(2 \pi)^6 \hbar^3 \kB T} &\int_0^{+ \infty}  \d \omega \, \left( - \frac{\partial \nB(\omega)}{\partial \omega} \right) \cdots \\ & \cdots \im{\frac{\epsilon(\omega) -1}{\epsilon(\omega) + 1} } \im{ \alpha(\omega) },
\end{split}
\end{equation}
where $\nB(\omega)$ is the Bose distribution, defined as $\nB(\omega)= 1/(e^{\hbar\omega/\kB T} -1)$.
In the classical limit $\hbar \to 0$, and considering an ideal metal of constant conductivity $\sigma \rightarrow \infty$,  Eq.~\eqref{fif polarizable} reduces to:
\begin{equation}
	\gamma_{\rm pol}^{\rm classical} = \frac{3 \kB T \epsilon_0 \alpha_T}{2 h^5 \sigma}, 
\end{equation}
where $\alpha_T$ is the classical polarizability of a rotating dipole moment $\mathbf{p}$, such that $\langle p_i^2 \rangle_{\rm classical} = \kB T \alpha_T$ for $i =x,y,z$.

We will justify in the next paragraph that the quantity $(\epsilon(\omega) -1)/(\epsilon(\omega) + 1)$ identifies with the local surface response function of the metal, namely $g_\s(q \rightarrow 0, \omega)$. The result in Eq.~\eqref{fif polarizable} thus contains the main features, except for the geometry, of the van der Waals or Casimir friction that we will discuss in the following: friction arises from a dynamical coupling between charge fluctuations in the two interacting bodies. However, we shall first analyze in more detail the physical meaning of surface response functions, which will be important ingredients in that discussion.   

\subsection{From charge susceptibilities to surface response functions}\label{section surface response fct}

We found in the previous paragraph that the fluctuation-induced friction force acting on a charged particle moving above a semi-infinite solid is naturally expressed in terms of the surface response function $g_\s$ of the solid, as shown in Eq.~\eqref{electronic friction}. Let us recall that the semi-infinite solid (in the half-space $z < 0$) responds to an external potential $\phi_{\rm ext}$ with an induced charge $\delta n$, according to a linear response function $\chi_\s$ (called \emph{charge susceptibility} or \emph{density response function}): 
\begin{equation}
	\delta n(\r,t) = \int_{-\infty}^{+\infty} \d t' \int \d \r' \, \chi_\s (\r, \r', t-t') \phi_{\rm ext}(\r', t'),
\end{equation}
The \emph{surface response function} is defined as \cite{liebschdynamicalscreening1987,pitarketheorysurface2007}: 
\begin{equation}\label{surface response function def}
	g_\s(\q, \omega) = - \frac{e^2}{2 \epsilon_0 q}\int_{-\infty}^0 \d z \, \d z' \, e^{q(z+z')} \chi_\s(\q, z, z', \omega). 
\end{equation}
We now discuss its physical interpretation and explain how it can be evaluated starting from the solid's electronic structure \cite{kavokinefluctuationinducedquantum2022}. 

Consider that the semi-infinite medium (occupying the lower half-space) is subject to an external evanescent plane wave at frequency $\omega$, which takes the form $\phi_\ext(\boldsymbol{\rho},z,\omega)=\phi_0 e^{i\q \cdot \boldsymbol{\rho}}e^{qz}$ in real space, and after taking its 2D Fourier transform, $\phi_\ext(\q, z, \omega) = \phi_0 e^{qz}$. The solid responds to this potential with an induced charge density
\begin{equation}
	\delta n (\q,z, \omega) = \phi_0 \int_{-\infty}^0 \d z' \, \chi_\s(\q, z, z', \omega) \, e^{qz'}.
\end{equation}
which in turn generates a potential $\phi_\ind$ outside the medium, which can be written at a distance $z$ above the interface as:
\begin{equation}
\begin{aligned}
	&\phi_\ind (q, z, \omega) = \int_{-\infty}^0 \d z'' \, V_\q(z-z') \, \delta n(\q, z', \omega) \\ &= \phi_0 \int_{-\infty}^0 \d z' \, \d z'' \, \chi_\s(\q, z', z'', \omega) \, e^{qz''} \, \frac{e^2}{2 \epsilon_0 q}e^{-q(z-z')} , 
\end{aligned}
\end{equation}
where we recall that $V_\q(z)$ stands for the 2D Fourier transform of the 3D Coulomb potential. We thus notice that the induced potential outside the medium is related to the external potential through:
\begin{equation}
	\phi_\ind(\q, z, \omega) = -g_\s(\q, \omega) \, \phi_0 e^{-qz}.
\end{equation}
from which we interpret the surface response function as a reflection coefficient for evanescent plane waves at the interface. Specifically, 
\begin{equation}
	\phi_\ind(\q, z= 0, \omega) = -g_\s(\q, \omega) \phi_\ext(\q, z=0, \omega). 
\end{equation}

\begin{figure}
	\begin{center}
		\includegraphics[width=\columnwidth]{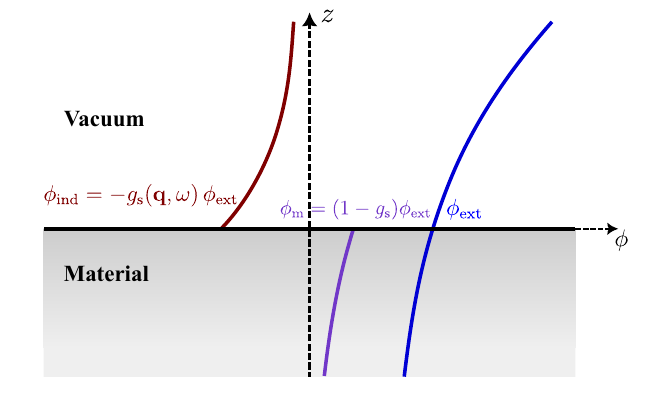}
		\caption{\textbf{Evanescent wave potentials at a single interface.} The external potential applied to the material is represented in blue. The material responds with an induced charge density, which produces an induced potential $\phi_{\rm ind}$ above the surface, and contributes to the screened potential $\phi_{\rm m}$ inside the material. The surface response function $g_\s(\q, \omega)$ plays the role of a reflection coefficient for the external potential on the interface. At the interface, continuity relation imposes $\phi_\ext + \phi_\ind = \phi_{\rm m}$.}\label{fig response function}
	\end{center}
\end{figure}

Based on this interpretation, the surface response function can be readily evaluated under the assumption of a local dielectric response -- which holds for any medium in the long wavelength limit ($q \to 0$) -- by simply enforcing boundary conditions. Consider indeed a semi-infinite medium characterized by a local dielectric function $\epsilon(\omega)$ and an evanescent plane wave at frequency $\omega$ impinging on the interface. In this case, according to the local response assumption, there can be no induced charge inside the medium, except at its surface, so that the potential $\phi_{\rm m}$ inside the material obeys the Laplace equation. Since it must vanish at $-\infty$, it takes the form $\phi_{\rm m}(\q, z, \omega) = \phi_{\rm m}e^{qz}$. Outside the medium, the potential is given by the sum of the external and induced potentials. With the Laplace equation holding outside the medium as well, the outside potential reads:
\begin{equation}
	\phi(\q, z, \omega) = \phi_\ext e^{qz} + \phi_\ind e^{-qz}.
\end{equation}
The different components of the potentials are depicted in figure \ref{fig response function}. Enforcing continuity of the potential and of the displacement field $\D = - \epsilon_0\epsilon(\omega) \nabla \phi$ at the interface yields:
\begin{equation}
	\phi_\ext + \phi_\ind = \phi_{\rm m} \quad \mathrm{and} \quad \phi_\ext - \phi_\ind = \epsilon(\omega) \phi_{\rm m}.
\end{equation}
Finally, we identify the surface response function in the long-wavelength limit as:
\begin{equation} \label{q0 limit}
	g_\s(q \rightarrow 0, \omega) = - \frac{\phi_{\rm ind}}{\phi_{\rm ext}} = \frac{\epsilon(\omega) -1}{\epsilon(\omega) +1}.
\end{equation}

Beyond the long-wavelength limit, the determination of the surface response function $g_\s(\q, \omega)$ is more involved, as it requires the knowledge of the density response function $\chi_\s(\q, z, z', \omega)$ for the semi-infinite medium. If the medium is a liquid or a dielectric solid, the density response function can be determined from molecular dynamics simulations, as a few-nanometer slab will already have semi-infinite-like interfacial response properties \cite{coquinotcollectivemodes}. In the case of a solid with delocalized electrons, however, a very thick slab may be needed to recover the interfacial response properties of the semi-infinite medium, leading to prohibitively large unit cell sizes in DFT calculations. For instance, more than a hundred layers were shown to be necessary in converging the plasmonic properties in heterostructures of graphene and transition metal dichalcogenides \cite{lavorprobingstructure2020}.

In the simplest perturbative treatment that takes into account electron-electron interactions (self-consistent Hartree or random phase approximation), the interacting density response function $\chi_\s$ can be deduced from the non-interacting one $\chi_\s^0$ by solving the following Dyson equation:
\begin{equation}\label{dyson response function}
\begin{split}
	&\chi_\s  (\q, z,z',\omega) = \chi_\s^0(\q, z,z', \omega)+ \cdots \\ & \cdots \int \d z_1 \, \d z_2 \, \chi_\s^0(\q, z,z_1, \omega) V_\q(z_1 -z_2) \chi_\s(\q, z_2,z',\omega). 	
\end{split}
\end{equation}
Nonetheless, the challenge of computing the non-interacting density response function $\chi_\s^0$ in the semi-infinite geometry remains. A possible approximation is to assume that the electronic excitations in the semi-infinite medium behave as if the electrons in the bulk were simply reflected specularly at the boundary, neglecting quantum interference effects. Concretely, this \emph{specular reflection approximation} relates the semi-infinite response function $\chi_\s^0$ to its bulk counterpart $\chi_{\s, \rm B}^0$, by simply "folding" $\chi_{\s, \rm B}^0$ with mirror symmetry along the surface:
\begin{equation}\label{specular reflection}
	\chi_\s^0(\q, z, z', \omega) = \chi_{\s, \rm B}^0(\q, z-z', \omega) + \chi_{\s, \rm B}^0(\q, z+z', \omega).
\end{equation}
The non-interacting bulk response function can be evaluated as long as the eigenstates and eigenenergies of the non-interacting bulk system are known: these are readily accessible within DFT, or simplified jellium or tight-binding models. The textbook result is \cite{rammer2007}:
\begin{equation}
\begin{split}
\chi_{\s, \rm B}^0(\q,& q_z, \omega) = \sum_{\nu, \nu'} \int_{\rm FBZ} \frac{\d ^3\k}{4 \pi^3} \, \vert \langle \k+\q, \nu \vert e^{i\q \cdot \r} \vert \k, \nu' \rangle \vert^2 \cdots \\ & \qquad \cdots \frac{\nF(E_\nu(\k+\q))-\nF(E_{\nu'}(\k))}{E_\nu(\k+\q)-E_{\nu'}(\k)-\hbar (\omega + i\delta)},	
\end{split}
\label{eq:chi0}
\end{equation}
where the states are labelled by a band index $\nu$ and a wavevector $\k$ within the first Brillouin zone, and $\delta \rightarrow 0$. The bulk non-interacting response function being known, combining equations \eqref{dyson response function} and \eqref{specular reflection} allows us to express the surface response function as:
\begin{equation}\label{general surface response function}
	g_\s(\q, \omega) = \frac{1 -q \ell_\q(\omega)}{1 + q \ell_\q(\omega)}, 
\end{equation}
with $$ \ell_\q(\omega) = \frac{2}{\pi} \int_0^\infty \frac{\d q_z}{(q^2 + q_z^2) \epsilon(\q, q_z, \omega)},$$
where $ \epsilon(\q, q_z, \omega) = 1 - \frac{e^2}{\epsilon_0 (q^2 + q_z^2)}\chi_{\s, \rm B}^0(\q, q_z, \omega)$ is the bulk system's dielectric response function. We refer to \cite{ritchie1966234, griffinharris, kavokinefluctuationinducedquantum2022} for the details of the derivation of formula \eqref{general surface response function}.

Before closing this section, we note that analogues of the surface response function, that conveniently pack dielectric response properties, can be defined not only in the semi-infinite medium geometry, but also in other geometries of relevance in nanofluidics: for instance, two-dimensional slits \cite{coquinotcollectivemodes} and one-dimensional tubes \cite{gispertelectrostaticscreeningnanotubes2025}.

\subsection{Van der Waals and Casimir friction}

\begin{figure}
	\begin{center}
		\includegraphics[width=\columnwidth]{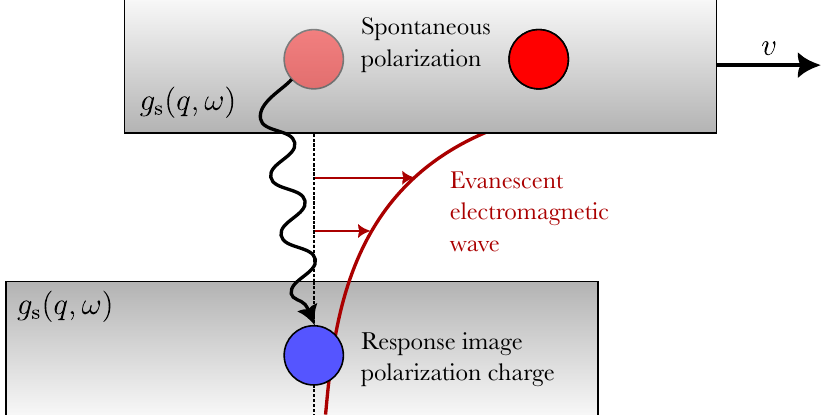}
		\caption{\textbf{Schematic of van der Waals friction.} Spontaneous charge fluctuations inside one of the solid bodies, which induces as a response an image charge in the other body. Resulting is a net van der Waals friction force parallel to the solid relative motions. For close enough separations between solids, evanescent electromagnetic waves dominate over propagative ones in van der Waals friction.} \label{fig van der Waals friction}
	\end{center}
\end{figure}

As discussed in section \ref{section Lifschitz theory}, two extended solid bodies separated by a vacuum gap are subject to an attractive van der Waals or Casimir force due to the coupling between spontaneous charge fluctuations inside the two bodies. Now, if the two bodies slide relative to each other, the fluctuating charge distributions cannot adjust instantaneously to the translational motion. This leads to a net component of the force that is parallel to the interface and opposes the motion: such a friction force is called \emph{van der Waals friction} or \emph{Casimir friction}, as illustrated in figure \ref{fig van der Waals friction}. In the same way that the van der Waals attraction between two solid bodies cannot be obtained by summing up two-atom potentials, the van der Waals friction cannot be directly obtained from the dielectric or electronic friction formulae discussed above. It requires, in fact, a non-equilibrium generalization of Lifschitz's theory, which was only achieved in the 1990's through the pioneering works of Pendry, Volokitin and Persson \cite{pendry1997, volokitin1999}. 

As for the static van der Waals force, those derivations focus on the fluctuating electromagnetic field in the vacuum gap between the two bodies. Given the fields $(\E(t), \mathbf{B}(t))$, the friction force can be deduced from the Maxwell stress tensor:
\begin{equation}\label{Maxwell stress tensor}
	\sigma_{ij} =  \epsilon_0 \left( E_i E_j -\frac{1}{2}\delta_{ij} E^2 \right) + \frac{1}{\mu_0}\left(B_i B_j - \frac{1}{2} \delta_{ij} B^2 \right),
\end{equation}
after averaging over equilibrium fluctuations.

Pendry's derivation tackles the case $T = 0$, where the fluctuations are purely quantum, and in the absence of propagation effects. He computed the Maxwell stress tensor directly out of equilibrium, taking into account the relative motion as a Doppler shift in the boundary conditions. Persson and Volokitin generalized Pendry's result to include finite temperature and propagation effects. In their initial derivation \cite{volokitin1999}, they relied on the phenomenological Rytov theory of "fluctuating electromagnetism", where a fluctuating current density is introduced into the classical Maxwell's equations, and the relative motion is taken into account as a boundary condition, as in Pendry's computation. They later proposed a rigorous quantum field theory derivation based on the coarse-grained Lifshitz formalism \cite{volokitin2006}. Since Lifshitz theory assumes thermal equilibrium, this approach only yields the friction coefficient $\lambda$ to linear order in velocity, through the Green-Kubo relation: 
\begin{equation}
	\lambda = \frac{1}{\mathcal{A}\kB T}  \int_0^{+\infty} \d t \langle F_x(t) F_x(0) \rangle_{\rm eq} ,
\end{equation}
where the lateral force $F_x$ is computed from the Maxwell stress tensor, and the average is taken over all equilibrium thermal and quantum fluctuations, $\mathcal{A}$ being the surface area. In both computations, the solid bodies are described in terms of their dielectric functions, which are in principle allowed to be nonlocal both in space and in time. As discussed in the previous section, the dielectric function is related to the reflection coefficient of electromagnetic waves at the solid boundaries, so that the two solid bodies may be equivalently described by their surface response functions $g_1(\q, \omega)$ and $g_2(\q, \omega)$. The fluctuating electromagnetic field between the two media decomposes into propagating waves (associated with thermal radiation at long distances) and evanescent waves, which both contribute to the friction coefficient, respectively, as:
\begin{equation}
\begin{split}
	\lambda^{\rm rad} &= \frac{\hbar^2}{32 \pi^2 \kB T} \int_0^{\infty} \frac{\d \omega}{\sinh^2(\hbar \omega/2\kB T)} \int_0^{\omega/c} \d q \, q^3 \cdots \\  \cdots & \frac{(1 - \vert g_1(q, \omega) \vert^2)(1- \vert g_2(q, \omega) \vert^2 )}{\vert 1 - e^{-2 id\sqrt{(\omega/c)^2 - q^2}} g_1(q, \omega) g_2(q, \omega) \vert^2},
\end{split}
\end{equation}
\begin{equation}\label{evanescent}
\begin{split}
	\lambda^{\rm evan}&= \frac{\hbar^2}{8 \pi^2 \kB T} \int_0^\infty \frac{\d \omega}{\sinh^2(\hbar \omega/2\kB T)} \int_{\omega/c}^\infty \d q \, q^3 \cdots \\   \cdots & \frac{e^{-2d \sqrt{q^2 - (\omega/c)^2}} \im{g_1(q, \omega)} \im{g_2(q, \omega)}}{\vert 1 - e^{-2d \sqrt{q^2 - (\omega/c)^2}} g_1(q, \omega) g_2(q, \omega) \vert^2}.
\end{split}
\end{equation}
In the limit where the distance between the two bodies is small compared to their typical absorption wavelengths, one may neglect propagation effects, and the most important contribution comes from evanescent waves, so that the van der Waals friction force reduces to $\F = - \lambda_{\rm vdW} \mathcal{A} \v$ with:
\begin{equation}
\begin{split}
	\lambda_{\rm vdW} &= \frac{\hbar^2}{8 \pi^2 \kB T} \int_0^\infty \frac{\d \omega}{\sinh^2(\hbar \omega/2\kB T)} \int_0^\infty \d q \, q^3 \cdots \\ & \qquad \cdots \frac{e^{-2qd} \im{g_1(q, \omega)} \im{g_2(q, \omega)}}{\vert 1 - e^{-2qd} g_1(q, \omega) g_2(q, \omega) \vert^2}.	
\end{split}
	\label{eq:QF_persson}
\end{equation}
It is illuminating to compare this result to the one obtained for electronic friction in Eq.~\eqref{electronic friction}. The electronic friction of an ion on a metal surface was determined by the metal's surface response function at the typical frequency of the ion motion. Here, since both solid bodies are fluctuating (neither of them carries a mean charge), the van der Waals friction is determined by the overlap of their respective surface response functions in frequency-momentum space. 

We note that the fluctuating electromagnetic field that supports momentum transfer between the two bodies in the form of van der Waals friction can also transport a net energy flux if the two bodies are not at the same temperature. This radiative heat transfer can directly be computed from the Poynting vector $\mathbf{S} = 1 / \mu_0 \langle \E \times \mathbf{B} \rangle$ when the fluctuating fields are known \cite{volokitin2008}. Similarly to the van der Waals friction, this radiative heat transfer is significantly increased in the near-field limit, where evanescent waves dominate \cite{volokitin2007, biehsNearfieldRadiativeHeat2021}.

\vspace{.5cm}

In this Section, we sought to provide a bird's eye view of the literature landmarks in the field of fluctuation-induced forces -- from the famous van der Waals force to more exotic fluctuation-induced friction effects. As the literature on these topics spans half a century, we did not strive for exhaustiveness; rather, we sought to outline the theoretical ideas and tools that underlie the main results of the field. Especially in the case of forces between two extended bodies, the prevalent theoretical approaches are based on the picture of two solids separated by a vacuum gap that supports a fluctuating electromagnetic field. This approach does not allow for easy intuition in the case where one of the bodies is a solid and the other is a liquid: there is essentially no vacuum gap at a solid-liquid interface. However, fluctuation-induced friction forces must also exist at such interfaces. While Eq.~\eqref{eq:QF_persson} has been applied to solid-liquid systems in a few instances \cite{volokitinVanWaals2008}, it is a priori not clear where its limits of validity are, nor how it connects to more usual descriptions of the liquid-solid interface. 

In the rest of this Article, we will "rediscover" fluctuation-induced forces in the context of nanoscale liquid-solid interfaces. We will first present a simplified picture that provides physical intuition (Sec. IV), and then expand into a microscopic description based on the Keldysh formalism of quantum field theory, which provides the rigorous foundation for fluctuation-induced and quantum effects in nanofluidic transport.

\section{Fluctuation effects in liquid-solid friction}
\label{sec4}

Fluctuations are a key determinant of friction in the solid and gas phases: solid-state charge fluctuations are at the origin of van der Waals or Casimir non-contact friction between solids and of electronic friction for molecules on metal surfaces. Yet, a liquid-solid analogue of van der Waals - Casimir friction has been proposed only recently~\cite{kavokinefluctuationinducedquantum2022} and termed \emph{fluctuation-induced quantum friction}. In this section, we introduce an intuitive theoretical approach to the effect that highlights its physical origin and provides some quantitative estimates for model systems. We discuss its most important practical consequences in Section V, before turning to the more powerful quantum theoretical approach in Section VI. 

\subsection{Liquid-solid friction in nanofluidics}

We first highlight the crucial role of liquid-solid friction in nanofluidics, regardless of its microscopic mechanism. In macroscopic hydrodynamics, no-slip boundary conditions are typically imposed for viscous flows at liquid-solid interfaces: the flow velocity is assumed to vanish at a solid wall ($\mathbf{v}|_{\rm wall} = 0$). This is, in fact, an approximation to the more general Navier partial slip boundary condition, which imposes that the viscous stress at the wall (viscosity times tangential velocity gradient) balances the liquid-solid friction force: 
\begin{equation}
\eta \left. \frac{\partial v}{\partial n} \right|_{\rm wall} = \lambda v|_{\rm wall}, 
\end{equation}
where $\bf n$ is the normal to the wall pointing into the fluid. Under a linear response assumption, the liquid-solid friction force is proportional to $v_{\rm wall}$ (which is then non-zero and is called the slip velocity); the proportionality coefficient is, by definition, the liquid-solid friction coefficient. Its SI unit is $\rm N \cdot s \cdot m^{-3}$ (force per unit velocity and per unit area). 

Let us consider for concreteness the flow of water through a slit-like channel of length $L$, width $W$ and height $h \ll L, W$ on the $1 - 100~\rm nm$ scale under a pressure drop $\Delta P$. Under experimentally achievable pressure drops $\Delta P \lesssim 1~\rm bar$, the corresponding Reynolds number is much smaller than unity: no turbulence is expected in nanoscale water flows. Under a no-slip boundary condition, the laminar flow profile is given by the Hagen-Poiseuille law: 
\begin{equation}
v_{\rm no-slip}(z) = \frac{\Delta P}{2 \eta L} \left( \frac{h^2}{4} - z^2 \right),
\end{equation}
assuming that the channel midplane is at $z = 0$. The effect of the partial slip boundary condition is to shift this parabolic velocity profile by the slip velocity: 
\begin{equation}
v(z) = v_{\text{no-slip}}(z) + \frac{6 b}{h} \cdot \overline{v_{\text{no-slip}}}, 
\label{v_with_slip}
\end{equation}
where $\overline{v_{\text{no-slip}}} = \Delta P h^2 / (12 \eta L)$ is the average no-slip velocity and $b = \eta / \lambda$ is called the slip length, typically in the $1-100~\rm nm$ range. As is apparent from Eq.~\eqref{v_with_slip}, this effect is important for channels of nanoscale dimensions; at macroscopic scales, no-slip (that is $b = 0$, or $\lambda \to \infty$) is a good approximation. Somewhat counter-intuitively, assuming an infinite liquid-wall friction coefficient is equivalent to assuming that no energy is being dissipated by the flow through liquid-solid friction. Indeed, interfacial energy dissipation requires a non-zero slip velocity: in a similar way, no energy can be dissipated by an electrical resistance if it is not traversed by any current. In a macrosopic channel, most of the energy is dissipated in the bulk through viscosity (the "internal friction" of the fluid); conversely, in a nanoscale channel, non-negligible dissipation may occur at the channel wall. 

\begin{figure}
	\begin{center}
		\includegraphics[width=\columnwidth]{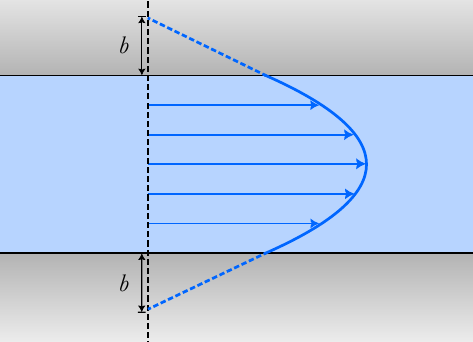}
		\caption{\textbf{Partial slip boundary condition in a slit geometry.} The slip length $b=\eta /\lambda$ corresponds to the distance from the interface to the point inside the wall where the linear extrapolation of the parabolic velocity profile vanishes.}\label{fig slip length}
	\end{center}
\end{figure}

Due to the factor 6 in the second term of Eq.~\eqref{v_with_slip}, the slip velocity often dominates the velocity profile in nanoscale channels, even with moderate slip lengths. As an example, let us consider a water flow in a channel of height $h = 1~\rm nm$, with a small slip length $b = 5~\rm nm$ at the walls. Then, the slip term in Eq.~\eqref{v_with_slip} exceeds the bulk Poiseuille term by a factor of 30, and
the velocity profile is well-approximated by a plug flow: 
\begin{equation}
v(z) = \frac{\Delta P}{2\eta L} \, bh = \frac{h \Delta P}{2 \lambda L}. 
\label{plug} 
\end{equation}
In a channel of height $h = 1~\rm nm$, the Navier-Stokes equation is at its limit of validity. However, it is important to note that the viscosity drops out of the result in Eq.~\eqref{plug}. In fact, this result could be obtained by enforcing global force balance on a rigid water slab moving at velocity $v$, subject to a pushing force $h W \Delta P /L$ and a friction force $2 \lambda v L W$. The thickness of this slab corresponds to about 3 layers of water molecules. The friction coefficient between two of these layers can be estimated as $\lambda_{\rm ww} \sim \eta / \delta \approx 3 \times 10^6 ~ \rm N \cdot s \cdot m^{-3}$, where $\delta \sim 0.3~\rm nm$ is the size of a water molecule. This is an order of magnitude larger than the water-solid friction coefficient corresponding to a slip length $b = 5~\rm nm$ ($\lambda \approx 2 \times 10^5~\rm N \cdot s \cdot m^{-3}$). It is thus intuitive that, under a pressure drop, the water slab will slide through the channel as a rigid body, without its constituent layers sliding relative to each other. The Navier-Stokes equation may not adequately describe shear (or the viscosity may be very different from the bulk viscosity) in 1 nm confinement, but this will not influence the flow rate under a pressure drop, which is ultimately determined by the liquid-solid friction coefficient. In such a situation, the friction coefficient $\lambda$ is the physically meaningful quantity, rather than the slip length $b$. However, $\lambda$ and $b$ are often used interchangeably, with $b$ being computed by convention using the bulk viscosity. 

Thus, in many nanofluidic systems -- and, particularly, single-digit nanochannels -- fluid transport is controlled by liquid-solid friction. Based on similar arguments, friction is also found to play a governing role in ion transport. Under an applied voltage, the ionic current through a nanochannel can be decomposed into two contributions: the electrophoretic contribution, which accounts for ions being dragged relative to the solvent by the electric field, and the electro-osmotic contribution, which corresponds to the advection of ions by the electro-osmotic flow. In nanometer-scale channels, the latter contribution dominates even with moderate slippage and surface charge \cite{kavokinefluidsnanoscalecontinuum2021}. Then, the ionic current under a voltage drop $\Delta V$ is
\begin{equation}
 I = \frac{2 e^2 W \Sigma^2}{\lambda} \frac{\Delta V}{L}, 
\end{equation}
where $\Sigma$ is the surface charge density (in multiples of the elementary charge per area). Here again, the solvent viscosity and ionic mobilities have dropped out to leave the friction coefficient $\lambda$ as a single governing parameter. 

Overall, understanding nanofluidic transport at single-digit scales hinges on understanding liquid-solid friction
-- at least in the case of long channels, where $L \gg h, b$. In short channels, entrance effects play a non-negligible role in determining the transport properties \cite{kavokinefluidsnanoscalecontinuum2021}. Entrance effects need to be analyzed with particular care in the case of water flows through carbon nanotubes, where effective slip lengths $b \gtrsim 1~ \rm \mu m$ have been observed \cite{alurufluidselectrolytesconfinement2023}. 

\subsection{Roughness-induced friction with a static wall}
\label{ssec:RI-friction}
We now discuss the microscopic origins of liquid-solid friction. The first microscopic theories of hydrodynamic friction date back more than 20 years \cite{bocquet1994,barrat1999}, and their predictions have already been reviewed elsewhere \cite{bocquet2007,kavokinefluidsnanoscalecontinuum2021}. These works were based on the picture of a solid acting on the liquid as a static, but spatially inhomogeneous potential, which accounted for the solid's surface roughness, possibly at the atomic scale. The main qualitative features of their findings can be summarized in the scaling expression 
\begin{equation}
\lambda_{\rm RI} \sim V_{\s \l}^2 \times S_{\l}(q_{\parallel}) \times \tau_{q_{\parallel}}. 
\label{lambdacl_scaling}
\end{equation}
This result is obtained under the assumption that the interaction potential between the solid and the liquid is periodic with a wavevector $q_{\parallel}$. $V_{\s\l}$ represents the amplitude of the potential, $S_{\l}(\mathbf{q}) = \langle \rho(\mathbf{q}) \rho(-\mathbf{q}) \rangle / \mathcal{A}$ is the density structure factor in the interfacial liquid layer ($\mathcal{A}$ is the surface area) and $\tau_{q_{\parallel}}$ is a microscopic relaxation time of the liquid density fluctuations at wavevector $q_{\parallel}$. Eq.~\eqref{lambdacl_scaling} captures the main qualitative determinants of the roughness-induced (RI) contribution to friction, which has also been called \emph{classical friction}, to distinguish it from the fluctuation-induced quantum friction introduced in the following \cite{kavokinefluctuationinducedquantum2022}. RI friction is sensitive to the strength of solid-liquid interaction $V_{\s\l}$: intuitively, the closer the interfacial liquid layer to the solid surface, the more it "feels" its roughness. The scaling of $\lambda_{\rm RI}$ with $V_{\s\l}^2$ establishes the link between friction and hydrophobicity \cite{huang2008}. Indeed, Young's law for the liquid-solid contact angle $\theta$ yields the scaling $1 + \cos \theta \sim V_{\s\l}$, so that $\lambda_{\rm RI} \sim (1 + \cos \theta)^2$. This relationship is well verified in both experiments and molecular dynamics simulations of water on various materials \cite{bocquetnanofluidicsbulkinterfaces2010}, although with some notable exceptions, such as the carbon-based materials already discussed above. Furthermore, the presence of the water structure factor in Eq.~\eqref{lambdacl_scaling} suggests that friction is influenced by the commensurability of the liquid and solid interfacial structures: friction is significant if the liquid has a large structure factor at the wavevector of the solid's surface roughness. A strong commensurability effect was demonstrated in molecular dynamics simulations of water flows through narrow (sub-10-nm) carbon nanotubes: classical friction was shown to nearly vanish in the narrowest tubes due to a curvature-dependent shift of the surface roughness wavevector, in addition to curvature-induced smoothening \cite{falk2010}. 

\begin{figure}
	\begin{center}
		\includegraphics[width=\columnwidth]{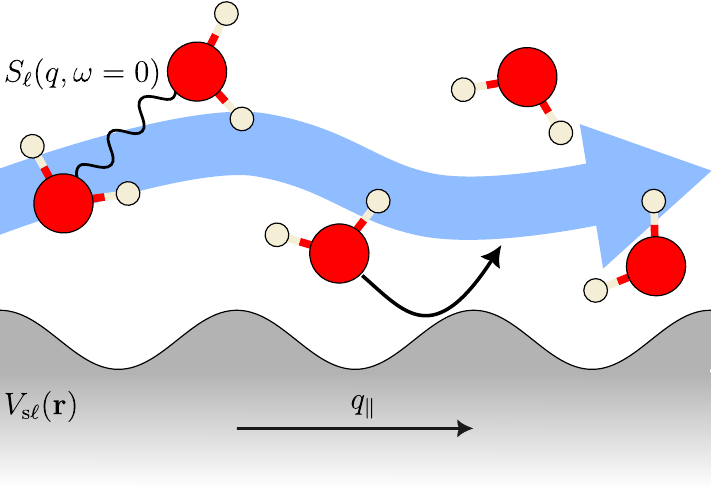}
		\caption{\textbf{Mechanism of roughness-induced friction.} The solid is modeled as a static periodic potential $V_{\s\l}(\r)$ with wavevector $q_{\parallel}$. This potential induces a density perturbation inside the liquid at the same wavevector, and friction arises from the solid pushing on this perturbation as the liquid flows. The amplitude of the perturbation is controlled by the liquid's structure factor $S_\l(q,\omega = 0)$.} \label{fig_RI_friction}
	\end{center}
\end{figure}

The analytical formula underlying the scaling expression in Eq.~\eqref{lambdacl_scaling} is usually derived through the Green-Kubo relation, considering equilibrium dynamics of the solid-liquid system \cite{bocquetGreenKuboRelationship2013}. Here, we provide a derivation based on a system in the presence of flow~\cite{kavokinefluctuationinducedquantum2022}, which will allow us to compare the physical mechanisms of roughness-induced and fluctuation-induced friction on equal footing. We thus consider a semi-infinite liquid occupying the half-space $z>0$, and flowing with a steady-state velocity field $\v$, in contact with a solid occupying the half-space $z<0$. Given the liquid's fluctuating particle density $\rho(\r, t)$, we wish to compute the force exerted by the solid on the liquid: 
\begin{equation}
\langle \mathbf{F}(t) \rangle = -\int \d \r \langle \rho(\r, t) \rangle \nabla V_{\s\l}(\r), 
\label{FCl_def}
\end{equation}
where $V_{\s\l}$ is a potential of the Lennard-Jones type and $\langle \cdot \rangle$ is the average over density fluctuations. Since we are looking for the component of the force that is parallel to the interface, we may assume without loss of generality that $V_{\s\l}$ has been subtracted with its component that is uniform along the interface: $\int \d \r_{\parallel} \, V_{\s\l}(\r_{\parallel}, z) = 0$. We then decompose the liquid density as 
\begin{equation}
\rho(\r, t) = \rho_0(\r, t) + \delta \rho(\r, t),
\end{equation}
where $\rho_0(\r ,t)$ is the density that the liquid would have if the solid was perfectly smooth; $\delta \rho$ is the perturbation induced by the solid's roughness. Only the latter contributes to friction force:
\begin{equation}
\langle \mathbf{F}(t) \rangle = -\int \d \r \langle \delta \rho(\r, t) \rangle \nabla V_{\s\l}(\r).
\label{FCl_def2}
\end{equation}
The potential $V_{\s\l}$ is short-ranged: it barely reaches beyond the interfacial liquid layer. It is then reasonable to assume that on that scale, the flow velocity is uniform (that is, independent of $z$). This allows us to go to the reference frame where the liquid's center of mass is at rest, writing
\begin{equation}
\langle \mathbf{F}(t) \rangle = -\int \d \r \langle \delta \rho(\r, t) \rangle_{\v = 0} \nabla V_{\s\l}(\r + \v t).  
\end{equation}
$\delta \rho$ now describes the response of an equilibrium liquid to the time-dependent perturbation $V_{\s\l}(\r + \v t)$: within linear response theory, there exists a response function $\chi$ such that 
\begin{equation}
\delta \rho (\r, t) = \int \d \r' \int \d t' \, \chi_\l(\r, \r', t - t') V_{\s\l}(\r' + \v t').  
\end{equation}
The response function $\chi_\l(\r, \r', t - t')$ depends only on the properties of the system with no surface roughness, so that we may introduce its Fourier transform along the interface, $\chi_\l(\q, z, z', t - t')$ (see Appendix A for Fourier transform conventions). Introducing in the same way the Fourier transform of $V_{\s\l}$ along the interface, $V_{\s\l} (\q, z)$, Eq.~\eqref{FCl_def2} transforms into
\begin{equation}
\begin{split}
\langle \mathbf{F}(t) \rangle &= \frac{1}{(2 \pi)^2} \int \d \q \d z \d z' \, (i \q) V_{\s\l}(\q, z') V_{\s\l}(-\q, z) \cdots \\ & \qquad \cdots \int \d t' \, e^{-i \q \v (t - t')} \chi_\l(\q, z, z', t - t'). 	
\end{split}
\end{equation}
Recognizing now the Fourier transform of $\chi_\l$ with respect to time, we obtain 
\begin{equation}
\begin{split}
\langle \mathbf{F}(t) \rangle = \frac{1}{(2 \pi)^2} \int \d \q & \d z \d z' \, (i \q)  V_{\s\l}(\q, z') V_{\s\l}(-\q, z) \cdots \\ & \cdots \chi_\l(\q, z, z', \omega = -\q \v). 	
\end{split}
\end{equation}
As the friction force must be real, only the imaginary part of $\chi_\l$ should contribute to the integral: this is indeed the case since $\chi_\l(\q, -\omega) = \chi_\l(\q, \omega)^*$ (Appendix A). The friction coefficient $\lambda_{\rm RI}$ is defined by $\lambda_{\rm RI} = - (\mathbf{F}\cdot \v)/(\mathcal{A} v^2)$. Thus, expanding $\chi$ to linear order in $\v$, we obtain 
\begin{equation}
\begin{split}
\lambda_{\rm RI} = \frac{1}{(2 \pi)^2} \int &\d \q \d z \d z' \,  \frac{(\q \cdot \v)^2}{v^2} \frac{V_{\s\l}(\q, z') V_{\s\l}(-\q, z) }{\mathcal{A}} \cdots \\ & \cdots \left(- \partial_{\omega} \mathrm{Im}[\chi_\l(\q, z, z', \omega)]|_{\omega = 0} \right). 	
\end{split}
\end{equation}
Finally, we note that since the potential $V_{\s\l}$ is periodic along the interface, its Fourier transform decomposes over reciprocal lattice wavevectors: $V_{\s\l}(\q, z) = (2 \pi)^2 \sum_{\G} V_{\G}(z) \delta(\q - \G)$, so that
\begin{equation}
 V_{\s\l}(\q, z') V_{\s\l} (-\q, z) = (2\pi)^2 \mathcal{A} \sum_{\G} V_{\G}(z') V_{-\G}(z) \delta(\q - \G).
\end{equation}
Thus, 
\begin{equation}
\begin{split}
	\lambda_{\rm RI} = \sum_{\G} \int & \d z \d z' \,  \frac{(\G \cdot \v)^2}{v^2} V_{\G}(z') V_{-\G}(z) \cdots \\ &\cdots \left(- \partial_{\omega} \mathrm{Im}[\chi_\l(\G, z, z', \omega)]|_{\omega = 0} \right). 
\end{split}
\label{eq:result_RI_chi}
\end{equation}
Alternatively, one may introduce the liquid's structure factor, defined as the correlation function of the equilibrium density fluctuations: 
\begin{equation}
S_\l(\r, \r', t- t') = \langle \rho_0(\r, t) \rho_0(\r', t') \rangle_{\v = 0}. 
\end{equation}
It is related to the response function $\chi_\l$ via the fluctuation-dissipation theorem (FDT, see Appendix C): 
\begin{equation}
S_\l(\r, \r', \omega) = - \frac{2 \kB T}{\omega} \im{\chi_\l(\r, \r', \omega)},  
\label{eq:FDT_RI}
\end{equation}
so that Eq.~\eqref{eq:result_RI_chi} can be recast as
\begin{equation}
\begin{split}
	\lambda_{\rm RI} = \frac{1}{2 \kB T}\sum_{\G} \int & \d z \d z' \,  \frac{(\G \cdot \v)^2}{v^2} V_{\G}(z') V_{-\G}(z) \cdots \\ &\cdots S_\l(\G, z, z', \omega = 0). 
\end{split}
\label{eq:result_RI_S}
\end{equation}
Eqs.~\eqref{eq:result_RI_chi} and \eqref{eq:result_RI_S} are the most compact and general results for the RI friction coefficient. The derivation pinpoints the underlying microscopic mechanism: friction arises because the solid's periodic potential is pushing on the density perturbation it has induced in the liquid (see Fig.~\ref{fig_RI_friction}). 

To make the qualitative picture even sharper, let us first note that the solid-liquid interaction potential is short-ranged, so that the integrals over $z$ may be restricted to the first liquid layer,
separated from the solid by a microscopic distance $z_1$. We then define 
\begin{equation}
\chi (\G, \omega) = \int_{\rm 1^{\rm st}~layer} \d z \d z' \chi_\l(\G, z, z', \omega), 
\end{equation}
and the friction coefficient becomes 
\begin{equation}
\lambda_{\rm RI} \approx \sum_{\G}  \frac{(\G \cdot \v)^2}{v^2} |V_{\G}(z_1)|^2  \left(- \partial_{\omega} \mathrm{Im}[\chi_\l(\G, \omega)]|_{\omega = 0} \right). 
\end{equation}
Furthermore, it can be assumed that the Fourier components of the potential decay quickly, so that only the shortest reciprocal lattice vectors (of norm $q_{\parallel}$) need to be considered:
\begin{equation}
\lambda_{\rm RI} \approx \alpha q_{\parallel}^2 |V_{q_{\parallel}}(z_1)|^2  \left(- \partial_{\omega} \mathrm{Im}[\chi_\l(q_{\parallel}, \omega)]|_{\omega = 0} \right),  
\label{eq:lambdaRI_approx}
\end{equation}
where $\alpha$ is a geometric factor of order 1. Now, consider for a moment that the amplitude in the liquid's density mode at wavevector $q_{\parallel}$ follows damped harmonic oscillator dynamics, with mass $m$, spring constant $k$ and friction coefficient $\mu$. Then, the response function has the explicit expression 
\begin{equation}
\chi_\l(q_{\parallel}, \omega) = \frac{1}{m \omega^2 - k + i \mu \omega}.   
\end{equation}
One identifies $\partial_{\omega} \im{\chi_\l(q_{\parallel}, \omega)}|_{\omega = 0} = - \mu \chi_\l(q_{\parallel})^2$, where $\chi(q_{\parallel}) \equiv \chi(q_{\parallel}, \omega = 0) = -1/k$ is the static value of the response function. This allows recasting the friction coefficient as 
\begin{equation}
\lambda_{\rm RI} \times v^2 \approx \alpha [q_{\parallel} v \times V_{q_{\parallel}}(z_1) \chi_\l(q_{\parallel})]^2 \times \mu.
\end{equation} 
As the liquid flows with velocity $v$, the solid's roughness at wavevector $q_{\parallel}$ drives a density oscillation in the liquid with amplitude $V_{q_{\parallel}} (z_1) \chi(q_{\parallel})$ and frequency $q_{\parallel} v$. We find that $\lambda_{\rm RI} \times v^2$ corresponds precisely to the power per unit surface dissipated by this density oscillation. 

In the general case, since the response function satisfies the Kramers-Kronig relation (see Appendix A), one can parametrize $\partial_\omega \im{\chi_\l(q_{\parallel})} |_{\omega = 0} = \chi_\l(q_{\parallel}) \tau_{q_{\parallel}}$, where $\tau_{q_{\parallel}}$ is a relaxation time of the liquid density in mode $q_{\parallel}$ ($\tau_{q_{\parallel}} = \mu/k$ in the harmonic oscillator model).  Using the FDT in Eq.~\eqref{eq:FDT_RI}, the static response function $\chi_\l(\q)$ can be expressed as
\begin{equation}
\chi_\l(\q) = \frac{1}{2 \pi \mathcal{A} \kB T} \langle \rho_0 (\q) \rho_0(-\q) \rangle_{\v = 0}, 
\end{equation}
where $\rho_0(\q) = \int_{\rm 1^{\rm st}~layer} \d z \, \rho_0(\q, z)$. One defines the dimensionless static structure factor of the interfacial liquid as a sum over particles $i$ and $j$: 
\begin{equation}
%\begin{split}
\bar S_{\l}(\q) = \frac{1}{N} \left \langle \sum_{i, j}^{1^{\rm st} ~\rm layer} e^{i \q (\r_i - \r_j)} \right \rangle = \frac{\langle \rho_0(\q) \rho_0(-\q) \rangle_{\v = 0}}{\rho_1 \cal A}, 	
%\end{split}
\end{equation}
where $N$ is the number of particles in the layer. The latter can be expressed as $N = \rho_1 \mathcal{A}$, where $\rho_1$ is the interfacial liquid density. This yields the approximate result   
\begin{equation}
\lambda_{\rm RI} \approx \frac{\alpha \rho_1 }{2\pi k_{\rm B} T} q_{\parallel}^2 |V_{\mathbf{q}_{\parallel}}(z_1)|^2 \bar S_{\l}(q_{\parallel}) \tau_{q_{\parallel}},  
\label{scaling2}
\end{equation}
which recovers the scaling expression anticpated in Eq.~\eqref{lambdacl_scaling}, and highlights the role played by liquid-solid commensurability in RI friction. The power dissipated in the density oscillation at $q_{\parallel}$ increases with the oscillation amplitude, which is controlled by the static response function $\chi_\l(q_{\parallel})$, or, through the fluctuation-dissipation theorem, by the structure factor $\bar S_\l(q_{\parallel})$. As the name suggests, the structure factor quantifies the liquid's molecular structure: it shows peaks at harmonics of $q = 2\pi/ \sigma$, where $\sigma$ is the molecular size \cite{falk2010}. Therefore, better liquid-solid commensurability leads to a stronger density perturbation at $q_{\parallel}$, and greater energy dissipation as the perturbation is driven by the flow.

\subsection{Fluctuation-induced quantum friction: a "call and response" picture}\label{Fluctuation-induced quantum friction}

Fluctuation-induced (FI) friction arises if not only the liquid, but also the solid, is able to fluctuate. These fluctuations can typically be decomposed into density fluctuations and charge fluctuations, which are important for a polar liquid such as water. FI friction can arise either through the Lennard-Jones interaction of density fluctuations, or through the Coulomb interaction of charge fluctuations -- we focus on the latter in the rest of this Article. A solid's charge fluctuations are largely the result of its electrons' quantum dynamics -- the term "quantum friction" thus refers to liquid-solid friction involving the solid's electronic fluctuations.  

A rigorous theory of liquid-solid friction involving the solid's electron dynamics needs to be based on a quantum formalism, which will be outlined in Sec. VI. However, the qualitative mechanism of FI friction does not rely on quantum mechanics: friction can arise from purely classical thermal fluctuations. We illustrate this mechanism by considering a liquid and a solid described by their classical fluctuating charge densities $n_\l$ and $n_{\rm s}$, coupled via Coulomb interactions. We decompose
%for which we assume generalized Langevin dynamics at temperature $T$: 
\begin{equation}
\left \{
\begin{array}{l}
n_\l (\r, t) = n_\l^0(\r, t) + \delta n_\l (\r,t) \\
n_\s (\r, t) = n_\s^0(\r, t) + \delta n_\s (\r,t) 
\end{array}
\right . , 
\label{nsw_dynamics}
\end{equation}
where $n_{\s, \l}^0$  are the charge densities in the absence of Coulomb interactions between the media (describing spontaneous charge fluctuations), and the $\delta n_{\s, \w}$ describe fluctuations induced by the interactions. In the framework of linear response theory, there exist response functions $\chi_{\s, \l}$ such that: 
\begin{equation}
\begin{split}
\delta n_{\s, \l} (\r, t) = \int \d \r' & \d t' \, \chi_{\s, \l}(\r, \r', t-t') \cdots \\ & \cdots \int \d \r_0 V(\r' - \r_0) n_{\l, \s}^0 (\r_0, t'), 
\label{nsw_linear_response}
\end{split}
\end{equation}
where $V(\r) = e^2 / (4 \pi \epsilon_0 r)$ is the Coulomb potential. Eqs.~\eqref{nsw_dynamics}-\eqref{nsw_linear_response} effectively define generalized Langevin dynamics for the charge densities.

\begin{figure}
	\begin{center}
		\includegraphics[width=\columnwidth]{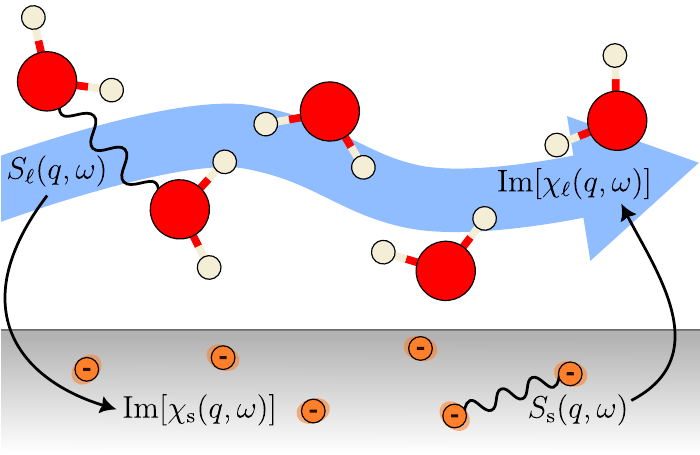}
		\caption{\textbf{Mechanism of fluctuation-induced friction.} Fluctuation-induced friction relies on a "call and response" mechanism: a spontaneous charge fluctuation in the liquid (governed by the structure factor $S_\l$) induces an excitation in the solid (determined by the response function $\chi_\s$), and vice versa. Energy dissipation occurs both in the liquid and the solid, and is related to the retardation in the response of each medium to the fluctuations of the other.}\label{fig fi friction}
	\end{center}
\end{figure}

The force exerted by the solid on the liquid is 
\begin{equation}
\langle \mathbf{F}(t) \rangle = -\int \d \r_\s \d \r_\l \langle n_\s (\r_\s, t) \nabla_\l V(\r_\l - \r_\s) n_\l(\r_\l, t) \rangle. 
\end{equation}
We assume that both media are locally neutral on average. The non-interacting densities being uncorrelated, $\langle n_\s^0(\r_\s, t) n_\l^0 (\r_\l, t) \rangle = 0$, and thus, to lowest order in the solid-liquid interaction, 
\begin{equation}
\begin{split}
&\langle \mathbf{F}(t) \rangle = -\int \d \r_\s \d \r_\l \langle \delta n_\s (\r_\s, t) \nabla_\l V(\r_\l - \r_\s) n_\l^0(\r_\l, t) \rangle \\
&-\int \d \r_\s \d \r_\l \langle  n_\s^0 (\r_\s, t) \nabla_\l V(\r_\l - \r_\s) \delta n_\l(\r_\l, t) \rangle. 
\end{split}
\label{Fq2terms}
\end{equation}
This expression contains a first intuition of the "call and response" mechanism behind FI friction. The first term corresponds to energy dissipation due to a spontaneous charge fluctuation in the liquid inducing an excitation in the solid. The second term corresponds to the symmetric process, where a spontaneous charge fluctuation in the solid induces an excitation in the liquid, as represented on figure \ref{fig fi friction}. Thus, energy is dissipated both in the liquid and in the solid. This is in contrast to RI friction, which corresponds to a static modulation of the solid inducing energy dissipation in the liquid only. 

To evaluate the correlators in Eq.~\eqref{Fq2terms}, one proceeds by using Eq.~\eqref{nsw_linear_response}: 
\begin{equation}
\begin{split}
\langle \mathbf{F}(t) \rangle = &-\int \d \r_\s \d \r_\s' \d \r_\l \d \r_\l' \d t' \chi_{\rm s}(\r_\s, \r_\s', t - t') \cdots \\ &\cdots \nabla_\l V(\r_\l - \r_\s) V(\r_\l' - \r_\s')  \langle  n_\l^0(\r_\l, t) n_\l^0 (\r_\l', t') \rangle \\
&-\int \d \r_\s \d \r_\s' \d \r_\l \d \r_\l' \d t' \chi_{\l}(\r_\l, \r_\l', t - t') \cdots \\ & \cdots \nabla_\l V(\r_\l - \r_\s) V(\r_\l' - \r_\s')  \langle  n_\s^0(\r_\s, t) n_\s^0 (\r_\s', t') \rangle.  \\
\end{split}
\label{Fqchi}
\end{equation}
At this point, we need to account explicitly for the presence of a flow field in the liquid. In a similar way to the case of RI friction, we assume that the flow velocity is uniform over the spatial scale of solid-liquid interactions, and that the spontaneous charge fluctuations in the flowing liquid are the same as in the equilibrium liquid, advected at velocity $\v$: $n^0_{\l}(\r, t) = n^0_{\l, \v = 0}(\r - \v t, t)$. Thus, in the first term of Eq.~\eqref{Fqchi}, we can replace the $n_\l^0$ factors accordingly. To determine the correct replacement in the second term, we may picture the system in the reference frame of the liquid's center of mass, in which the solid is moving with velocity $-\v$. We thus replace $n^0_{\s}(\r, t) \mapsto n^0_{\s, \v = 0}(\r + \v t, t)$. The remaining equilibrium correlators amount to structure factors:  
\begin{equation}
\begin{split}
&\langle \mathbf{F}(t) \rangle = -\int \d \r_\s \d \r_\s' \d \r_\l \d \r_\l' \d t' \chi_{\rm s}(\r_\s, \r_\s', t - t') \cdots \\ &\cdots \nabla_\l V(\r_\l - \r_\s) V(\r_\l' - \r_\s')  S_\l(\r_\l - \v t, \r_\l' - \v t', t - t') \\
&-\int \d \r_\s \d \r_\s' \d \r_\l \d \r_\l' \d t' \chi_{\l}(\r_\l, \r_\l', t - t') \cdots \\ & \cdots \nabla_\l V(\r_\l - \r_\s) V(\r_\l' - \r_\s')  S_\s (\r_\s + \v t, \r_\s' + \v t', t - t').  \\
\end{split}
\label{Fqchi_v0}
\end{equation}
Obtaining a closed-form expression for the fluctuation-induced friction coefficient is then a matter of carrying out Fourier transforms parallel to the interface and using the FDT relating structure factors to response functions: 
\begin{equation}
S_{\l,\s}(\r, \r', \omega) = - \frac{2 \kB T}{\omega} \im{\chi_{\l,\s}(\r, \r', \omega)}. 
\label{eq:FDT_FI}
\end{equation}
The calculation is detailed in Appendix B. The $\chi_{\s, \l}$ end up conveniently bundled as surface response functions, as introduced in Sec. III.C: 
\begin{equation}
g_\l(q, \omega) = -\frac{e^2}{2 \epsilon_0 q} \int_0^{+\infty} \d z \d z' e^{-q(z + z')} \chi_\l(q, z, z', \omega), \label{surface response function water}
\end{equation}
\begin{equation}
g_\s(q, \omega) = -\frac{e^2}{2 \epsilon_0 q} \int_{-\infty}^{0} \d z \d z' e^{q(z + z')} \chi_\s(q, z, z', \omega),\label{surface response function solid}
\end{equation}
The decomposition into surface response functions is possible if one can define a boundary separating the solid and the liquid: $n_\l$ vanishes for $z < 0$ and $n_\s$ vanishes for $z>0$. The definition of this boundary will be discussed in more detail in Sec. IV.D. Ultimately one obtains for the friction coefficient $\lambda_{\rm FI} = - \langle \mathbf{F} \cdot v \rangle /(\mathcal{A} v^2)$
\begin{equation}
\lambda_{\rm FI} = \frac{k_{\rm B} T}{2 \pi^2} \int_0^{+\infty} \d q \, q^3 \int_0^{+\infty} \frac{\d \omega}{\omega^2} \, \mathrm{Im} [g_{\l}(q, \omega)] \mathrm{Im} [g_{\s}(q, \omega)].  
\label{lambdaQ_cl}
\end{equation}
The FI friction coefficient takes the form of an integral over modes $(q, \omega)$. The factors $\mathrm{Im} [g_{\l,\s}(q,\omega)]$ represent the spectral density of charge fluctuations in the liquid and the solid: FI friction relies on a spectral matching condition between the two media's charge fluctuations. Due to the factor $1/\omega^2$, lower-frequency modes make a larger contribution: indeed, the lower the frequency of the mode, the larger its spontaneous fluctuations at a given temperature. 

To sharpen the qualitative mechanism behind FI friction, let us consider for simplicity that the liquid and the solid exhibit charge fluctuations at a single wavevector $\q_{\parallel}$. Let us also restrict the $z$ integrals to a single layer in either the liquid or the solid, with a distance $z_1$ between the layers: this is, in general, not a valid approximation in the case of long-range Coulomb interactions, but it does not affect the qualitative discussion. This yields an approximate expression for FI friction that is similar to Eq.~\eqref{eq:lambdaRI_approx} for RI friction: 
\begin{equation}
	\begin{split}
\lambda_{\rm FI} \approx \alpha \kB T &q_{\parallel}^2 |V_{q_{\parallel}}(z_1)|^2 \dots \\ \dots &\int_0^{+\infty} \frac{\d \omega}{\omega^2} \, \mathrm{Im} [\chi_{\l}(q_{\parallel}, \omega)] \mathrm{Im} [\chi_{\s}(q_{\parallel}, \omega)], 
	\end{split}
\label{eq:lambdaFI_approx} 
\end{equation}
with $\alpha$ a geometric factor of order 1. Let us now assume that the charge density in mode $q_{\parallel}$ follows damped harmonic oscillator dynamics for both the liquid and the solid. Then, 
\begin{equation}
\chi_{\l,\s}(q_{\parallel}, \omega) = \frac{1}{m_{\l,\s} \omega^2 - k_{\l,\s} + i \mu_{\l,\s} \omega},  
\end{equation} 
and we denote $\omega_{\l,\s} = \sqrt{k_{\l,\s}/m_{\l,\s}}$ the oscillator frequency. We further consider the overdamped limit $\tau_{\l, \s} = \mu_{\l,\s}/k_{\l,\s} \gg 1/\omega_{\l,\s}$. 
Then, the frequency integral in Eq.~\eqref{eq:lambdaFI_approx} can be computed analytically: 
\begin{equation}
\lambda_{\rm FI} \approx \frac{\pi \alpha}{2}  \kB T q_{\parallel}^2 |V_{q_{\parallel}}(z_1)|^2 \frac{1}{k_\s k_\l}\frac{\tau_\l \tau_\s}{\tau_\l + \tau_\s}.   
\end{equation}
If the solid is much slower than the liquid ($\tau_\s \gg \tau_\l$), then the power per unit surface dissipated by FI friction can be cast as 
\begin{equation}
\lambda_{\rm FI} v^2 
= \frac{\pi \alpha}{2}
\left[
q_{\parallel} v \times
\chi_{\l}(q_{\parallel})
V_{q_{\parallel}}(z_1)
\delta n_{\s}
\right]^2
\mu_{\l}. 
\label{eq:FI_power_liquid}
\end{equation}
Similarly, if the liquid is much slower than the solid ($\tau_\l \gg \tau_\s$), the dissipated power reads 
\begin{equation}
\lambda_{\rm FI} v^2 
= \frac{\pi \alpha}{2}
\left[
q_{\parallel} v \times
\chi_{\s}(q_{\parallel})
V_{q_{\parallel}}(z_1)
\delta n_{\l}
\right]^2
\mu_{\s}. 
\label{eq:FI_power_solid}
\end{equation}
Here $\chi_{\l,\s}(q_{\parallel})$ stand for the static susceptibilities $\chi_{\l,\s}(q_{\parallel}, \omega = 0) = -1/k_{\l,\s}$ and $\delta n_{\l,\s} = \sqrt{\kB T/ k_{\l,\s}}$ are the typical amplitudes of the spontaneous density fluctuations at temperature $T$. Thus, in the slow solid limit, FI friction looks like RI friction of the liquid on a static charge density modulation $\delta n_\s$ in the solid. The right hand side in Eq.~\eqref{eq:FI_power_liquid} represents the power dissipated by an oscillation in the liquid charge density driven with amplitude $\chi_{\l}(q_{\parallel}) V_{q_{\parallel}}(z_1) \delta n_{\s}$ and at frequency $q_{\parallel} v$. Conversely, in the slow liquid limit (Eq.~\eqref{eq:FI_power_solid}), FI friction looks like RI friction of the solid on a static charge density modulation $\delta n_\l$ in the liquid. But the similarity with RI friction is only formal: indeed, $\delta n_{\l,\s}$ are not actually static modulations, but rather typical amplitudes of thermal fluctuations, that may be occuring at frequencies that are much larger than $q_{\parallel} v$. In fact, $q_{\parallel} v$ is not the actual frequency at which the density perturbations are driven, but rather the Doppler frequency shift between modes $\q_{\parallel}$ and $-\q_{\parallel}$. It is their combined effect that amounts to an effective driving frequency $q_{\parallel} v$ in the particular case where the timescales of the liquid and the solid are well-separated. 
%As opposed to RI friction, that only involves dynamics of the liquid density at frequencies $\omega \sim q_{\parallel} v$, FI friction can involve dynamics at any frequency and wavevector. 
In the general case, a quantitative discussion of FI friction requires a detailed analysis of the charge fluctuation spectra of the two media, that we will carry out in Sec. IV.D. 
   
As opposed to RI friction, that only involves dynamics of the liquid density at frequencies $\omega \sim q_{\parallel} v$, FI friction can involve dynamics at any frequency. In the case of modes with frequencies $\omega \gtrsim k_{\rm B} T / \hbar \approx 6~\rm THz$, a classical description for the underlying dynamics is no longer accurate: indeed, the zero-point energy of thes modes is not negligible compared to temperature. Starting from the quantum analogue of the Langevin dynamics in Eq.~\eqref{nsw_dynamics} (see Sec. VI), one obtains a quantum friction coefficient 
\begin{equation}
\begin{split}
	\lambda_{\rm Q} = \frac{\hbar^2}{8 \pi^2 k_{\rm B} T} &\int_0^{+\infty} \d q \, q^3 \int_0^{+\infty} \frac{\d \omega}{\mathrm{sinh}^2[\hbar \omega / (2 k_{\rm B}T)]} \cdots \\ & \quad \cdots  \mathrm{Im} [g_\l(q, \omega)] \mathrm{Im} [g_\s(q, \omega)].  
\end{split}
\label{lambdaQ_q}
\end{equation}
In the limit $\hbar \to 0$, Eq.~\eqref{lambdaQ_q} reduces to Eq.~\eqref{lambdaQ_cl}. It is important to note, however, that while the classical limit can be a good approximation for this formula, the quantum nature of the electron dynamics is usually crucial for the solid's response function $g_\s$. We thus delay discussing in detail the quantum character of fluctuation-induced friction until Sec. \ref{how_quantum}. 

Finally, we note that the results in Eqs.~\eqref{lambdaQ_cl} and \eqref{lambdaQ_q} are obtained to lowest non-zero (second) order in the solid-liquid interaction. In the framework of the Langevin dynamics described by Eq.~\eqref{nsw_dynamics} (or its quantum analogue), interactions can in fact be taken into account to arbitrary order yielding the result
\begin{equation}\label{lambdaQ_complete}
\begin{split}
\lambda_{\rm Q} = \frac{\hbar^2}{8 \pi^2 k_{\rm B} T} &\int_0^{+\infty} \d q \, q^3 \int_0^{+\infty} \frac{\d \omega}{\mathrm{sinh}^2[\hbar \omega / (2 k_{\rm B}T)]} \cdots \\ & \cdots \frac{\mathrm{Im} [g_{\l}(q, \omega)] \mathrm{Im} [g_{\s}(q, \omega)]}{| 1 - g_\l(q,\omega)g_\s(q, \omega)|^2}. 
\end{split}
\end{equation}
This is the most general expression for the liquid-solid quantum friction coefficient. The denominator represents the contribution of multiple excitation processes, where, for example, a spontaneous fluctuation of the solid induces a perturbation in the liquid, that in turn perturbs the solid, and so on. These processes are often quantitatively important and need to be taken into account when evaluating the friction coefficient.

\begin{figure*}
	\begin{center}
		\includegraphics[width=\textwidth]{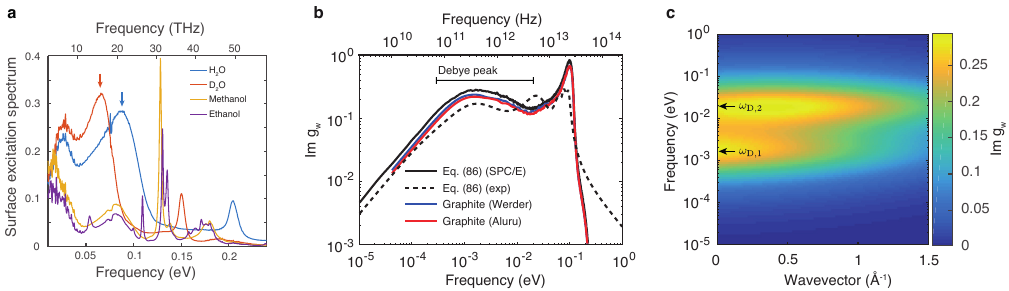}
		\caption{\textbf{Surface response functions of polar liquids.} (a) Surface excitation spectra at zero momentum, $\im{g_\w(0, \omega)}$, for water, heavy water, methanol and ethanol, as obtained with infrared spectroscopy. Reproduced from \cite{yuelectroncooling2023}. (b) Surface excitation spectrum of water on a graphene surface at zero momentum, as obtained from a compilation of experimental data and MD simulations with different interaction parameters. Adapted from \cite{kavokinefluctuationinducedquantum2022} (c) Model for the surface excitation spectrum of water on graphene, as defined by Eq.~\eqref{water spectrum}. Reproduced from \cite{kavokinefluctuationinducedquantum2022}.}\label{fig:spectrum liquid}
	\end{center}
\end{figure*}

\subsection{Quantitative estimates for model systems}

The result in Eq.~\eqref{lambdaQ_q} highlights that FI friction is determined by the matching of charge fluctuation spectra between the liquid and the solid. Significant friction is obtained when there is an overlap in the typical frequencies and momenta of the excitation spectra $\im{g_\l(q, \omega)}$ and $\im{g_\s(q, \omega)}$. Furthermore, due to the exponential behavior of the factor $\sinh^2[\hbar \omega /(2\kB T)]$ that appears in Eq.~\eqref{lambdaQ_q}, only the low energy modes, with frequencies smaller or on the order of the thermal frequency $\kB T/\hbar \simeq 6$ THz, can contribute significantly to fluctuation-induced friction, in contrast to the static van der Waals force discussed in section \ref{section Lifschitz theory}. We will now discuss the characteristic features of the surface excitation spectra for typical liquids and solids in the thermal frequency range, and obtain quantitative estimates of the FI contribution to friction. 

\subsubsection{Liquid excitations} 
If the liquid is polar -- that is, if the molecules have a net dipole moment -- then it exhibits significant charge fluctuations at thermal frequencies. In the long wavelength limit, the corresponding surface response function can be obtained from the liquid's bulk dielectric function: 
\begin{equation}
	g_\l(q\rightarrow 0, \omega) = \frac{\epsilon_\l(\omega)-1}{\epsilon_\l(\omega)+1}.
\end{equation}
The surface response function obeys a sum rule (Kramers-Kronig relation) $\int_0^{\infty} \d \omega \, \im{g_\l(q, \omega)}/\omega = (\pi/2)\, g_\l(q, 0)$. Together with the above equation, it implies that the larger the static dielectric constant of the liquid, the larger the spectral density $\im{g_\l}$.
Experimental surface excitation spectra at zero momentum, $\im{g_\l(0, \omega)}$, are shown for different polar liquids in Fig.~\ref{fig:spectrum liquid}.a. These spectra typically present a broad feature at GHz to THz frequencies called a Debye mode, which arises from the collective relaxation of molecular dipoles. In the case of water, there is also a feature around 20 THz (100 meV) -- the libration mode, which corresponds to hindered molecular rotations. Collectively, one may refer to a liquid's charge excitations as \emph{hydrons}, by analogy with phonons in solid-state systems \cite{yuelectroncooling2023,coquinotmomentumtunnelling2025}. The hydron modes behave essentially as broadened optical phonons, with weak dispersion in momentum space. Thus, as the simplest model for the momentum dependence of the surface response function, one may take
\begin{equation}
g_\l(q, \omega) = g_\l(0, \omega) e^{-2 q d},
\end{equation}
where $d$ is the distance (typically on the few-angstrom scale) between the interfacial molecular layer in the liquid and the onset of the solid's electronic density. If the solid's conduction electrons can be assigned to Wannier orbitals that are well-separated from the interfacial liquid (e.g., water on transition metal dichalcogenides), the distance $d$ can be defined unambiguously. If the electronic orbitals come into direct contact with the interfacial liquid, then MD simulations are required to model the momentum dependence more accurately. For water on graphene/graphite, \cite{kavokinefluctuationinducedquantum2022} proposed an analytical representation of the surface response function, based on a fit to force-field MD simulations of SPC/E water with the interaction parameters of \cite{wuGraphiticCarbonwater2013} (Fig.~\ref{fig:spectrum liquid}.b): 
\begin{equation}\label{water spectrum}
	g_\l(q, \omega) = \frac{f_1(q)}{1 - i\omega/\omega_{D,1}} + \frac{f_2(q)}{1 - i\omega/\omega_{D,2}},
\end{equation}
with
\begin{equation}
	\begin{split}
\quad f_1(q) &= \frac{g_\l(q,0)}{2}\, e^{-q/q_0}, \\
	f_2(q) &= \frac{g_\l(q,0)}{2}\, (2-e^{-q/q_0}), \\
	g_\l(q,0) &= \exp \left[ a+a' \{ 1+ (q/q_1)^\alpha\}^{1/\alpha} \right].
\end{split}
\end{equation}
This is the sum of two Debye functions, that provide a coarse description of the actual Debye and libration modes. The momentum dependence consists in spectral weight transfer between the modes, and a decay of the overall amplitude, which becomes exponential at high momenta -- accounting for the contact repulsion between the liquid and the solid (Fig.~\ref{fig:spectrum liquid}.c). The parameters are $\hbar\omega_{D,1} = 1.5$ meV, $\hbar\omega_{D,2} = 20$ meV,
$q_0 = 3.12 \, \mathring{\A}^{-1}$, $q_1 =1.79 \, \mathring{\A}^{-1}$, $a=3.38$, $a'=-3.41$, and $\alpha=2.4$. While developed for a semi-infinite slab of water, this model has been shown to describe the water response reasonably well down to $\sim 1~\rm nm$ confinement between two graphene sheets \cite{coquinotcollectivemodes}.

\subsubsection{Solid excitations, quantum friction estimates.} 

Charge dynamics in solids arise from either electronic excitations or optical phonons. However, optical phonons, which are associated with out-of-phase vibrations of the lattice ions within a unit cell, typically occur at higher-than-thermal frequencies and thus do not significantly contribute to FI friction \cite{kavokinefluctuationinducedquantum2022}. To have electronic excitations at thermal frequencies, a solid must be electrically conducting -- these are then excitations of the free electrons that contribute to conduction.

\begin{figure}
	\begin{center}
		\includegraphics[width=\columnwidth]{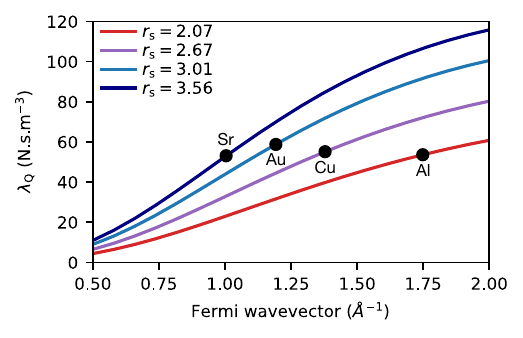}
		\caption{\textbf{Quantum friction on different metals.} Fluctuation-induced quantum friction is computed from Eq.~\eqref{lambdaQ_complete} for solids described by a jellium model (Eq.~\eqref{spectrum jellium}) and water (Eq.~\eqref{water spectrum}), as a function of the Fermi wavevector and for different values of the Wigner-Seitz radius (Eq.~\eqref{Wigner Seitz radius}). Values of quantum friction on aluminum, copper, gold and strontium are specified. }\label{fig lambda Q metals}
	\end{center}
\end{figure}

\begin{figure*}
	\begin{center}
		\includegraphics[width=\textwidth]{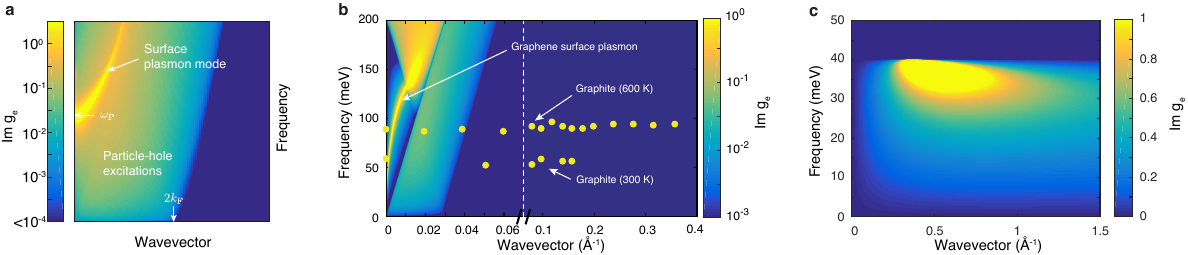}
		\caption{\textbf{Surface response function of electronic systems.} The surface excitation spectrum $\im{g_\s(q, \omega)}$ is represented for different solid models. (a) Semi-infinite jellium (radius parameter $r_s = 5$), reproduced from \cite{kavokinefluctuationinducedquantum2022}. (b) Doped graphene ($E_{\rm F} = 0.1$ eV) together with experimental points for graphite surface plasmon, as obtained in \cite{laitenbergerplasmon} at 600 K and \cite{portail1999dynamical} at 300 K, reproduced from \cite{kavokinefluctuationinducedquantum2022}. (c) Local susceptibility model of the graphite surface response function, as given by Eq.~\eqref{eq:local susceptibility model}, with the local susceptibility obtained from a 1D tight-binding description, with $n_\s = 2.3 \times 10^{12}$ cm$^{-2}$, reproduced from \cite{kavokinefluctuationinducedquantum2022}.}\label{fig:solid spectra}
	\end{center}
\end{figure*}

\paragraph{Jellium model.} For a conventional metal (such as gold, copper, etc.) the electronic excitations are described at the simplest level within the jellium model. In this model, the atomic nuclei and core electrons are reduced to a uniform positive background, in which the conduction electrons behave as free electrons, with parabolic dispersion $E(\q) = \hbar^2q^2/(2m^*)$. The model is parametrized by the effective band mass $m^*$ and the electron density $n_\e$, that fixes the Fermi level $E_{\rm F} = (\hbar^2/(2m^*)) \times (3 \pi^2 n_\e)^{2/3}$. Equivalently, the parameters can be specified as the Fermi wave-vector $\kF = (3 \pi^2 n_\e)^{1/3}$ and the Wigner-Seitz radius parameter 
\begin{equation}\label{Wigner Seitz radius}
	r_\s = \left( \frac{9 \pi}{4} \right)^{1/3}\frac{(m^*/m_\e)}{\kF a_0 },
\end{equation}
where $a_0$ is the Bohr radius and $m_\e$ the electron mass. The surface response of the semi-infinite jellium can be deduced from the charge susceptibility of the bulk jellium following the specular reflection procedure, as detailed in section \ref{section surface response fct}. A typical spectrum resulting from this procedure is shown in figure \ref{fig:solid spectra}.a. It decomposes into an incoherent particle-hole continuum and a collective surface plasmon mode. In conventional metals, the plasmon frequency is on the order of electronvolts, so that only particle-hole excitations may contribute to FI friction. In this case, the following phenomenological expression may be used for the surface excitation spectrum \cite{kavokinefluctuationinducedquantum2022}:
\begin{equation}\label{spectrum jellium}
\begin{split}
	g_\s(\q, \omega)&\simeq e^{-q/(C[r_\s]\kF)} \\ & \quad + i A[r_\s]  \frac{\omega}{E_{\rm F}} \frac{q}{\kF} e^{-q/(B[r_s]\kF)} \, \theta(2 \kF -q), 
\end{split}
\end{equation}
where the function $A[r_\s] \simeq 4.35 \times r_\s^{-0.9}$, and $B[r_\s]$ and $C[r_\s]$ are tabulated for a range of $r_\s$ values in \cite{kavokinefluctuationinducedquantum2022}. Fig. \ref{fig lambda Q metals} shows the QF coefficient of water on a metal as a function of $\kF$ and $r_\s$, as computed from Eq.~\eqref{lambdaQ_complete}, using the jellium spectrum in Eq.~\eqref{spectrum jellium} and the water spectrum in Eq.~\eqref{water spectrum}. For gold, $r_\s = 3.01$ and $\kF = 1.20 \, \mathring{\A}$ \cite{kittel2018introduction}, yielding $\lambda_{\rm Q} \simeq 60 ~ \rm N\cdot s \cdot m^{-3}$. For water on gold nanoparticles \cite{collismeasurementnavierslip2021} the total hydrodynamic friction was measured at $\lambda \sim 10^5 \, \rm N\cdot s \cdot m^{-3}$. While this measurement employed citrate-passivated nanoparticles, it is unlikely that a pristine gold surface would show lower friction due to its hydrophilic nature. Thus, QF is expected to be a rather small correction to hydrodynamic friction on metal surfaces. Nevertheless, the value $\lambda_{\rm Q} \simeq 60 ~ \rm N\cdot s \cdot m^{-3}$ is significant when it comes to hydro-electronic cross-coupling effects, as we will discuss in section \ref{section v}. 

\paragraph{2D materials.} For metallic 2D materials such as graphene, the problem of determining the surface response function $g_\s(q, \omega)$ reduces to computing the density response function $\chi_\s(q, \omega)$. Assume that the 2D material is in the plane $z = 0$ and is subject to an evanescent wave external potential $\phi_{\rm ext}(\r, \omega) = e^{i \q \r}e^{q z} \phi_{\rm ext}(\q, \omega)$. If no out-of-plane polarization is allowed, then the induced charge density in the 2D plane is 
\begin{equation}
\delta n(q, \omega) = \chi_\s(q, \omega) \phi_{\rm ext}(q, \omega), 
\end{equation}
and the induced potential above the 2D material is $\phi_{\rm ind}(q, z) = v_q e^{-q z} \delta n(q, \omega)$, with $v_q = e^2/(2 \epsilon_0 q)$, so that by definition, the surface response function is 
\begin{equation}
g_\s(q, \omega) = -v_q \chi_\s(q, \omega). 
\label{eq:g_2D}
\end{equation}
This expression strictly holds for wavevectors $q$ such that $1/q$ is much larger than the lattice spacing, and can be corrected with orbital form factors and local field terms for larger $q$ \cite{kavokinefluctuationinducedquantum2022, Tudorovskiy2010}. In the case of graphene, the density response function can be computed analytically at zero temperature, and at finite temperature by straightforward numerical integration \cite{hwangDielectricFunction2007,wunschDynamicalPolarization2006}. A typical excitation spectrum of doped graphene is shown in Fig.~\ref{fig:solid spectra}.b. Plugging the analytical $T = 0$ result into Eq.~\eqref{lambdaQ_complete} yields a numerical estimate for the QF coefficient of water on graphene: 
\begin{equation}
\lambda_{\rm Q} \approx 10^{-3} \times \left( \frac{n_\s}{n_0} \right)^{3/2} ~\rm N \cdot s \cdot m^{-3}, 
\label{eq:lambdaQ_graphene}
\end{equation}
where $n_\s$ is the graphene charge carrier density and $n_0 = 10^{12}~\rm cm^{-2}$. 
Thus, for a typical electronic density ($n_\s \sim 5 \times 10^{12} ~\rm N \cdot s \cdot m^{-3}$), 
$\lambda_{\rm Q} \sim 10^{-2} ~\rm N \cdot s \cdot m^{-3}$, 
which would correspond to a 10 cm slip length in the absence of other sources of friction. 
Thus, QF is predicted to be a small correction to the friction of water on graphene; nevertheless, as in the case of metals, it is significant from the point of view of hydro-electronic couplings (Sec. V).  

\paragraph{Drude plasmon model.} The jellium model does not provide an adequate description of the electronic excitations for metallic systems where the band structure cannot be reduced to a single parabolic band. In particular, this is the case for systems where the bandwidth is comparable to the plasmon frequency that would be expected within the jellium model: the finite bandwidth then results in a flattening of the plasmon dispersion \cite{stauberquasiflatplasmonic2016,lewandowskiintrinsicallyundamped2019,khalijiplasmonsscreening2020,dajornadauniversalslow2020}. The plasmon may then reach frequencies that are low enough to contribute to QF with water. This type of mechanism is likely involved in explaining the nearly flat dispersion of graphite's low-energy surface plasmon observed experimentally at around $\omega_\s \sim 50~\rm meV$ \cite{laitenbergerplasmon,portail1999dynamical}. The simplest phenomenological description one can adopt for such a mode is a Drude plasmon model \cite{pitarketheorysurface2007}: 
\begin{equation}
g_\s(q, \omega) = \frac{\omega_\s^2}{\omega_\s^2 - \omega^2 -  i \gamma \omega} \, \theta (q_{\rm max} - q), 
\label{eq:Drude_plasmon}
\end{equation}
which introduces, in addition to the surface plasmon frequency $\omega_\s$, the width $\gamma$ and a cutoff wavevetor $q_{\rm max}$, above which the plasmon decays or disperses to higher frequencies ($\theta$ stands for the Heaviside function). With $\gamma = 50~\rm meV$ and $q_{\rm max} = 0.4~ \mathring{\A}^{-1}$, Eq.~\eqref{lambdaQ_complete} predicts a QF coefficient $\lambda_{\rm Q} \approx 5 \times 10^3 ~\rm N \cdot s \cdot m^{-3}$ for water on graphite, that would amount amount to a slip length $b \sim 200~\rm nm$ in the absence of RI friction. Note that this prediction is very sensitive to the value of $q_{\rm max}$ (see below), and $q_{\rm max} = 0.4~ \mathring{\A}^{-1}$ represents the largest wavevector that could be accessed in electron energy loss experiments \cite{laitenbergerplasmon}. With $q_{\rm max} = 0.8~ \mathring{\A}^{-1}$, Eq.~\eqref{lambdaQ_complete} predicts $\lambda_{\rm Q} \approx 3 \times 10^4~\rm N \cdot s \cdot m^{-3}$, corresponding to a slip length $b \approx 30~\rm nm$ -- on the order of the exprimentally measured slip lengths for water on graphite and in large multiwall carbon nanotubes \cite{maalimeasurementsliplength2008a,secchimassiveradiusdependent2016}. This led \cite{kavokinefluctuationinducedquantum2022} to propose that QF may be a significant -- if not the dominant -- source of friction for water on graphite. 

\begin{figure}
	\begin{center}
		\includegraphics[width=\linewidth]{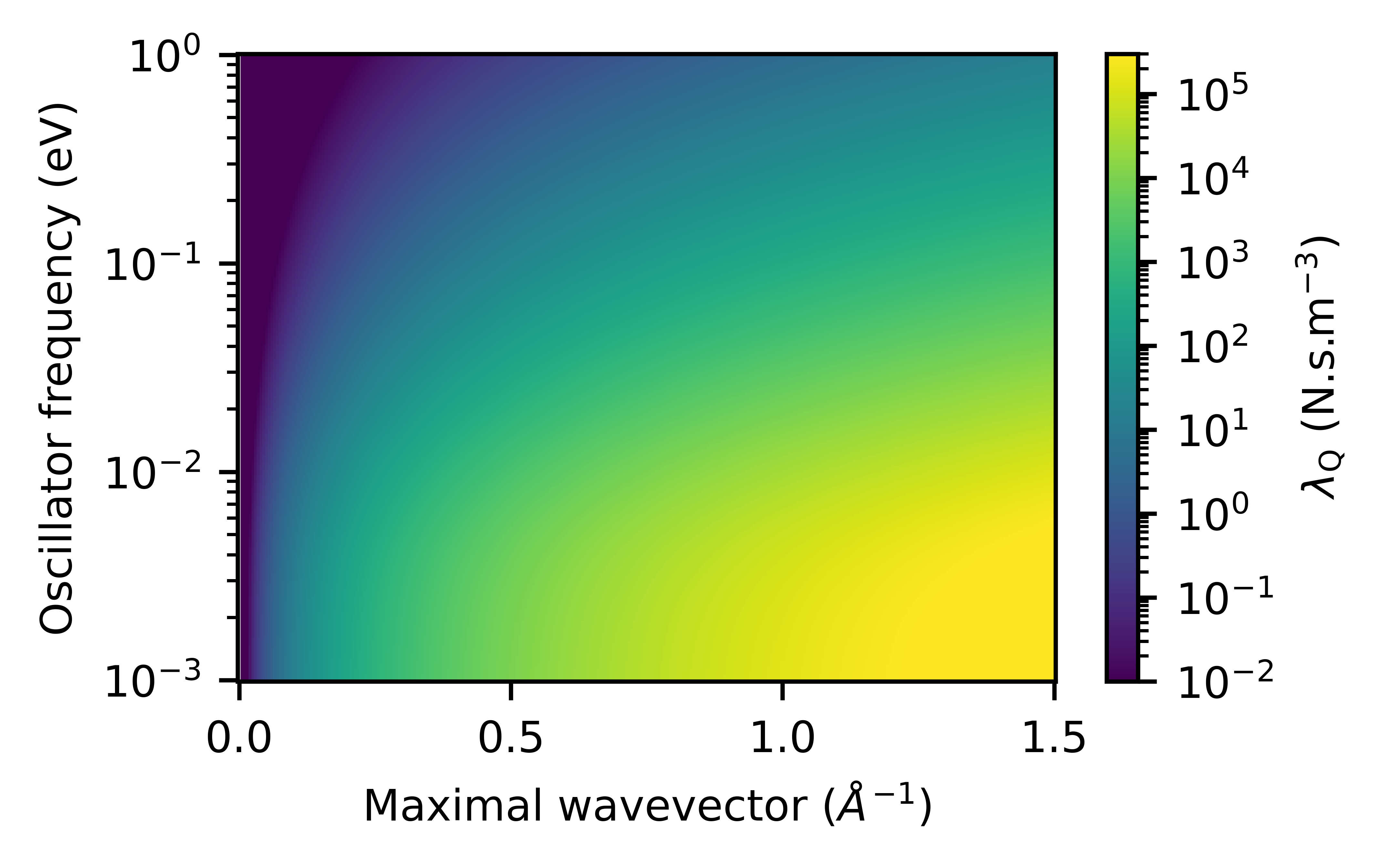}
		\caption{\textbf{"Rule of thumb" for fluctuation-induced friction.} Fluctuation-induced friction was computed from Eq.~\eqref{eq:rule of thumb} with the simplified water spectrum from Eq.~\eqref{eq:simplified water} and a single frequency solid as in Eq.~\eqref{simple solid spectrum}. The higher $q_{\rm max}$ and the lower $\omega_{\rm p}$, the higher the friction. }\label{fig rule of thumb}
	\end{center}
\end{figure}

For $\gamma \to 0$, Eq.~\eqref{eq:Drude_plasmon} reduces to 
\begin{equation}\label{simple solid spectrum}
	\im{g_\s(q, \omega)} = \frac{\pi}{2} \omega \, \delta(\omega \pm \omega_\s) \, \theta(q_{\rm max} - q).
\end{equation}
This expression is useful to derive a rule-of-thumb expression of the FI friction coefficient. If we further reduce water to a single Debye peak, 
\begin{equation}\label{eq:simplified water}
	\im{g_\l(q, \omega)} \simeq \eta_\l \frac{\omega \omega_0}{\omega^2 + \omega_0^2} e^{-2 q d},
\end{equation}
with $\eta_\l \approx 1.7$ and $\omega_0 \approx 5~\rm meV$, and use the first order expression for the FI friction coefficient in Eq. \eqref{lambdaQ_cl}, we obtain 
% \begin{equation}\label{eq:rule of thumb}
% \begin{split}
% 		\lambda_{\rm FI} &\simeq \frac{\kB T \eta_\l}{64 \pi d^4} \frac{\omega_0}{\omega_\s^2 + \omega_0^2} \int_0^{2dq_{\rm max}} \d u \, u^3 \,e^{-u} \\ & \simeq \frac{5 \cdot 10^3 \; \rm N \cdot s \cdot m^{-3}}{1 + (\omega_\s/\omega_0)^2},
% \end{split}  
% \end{equation}
\begin{equation}
\lambda_{\rm FI} \, [\mathrm{N \cdot s \cdot m^{-3}}] \simeq \frac{1.8 \times 10^5 \times \varphi(d q_{\max})}{d[\mathring{\A}]^4 (1 + (\omega_\s/\omega_0)^2) }, 
\label{eq:rule of thumb}
\end{equation}
where $\varphi(x) = (1/4)\int_0^{2x} \d u \, u^3 \,e^{-u}$. One has $\varphi(x) \approx x^4$ as $x \to 0$ and $\varphi(x) \to 3/2$ as $x \to \infty$. 
 This expression, which can be used for a first rough estimate on an unknown material, highlights that FI friction is not a resonance effect: rather, the lower the frequency of the solid's excitations the stronger the friction. The momentum also plays a key role: the higher the momentum $q$ at the frequency matching, the stronger its contribution to the friction force -- as illustrated in Fig.~\ref{fig rule of thumb}. Shorter wavelength fluctuations are more efficient at transferring momentum between the two media, as can be seen from the $q^3$ factor in Eq.~\eqref{lambdaQ_q}. This result is intuitive when FI (quantum) friction is interpreted in terms of the tunneling of elementary excitations between the liquid and the solid, as discussed in Sec. \ref{section quasiparticle picture}. Since each elementary excitation carries a quasi-momentum $\hbar \q$, more momentum is transferred per tunneling event when excitations are at higher $\q$. 

\paragraph{Drude oscillator model.} The Drude plasmon model is not to be confused with the Drude oscillator model -- a model originally developed to describe atomic polarizability in solids \cite{tholemolecularpolarizabilitiescalculated1981}, and has recently been extended to account for a plasmon-like excitation in the framework of MD simulations \cite{buiclassicalquantumfriction2023} -- see Sec. VII. The model considers explicitly the solid's crystal lattice and assigns to each lattice site an "electron" of mass $m$ and charge $-Q$, that is elastically bound to a fixed counter-charge $+Q$ with a spring of stiffness $k$. Thus, the model effectively assigns to each lattice site a polarizability $\alpha$, which relates the dipole moment of an oscillator $\mathbf{p}$ to the local electric field $\mathbf{E}$: $\mathbf{p} = \alpha \mathbf{E}$. In the frequency domain, 
\begin{equation}
\alpha(\omega) = \frac{Q^2}{m}\frac{\omega_0^2}{\omega_0^2 - \omega^2 - i \gamma \omega}. 
\end{equation} 
where $\omega_0 = \sqrt{k/m}$ and $\gamma$ is a damping rate. 
The simulations by \textcite{buiclassicalquantumfriction2023} implemented a two-dimensional lattice of Drude oscillators. 
In this setting, the surface response function is $q$-dependent to leading order in $q$. Let us first consider the case where the oscillators can only polarize in the plane of the 2D lattice. In the absence of interaction between oscillators, and for wavevectors $q$ such that $1/q$ is much larger than the lattice spacing, the density response function of the lattice is \cite{coquinotmomentumtunnelling2025}
\begin{equation}
\chi^0_\s(q, \omega) = n_\s q^2 \alpha(\omega), 
\end{equation}
where $n_\s$ is the surface density of oscillators. Coulomb interactions are taken into account by self-consistently enforcing that the oscillators respond to the total local potential, that comprises both the external potential $\phi_{\rm ext}$ and the induced potential, via their non-interacting response function: 
\begin{equation}
\delta n_{\rm ind} = \chi^0_\s  (\phi_{\rm ext} + v_q \delta n_{\rm ind}). 
\label{eq:sc_RPA}
\end{equation}
The interacting response function $\chi_\s$ satisfies $\delta n_{\rm ind} = \chi_\s \phi_{\rm ext}$, so that, explicitly 
\begin{equation}
\chi_\s(q, \omega) = \frac{\chi^0_\s (q, \omega)}{1 - v_q \chi^0_\s(q,\omega)}. 
\end{equation}
For a generic electronic system, Eq.~\eqref{eq:sc_RPA} defines the self-consistent Hartree ot random phase approximation (RPA) -- which is exact for a harmonic system (see Sec. VI). 
The corresponding surface response function reads, according to the prescription in Eq.~\eqref{eq:g_2D},  
\begin{equation}
	\begin{split}
g_\s(q, \omega) &= -v_q \chi_\s(q, \omega)  \\
&  = \frac{\omega_{\rm p}^2(q)}{\omega_0^2 + \omega_{\rm p}^2(q) - \omega^2 - i\gamma\omega}, 
	\end{split}
\end{equation} 
with $\omega_{\rm p}(q)^2 = q n_\s Q^2 / (2 m \epsilon_0)$. If the oscillators can also polarize perpendicular to the lattice, the surface response function acquires a second mode, with opposite dispersion: 
\begin{equation}
	\begin{split}
g_\s(q, \omega) = &\frac{\omega_{\rm p}^2(q)}{\omega_0^2 + \omega_{\rm p}^2(q) - \omega^2 - i\gamma\omega} \\
+ &\frac{\omega_{\rm p}^2(q)}{\omega_0^2 - \omega_{\rm p}^2(q) - \omega^2 - i\gamma\omega}.
	\end{split}
\end{equation}
In practice, the parallel and perpendicular modes may show an additional splitting due to the lattice geometry \cite{buiclassicalquantumfriction2023}. If one wishes to describe a weakly dispersing mode, the parameters of the oscillators need to be chosen so that $\omega_{\rm p}(q) \ll \omega_0$ for relevant $q$ values. Then, 
\begin{equation}
g_\s(q, \omega) \approx \frac{2 \omega_{\rm p}(q)^2}{\omega_0^2} \frac{\omega_0^2}{\omega_0^2 - \omega^2 - i \gamma \omega}. 
\end{equation}
Thus, in contrast to the Drude plasmon model in Eq.~\eqref{eq:Drude_plasmon}, where the oscillator strength is 1 is order to satisfy the f-sum rule for the electron gas \cite{pitarketheorysurface2007}, the 2D Drude oscillator model rather describes an optical-phonon-like excitation, with oscillator strength $2 \omega_{\rm p}(q)^2 /\omega_0^2 < 1$. 
%At a given $q$, the 2D Drude oscillator model underestimates the amplitude of the surface response function compared to the Drude plasmon model by a factor $2 \omega_{\rm p}(q)^2 / \omega_0^2$. 

For the sake of completeness, we now briefly discuss the response properties of a 3D lattice of Drude oscillators. In this case, the Clausius-Mossotti relation \cite{rysselbergheRemarksConcerning1932} gives the dielectric function in the long-wavelength limit in terms of a single oscillator's polarizability: 
\begin{equation}
\frac{\epsilon(\omega) - 1}{\epsilon(\omega) + 2} = \frac{n_v \alpha(\omega)}{3 \epsilon_0},  
\end{equation}
where $n_v$ is the volume number density of oscillators. This yields explicitly for the dielectric function 
\begin{equation}
\epsilon(\omega) = 1 + \frac{\omega_{\rm p}^2}{\tilde \omega_0^2 - \omega^2 - i \gamma \omega}, 
\end{equation}
with $\omega_{\rm p}^2 = n Q^2 / (m \epsilon_0)$ and $\tilde \omega_0^2 = \omega_0^2 - \omega_{\rm p}^2/3$. The resonance in the dielectric function is shifted from the bare frequency $\omega_0$ due to Coulomb interactions between the oscillators. The corresponding long-wavelength expression for the surface response function of a semi-infinite lattice of Drude oscillators is 
\begin{equation}
g_\s(\omega) = \frac{\omega_{\rm p}^2/2}{\tilde{\omega}_0^2 + \omega_{\rm p}^2/2 - \omega^2 - i\gamma\omega}, 
\end{equation}
which reduces to the Drude plasmon model in the limit $\tilde \omega_0 = 0$, with the surface plasmon frequency $\omega_\s = \omega_{\rm p} / \sqrt{2}$. However, this limit can hardly be implemented in practice (say, in the framework of MD simulations) because it places the system right at a ferroelectric instability. As in the 2D case, the 3D Drude oscillator model rather describes an optical-phonon-like excitation, with finite oscillator strength -- but that remains momentum-independent to leading order in $q$. Indeed, the surface response function can be recast as 
\begin{equation}
g_\s(\omega) = \frac{\omega_{\rm p}^2}{2\omega_0^2 +  \tilde \omega_{\rm p}^2/3} \frac{\tilde \omega_\s^2}{\tilde \omega_\s^2  - \omega^2 - i\gamma\omega}, 
\end{equation}
with $\tilde \omega_\s^2 =  \omega_0^2 + \omega_{\rm p}^2/6$. The oscillator strength is $\omega_{\rm p}^2/(2\omega_0^2 +  \tilde \omega_{\rm p}^2/3) < 1$. 
%Thus, if the model is used to describe a surface plasmon mode, then the response function is underestimated by a factor $\omega_{\rm p}^2/(2\omega_0^2 + \omega_{\rm p}^2/3)$ in the long-wavelength limit. 

\begin{figure}
	\begin{center}
		\includegraphics[width=\columnwidth]{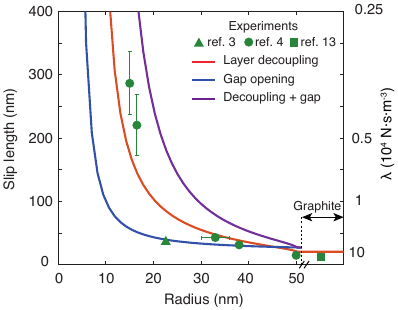}
		\caption{\textbf{Quantum friction of water in CNTs.} Water slip length ($b = \eta/\lambda$, with $\eta$ the water viscosity) in a multi-wall CNT as a function of inner tube radius. The green symbols are experimental data from \cite{secchimassiveradiusdependent2016} and the full lines are theoretical predictions, corresponding to different models for the radius dependence and the carrier density. The error bars correspond to the experimental uncertainty, which has been estimated in \cite{secchimassiveradiusdependent2016}. Reproduced from \cite{kavokinefluctuationinducedquantum2022}. }\label{fig graphene graphite}
	\end{center}
\end{figure}

\paragraph{Local susceptibility model.} At its root, the Drude oscillator model describes a material through a local relation between polarization and electric field, which is the hallmark of a dielectric. For a metal-like system, it is more natural to assume a local relation between induced charge and applied potential -- that is, a local non-interacting charge susceptibility. This assumption leads, for instance, to Thomas-Fermi screening in the static limit. In the case of a 2D system, the corresponding expression for the surface response function is straightforward. Consider indeed a 2D lattice where the induced charge density on a site is related to the applied potential according to $\delta n (\omega) = \chi(\omega) \phi_{\rm ext}(\omega)$. Then, in the absence of interaction between the sites, the surface response function is 
\begin{equation}
g_\s^0(q, \omega) = -n_\s v_q \chi (\omega),  
\end{equation}
where $n_\s$ is the site density. Accounting for the mutual interactions at the self-consistent Hartree level as in Eq.~\eqref{eq:sc_RPA}, the full surface response function becomes
\begin{equation}\label{eq:local susceptibility model}
g_\s(q, \omega) = \frac{n_\s v_q \chi(\omega)}{n_\s v_q \chi(\omega) - 1}. 
\end{equation}
This type of model was used by Kavokine \emph{et al.} to describe the radius dependence of the low-energy surface plasmon in multiwall CNTs \cite{kavokinefluctuationinducedquantum2022}. The local susceptibility $\chi(\omega)$ was obtained from a 1D tight-binding description of electron dynamics perpendicular to the tube layers, and $n_\s$ was used as a parameter to quantify the number of electrons that participate in the collective mode. A typical spectrum associated with this independent chain model is represented in Fig.~\ref{fig:solid spectra}c. $n_\s$ is expected to decrease with decreasing tube radius. Indeed, in larger tubes, the layers are well aligned, so the electronic response is expected to be similar to that of graphite. In smaller tubes, the layers misalign, eventually leading to a graphene-like electronic response that contains no low-energy plasmon. With reasonable assumptions on the radius dependence of $n_\s$, Kavokine \emph{et al.} could reproduce quantitatively the experimentally measured slip lengths for water in CNTs (Fig.~\ref{fig graphene graphite}). At present, graphite and multiwalled carbon nanotubes, due to their atomically smooth surfaces and distinctive electronic excitations, appear to be the most promising material systems for exhibiting high quantum friction.

\subsection{How quantum is quantum friction?} \label{how_quantum}

We now discuss in detail the quantum character of fluctuation-induced friction at liquid-solid interfaces. FI friction in itself is not a quantum effect: friction can arise from purely classical thermal fluctuations, as well as from quantum fluctuations. The question is then whether the specific fluctuations that are relevant at liquid-solid interfaces are amenable to a classical description. 

\begin{figure}
	\begin{center}
		\includegraphics[width=\columnwidth]{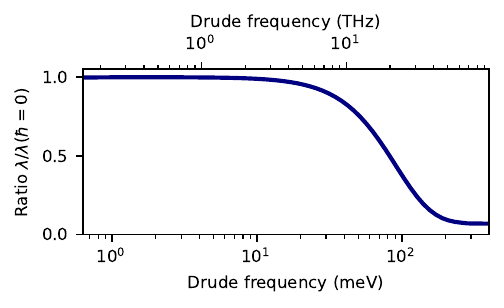}
		\caption{\textbf{Importance of explicit quantum corrections to fluctuation-induced friction.} We computed the fluctuation-induced friction in the classical limit ($\lambda(\hbar=0)$, according to Eq.~\eqref{lambdaQ_cl}) and with quantum corrections ($\lambda$, Eq.~\eqref{lambdaQ_q}), with the water spectrum given by \eqref{water spectrum}, and a solid modeled by a narrow Drude plasmon as in \eqref{eq:Drude_plasmon}, with $q_{\rm max} = 0.4$ Angstrom$^{-1}$ and $\gamma = 1$ meV. The ratio $\lambda / \lambda(\hbar = 0)$. Quantum corrections are important when the friction involves fluctuations above thermal frequencies that have non-negligible zero-point energy.} \label{fig comparison classical quantum}
	\end{center}
\end{figure}

If the fluctuations are described by a set of harmonic modes, then the classical description breaks down for modes with frequencies $\hbar \omega \gtrsim \kB T$. This is embodied by the factor $\sinh^2[\hbar \omega/(2 \kB T) ]$ in Eq.~\eqref{lambdaQ_q} for the FI friction coefficient, which describes the quantum correction to the thermal population of high-frequency modes. Figure \ref{fig comparison classical quantum} shows the ratio between the friction coefficient $\lambda$ including quantum corrections, as computed from Eq.~\eqref{lambdaQ_q}, and its classical counterpart $\lambda(\hbar=0)$, as computed from Eq.~\eqref{lambdaQ_cl}. The liquid is modeled by the surface response function in Eq.~\eqref{water spectrum}, and the solid is described with a Drude plasmon at frequency $\omega_0$, as given by Eq.~\eqref{eq:Drude_plasmon}. The classical and quantum expressions indeed agree when $\omega_0 \lesssim \kB T$, while for $\omega_0 \gtrsim \kB T$ the classical expression significantly overestimates the friction.

The situation becomes more complex if the solid's charge fluctuations arise from free electron dynamics. Then, even at lower-than-thermal frequencies, these fluctuations may require a quantum description. As an example, consider the FI friction of water with a 2D electron gas (2DEG).  For this system, within a mean-field (RPA) treatment of electron-electron interactions, the surface response function can be written as 
\begin{equation}\label{interacting g}
	g_\s(q, \omega, T) = \frac{-v_q\chi_0(q, \omega, T)}{1- v_q \chi_0(q, \omega, T)},
\end{equation}
where $\chi_0(q, \omega, T)$ denotes the non-interacting density response function of the 2DEG. The latter can be computed with varying degrees of accuracy to assess the importance of quantum effects. In the fully quantum treatment, one starts from the non-interacting response function at zero temperature (the Lindhard function) \cite{mihaila2011lindhard}:  
\begin{equation}
	\begin{split}
		\re{\chi_0(q, \omega)}&= - \frac{m_\e}{\pi \hbar^2} \Bigg\{ 1 + \frac{\kF}{q} \Big[ \mathrm{sgn}(\nu_- ) \theta(\nu_-^2-1) \sqrt{\nu_-^2 - 1}  \\ & \qquad \qquad \quad \cdots - \mathrm{sgn}(\nu_+) \sqrt{\nu_+^2 -1} \Big] \Bigg\}, 
	\end{split}
\end{equation}
\begin{equation}
	\begin{split}
		\im{\chi_0(q, \omega)}&= - \frac{m_\e}{\pi \hbar^2} \frac{\kF}{q} \Big[ \theta(1 - \nu_-^2) \sqrt{1 - \nu_-^2}  \\ &   \qquad \dots - \theta( 1 - \nu_+^2) \sqrt{1 - \nu_+^2} \Big],
	\end{split}
\end{equation}
with $\nu_\pm = \frac{m_\e \omega}{\hbar \kF  q } \pm \frac{q}{2 \kF}$, $\kF$ being the Fermi wavevector and $m_\e$ the electron mass. It is then extended to finite temperature according to 
\cite{maldagueManybodyCorrections1978}: 
\begin{equation}\label{finite T Lindhard}
	\chi_0^{\rm Q}(q, \omega, T) = \int_0^\infty \dd E \, \frac{\big[ \chi_0(q, \omega) ]_{T=0, E_{\rm F} = E}}{4 \kB T \, \cosh^2[ (E-\mu(T))/(2\kB T)]}. 
\end{equation}

In the classical limit, the 2DEG becomes a collisionless plasma, whose density dynamics are described by the Vlasov equation
\cite{vlasov1968vibrational}:
\begin{equation}\label{eq:Vlasov}
	\partial_t \delta n_\p + \frac{\p}{m} \cdot \nabla_\r \delta n_\p - (\nabla_\r V)\cdot (\nabla_\p n_\p^0) = 0,
\end{equation}
where $\delta n_\p(\r,t) = n_\p(\r, t) - n_\p^0$ is the deviation of the density of electrons (of a given spin) with momentum $\p$ with respect to the equilibrium density $n_\p^0$, and $V(\r,t)$ an external potential. The equilibrium distribution is the Boltzmann distribution: $n_{\p}^0 = e^{-\beta( \p^2 / (2m) - \mu)}$, with the chemical potential $\mu$ fixed so that the total electron density is $n_\e$. The density response function of the collisionless plasma is 
\begin{equation}\label{eq:classical 2DEG}
	\chi_0^{\rm cl}(q, \omega, T) = - \frac{n_\e}{\kB T} \Bigg[ 1 + \Big( \frac{\omega}{\sqrt{2} q v_T} + i 0^+ \Big) \, Z\Big( \frac{\omega}{\sqrt{2} q v_T} + i 0^+ \Big) \Bigg],
\end{equation}
where $v_T = \sqrt{\kB T/m_\e}$ and
\begin{equation}Z(\zeta) = \frac{1}{\sqrt{\pi}} \int_{-\infty}^{+ \infty} \dd x \, \frac{e^{-x^2}}{x- \zeta} \end{equation}
is the plasma dispersion function.

A semiclassical description of the 2DEG density response is obtained by replacing the Boltzmann distribution by the Fermi-Dirac distribution: $n_{\p}^0 = 1 / (e^{\beta( \p^2 / (2m) - \mu)} + 1)$. The Vlasov equation then becomes the collisionless Boltzmann equation, which yields for the density response function, in the low-temperature limit, 
\begin{equation}\label{eq:semi-classical 2DEG}
	\chi_0^{\rm semi-cl}(q, \omega, T) = \frac{m_\e}{2 \pi \hbar^2} \left[ - 1 + \frac{\omega+i 0^+}{\sqrt{ (\omega+i0^+)^2 - (q v_{\rm F})^2}} \right].
\end{equation} 

\begin{figure}
	\begin{center}
		\includegraphics[width=\columnwidth]{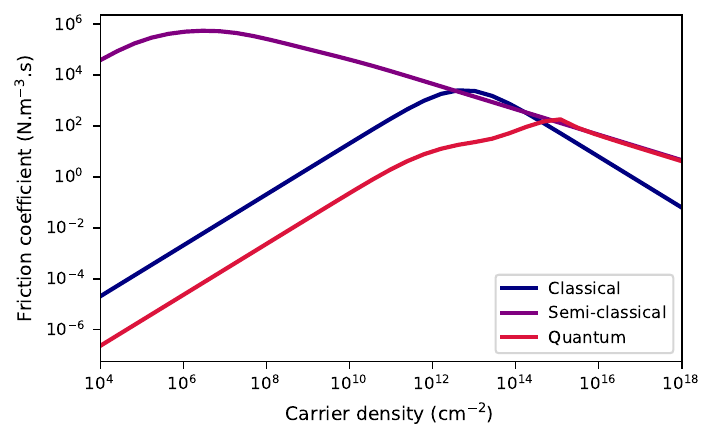}
		\caption{\textbf{Quantum effects in the fluctuation-induced friction of water on a 2D electron gas.} Fluctuation-induced friction is computed from Eq.~\eqref{lambdaQ_complete} for water on a 2D electron gas, with its surface response described at either the quantum (Eq.~\eqref{finite T Lindhard}), semi-classical (Eq.~\eqref{eq:semi-classical 2DEG}) or classical (Eq.~\eqref{eq:classical 2DEG}) level.} \label{fig quantum v classical 2DEG}
	\end{center}
\end{figure}

The water-2DEG FI friction coefficient, computed according to Eq.~\eqref{lambdaQ_complete}, is plotted in Fig. \ref{fig quantum v classical 2DEG} as a function of the 2DEG's carrier density $n_\e$, related to its Fermi wavevector through $\kF = \sqrt{2 \pi n_\e}$, within each of the descriptions of the 2DEG's electronic response (Lindhard, Boltzmann, Vlasov). One notices that the classical and quantum results differ significantly, regardless of the $n_\e$ value. This is indeed expected, since the Boltzmann and Fermi distributions are very different in the considered electron density range. However, the difference in the equilibrium distribution functions is not the sole origin of the discrepancy. Indeed, the semiclassical description, which accounts for the Fermi distribution, agrees with the full quantum result only for the largest densities. Compared to the exact Lindhard function, the Boltzmann description misses quantum interference effects in the scattering of electrons on the external potential \cite{friedelXIVDistribution1952}: these are therefore important in the water-2DEG FI friction. Indeed, water exhibits thermal density fluctuations down to molecular length scales, which perturb the 2DEG at wavevectors up to $q_{\rm max} \sim 1 ~\rm A^{-1}$. Unless the 2DEG has very high density, $q_{\rm max} \gtrsim \kF$: the interaction with water scatters electrons across the Fermi surface. Quantum interference effects are important in such scattering events, even if electrons transition between states that are very close in energy. 

We conclude that FI liquid-solid friction has a significant quantum character if it involves: i) fluctuations at higher-than-thermal energies, and ii) electronic fluctuations at around the Fermi wavelength. Since the first case is relatively rare (and corresponds to situations where FI friction is a small effect), we reserve the term \emph{quantum friction} for referring to FI friction between a liquid and an electronic system. 

As a last remark, we note that the FI friction between a liquid and a solid is intrinsically "more quantum" than the non-contact van der Waals friction between two solids, discussed in Sec. III. Indeed, in the solid-solid case, the largest excitation wavevectors that can contribute to friction are $q \sim 1/d$, where $d$ is the size of the vacuum gap between the solids. For any appreciable $d$, this condition imposes $q \ll \kF$, and the semi-classical description is then valid for the electronic excitations. In the liquid-solid case, there is essentially no vacuum gap: there is then a significant contribution from excitations at $q \sim \kF$, which require a fully quantum description.

\section{The hydro-electronic transport matrix}\label{section v}

\begin{figure}
	\begin{center}
		\includegraphics[width=0.9\columnwidth]{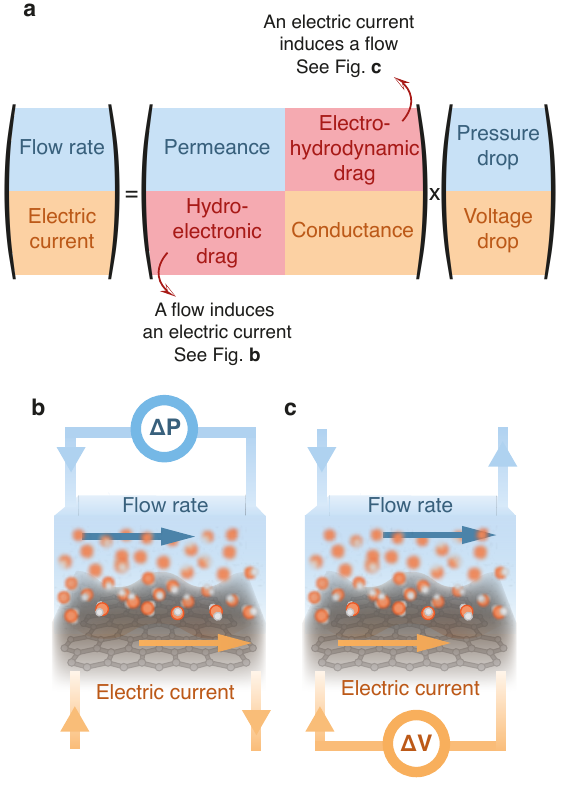}
		\caption{\textbf{Transport matrix and hydro-electronic cross-couplings.} (a) Sketch of the transport matrix, with hydro-electronic cross-couplings appearing in red. (b) Sketch of the hydro-electronic drag effect. (c) Sketch of the electro-hydrodynamic drag effect. Reproduced from \cite{coquinothydroelectricenergy2024}.} \label{fig2prx}
	\end{center}
\end{figure}

As discussed in the previous section, the coulombic coupling between a liquid's charge fluctuations and electronic excitations in the confining wall leads to an additional contribution to liquid-solid friction, termed quantum friction (QF). 
QF effectively couples the liquid's mass flow to the electronic current inside the wall.
Formally, and under the assumption of small driving forces, the ensuing physics can be captured in the form of a 
transport matrix $\mathbf{L}$ that couples the fluid and the electronic degrees of freedom across the liquid-solid interface:
\begin{equation}\label{transport matrix eq 0}
	\begin{pmatrix}
		Q \\ I 
	\end{pmatrix} = \mathbf{L} \begin{pmatrix}
		\Delta P \\ \Delta V
	\end{pmatrix} \quad \mathrm{with} \quad \mathbf{L} = \begin{pmatrix}
		\mathcal{L} & \mu_{\h \e} \\ \mu_{\h \e} & G_\e 
	\end{pmatrix}.
\end{equation} 
Here the driving forces are $\Delta P$, the pressure drop applied to the fluid and $\Delta V$ the voltage drop on the solid, while the fluxes are the flow rate $Q$ and electronic current $I$.
The coefficients of the matrix, which is symmetric due to the Onsager reciprocity principle, are the hydrodynamic permeance $\mathcal{L}$, the electronic conductance $G_\e$, and the hydro-electronic mobility $\mu_{\h \e}$. The matrix $\mathbf{L}$, depicted schematically in Fig.~\ref{fig2prx}.a, formalizes the specific transport properties of the liquid-electron interface, that arise as a consequence of QF. In this Section, we discuss these transport properties within a simplified model where explicit expressions can be provided for the coefficients of the transport matrix. In particular, we provide quantitative estimates for the coupled liquid-electron transport phenomena: the induction of an electronic current by a liquid flow (hydro-electronic drag, Fig.~\ref{fig2prx}.b) and the induction of a liquid flow by an electronic current (electro-hydrodynamic drag, Fig.~\ref{fig2prx}.c).

\subsection{Hydro-electronic cross-couplings}

In Sec. IV, we identified two hydrodynamic friction mechanisms. Roughness-induced (RI) friction transfers momentum from the moving fluid to the solid's internal vibrational excitations (phonons), while fluctuation-induced quantum friction (QF) transfers momentum directly to the solid's electrons. Since the electrons cannot relax this momentum instantaneously, they will accumulate some momentum and therefore carry a net current under the influence of the liquid flow. This effect has been called \emph{hydro-electronic drag} \cite{coquinothydroelectricenergy2024}, or simply Coulomb drag when there is no ambiguity with other mechanisms. Hydro-electronic drag is indeed the liquid-solid analogue of the Coulomb drag effect in condensed matter systems \cite{narozhnyCoulombDrag2016}.

Hydro-electronic drag is indeed not the only mechanism for flow-induced electronic current generation. In particular, if the liquid contains ions, and the solid bears a surface charge $\Sigma$, then a flow with velocity $\v_\l$ will drag the counter-ions and induce an ionic streaming current $\mathbf{j}_{\rm sc} = - \Sigma \v_\l$. If the solid is metallic, the moving ions may in turn drag the free electrons through electronic friction, leading to an electronic current in the solid.  
Moreover, the possibility of a phonon-mediated electronic current generation has been investigated \cite{kralnanotubeelectron2001,coquinotquantumfeedback2023}. There, the liquid flow drives a phonon wind as a result of RI friction, which then induces an electronic current through electron-phonon coupling. In this Section, for simplicity, we choose not to consider all these effects and focus only on the direct coupling between the liquid and electronic charge fluctuations. 

Hydro-electronic drag cannot be rigorously described within the Langevin equation formalism outlined in previous section, and requires in principle the quantum field theory framework of Sec. \ref{section quantum theory}, which gives access to the non-equilibrium state of the solid under liquid flow. In the following, we nonetheless propose an approximate treatment of the effect based on momentum conservation arguments. 
As a starting point, we consider a semi-classical Drude model for electronic conduction in the solid. Within this model, the solid's electrons are independent and each follow a Newton's equation of motion:
\begin{equation}
	m_\e \frac{\d \v_\e}{\d t} = - \frac{m_\e}{\tau_\e^0} \v_\e + \mathbf{F}_{\rm ext},
\end{equation}
where $m_\e$ is the mass of an electron and $\F_{\rm ext}$ is an external force. All the momentum relaxation processes (electron scattering on phonons, impurities, other electrons) are described with a single relaxation time $\tau_\e^0$: in particular, we neglect the momentum dependence of the various scattering processes. 

Let us denote by $n_\e$ the free electron density inside the solid, and $\delta$ its thickness. Then, the momentum flux going out of the electronic system per unit area is $\lambda_\e^0 v_\e$, with $\lambda_\e^0 = n_\e m_\e \delta / \tau_\e^0$. Note that the friction coefficient $\lambda_\e^0$ can still be defined for a 2D solid such as graphene: then $n_\e \delta$ becomes the surface electron density. If the solid is in contact with a flowing liquid with imposed interfacial velocity $v_\l$, then the momentum flux going into the electronic system as a result of QF is $\lambda_{\rm Q} (v_\l - v_\e)$. Indeed, the derivation in Sec. IV can be generalized to the situation where an electric current is established inside the solid by going to the reference frame where the electrons have no drift velocity: this immediately yields a QF force proportional to $v_\l - v_\e$. In the steady state, and in the absence of applied voltage in the solid, the incoming and outgoing momentum fluxes must be equal, so that $v_\e$ and $v_\l$ must be related according to 
\begin{equation}
v_\e = \frac{\lambda_{\rm Q}}{\lambda_\e^0 + \lambda_{\rm Q}} v_\l. 
\end{equation}
This relation provides a first estimate of hydro-electronic (HE) drag in terms of the liquid's slip velocity. Since quantitative estimates for $\lambda_{\rm Q}$ have been discussed in detail in Sec. IV.D., it is now useful to discuss the order of magnitude of $\lambda_\e^0$. For a thin layer of metal such as gold, $\lambda_\e^0$ can be deduced from the conductivity $\sigma = n_\e e^2 \tau_\e^0/m_\e \approx 4 \times 10^7 ~\rm S/m$, with $n_\e \approx 6 \times 10^{28} ~\rm m^{-3}$ and $m_\e$ close to the bare electron mass. This yields 
\begin{equation}
\lambda_\e^0 \approx 2 \times \delta (\mathrm{nm}) \times 10^3 ~\rm N \cdot s \cdot m^{-3}
\end{equation}
Given $\lambda_{\rm Q} \approx 60~\rm N \cdot s \cdot m^{-3}$ for the water-gold system, we estimate that the electronic drift velocity due to HE drag should be on the order of one percent of the slip velocity at the interface between water and a few-nm-thick gold layer. Another system of interest is graphene, where $\lambda_\e^0$ is best expressed in terms of the electron mobility $\mu_\e$: $\lambda_\e^0 = e n_\s / \mu_\e$, where $n_\s$ is the electron density per unit surface. Taking $n_\s = 5 \times 10^{12}~\rm cm^{-2}$, this yields 
\begin{equation}
\lambda_\e^0 = \frac{80}{\mu_\e [\rm cm^2 / (V \, s)]} ~ \rm N \cdot s \cdot m^{-3}. 
\label{eq:lambdae_graphene}
\end{equation}
With typical mobilities in high-quality graphene exceeding $10^4 ~\rm cm^2 / (V \, s)$, we estimate that $\lambda_\e^0 \ll \lambda_{\rm Q}$ despite the small water-graphene quantum friction $\lambda_{\rm Q} \approx 10^{-2} ~\rm N \cdot s \cdot m^{-3}$. Thus, when water is flowing on graphene, one may expect the induced electronic velocity to be on the order of the water slip velocity. 

According to the Onsager reciprocity principle, if there is HE drag, there must be a symmetric effect, whereby an electric current driven through the solid induces a liquid flow. We call this effect electro-hydrodynamic (EH) drag, adopting the symmetric nomenclature of \textcite{herrerofluidselectrostaticallyactive2026} (the names \emph{quantum osmosis} \cite{coquinothydroelectricenergy2024} or \emph{electronic pumping} \cite{coquinotElectronElectrolyte2026} have been used in previous works). 
 A first estimate of EH drag can be obtained using a momentum balance argument as above. Assuming that the liquid flows through a channel whose walls support an electric current with drift velocity $v_\e$, then the momentum flux going into the liquid per unit area is $\lambda_{\rm Q} (v_\e - v_\l)$. The momentum flux going out of the liquid corresponds to the roughness-induced frictional stress, $\lambda_{\rm RI} v_\l$. In the steady state, and in the absence of applied pressure, the incoming and outgoing fluxes must be equal so that 
\begin{equation}
v_\l = \frac{\lambda_{\rm Q}}{\lambda_{\rm RI} + \lambda_{\rm Q}} v_\e. 
\end{equation}
In most cases, as discussed in Sec. IV.D., $\lambda_{\rm Q} \ll \lambda_{\rm RI}$ so that the slip velocity will be a small fraction of the electronic drift velocity. However, electronic drift velocities can be much larger than typical hydrodynamic velocities: for instance, drift velocities of up to $\sim 10^5~\rm m/s$ have been achieved in graphene \cite{schmittMesoscopicKleinSchwingerEffect2023}. The water-graphene interface thus appears as a promising system for observing EH drag -- an effect, which, to our knowledge, has not been directly observed so far.

\begin{figure}
	\begin{center}
		\includegraphics[width=\columnwidth]{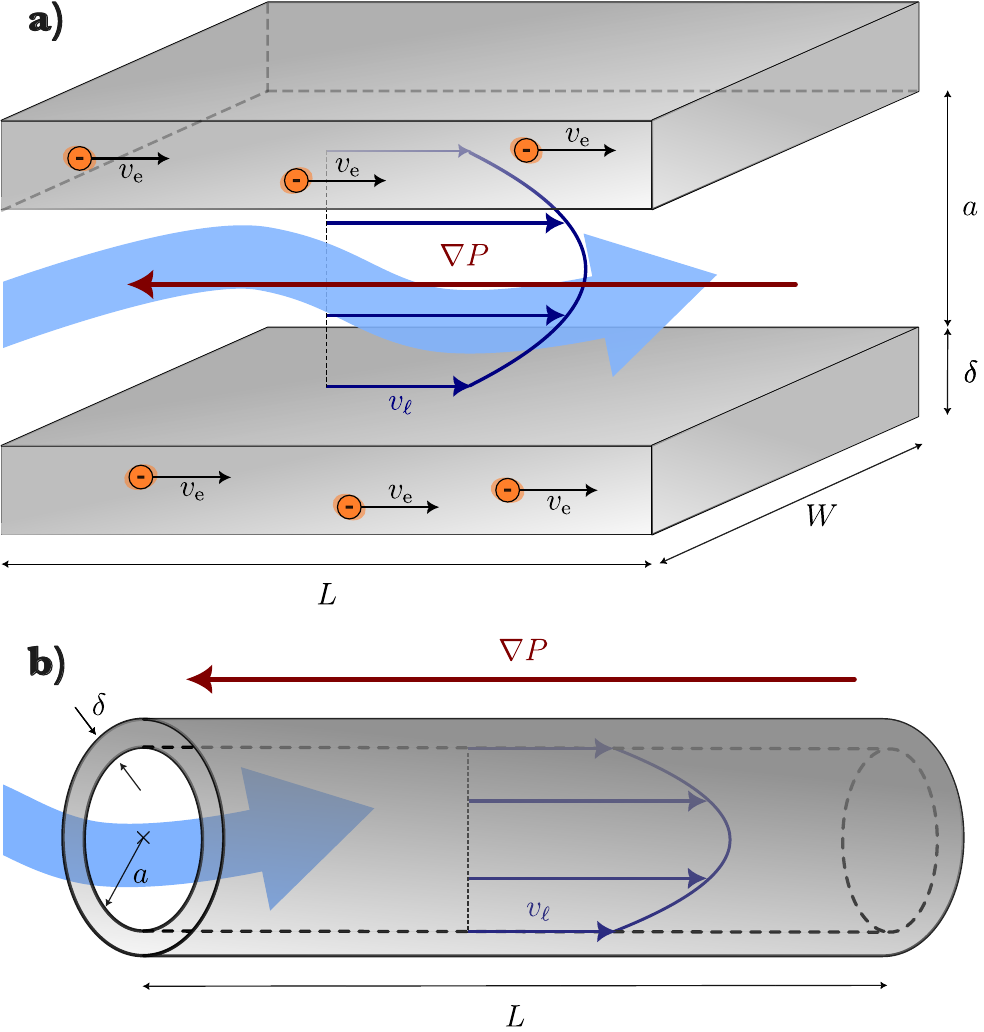}
		\caption{\textbf{Prototypical systems for the study of hydro-electronic cross-couplings.} (a) Slit-like channel of height $a$, width $W$ and length $L$ with electrically-conducting walls of thickness $\delta$. (b) Nanotube with radius $a$, length $L$, and electrically conducting wall with thickness $\delta$. In both geometries, the pressure drop is denoted $\Delta P$. The conducting walls are connected to an external circuit with negligible load resistance.} \label{fig1section5}
	\end{center}
\end{figure}

\subsection{Expression of the hydro-electronic transport matrix}

In the previous section, we introduced momentum balance relations between the electronic drift velocity and the liquid slip velocity, which gave us a first intuition of coupled hydro-electronic transport. However, it is typically the total flow rate through a channel, rather than the slip velocity at the walls, that can be accessed experimentally. To go beyond this limitation (and to fully describe transport under combined voltage and pressure driving), we now establish the general expression of the hydro-electronic transport matrix -- within the same Drude model framework for the electrons as above \cite{coquinothydroelectricenergy2024}. 

To write out the transport matrix, we need to specify the channel geometry. We will consider both a slit-like and a tubular channel, as pictured in Fig.~\ref{fig1section5}. The slit channel has height $a$, width $W$ and length $L$, with $L, W \gg a$. The tubular channel has radius $a$ and length $L$. For both channels, the walls are electronic conductors of thickness $\delta \ll a$, connected to an external circuit, so that they can support a voltage drop $\Delta V$ and a steady-state electric current $I$. We denote as $\lambda_{\rm RI}, \lambda_{\rm Q}$ the hydrodynamic friction coefficients at the walls; the flow rate and pressure drop in the liquid are denoted as $(Q, \Delta P)$. In order to write equations that apply to both geometries, we will denote as $\mathcal{A}$ the channel cross-sectional area, and as $2 W$ the length of the active interface ($W = \pi a $ for the tube). 

The global momentum balance equations for the liquid and for the electrons read respectively as 
\begin{align}\label{eq dynamical hydroelectonic}
	&2W L \times \big[\lambda_{\rm RI} v_\l + \lambda_{\rm Q} (v_\l- v_\e) \big]= \mathcal{A} \times \Delta P, \\
	&2 W L \times \big[\lambda_\e^0 v_\e -\lambda_{\rm Q} (v_\l- v_\e) \big]=2 W L \delta \times n_\e (-e) \, \frac{\Delta V}{L}.
\end{align}
The solution of this coupled linear system reads in matrix form:
\begin{equation} \label{eq:v_matrix}
	\begin{split}
	& \begin{pmatrix} v_\l \\ v_\e\end{pmatrix} = \frac{1}{1 - \Lambda_{\l \e}} \begin{pmatrix}
		\lambda_\l^{-1} & \lambda_{\rm Q}/(\lambda_\l \lambda_\e) \\ \lambda_{\rm Q}/(\lambda_\l \lambda_\e)& \lambda_\e^{-1} 
	\end{pmatrix} \dots
	\\
	& \qquad \dots 
	\begin{pmatrix}
		\mathcal{A} \Delta P /(2WL) \\ (-e) \delta n_\e \Delta V / L
	\end{pmatrix},
\end{split}
\end{equation}
where we define $\lambda_\l = \lambda_{\rm RI} + \lambda_{\rm Q}$, $\lambda_\e = \lambda_\e^0 + \lambda_{\rm Q}$ and $\Lambda_{\l \e} = \lambda_{Q}^2 / (\lambda_\l \lambda_\e) <1$. The liquid flow rate is expressed in terms of the pressure drop and slip velocity $v_{\l}$ according to the Hagen-Poiseuille law: 
\begin{equation}\label{debit}
	Q= \frac{\mathcal{A} a^2}{\alpha \eta} \frac{\Delta P}{L} + \mathcal{A} v_\l, 
\end{equation}
with $\alpha = 8$ ($\alpha = 12$) for the tube (slit) geometry. 
The electronic current in the wall is written in terms of the drift velocity $v_\e$ as 
\begin{equation} \label{electrical current}
I =  2 W \delta \times (-e) n_\e v_\e. 
\end{equation}
Combining the matrix solution \eqref{eq:v_matrix} with Eqs. \eqref{debit} and \eqref{electrical current}, one obtains the full transport matrix $\mathbf{L}$ introduced in \eqref{transport matrix eq 0}:
\begin{equation}\label{transport matrix eq}
	\begin{pmatrix}
		 Q \\ I 
	\end{pmatrix} = 
	\begin{pmatrix}
		\mathcal{L} & \mu_{\h \e} \\ \mu_{\h \e} & G_\e 
	\end{pmatrix}\times
	\begin{pmatrix}
		\Delta P \\ \Delta V
	\end{pmatrix} 
\end{equation} 
where the expressions for the permeance $\mathcal{L}$, the conductance $G_\e$ and the hydro-electronic mobility $\mu_{\h \e}$ are given by:
\begin{align}
	\mathcal{L} &= \frac{ \mathcal{A} a^2}{\alpha \eta L } \left( 1 + \frac{1}{1- \Lambda_{\l \e}}\frac{\beta b}{a} \right), \quad \mathrm{with}\,\, b = \frac{\eta}{\lambda_\l }, \\
	G & = \frac{1}{1- \Lambda_{\l \e}} \times \frac{2W \delta}{L} \frac{n_\e e^2 \tau_\e}{m_\e}, \quad \mathrm{with} \,\, \tau_\e = \frac{\delta n_\e m_\e}{\lambda_\e}, \\
	\mu_{\h \e} &= \frac{\delta \mathcal{A} n_{\e}(- e)}{L} \times \frac{\lambda_{\rm Q}}{\lambda_{\rm RI}  \,\lambda_\e^0 + (\lambda_{\rm RI}  + \lambda_\e^0) \lambda_{\rm Q}}. \label{Cle}
\end{align}
The geometric factor $\beta$ is 4 for the tube and $6$ for the slit. 

We may now discuss the expressions of the transport coefficients. The cross-mobility $\mu_{\h \e}$ vanishes in the absence of QF, and saturates to $\mu_{\lim} = -\frac{\mathcal{A} \delta \,  n_\e e}{L} \times \frac{1}{\lambda_{\rm RI}  + \lambda_\e^0}$ as $\lambda_{\rm Q} \rightarrow + \infty$. This saturation accounts, e.g., for the impossibility that flow-driven electrons go faster than the flow, in the absence of applied voltage drop. $\mu_{\h \e}$ is negative because we assumed that the charge carriers are negative electrons: in this case, the imposed flow and induced current (or the imposed current and induced flow) have opposite directions. The direct transport coefficients -- the permeance and the conductance -- are both enhanced by the hydro-electronic coupling, up to diverging when $\Lambda_{\l \e} \to 1$. This behavior can be understood as follows: from the liquid's point of view, turning on the hydro-electronic coupling amounts to going from an open-circuit configuration, where the electrons in the wall remain static, to a closed-circuit configuration, where they can flow. If the channel permeability is mostly limited by QF (large $\Lambda_{\l \e}$), then closing the circuit removes the most important source of dissipation for the liquid flow: the interfacial liquid and the electrons can then drift at the same velocity, generating no friction. A similar picture holds if one takes the point of view of the electrons: then, switching on the hydro-electronic coupling amounts to going from a closed to an open fluidic channel. 

\begin{figure}
\centering
\includegraphics{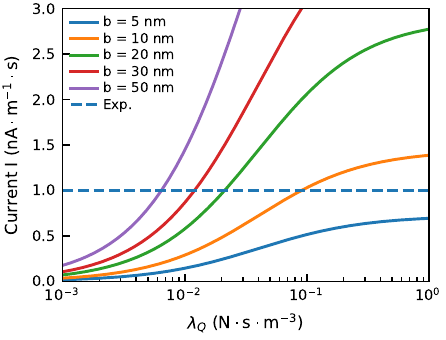}
\caption{\textbf{Hydro-electronic drag: comparison to experiment}.The hydro-electornic drag current, as predicted by Eq.~\eqref{transport matrix eq}, is plotted as a function of water-graphene QF coefficient and for different values of the slip length $b$, in the experimental geometry of \cite{takedaenhancingelectricitygeneration2025}. The slit-like channel dimensions are $(W, a) = \rm (2.1~mm, 0.7~mm)$ and the graphene parameters are $n_\e = 5 \times 10^{12} ~\rm cm^{-2}$ and $\mu = 2000 ~\rm cm^2 \cdot V^{-1} \cdot s^{-1}$.}
\label{fig:takeda}
\end{figure}

The hydro-electronic transport matrix may be used to provide explicit estimates for flow-induced electric currents and current-induced liquid flows, that can be compared to experiment. Recently, Takeda \emph{\emph{et al.}} fabricated slit-like channels with dimensions $(W, a) = \rm (2.1~mm, 0.7~mm)$, where the bottom wall, made of silicon oxide, was covered with graphene via chemical vapor deposition (CVD). Upon flowing deionized water through the channel, they observed an electric current proportional to the average flow velocity $\bar v_\l$: $I/\bar v_\l \approx 1 ~\rm nA\cdot m^{-1} \cdot s$ \cite{takedaenhancingelectricitygeneration2025}. Qualitatively, an electric current going in the direction of the water flow is consistent with a Coulomb drag mechanism acting on positive hole charge carriers, as expected in CVD graphene on $\rm SiO_2$. 
Quantitatively, we show in Fig.~\ref{fig:takeda} the prediction of Eq.~\eqref{transport matrix eq} for the geometry of Takeda \emph{\emph{et al.},} where we have assumed a typical hole density $n_\e = 5 \times 10^{12} ~\rm cm^{-2}$ and mobility $\mu = 2000 ~\rm cm^2 \cdot V^{-1} \cdot s^{-1}$ for CVD graphene.\footnote{The predicted current was divided by 2 to account for the fact that the device has only one "active" wall. This approximate treatment is valid if $\lambda_{\rm Q} \ll \lambda_{\rm RI}$, as is the case for water on graphene.} These values correspond to an intrinsic friction coefficient of the hole charge carriers $\lambda_\e^0 = 4 \times 10^{-2} ~\rm N \cdot s \cdot m^{-3}$. Given the predicted QF coefficient $\lambda_{\rm Q} \approx 10^{-2}~\rm N \cdot s \cdot m^{-3}$ at the assumed hole density (Eq.~\eqref{eq:lambdaQ_graphene}), the HE drag mechanism is consistent with the measurements of Takeda \emph{\emph{et al.}} if one assumes a small slip length $b \approx 30~\rm nm$ at the water-graphene interface. 
It is also of interest to estimate the expected EH drag velocity in the same geometry\footnote{We assume for simplicity that the channel has the same length as the graphene-covered area.}. 
From the experimental data, $\mu_{\h \e} \approx 1 ~\rm \mu A / bar$. CVD graphene can sustain a maximum current density on the order of $35~\rm A/m$ \cite{belotcerkovtcevaExtremeCurrent2025}, corresponding to a drift velocity $v_\e \approx 2~\rm km/s$ at $n_\e = 5 \times 10^{12}~\rm cm^{-2}$ carrier density. In the experimental geometry, such a current density is achieved by applying $\Delta V \approx 130~\rm V$ to the sample. The corresponding flow rate is $Q = \mu_{\h \e} \Delta V /2 \approx 0.65~\rm \mu L / s$; the factor of 2 accounts for there being only one active wall. This amounts to an average flow velocity $\bar v_\e \approx 0.5 ~\rm mm/s$ in the channel. This is a measurable velocity, comparable to what could be achieved with an electroosmotic pump at similar voltage. Yet, contrary to electroosmosis, the direct electronic pumping mechanism base on QF does not rely on dissolved ions, nor does it require faradaic electrodes held at high voltages beyond the water electrochemical window. 

\subsection{Practical implications}

We now discuss some of the practical consequences of nanoscale hydro-electronic couplings. We will first introduce the flow tunneling phenomenon \cite{coquinotmomentumtunnelling2025}, and then discuss the potential of HE drag as a mechanism for nanoscale hydroelectric energy conversion \cite{coquinothydroelectricenergy2024}. 

\subsubsection{Flow tunneling}

\begin{figure}
	\begin{center}
		\includegraphics[width=\columnwidth]{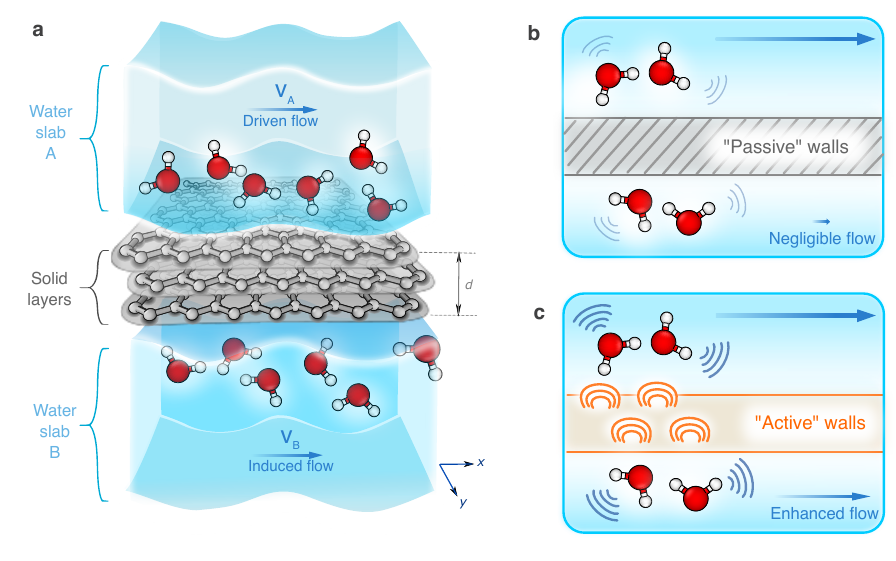}
		\caption{\textbf{Principle of flow tunneling and role of the solid wall.} (a) Schematic of the prototypical flow tunneling system. Two water slabs $A$ and $B$ are separated by a solid of thickness $d$. A flow of velocity $\v_A$ is imposed in the slab $A$, inducing a flow of velocity $\v_B$ in the slab $B$. (b) When the wall is "passive", i.e. when the hydro-electronic mobility vanishes, the induced flow is negligible. (c) When the wall is "active", with important hydro-electronic mobility, there is an induced flow across the wall. Reproduced from \cite{coquinotmomentumtunnelling2025}.}\label{fig1flowtunneling}
	\end{center}
\end{figure}

Flow tunneling is a direct implication of hydro-electronic coupling in a two-interface system as depicted in Fig.~\ref{fig1flowtunneling}. Consider two liquid slabs separated by a conducting solid wall. Then, a flow imposed in the top slab will induce an electric current in the wall through HE drag, which in turn will induce a flow in the bottom slab through EH drag. The flow is effectively transmitted through the solid wall. 

At the simplest level, this effect can be analyzed in the same momentum balance framework as above. For a straightforward analysis, we assume plug flow in both liquid slabs with velocities $\v_A$ and $\v_B$, and an electron drift velocity $\v_\e$ that is uniform across the solid. Then, the momentum balance equations for the electrons and for slab $B$ respectively read: 
\begin{align}
	0 &= - \lambda_\e^0 \v_\e - \lambda_{\rm Q}(\v_\e - \v_A) - \lambda_{\rm Q} (\v_\e - \v_B), \\
	0 &= - \lambda_{\rm RI} \v_B - \lambda_{\rm Q} (\v_B - \v_\e).
\end{align}
Eliminating the electronic velocity yields 
\begin{equation}\label{eq flow tunneling}
	\v_B = \frac{ \lambda_{\rm Q}^2/(\lambda_\e^0 + 2 \lambda_{\rm Q})}{\lambda_{\rm RI}  + \lambda_{\rm Q} (\lambda_\e^0 + \lambda_{\rm Q})/(\lambda_\e^0 + 2 \lambda_{\rm Q})} \, \v_A.
\end{equation}
The "tunneling efficiency" $v_B/v_A$ depends on the thickness $\delta$ of the solid wall through the electronic friction coefficient $\lambda_\e^0 \propto \delta$. For a single layer of high-mobility graphene ($n_{\e} = 5 \times 10^{12}~\rm cm^{-2}$, $\mu = 10^5~\rm cm^2/(V \cdot s)$), according to Eqs.~\eqref{eq:lambdaQ_graphene} and \eqref{eq:lambdae_graphene} $\lambda_{\rm Q} / \lambda_{\e}^0 \approx 10$ so that the tunneling efficiency through a few-layer graphene wall will not depend on the number $N$ of layers up to $N \sim 10$, and then decrease roughly as $1/N$. The maximum tunneling efficiency, achieved for $N \lesssim 10$, is 
\begin{equation}
\left( \frac{v_B}{v_A} \right)_{\rm max} = \frac{\lambda_{\rm Q}}{2 \lambda_{\rm RI} + \lambda_{\rm Q}}. 
\end{equation}
Assuming $\lambda_{\rm Q} = 10^{-2}~\rm N \cdot s \cdot m^{-3}$ and $\lambda_{\rm RI} \sim 10^4~\rm N \cdot s \cdot m^{-3}$ for suspended graphene, this evaluates to a negligibly small $(v_B/v_A)_{\rm max} \approx 5 \times 10^{-7}$. However, the case of few-layer graphene is subtle, because, as the number of layers is increased, the system is expected to transition from graphene-like to graphite-like electronic properties \cite{kavokinefluctuationinducedquantum2022}: for graphite, $\lambda_{\rm Q} \gg \lambda_{\rm RI}$ and $(v_B/v_A)_{\rm max} \approx 1$. It is not known exactly when this transition occurs, but it can be expected to take 10 -- 100 layers \cite{lavorprobingstructure2020}. At the same time, as the number of layers is increased, it becomes wrong to assume that the electron drift velocity is uniform across the solid: eventually, the propagation of the drift velocity is limited by the electron mean free path. The interplay between ballistic and diffusive momentum propagation in flow tunneling was studied in the framework of a multilayer Drude model by \cite{coquinotmomentumtunnelling2025}; to go towards more accurate descriptions, experimental benchmarks are now crucially required. Fundamentally, flow tunneling challenges the applicability of hydrodynamics to systems with walls reaching nanoscale thickness, as the effect cannot be packed into a boundary condition, e.g. an effective slip length. It is of particular interest to asses its importance in water transport through lamellar membranes.

\subsubsection{Nanoscale hydro-electricity}

Hydro-electronic drag provides a mechanism for directly converting hydraulic energy into electric energy at a liquid-solid interface -- a concept referred to as \emph{hydronic energy}. We now summarize the main results of \cite{coquinothydroelectricenergy2024}, who analyzed this mechanism in terms of conversion efficiency. 

\begin{figure}
	\begin{center}
		\includegraphics[width=\columnwidth]{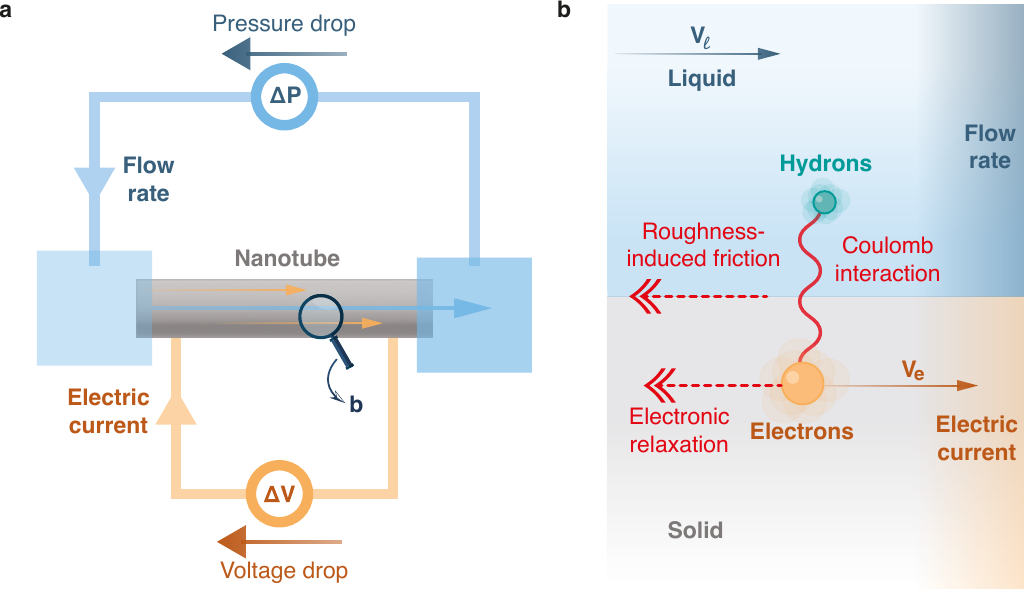}
		\caption{\textbf{Hydronic energy.} (a) Sketch of the model system: a liquid flows through a nanotube connected to an external electrical circuit. A fraction of the energy used to pump the fluid through the tube is recovered as electricity. (b) Schematic of the momentum transfer and relaxation processes governing the energy conversion efficiency. Adapted from \cite{coquinothydroelectricenergy2024}.}\label{figure hydroelectronic drag}
	\end{center}
\end{figure}

Let us consider a single-nanotube geometry as depicted in Fig.~\ref{figure hydroelectronic drag}.a and assume that the nanotube is now connected to an external load resistance $R_L$. Then, if a current $I$ flows through the circuit as a result of HE drag, a voltage drop $\Delta V$ builds up according to $\Delta V = - R_L I $. Rewriting the transport equations \eqref{transport matrix eq} with $\Delta P$ an imposed pressure drop and $\Delta V$ constrained as above, the flow-induced electric current is deduced as:
\begin{equation}
	I = \frac{I_{\rm max}}{1 + G R_L} \quad \mathrm{with} \quad I_{\rm max} = \mu_{\h \e} \Delta P. 
\end{equation}
The electrical power delivered to the resistance is $\mathcal{P}_\e = R_L I^2$, which is maximized when $R_L = 1/G$.
The corresponding flow rate is $Q = \mathcal{L} \Delta P - R_L \mu_{\h \e} I $, and the mechanical power required to maintain it is $\mathcal{P}_\h = Q \, \Delta P$.
The hydro-electric conversion efficiency is then defined as 
\begin{equation}
	\gamma = \frac{\mathcal{P}_\e}{\mathcal{P}_\h} = \frac{\mathcal{I}(1 - \mathcal{I})}{\mathcal{I} + 1/Z},
\end{equation}
where $\mathcal{I} = I/I_{\rm max}$ and $Z$ is a dimensionless figure of merit, defined as:
\begin{equation}
\begin{split}
		Z &= \frac{\Lambda_{\l \e}}{1 - \Lambda_{\l \e}} \frac{1}{1 + \frac{a}{4b}} \\ &= \frac{\lambda_{\rm Q}^2}{\lambda_\e^0 \lambda_{\rm RI} + \lambda_{\rm Q}(\lambda_\e^0 + \lambda_{\rm RI})} \frac{1}{1 +  a/4b}.
\end{split}
\end{equation}
For a given value of $Z$, the maximum efficiency is achieved for $\mathcal{I} = 1 / \sqrt{1+Z}$ and reaches:
\begin{equation}
	\gamma_{\rm max} = \frac{Z}{(1 + \sqrt{1+Z})^2}.
\end{equation}
The efficiency is 0 when $Z \to 0$; when $Z \to \infty$, the maximum efficiency goes to $1$, corresponding to lossless energy conversion. We note that, by Onsager symmetry, we would have arrived at the same expression for the maximum efficiency had we considered "electro-hydraulic" energy conversion and defined $\gamma = \mathcal{P}_\h / \mathcal{P}_\e$. 

The figure of merit essentially compares the rate of liquid electron momentum transfer ($\lambda_{\rm Q}$) to the rate of momentum loss in the liquid-electron system. Momentum loss originates from roughness-induced friction ($\lambda_{\rm RI}$), electron scattering on impurities or phonons $\lambda_\e^0$ and viscous losses in the bulk of the channel ($a / (4 b)$). Interestingly, there is no evident contradiction between maximizing interfacial momentum transfer and minimizing momentum loss, so that it should be possible in principle to obtain a large hydro-electronic figure of merit. 

\begin{figure}
	\begin{center}
		\includegraphics[width=\columnwidth]{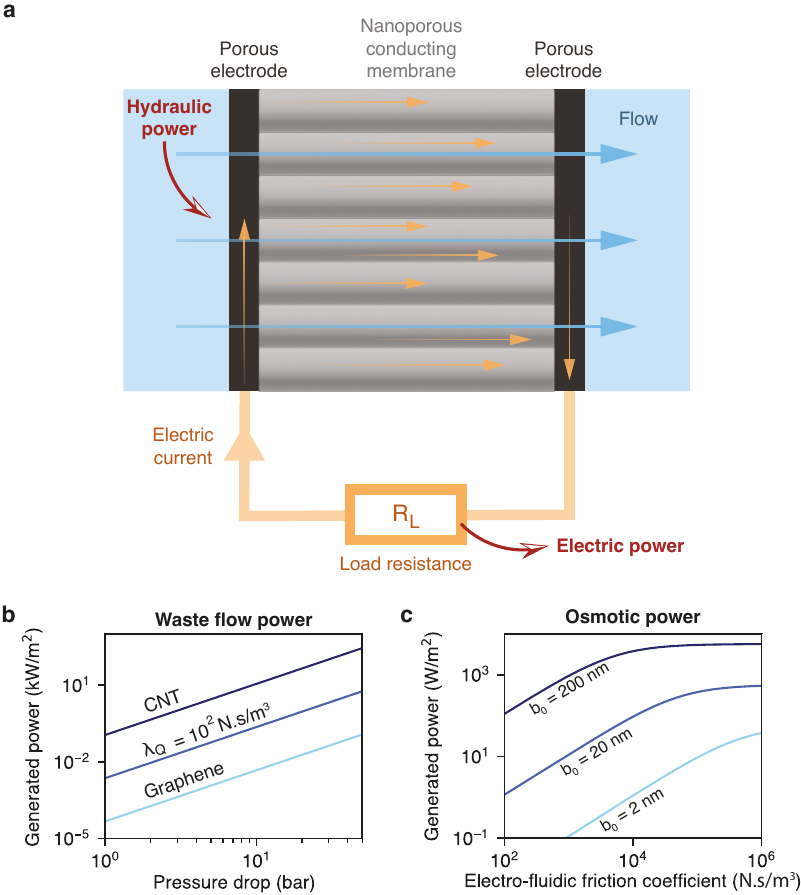}
		\caption{\textbf{Concept of a large-scale hydronic generator for waste flow recovery.} (a) Schematic of a hydronic generator based on a nanoporous conducting membrane placed between two porous electrodes. (b) Electrical power recovered per unit surface as a function of waste flow pressure, for different membrane materials. (c) Power produced by pressure-retarded osmosis through a semipermeable and electrically conducting membrane. Adapted from \cite{coquinothydroelectricenergy2024}.}\label{fig4prx}
	\end{center}
\end{figure}

Let us now analyze the relative importance of the momentum loss mechanisms. First, viscous dissipation in the channel can be avoided if the channel is sufficiently narrow and/or has sufficiently large slippage ($a \ll 4 b$). This is typically the case for carbon nanotubes or slit nanochannels with graphite walls. Then, in most cases -- and especially for carbon-based materials -- the intrinsic electronic dissipation is weak, and certainly weaker than RI friction ($\lambda_\e^0 \ll \lambda_{\rm RI}$). Then, one has a simplified expression for the figure of merit: 
\begin{equation}
	Z \approx \frac{\lambda_{\rm Q}^2}{\lambda_{\rm RI}(\lambda_\e^0 + \lambda_{\rm Q})}. 
\end{equation}
Thus, the key condition for efficient energy conversion is $\lambda_{\rm Q} \gtrsim \lambda_{\rm RI}$. One can expect this to be realized in multiwall CNTs or slit channels with graphite walls, where $Z$ could reach up to $\sim 10$ based on the estimates of Sec. IV. It could thus be of interest to build large-scale hydroelectric conversion devices -- termed \emph{hydronic generators} \cite{coquinothydroelectricenergy2024} -- based on those materials, as depicted schematically in Fig. \ref{fig4prx}a. Such devices could serve to recover energy from \emph{waste flows} -- flows that have too little intertia to spin a turbine -- that are a by-product of membrane-based filtration processes such as water desalination (Fig. \ref{fig4prx}b). 

It is important to note that a high energy conversion efficiency is not a requirement for observable HE drag currents or EH drag flows. For instance, in the experimental configuration of \cite{takedaenhancingelectricitygeneration2025}, $Z \sim 10^{-6}$ without viscous losses (and $Z \sim 10^{-10}$ with viscous losses). Similarly, the power per unit surface produced in a hydronic generator can be significant despite the conversion efficiency being low -- this feature may be exploited for osmotic energy harvesting via pressure retarded osmosis (PRO). 
When sea water and fresh water are separated by a semi-permeable membrane, an osmotic pressure difference $\Delta \Pi \approx 30~\rm bar$ builds up across the membrane. This pressure difference drives a spontaneous flow, whose kinetic energy can be harvested using a turbine. Given an electrically-conducting semi-permeable membrane, the energy could be harvested instead using HE drag directly at the membrane; only moderate values of $Z$ would then be required for the generated power per unit area to exceed the industrial relavance thershold of $5 ~\rm W/m^2$ \cite{sirianewavenues2017}, as shown in Fig. \ref{fig4prx}c. 
Such a concept for osmotic energy harvesting was recently realized with a chitosan-alginate biopolymer membrane that was "doped" with graphene platelets \cite{songIonConveying2026}. An electric current through the membrane could indeed be collected under osmotic flow, with a power density reaching $24~\rm W/m^2$, exceeding by nearly a factor of 5 the power density achieved with conventional reverse electrodyalisis using the same membrane. Given the complexity of the membrane system and the presence of ions in high concentrations, the exact mechanisms at play remain to be investigated.

\section{Quantum theory of nanoscale liquid-solid interfaces}\label{section quantum theory}

In general, fluctuation-induced friction can arise from quantum or classical thermal fluctuations. For liquids at room temperature, charge or mass density fluctuations are well described by classical stochastic dynamics: for instance, the experimental charge fluctuation spectrum of water is well reproduced by classical molecular dynamics simulations up to frequencies $~\sim 100~\rm meV$ or $25~\rm THz$ \cite{carlson2020, bonthuis2012}. However, the microscopic description of low-energy electronic excitations in solids (including at energies $\hbar \omega \ll k_{\rm B} T$) is intrinsically based on quantum mechanics. Solid-liquid interfaces can be viewed as a hybrid quantum-classical system, where classical and quantum theoretical descriptions collide. In \cite{kavokinefluctuationinducedquantum2022}, the clash was resolved by taking the description entirely to the quantum side: conceptually, this must be possible, since classical mechanics can be obtained as a limiting case of quantum mechanics. In the following, we reformulate the operator formalism of \cite{kavokinefluctuationinducedquantum2022} in a field theory language, which emphasizes the continuous transition between quantum and classical dynamics. In the fully classical limit, this formalism essentially becomes the stochastic density functional theory used by \textcite{deanNonequilibriumTuning2016} in the study of the thermal Casimir effect. 

Quantum field theory is well-established as a formalism in condensed-matter physics. Thus, a quantum field theory of liquid-solid interfaces benefits from well-rounded analytical methods: in particular, diagrammatic perturbation theory. In this section, we illustrate how Feynman diagrams are a powerful tool for understanding fluctuation-induced friction and coupled hydro-electronic transport effects. We also illustrate how the quantum concept of quasiparticles can illuminate fluctuation effects even in a purely classical system.

\subsection{Keldysh action: general idea}

In the framework of statistical mechanics, any observable of a many-body system in thermal equilibrium can be computed as a statistical average over the configurations of its microscopic degrees of freedom. The Keldysh formalism \cite{keldysh1965}  generalizes this idea to non-equilibrium many-body quantum systems. Given some microscopic degrees of freedom $\{ \phi (\r, t) \}$, it constructs an action $\calS[ \{ \phi(\r, t) \}]$, such that the mean value of any observable $\mathcal{O}(\{\phi(\r, t_0) \})$ can be evaluated as 
\begin{equation}
\langle \mathcal{O} \rangle = \int [D \{\phi \}] \mathcal{O}(\{\phi(\r, t_0) \}) e^{i\calS[ \{ \phi(\r, t) \}]}. 
\end{equation}
Here $\int [D \{\phi \}]$ represents a path integral over all configurations $\phi(\r, t)$. 
The time $t$ is sampled on a contour $\cal C$ -- the Keldysh contour, going from $-\infty$ to $+ \infty$ and then back to $-\infty$. 
The Keldysh formalism is derived from first principles in several recent textbooks \cite{kamenev2011, rammer2007, altlandcondensedmatter2023}. It is, in a sense, "a theory of everything" for many-body systems: other descriptions of many-body dynamics, such as the Langevin, Fokker-Planck or Boltzmann equations, can be derived from Keldysh field theory under suitable approximations.

\subsection{Keldysh action of interacting electrons}

The microscopic degrees of freedom associated with the electrons in solids are their creation and annihilation fields, $\psibar_{\sigma}(\r,t)$ and $\psi_{\sigma}(\r,t)$, respectively; $\sigma = \uparrow, \downarrow$ is the spin index. 
In the absence of interaction with the liquid, the Keldysh action describing their dynamics can be  written as 
\begin{equation}
\begin{split}
\S_\e &= \int_{\cal C} \frac{\d t}{\hbar} \int \d \r\sum_{\sigma}  \psibar_{\sigma}(\r, t)(i\hbar \partial_t - \hat H(\r)) \psi_{\sigma}(\r, t)\\
& - \frac{1}{2}  \int_{\cal C} \frac{\d t}{\hbar} \int \d \r \d\r' n_\e(\r, t)V(\r - \r') n_\e(\r', t). 
\end{split}
\label{Se}
\end{equation}
The first term is the action of free electrons, while the second accounts for electron-electron Coulomb interactions. Here $\hat H(\r)$ is the one-electron Hamiltonian in the position representation, $V(\r)$ is the Coulomb potential and $n_\e(\r,t) = \sum_{\sigma}[\psibar_{\sigma} \psi_{\sigma}](\r, t)$ is the electron density. It is important to note that this action describes electrons at a given temperature $T$ -- in the sense that, at $t \to - \infty$, when all interactions and non-equilibrium perturbations have been adiabatically switched off, the density matrix is the thermal density matrix $\hat \rho_0 = e^{-\hat H / \kB T}$. However, the temperature does not appear in the continuum notation: it hides in the prescription for inverting the operator $(i\hbar \partial_t - \hat H(\r))$ on the contour, which we will make explicit once we have mapped the integration to regular time. Note as well that the $\psi, \psibar$ are anticommuting Grassmann variables, with specific Gaussian integration rules, whereas their combination $n_\e(\r,t)$ is a regular number.

The forward and backward times can be transformed into a single real time at the cost of doubling the number of degrees of freedom. For any field $\phi(t)$ defined on the closed time contour we may distinguish the field $\phi^+(t)$ on the forward part and $\phi^-(t)$ on the backward part, and define the vector $\boldsymbol{\phi} = \left( \phi^+(t), ~ \phi^-(t) \right)$. Then, the action acquires a matrix structure 
\begin{equation}
	\begin{aligned}
\mathcal{S}_\e &= \frac{1}{\hbar} \int_{-\infty}^{+\infty} \d t \d t' \int \d \r \d \r' \sum_{\sigma} \left[ \overline{\boldsymbol{\psi}}_{\sigma} \cdot \mathbf{G}^{-1}_{0,\sigma} \cdot \boldsymbol{\psi}_{\sigma} \right]_{\r, t, \r', t'} \\
& - \frac{1}{2\hbar} \int_{-\infty}^{+\infty} \d t \int \d \r \d \r' \left[ \mathbf{n_\e \cdot V \cdot n_\e} \right]_{\r, \r', t}
	\end{aligned}
	\label{eq:Se}
\end{equation}

The $(+, -)$ basis of the "Keldysh space" is in fact not the most practical choice, since it does not make explicit the causal structure of correlation functions. For the fermionic fields, one defines \cite{kamenev2011, larkinnonlinearconductivity1975}: 
\begin{align}
	\psi^{1} = \frac{\psi^+ + \psi^-}{\sqrt2},\qquad \psi^{2} = \frac{\psi^+ - \psi^-}{\sqrt2},  \\
\psibar^1 = \frac{\psibar^+ - \psibar^-}{\sqrt2},\qquad
 \psibar^2 = \frac{\psibar^+ + \psibar^-}{\sqrt2}.
\end{align}
In this new $(1,2)$ basis, the matrix $\mathbf{G}_0$ takes the form
\begin{equation}
\mathbf{G}_0= \left(
		\begin{array}{cc}
			G_0^{\R} & G_0^{\K}  \\
			0 & G_0^{\A}
		\end{array}
	\right)
	.
\end{equation}
With the notation $x = (\r, t)$, $x' = (\r', t')$,
\begin{equation}
G_0^{\R,\A}(x,x') = \left[\delta(x - x') \left(i\hbar \partial_{t'} - \hat H(\r') \pm i0^+ \right) \right]^{-1},  
\label{eq:G0R}
\end{equation}
and the expression of $G_0^{\K}$ will be given below in Fourier space (Eq.~\eqref{eq:G0K}). Applying the rules of fermionic Gaussian integration, $\G_0$ can be expressed in terms of field correlators taken at $V = 0$: 
\begin{equation}
\mathbf{G}_0(x,x') = -\frac{i}{\hbar} \left(
	\begin{array}{cc}
		\langle \psi^1(x) \psibar^1(x') \rangle_0 & \langle \psi^1(x) \psibar^2(x') \rangle_0 \\
		\langle \psi^2(x) \psibar^1 (x') \rangle_0 & \langle \psi^2(x) \psibar^2(x') \rangle_0
	\end{array}
	\right), 
\end{equation}
Note that the correlator $\langle \psi^2(x) \psibar^1 (x') \rangle_0$ indeed vanishes; this can be traced back to the fact that the Keldysh contour folds back on itself at $+ \infty$, imposing that $\psi^+(+\infty) = \psi^-(+\infty)$ and $\psibar^+(+\infty) = \psibar^-(+\infty)$. 

One may similarly define an interacting Green's function matrix $\mathbf{G}$ in terms of the correlators taken with respect to the interacting action. The components $G^{\R}, G^{\A}, G^{\K}$ correspond indeed to physically meaningful Green's functions. The retarded Green's function $G^{\R}(x,x')$ represents the probability amplitude for an electron created at $(\r', t')$ to propagate until $(\r, t)$. It is thus non-zero only if $t > t'$. Similarly, the advanced Green's function $G^{\A}(x, x')$ represents the amplitude for an electron that is detected at $(\r', t')$ to have propagated from $(\r, t)$. It is thus non-zero only if $t' > t$. 
For bookkeeping, it is useful to remember that the creation (bar) field always comes second in the definition of the Green's function, and one writes $G(t_{\psi}, t_{\psibar}) = G(t_{\psi} - t_{\psibar})$ if there is time translation invariance. If $\hat H(\r)$ is also translationally invariant in space, the Fourier transform of Eq.~\eqref{eq:G0R} yields
\begin{equation}
G^{\R,\A}_0 (\textbf{q},\omega)=\frac{1}{\hbar\omega-\xi_\textbf{q} \pm i0^+}
\label{GReq}
\end{equation}
where $\xi_\textbf{q}=u_\textbf{q}-\mu$ is a single electron excitation energy: $u_\textbf{q}$ is the energy dispersion and $\mu$ is the chemical potential. Note that $G_0^{\R} (\q, \omega) = [G_0^{\A}(\q, \omega)]^*$; this relation also holds for the interacting Green's function at thermal equilibrium. Eq.~\eqref{GReq} shows that the retarded and advanced Green's functions contain information on the accessible electronic states. 

The Keldysh Green's function $G^{\K}$ is related at equilibrium to the retarded and advanced Green's functions via the fluctuation-dissipation theorem: 
\begin{equation}
\begin{split}
G^{\K}(\r, \r', \omega) &= \tanh\left( \frac{\hbar \omega}{2 \kB T} \right) [G^{\R} - G^{\A}](\r, \r', \omega) \\
&= 2i \, \tanh\left( \frac{\hbar \omega}{2 \kB T} \right) \im{G^{\R}(\r, \r', \omega)}  
\end{split} 
\label{eq:G0K}
\end{equation}
This introduces explicitly the temperature as a prescription for inverting $\mathbf{G}_0^{-1}$. For non-interacting electrons:
\begin{equation}
 G_0^{\K}(\textbf{q},\omega)=(2n_{\rm F}(\omega) -1) \times  2 i \pi \delta(\hbar\omega- \xi_{\q}),
\end{equation}
where $n_{\rm F}(\omega) = 1/(e^{\hbar\omega/\kB T} + 1)$ is the Fermi-Dirac distribution. Thus, the Keldysh Green's function carries information on the (possibly non-equilibrium) occupation of electronic states. 

For a real bosonic field such as the electron density $\n_\e$, we define the "real" field $n_\e$ and the "response" field $\tilde n_\e$ as: 
\begin{equation}
	n_\e = \frac{n^+ + n^-}{2},\qquad \tilde n_\e = \frac{n^+ - n^-}{\hbar}.  
\end{equation}
These definitions are not conventional for quantum field theory but they ensure a consistent classical limit, as detailed in the following. With these definitions, $\mathbf{V} = V(\r - \r') \boldsymbol{\sigma}_x$ (the first Pauli matrix), so that the electron-electron interaction term in the action is explicitly 
\begin{equation}
\mathcal{S}^{\rm int}_{\e\e} = - \int_{t, \r, \r'} n_\e(\r, t) V(\r-\r') \tilde n_\e(\r', t). 
\end{equation}
Note the expression of the density operators in terms of the fermionic operators:
\begin{align}
	&n_\e = \sum_{\sigma}\frac{\psibar^1_{\sigma} \psi^2_{\sigma} + \psibar^2_{\sigma} \psi^1_{\sigma}}{2}, \\
	 & \tilde n_\e = \sum_{\sigma} \frac{\psibar^1_{\sigma}\psi^1_{\sigma} + \psibar^2_{\sigma}\psi^2_{\sigma}}{\hbar}.  
	 \label{eq:npsi}
\end{align}
With this setup, one can in principle evaluate perturbatively any observable on the interacting electron system by expanding $e^{i \mathcal{S}_\e}$ in powers of $\mathcal{S}^{\rm int}_{\e\e}$ and applying Wick's theorem. However, before specifying the Feynman rules governing this expansion, we will describe the liquid dynamics and the electron-liquid interaction.

\subsection{Formal quantization of liquids: the hydron quasiparticle}

At energies close to room temperature, the relevant microscopic degrees of freedom in a liquid are molecular positions, bond lengths and angles. It is formally possible to write out a Keldysh action describing those degrees of freedom and their interactions. However, since interactions are strong in a liquid, the perturbation theory in those interactions is intractable -- even without yet considering any interactions with a solid. \cite{kavokinefluctuationinducedquantum2022} took instead a phenomenological approach, where the liquid is described by its fluctuating charge density $n_\w(\r, t)$. Under the assumption of Gaussian fluctuations, the dynamics are completely determined by the linear response function $\chi^{\R}_\w(\r,\r',t-t')$ of the mean charge density to an external potential, defined by: 
\begin{equation}
\langle n_\w(\r, t) \rangle = \int_{-\infty}^{+\infty} \d t' \int \d \r' \chi^{\R}_\w(\r, \r', t-t') \phi_{\rm ext}(\r', t').
\label{eq:linear_reponse_nw}
\end{equation}
$\chi_\w^{\R}$ is the only input required for the theory. Indeed, the assumption is that $n_\w(\r,t)$ is a Gaussian random process, which is fully specified by its mean value and two-point correlation function (or structure factor) $S_\w(\r, \r', t-t') =  \langle n_\w(\r, t) n_\w(\r', t') \rangle$. The latter is related to the linear response function via the fluctuation-dissipation theorem: 
\begin{equation}
S_\w(\r, \r', \omega) = -\frac{2 \kB T}{\omega} \im{\chi^{\R}_\w}(\r, \r', \omega). 
\label{eq:FDT_S_class}
\end{equation}
A dynamical equation for $n_\w$ may therefore be written as 
\begin{equation}
	\begin{split}
n_\w(\r,t) = &\int_{-\infty}^{t} \d t' \int \d \r' \chi^{\R}_\w(\r, \r', t-t') \phi_{\rm ext}(\r', t')\\ &+ \xi(\r,t), 
	\end{split}
\label{eq:nw_GLE}
\end{equation}
where we used that $ \chi^{\R}_\w(t-t')$ vanishes for $t'>t$ and $\xi(\r,t)$ is the Gaussian random process defined by
\begin{equation}
\langle \xi(\r,t) \xi(\r',t') \rangle = S_\w(\r, \r', t- t'). 
\end{equation}

The generalized Langevin description of the liquid charge dynamics in Eq.~\eqref{eq:nw_GLE} can be interfaced with the Keldysh description of electron dynamics if cast into a path integral form \cite{martinstatisticaldynamics1973}. We are thus looking for an action $\mathcal{S}_\w[n_\w]$ such that the mean value of any observable $\mathcal{O}[n_\w]$ can be computed as 
\begin{equation}
\langle \mathcal{O} \rangle = \int [D n_\w] \mathcal{O}[n_\w] e^{i \mathcal{S}_\w[n_\w]}. 
\end{equation}
To this end, we rewrite the dynamical equation as 
\begin{equation}
\int \d x' \, (\chi_\w^{\R})^{-1}(x,x') n_\w(x') = \phi_{\rm ext}(x) + \zeta(x), 
\label{eq:nw_zeta}
\end{equation}
where we have used the shorthand $\int \d x = \int \d \r \d t, \int \d x' = \int \d \r' \d t'$ and
\begin{equation}
\langle \zeta(x) \zeta(x') \rangle = \left[(\chi_\w^{\R})^{-1} \cdot S_\w \cdot (\chi_\w^{\A})^{-1} \right](x,x') \equiv C_\w(x,x'), 
\end{equation}
with $\chi_\w^{\A}(x,x') \equiv \chi_\w^{\R}(x',x)$. Since the noise trajectories $\zeta(x)$ are sampled from a Gaussian distribution, one has,
\begin{equation}
\langle \mathcal{O} \rangle = \int [D \zeta] \O[n_\w[\zeta]] e^{- \frac{1}{2}\int \d x \d x' \, [\zeta \cdot C_\w^{-1} \cdot \zeta](x,x') }, 
\end{equation}
where $n_\w[\zeta]$ is the solution of Eq.~\eqref{eq:nw_zeta} for a given noise realization $\zeta$. Introducting a functional $\delta$, 
\begin{equation}
	\begin{aligned}
\langle \mathcal{O} \rangle = &\int [D \zeta] [D n_\w] \O[n_\w] e^{- \frac{1}{2}\int \d x \d x' \, [\zeta \cdot C_\w^{-1} \cdot \zeta](x,x') } \dots \\
\dots \,  &\delta\left[\zeta(x) + \phi_{\rm ext}(x) - [(\chi_\w^{\R})^{-1} \cdot n_\w](x)\right]. 
	\end{aligned}
\end{equation}
Now, using the integral representation of the $\delta$ function,
\begin{equation}
	\begin{aligned}
\langle \mathcal{O} \rangle = &\int [D \zeta][D n_\w] [D \tilde n_\w] \O[n_\w] e^{- \frac{1}{2}\int \d x \d x' \, [\zeta \cdot C_\w^{-1} \cdot \zeta] } \dots \\
\dots \,  &e^{-i \int \d x \, \tilde n_\w (x) \left[\zeta(x) + \phi_{\rm ext}(x) - [(\chi_\w^{\R})^{-1} \cdot n_\w](x)\right]}. 
	\end{aligned}
\end{equation}  
Carrying out the Gaussian integral over $\zeta$, one reaches the desired result: 
\begin{equation}
	\begin{aligned}
\langle \mathcal{O} \rangle = &\int [D n_\w] [D \tilde n_\w] \O[n_\w] e^{- \frac{1}{2}\int \d x \d x' \, [\tilde n \cdot C_\w \cdot \tilde n](x,x') } \dots \\
\dots \,  &e^{i \int \d x \d x' \, [\tilde n_\w \cdot (\chi_\w^{\R})^{-1}\cdot n_\w](x,x')}  e^{-i \int \d x \, \tilde n_\w(x) \phi_{\rm ext}(x)}.
	\end{aligned}
\end{equation} 
This is the Martin-Siggia-Rose-Janssen-DeDominicis (MSRJD) representation of Eq.~\eqref{eq:nw_GLE}. The MSRJD action can be cast into a matrix form that bears a strong analogy with the electronic Keldysh action derived in Eq.~\eqref{Se}: 
\begin{equation}
	\begin{split}
\mathcal{S}_\w[n_\w, \tilde n_\w] = &\frac{1}{2} \int \d x \d x' \, \left[ \n_\w \cdot \boldsymbol{\chi}_\w^{-1} \cdot \n_\w \right] (x, x')\\ &- \int \d x \, \tilde n_\w(x) \phi_{\ext}(x)
	\end{split}
\label{eq:Sw}
\end{equation}
with $\n = (n_\w \, \tilde n_\w)$ and 
\begin{equation}
\boldsymbol{\chi} = \left(
	\begin{array}{cc}
		-iS_\w & \chi_\w^{\R} \\
		\chi_\w^{\A} & 0 
	\end{array}
\right). 
\end{equation}

However, the MSRJD action still describes classical dynamics, and to be compatible with the electronic Keldysh action it needs to be quantized. The quantization turns out to be formally straightforward: it amounts to replacing the classical FDT in Eq.~\eqref{eq:FDT_S_class} with its quantum version: 
\begin{equation}
S_\w(\r, \r', \omega) = -\hbar \, \coth \left(\frac{\hbar \omega}{2 \kB T}\right) \im{\chi^{\R}_\w}(\r, \r', \omega). 
\end{equation}
To clarify the physical meaning of this quantization procedure, we specialize to the case of a two-dimensional liquid layer. With translational invariance, one may Fourier-transform Eq.~\eqref{eq:nw_zeta} in space and in time to obtain independent algebraic equations for each of the Fourier modes (in the absence of external potential):
\begin{equation}
\frac{1}{\chi_\w^{\R}(\q, \omega)} n_\w(\q, \omega) =  \zeta(\q, \omega)
\end{equation}
The response function $\chi^{\R}$ can always be cast in the form
\begin{equation}
[\chi_\w^{\R}]^{-1}(\q, \omega) =  (k_{\q} + i \omega \gamma_{\q}(\omega) - m_{\q} \omega^2). 
\end{equation}
For instance, a constant $\gamma_{\q}$ corresponds to the common damped harmonic oscillator model for liquid charge fluctuations. A combination of damped harmonic oscillators may be represented with a frequency-dependent $\gamma_{\q}$. Fourier-transforming back to time, the dynamical equation for a mode $n_\w$ can be recast as 
\begin{equation}
\begin{split}
m_{\q} \partial^2_t n_\w(\q, t) = &- k_{\q} n_\w(\q, t) + \zeta(\q, t) \\
&- \int_{-\infty}^t \d t' \, \gamma_{\q} (t-t') \partial_t n(\q, t').
\end{split}
\end{equation}
Thus, the action in Eq.~\eqref{eq:Sw} describes the dynamics of a collection of independent modes $\q$. The amplitude $n_\w(\q, t)$ in each mode behaves as the position of a particle of mass $m_{\q}$, confined in a potential $(1/2)k_{\q}n_\w^2$, that undergoes non-Markovian Langevin dynamics with friction kernel $\gamma_{\q}(\omega)$.
 Microscopically, such dynamics arise if a particle is coupled to a bath of harmonic oscillators (Caldeira-Legett model): if $q_j$ represent the positions of harmonic oscillators with frequencies $\omega_j$ and masses $m_j$, and they couple to $n_\w$ via a force $\sum_j c_j q_j$, then their effect amounts to a non-Markovian friction with kernel 
\begin{equation}
\gamma(t) = \sum_j \frac{c_j^2}{m_j \omega_j^2} \cos(\omega_j t),
\end{equation}
plus the corresponding fluctuation term. An arbitrary friction kernel can be obtained by letting the number of oscillators go to infinity. Then, defining the bath spectral density 
$J(\omega)=\frac{\pi}{2}\sum_j \frac{c_j^{2}}{m_j\omega_j}\,\delta(\omega-\omega_j)$, 
\begin{equation}
\gamma(t) =\frac{2}{\pi}\int_{0}^{\infty} \d\omega \,\frac{J(\omega)}{\omega}\cos(\omega t). 
\end{equation}
Harmonic oscillators have a well-defined quantization procedure. Starting from a quantum version of the Caldeira-Legett model, one may construct the Keldysh action according to the standard procedure, obtaining indeed Eq.~\eqref{eq:Sw} with the classical FDT replaced with its quantum version. 

We note that the generalized Langevin description for liquid charge fluctuations does not necessarily rely on a Gaussian fluctuation assumption; it can be shown to hold on more general grounds using the Mori-Zwanzig projection framework. However, assuming Gaussian fluctuations has allowed us to represent the liquid by a collection of harmonic modes coupled to harmonic baths. The formal quantization of liquid dynamics then amounts to the quantization of all these harmonic oscillators. Elementary excitations in the harmonic modes can be represented as quasiparticles, dubbed \emph{hydrons} \cite{yuelectroncooling2023,kingquantumfrictionwater2023}. Hydrons bear analogy with phonons, with the important difference that they typically have very short lifetimes (very broad spectra) due to the harmonic bath coupling. 

\subsection{Electron-liquid action, observables and Feynman rules}

Adding together the electronic action in Eq.~\eqref{eq:Se} and the formally quantized liquid action in Eq.~\eqref{eq:Sw},  one obtains a Keldysh action for the electron-liquid system, which still misses, however, the electron-liquid interaction term. From the point of view of the liquid, the electrons produce an external potential $\phi_{\rm ext}(x) = -\int \d x' V(x - x') n_\e(x')$, corresponding to a term $ \int \d x \d x' \, \tilde n_\w(x) V(x - x') n_\e(x')$ in the liquid action. But from the point of view of the electrons, it is the liquid that produces an external potential, corresponding to a term $ \int \d x \d x' \, \tilde n_\e(x) V(x - x') n_\w(x')$ in the electronic action. Altogether, using the symmetry property $V(x-x') = V(x'-x)$ the Keldysh action for the coupled electron-liquid system reads 
\begin{equation}
\begin{aligned}
\calS[\boldsymbol{\psi}, &\bar{\boldsymbol{\psi}}, n_\w] =  \frac{1}{\hbar}\int \d x \d x' \sum_{\sigma} \, \left[ \bar{\boldsymbol{\psi}}_{\sigma} \cdot \mathbf{G}^{-1}_{0,\sigma} \cdot \boldsymbol{\psi}_{\sigma} \right] (x, x') \\
&+\frac{1}{2} \int \d x \d x' \, \left[ \n_\w \cdot \boldsymbol{\chi}_\w^{-1} \cdot \n_\w \right] (x, x') \\
& - \int \d x \d x' \, V(x - x')\tilde n_\e(x)  n_\e(x') \\
& + \int \d x \d x' \, V(x - x')[ \tilde n_\e(x) n_\w(x') + \tilde n_\w(x) n_\e(x')].
\end{aligned}
\label{eq:Sew}
\end{equation}
This action can describe the coupled system in the presence of any non-equilibrium perturbation, such as an external pressure that induces a liquid flow field $\mathbf{v}(\r)$ in the steady state. The flow field is taken into account by putting the equilibrium charge density fluctuations in a moving reference frame: $n_\w(\r,t) = n_\w^{\rm eq} (\r + \mathbf{v}t)$. In practice, when the lenghtscale on which $\v(\r)$ varies is much longer than the lengthscale of solid-liquid interactions (1-2 liquid layers), one may assume that the velocity $\v$ is constant and equal to the interfacial velocity. Then, the change of reference frame for the liquid density is equivalent to a Doppler shift in the correlation functions: 
\begin{equation}
\boldsymbol{\chi}_\w(\q, z, z', \omega) \mapsto \boldsymbol{\chi}_\w(\q, z, z', \omega - \q \cdot \v), 
\label{eq:doppler}
\end{equation}
where $\q$ is the wavevector parallel to the interface. 

The key observable to be evaluated is the force exerted by the solid on the liquid  
\begin{equation}
\begin{split}
\langle \mathbf{F} \rangle (t) &= \int \d \r_\e \d \r_\w \, \nabla_{\r_\w}V(\r_\e - \r_\w) \langle n_\w (\r_\w, t) n_\e(\r_\e, t) \rangle \\
&\equiv \int \d \r_\e \d \r_\w \, \nabla_{\r_\w} V(\r_\e - \r_\w) S_{\e\w} (\r_\e, \r_\w, 0),
\end{split}
\label{eq:Esw}
\end{equation}
where we have defined the electron-liquid structure factor $S_{\e\w}$. 
One may also be interested in the electronic current induced in the solid by the liquid flow. In terms of the fermionic operators $\hat \psi(\r,t), \hat \psi^{\dagger}(\r, t)$, the electric current operator is 
\begin{equation}
\hat{\textbf{j}}(\r, t)=-\frac{i\hbar e}{2m_\e}\sum_\sigma \left[\hat \psi(\r,t)\nabla_{\r} \hat\psi^{\dagger}(\r,t)-\hat\psi^{\dagger}(\r,t)\nabla_{\r} \hat \psi(\r,t)\right], 
\end{equation}
where $m_\e$ is the electron mass. In a steady-state translationally invariant system, this definition leads to the expression, in space dimension $d$ \cite{coquinotquantumfeedback2023}, 
 \begin{equation}
 \langle \mathbf{j} \rangle = 2i e \int \frac{\d \q \d \omega}{(2\pi)^{d+1}} (\nabla_{\q} \xi_{\q})   G^{\K} (\q,\omega), 
 \label{current0}
 \end{equation}
 assuming that the electrons behave as well-defined quasiparticles with dispersion $\xi_{\q}$. Thus, the relevant observables for the nanoscale liquid-solid interface can be obtained from two-point correlation functions in the field theory defined by Eq.~\eqref{eq:Sew}.

\begin{figure}
	\begin{center}
		\includegraphics{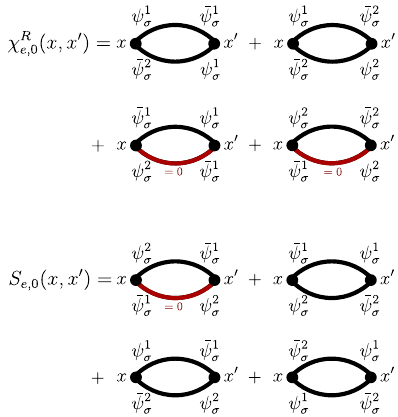}
		\caption{\textbf{Feynman diagrams for electron density propagator.} We represent the expansion of the non-interacting electron density propagators defined in Eq.~\eqref{eq:def chie} in terms of contractions of the fermionic operators (depicted by a plain line), after using Wick's theorem. Note that the contraction of the fermionic fields $\psi_\sigma^2$ and $\psibar_\sigma^1$ vanishes, as per the Feynman rules in Eq.~\eqref{eq:Ge0}. This yields Eqs.~\eqref{eq:chiR0} and \eqref{eq:exp Se0}.}\label{fig:diagrams electron density}
	\end{center}
\end{figure}

Observables can be evaluated perturbatively by expanding the path integral weight $e^{i\calS}$ in powers of the interaction terms and applying Wick's theorem, since without interactions the path integral is Gaussian. The expansion terms can be represented as Feynman diagrams, and the practical rules for their evaluation are as follows: 
\begin{itemize}
\item Every vertex brings a factor of $i$. Note the explicit expression of the liquid-electron vertex: 
\begin{equation}\label{eq:vertex feynman}
	\begin{split}
n_\w \tilde n_\e + \tilde n_\w n_\e = &\sum_{\sigma}\left[ \frac{n_\w \psibar^1_{\sigma} \psi^1_{\sigma}}{\hbar} + \frac{n_\w \psibar^2_{\sigma} \psi^2_{\sigma}}{\hbar}\right] \\+ &\sum_{\sigma}\left[\frac{\tilde n_\w \psibar^1_{\sigma} \psi^2_{\sigma}}{2} + \frac{\tilde n_\w \psibar^2_{\sigma} \psi^1_{\sigma}}{2}\right].
	\end{split}
\end{equation}
\item Contractions of electron fields yield the propagators: 
\begin{equation}
	\begin{aligned}
&\langle \psi_{\sigma}^1(x) \psibar_{\sigma}^1(x') \rangle_0 = i \hbar G_{0,\sigma}^{\R}(x,x') \\ &\langle \psi_{\sigma}^1(x) \psibar_{\sigma}^2(x') \rangle_0 = i \hbar G_{0,\sigma}^{\K}(x,x') \\
&\langle \psi^2(x) \psibar^1(x') \rangle_0 = 0 \\ &\langle \psi_{\sigma}^2(x) \psibar_{\sigma}^2(x') \rangle_0 = i \hbar G_{0,\sigma}^{\A}(x,x') \\
	\end{aligned}
	\label{eq:Ge0}
\end{equation}
\item Every loop of electron propagators brings a minus sign. 
\item Contractions of hydron fields yield the propagators: 
\begin{equation}
	\begin{aligned}
&\langle n_\w(x) \tilde n_\w(x') \rangle_0 = i\chi_\w^{\R}(x,x') = i\chi_\w^{\A}(x',x) \\
&\langle n_\w(x) n_\w(x') \rangle_0 = S_\w(x, x') \\ &\langle \tilde n_\w(x) \tilde n_\w(x') \rangle_0 = 0 
\label{eq:chiw0}
	\end{aligned}
\end{equation}
\item Integrals (and sums over spin) are taken over the coordinates of all vertices. 
\end{itemize}
For contractions of the electron fields that correspond to contractions of the electron density, it is convenient to use the electron density propagators: 
\begin{equation}\label{eq:def chie}
	\begin{aligned}
&\langle n_\e(x) \tilde n_\e(x') \rangle_0 = i\chi_{\e,0}^{\R}(x,x') = i\chi_{\e,0}^{\A}(x',x) \\
&\langle n_\e(x) n_\e(x') \rangle_0 = S_{\e,0}(x, x') \\ &\langle \tilde n_\e(x) \tilde n_\e(x') \rangle_0 = 0. \\
	\end{aligned}
\end{equation}
These can be obtained in terms of the bare electron Green's functions by using the expressions of the density fields in terms of the fermionic fields in Eq.~\eqref{eq:npsi} and the definitions of the Green's functions in Eq.~\eqref{eq:Ge0}. Diagrammatically, computing an electron density correlator corresponds to evaluating a "bubble" diagram formed by two electron Green's functions (Fig.~\ref{fig:diagrams electron density}). One obtains: 
\begin{equation}
	\begin{split}
\chi^{\R}_{\e,0}(x,x') = \frac{i \hbar}{2} \sum_\sigma &[G^{\R}_{0, \sigma}(x, x')G^{\K}_{0, \sigma}(x',x) \\ &+ G^{\K}_{0, \sigma}(x, x')G^{\A}_{0, \sigma}(x',x) ],
	\end{split}
\label{eq:chiR0} 
\end{equation}
\begin{align}
	&
\begin{aligned}
S_{\e,0}(x&,x') = -\frac{\hbar^2}{4}\sum_\sigma \Big[ G^{\K}_{0, \sigma}(x,x')G^{\K}_{0, \sigma}(x',x) \\
        & + G^{\R}_{0, \sigma}(x,x')G^{\A}_{0, \sigma}(x',x)  + G^{\A}_{0, \sigma}(x,x')G^{\R}_{0, \sigma}(x',x) \Big],
        \label{eq:exp Se0}
\end{aligned}\\
&
\begin{aligned}
\chi^{\A}_{\e,0}(x,x') &= \chi^{\R}_{\e,0}(x',x).
\end{aligned}
\label{eq:chiA0}
\end{align}
assuming that the bare Green's functions do not depend on spin. Using the expressions of the bare Green's functions in Eq.~\eqref{GReq}, 
\begin{equation}
\chi^{\R(\A)}_{\e,0}(\q, \omega) = 2\int \frac{\d \k}{(2\pi)^d} \frac{n_F(\xi_{\k + \q}) - n_F(\xi_{\k})}{\hbar \omega \pm i 0^+ + \xi_{\k} - \xi_{\k + \q}},
\end{equation}
which is exactly the expression given in Eq.~\eqref{eq:chi0}, with the matrix elements $| \langle \k + \q | e^{i \q \cdot \r} | \k \rangle |^2 = 1$. 

As one often considers systems in a non-equilibrium steady state, it is practical to apply Feynman rules directly in frequency space. Then, one assigns a frequency arrow to each propagator and imposes frequency conservation at the vertices. By convention, an arrow with frequency $\omega$ pointing towards a $\psibar$ field yields a Green's function $G(\omega)$. Similarly, for density fields, an $\omega$ arrow pointing towards a $\tilde n$ field yields a response function $\chi^{\R}(\omega)$. A normalized frequency integral ($\int \frac{\d \omega}{2 \pi}$) is then carried out over all internal frequencies. 

Since a liquid-solid system is not translationally invariant parallel to the interface, it is in general not possible to apply Feynman rules in momentum space. However, a simplification is still possible if one can define a flat dividing surface between the liquid and the solid. Placing the dividing surface at $z = 0$, so that the solid is at $z <0$ and the liquid at $z > 0$, we define the "surface densities" as 
\begin{align}
&\n_\w^{\circ}(\q, t) = \int_0^{+\infty} \d z \, e^{-q z}\, \n_\w(\q, z), \\
& \n_\e^{\circ}(\q, t) = \int_{-\infty}^{0} \d z \, e^{q z}\, \n_\e(\q, z). \\
\end{align}
Then, noting that 
\begin{equation}
	\begin{split}
V(q, z, z') &= \frac{e^2}{2 \epsilon_0 q} e^{-q | z- z'|} \\
&\equiv V(q) e^{-q | z- z'|},
	\end{split}
\end{equation}
the liquid-electron interaction term in the action \eqref{eq:Sew} can be recast as 
\begin{equation}
	\begin{split}
\calS_{\e\w}^{\rm int} = \int \frac{\d t \d \q}{(2\pi)^2} \, V(q) &[ \tilde n_\e^{\circ} (-\q, t) n_\w^{\circ}(\q, t) \\ &+ \tilde n_\w^{\circ} (-\q, t) n_\e^{\circ}(\q, t) ]
	\end{split}
\end{equation}
Thus, the perturbation series in the liquid-electron interaction can be expressed in terms of surface density correlators, that are none other than the surface response functions defined in Sec. III: 
\begin{equation}
	\begin{aligned}
&V(q) \langle n_{\e,\w}^{\circ} \tilde n_{\e,\w}^{\circ} \rangle_{\q, \omega} = -ig_{\e,\w}^{\R}(\q, \omega) = -ig_{\e,\w}^{\A}(\q, -\omega) \\
&V(q) \langle n_{\e,\w}^{\circ} n_{\e,\w}^{\circ} \rangle_{\q, \omega} \equiv -g_{\e,\w}^{\S}(\q,\omega)
	\end{aligned}
\end{equation}
with the FDT at equilibrium:
\begin{equation}
g^{\S}_{\e,\w}(\q, \omega) = - \hbar \, \coth \left(\frac{\hbar \omega}{2 \kB T}\right) \im{g^{\R}_{\e,\w}}(\q, \omega). 
\label{eq:FDT_g}
\end{equation}

\subsection{Quantum friction and van der Waals attraction from Keldysh formalism}

According to Eq.~\eqref{eq:Esw}, the evaluation of solid-liquid forces  reduces to the computation of the liquid-electron surface density correlator: 
\begin{equation}
\left[ \begin{array}{c}
	\langle F^z \rangle \\
	\langle \mathbf{F}^{\parallel} \rangle
\end{array}
	\right]
=  \mathcal{A} \int \frac{\d \q \d \omega}{(2\pi)^3} 
\left[ \begin{array}{c}
	 q  \\
	 i \q 
\end{array}
	\right]
 g^{\S}_{\e\w}(\q, \omega). 
\end{equation}
Diagrammatically, the perturbative expansion for $\mathbf{g}_{\e\w}$ can be represented as a self-consistent Dyson equation: 
\begin{equation}
\includegraphics[scale=0.8]{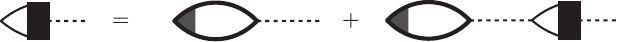}
\label{eq:dyson_ew}
\end{equation}
where the dashed line represents the liquid surface response function $\mathbf{g}_\w$ and the bubble represents the electron surface response function $\mathbf{g}_\e$. Note that the latter is not a bare response function: $\mathbf{g}_\e$ is completely irreducible with respect to the liquid-electron interaction (it cannot be made disconnected by cutting one hydron propagator), but fully renormalized within that constraint. A common choice is to renormalize $\mathbf{g}_\e$ by the electron-electron interactions in the random phase approximation (see Sec. V), which amounts diagramatically to another self-consistent Dyson equation:
\begin{equation}
\includegraphics[scale=0.8]{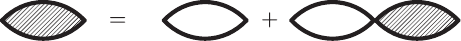}
\label{eq:RPA_Dyson}
\end{equation}
Applying the Feynman rules, Eq.~\eqref{eq:dyson_ew} becomes
\begin{align}
&g_{\e\w}^{\S} = (g_\e^{\R} \cdot g_\w^{\S} + g_\e^{\S} \cdot g_\w^{\A})(1 + g_{\e\w}^{\A}) + g_\e^{\R} \cdot g_\w^{\R} \cdot g_{\e\w}^{\S}
\label{eq:dyson_ew_S}
 \\
& g_{\e\w}^{\R,\A} = g_\e^{\R,\A} \cdot g_\w^{\R,\A} + g_\e^{\R,\A} \cdot g_\w^{\R,\A} \cdot g_{\e\w}^{\R,\A}.
\label{eq:dyson_ew_A}
\end{align}
Using the property $g^{\A}(\q,\omega) = g^{\R}(\q, \omega)^*$, the force components read 
\begin{equation}
\left[ \begin{array}{c}
	\langle F^z \rangle \\
	\langle \mathbf{F}^{\parallel} \rangle
\end{array}
	\right]
=  \mathcal{A} \int \frac{\d \q \d \omega}{(2\pi)^3} 
\left[ \begin{array}{c}
	 -q  \\
	 i \q 
\end{array}
	\right]
 \left(\frac{g_\w^{\R} \cdot g_\e^{\S}+g_\w^{\S} \cdot {g}_\e^{\A}}{|1-g_\w^{\R} \cdot g_\e^{\R}|^2} \right)_{\q, \omega}. 
 \label{eq:F_components}
\end{equation}

At equilibrium (in the absence of flow), with the FDT in Eq.~\eqref{eq:FDT_g}, it is only a matter of algebra to obtain the attractive van der Waals force between the solid and the liquid: 
\begin{equation}
\frac{\langle F^z \rangle}{\mathcal{A}} =  -\int \frac{\d \q \d \omega}{(2\pi)^3}  \frac{\hbar q}{\mathrm{tanh}(\hbar \omega/ 2 \kB T)}\left( \frac{\im{g_\w^{\R} \cdot g_\e^{\R}}}{| 1 - g_\e^{\R} g_\w^{\R}|^2} \right)_{\q, \omega}. 
\end{equation}
This result generalizes the Lifshitz formula in Eq.~\eqref{eq:vdw_lifshitz} to media with non-local dielectric response, and to arbitrary order in the Coulomb interaction between the media. 

In the presence of flow with interfacial velocity $\v = v \mathbf{e}_x$, the surface response response functions in Eq.~\eqref{eq:F_components} refer to the non-equilibrium steady state and in principle do not satisfy the FDT. For the liquid, we use the Doppler shift prescription in Eq.~\eqref{eq:doppler}, which results in a Doppler-shifted FDT. We will discuss in the following the non-equilibrium state of the solid under liquid flow; for now, we assume that the solid stays at equilibrium so that its response functions satisfy the FDT. Then, expanding to linear order in $\v$, the friction force reads \cite{kavokinefluctuationinducedquantum2022}
\begin{equation}
	\begin{split}
\frac{\langle F_x \rangle }{\mathcal{A}}  =  - v \frac{\hbar^2}{2 \kB T} &\int \frac{\d \q \d \omega}{(2\pi)^3} \frac{q_x^2}{\mathrm{\sinh}^2 (\hbar \omega / 2 \kB T)} \dots \\ \dots & \left( \frac{\im{g_\w^{\R}} \im{ g_\e^{\R}}}{| 1 - g_\e^{\R} g_\w^{\R}|^2} \right)_{\q, \omega}. 
	\end{split}
\label{eq:QF_result}
\end{equation}
The derivation of quantum friction from the Keldysh framework formally recovers the results obtained by Volokitin and Persson from Lifshitz-like theory (Sec. III). But, importantly, the Keldysh theory is fully microscopic -- the electrons are described at the level of their creation and annihilation operators -- while the Lifshitz framework is coarse-grained: the solid's dielectric function (or, equivalently, the surface response function $g_\e$) is treated as an input of the theory. Within the Keldysh theory, $g_\e$ can be renormalized by interactions with the liquid, allowing for the description of liquid-flow-induced electric currents and further hydron-electron correlation effects. Before discussing these effects in detail, we show how a Lifshitz-like theory emerges by taking the limit of Gaussian density fluctuations in the Keldysh action. 

\subsection{From Keldysh to Lifshitz-like theory}

Consider the electronic part of the partition function: 
\begin{equation}
Z_\e = \int [D \bpsibar] [D \bpsi] e^{i \S_\e [\bpsibar,\bpsi]}, 
\end{equation}
with the action 
\begin{equation}
\S_\e[\bpsi, \bpsibar] =  \bpsibar \cdot (\G_0^{-1} / \hbar) \cdot \bpsi - \tilde n_\e \cdot V \cdot n_\e. 
\end{equation}
Here, we use a compact notation where integration over the space and time coordinates of the fields that appear in the action is implied. For the quantities in bold, that represent vectors or matrices in Keldysh space, we imply summation over the Keldysh indices. In any bilinear combination of $\psibar_{\sigma}, \psi_{\sigma}$, we further imply summation over spin indices: $\bpsibar \cdot \mathbf{A} \cdot \bpsi \equiv \sum_\sigma \bpsibar_{\sigma} \cdot \mathbf{A} \cdot \bpsi_{\sigma}$. The density fields are expressed in terms of the fermionic fields according to Eq.~\eqref{eq:npsi}, which can be recast in matrix form as $n_\e = \bpsibar \cdot \N \cdot \bpsi$, $\tilde n_\e = \bpsibar \cdot \tilde \N \cdot \bpsi$, with $\tilde \N = \boldsymbol{1}/\hbar$ and $\N = \boldsymbol{\sigma}_x / 2$, $\boldsymbol{\sigma}_x$ being the first Pauli matrix. We decouple the density fields by introducing delta functions: 
\begin{equation}
\begin{split}
Z_\e = \int &[D \bpsibar] [D \bpsi] [D \n_\e] e^{i \left[ \bpsibar \cdot (\G_0^{-1} / \hbar) \cdot \bpsi - \tilde n_\e \cdot V \cdot n_\e\right] }\dots \\
\dots \, &\delta(n_\e - \bpsibar \cdot \N \cdot \bpsi)   \delta (\tilde n_\e - \bpsibar \cdot \tilde \N \cdot \bpsi). 
\end{split}
\end{equation}
Using Fourier representations for the delta functions, 
\begin{equation}
\begin{split}
Z_\e = \int &[D \bpsibar][D \bpsi][D \n_\e][D \boldsymbol{\phi}] \dots \\
& \exp \Bigl\{ i \Bigl[ 
\bpsibar \cdot \Bigl( \G_0^{-1}/\hbar 
- \tilde \phi\, \N - \phi\, \tilde \N \Bigr) \cdot \bpsi + \dots \\
&
+ \tilde \phi \cdot n_\e 
+ \phi \cdot \tilde n_\e 
- \tilde n_\e \cdot V \cdot n_\e
\Bigr] \Bigr\}.
\end{split}
\end{equation}
Carrying out the Gaussian integral over the fermionic fields and using the identity $\log \det (\dots) = \mathrm{Tr} \log (\dots)$ yields 
\begin{equation}
\begin{split}
Z_\e = \int &[D \n_\e][D \boldsymbol{\phi}] \dots \\
& \exp \Bigl\{ i \Bigl[
- \mathrm{Tr} \log \Bigl( \G_0^{-1}/\hbar
- \tilde \phi\, \N - \phi\, \tilde \N \Bigr) + \dots \\
&
+ \tilde \phi \cdot n_\e
+ \phi \cdot \tilde n_\e
- \tilde n_\e \cdot V \cdot n_\e
\Bigr] \Bigr\}.
\end{split}
\end{equation}
At this point, we have carried out a change of variable in the path integral, $(\bpsibar, \bpsi) \mapsto (\n_\e, \boldsymbol{\phi})$. Taking the limit of Gaussian density fluctuations corresponds to expanding the action $\S_\e[\n_\e, \boldsymbol{\phi}]$ around its saddle point $(\n_\e^{\star}, \boldsymbol{\phi}^{\star})$ to quadratic order. Letting $\n_\e = \n_\e^{\star} + \delta \n_\e$ and $\bphi = \bphi^{\star} + \delta \bphi$,
\begin{equation}
\begin{split}
&S_\e[\n_\e,\bphi_\e] \approx\;
S_\e[\n_\e^{\star},\bphi^{\star}] \dots  \\
& - \frac{1}{2}\,
\mathrm{Tr} \Bigl[
\hbar \G_0 \cdot
\bigl( \delta \tilde \phi\, \N + \delta \phi\, \tilde \N \bigr)
\cdot \hbar \G_0 \cdot
\bigl( \delta \tilde \phi\, \N + \delta \phi\, \tilde \N \bigr)
\Bigr] \\
& + \delta \tilde \phi \cdot \delta n_\e
+ \delta \phi \cdot \delta \tilde n_\e
- \delta \tilde n_\e \cdot V \cdot \delta n_\e .
\end{split}
\end{equation}
since any terms that are linear in $\delta n_\e, \delta \phi$ must vanish at the saddle point. Now, by expanding the trace, one may identify the electron density correlators as defined in Eqs.~\eqref{eq:chiR0}-\eqref{eq:chiA0}: 
\begin{equation}
\begin{split}
S_\e[\n_\e,\bphi_\e] =\;&
S_\e[\n_\e^{\star},\bphi^{\star}] \dots \\
& - \frac{1}{2}\,
\delta\bphi \cdot
\boldsymbol{\sigma}_x \boldsymbol{\chi}_{\e,0} \boldsymbol{\sigma}_x
\cdot \delta\bphi \\
& + \delta \tilde\bphi \cdot \boldsymbol{\sigma}_x \cdot \delta \n_\e
- \delta \n_\e \cdot V \boldsymbol{\sigma}_x \cdot \delta \n_\e .
\end{split}
\end{equation}
with 
\begin{equation}
\boldsymbol{\chi}_{\e,0} = \left(
	\begin{array}{cc}
		-iS_{\e,0} & \chi_{\e,0}^{\R} \\
		\chi_{\e,0}^{\A} & 0 
	\end{array}
\right). 
\end{equation}
Finally, carrying out the Gaussian integral over $\phi$, one obtains an effective action for the electronic density fluctuations: 
\begin{equation}
Z_\e = \int [D \delta \n_\e]\,  \exp \left\{\frac{i}{2} \delta \n_\e  \cdot \boldsymbol{\chi}_{\e, \rm RPA}^{-1} \cdot \delta \n_\e \right\}, 
\end{equation}
with 
\begin{equation}
\boldsymbol{\chi}_{\e, \rm RPA}^{-1} = \boldsymbol{\chi}_{\e,0}^{-1} - V \boldsymbol{\sigma}_x, 
\end{equation}
which is exactly the Dyson equation that defines the random phase approximation (Eq.~\eqref{eq:RPA_Dyson}). 

By adding the liquid's degrees of freedom into this electronic RPA effective action, one obtains an action for the electron-liquid system that has the physical content of Lifshitz theory: 
\begin{equation}
\begin{aligned}
\calS[\n_\e, n_\w] &=  \frac{1}{2} \int \d x \d x' \, \left[ \n_\e \cdot \boldsymbol{\chi}_{\e, \rm RPA}^{-1} \cdot \n_\e \right] (x, x') \\
&+\frac{1}{2} \int \d x \d x' \, \left[ \n_\w \cdot \boldsymbol{\chi}_\w^{-1} \cdot \n_\w \right] (x, x') \\
& + \int \d x \d x' \, V(x - x')[ \tilde n_\e(x) n_\w(x') + \tilde n_\w(x) n_\e(x')].
\end{aligned}
\label{eq:Sew_RPA}
\end{equation}
Here, both the liquid and the electrons are fully described by the two-point functions of their density fluctuations. The above derivation highlights that the electron density fluctuations can be assumed Gaussian if their two-point function is well-described within RPA. However, the action in Eq.~\eqref{eq:Sew_RPA} could be obtained phenomenologically, by treating the electrons exactly as we treated the liquid: assume that fluctuations are Gaussian and write down the corresponding action, with the constraints of causality and microscopic reversibility (FDT). In fact, Eq.~\eqref{eq:Sew_RPA} corresponds to the (quantized) MSRJD representation of the dynamical equations \eqref{nsw_dynamics}, that were used as the starting point for our discussion of liquid-solid friction in Sec. IV. The main limitation of Eq.~\eqref{eq:Sew_RPA} is its inability to capture the non-equilibrium state of the electronic system upon interaction with the liquid flow: namely, the flow-induced electronic current discussed in the next section. 

\subsection{Flow-induced electric current from electron-hydron self-energy}

\begin{figure}
	\begin{center}
		\includegraphics{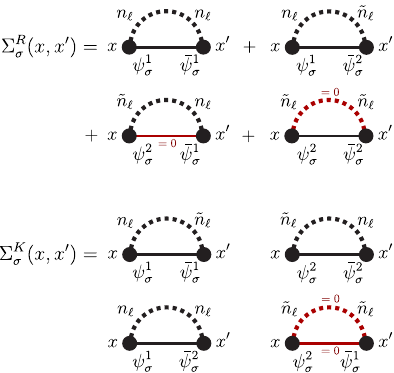}
		\caption{\textbf{Diagrammatic representation of the electron-hydron self-energy.} We represent the electron self-energy due to the Coulomb interaction with hydrons, as defined in Eq.~\eqref{dyson}, according to the Feynman rules for the interaction vertex given by Eq.~\eqref{eq:vertex feynman}. Full lines stand for the bare electronic Green's functions $\mathbf{G}_{0,\sigma}$, and dashed lines for bosonic liquid density contractions $D_\w$ as defined in Eq.~\eqref{eq:def Dw}.}\label{fig:electron self energy}
	\end{center}
\end{figure}

\begin{figure}
	\begin{center}
		\includegraphics[width=\columnwidth]{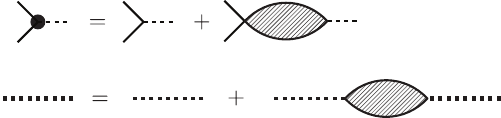}
		\caption{Diagrammatic definition of the renormalized surface response function $\tilde g_\w$ in Eq.~\eqref{eq:g_tilde} (bold dashed line including vertices, see also Fig.~\ref{fig:electron self energy}). The thin dashed line represents the bare surface response function $g_\w$.}\label{fig:sigma_diag}
	\end{center}
\end{figure}

The current transported by the electrons can be evaluated in terms of the Keldysh component of their non-equilibrium Green's function as per Eq.~\eqref{current0}. The perturbation series for the Green's function can alway be partially resummed by introducing a self-energy $\boldsymbol{\Sigma}$: 
\begin{equation}
\G_{\sigma} = \G_{\sigma,0} + \G_{\sigma,0} \cdot \boldsymbol{\Sigma}_{\sigma} \cdot \G_{\sigma}, 
\label{dyson}
\end{equation}
Within the trigonal (Larkin-Ovchinnikov) representation chosen here, the self-energy has the same matrix structure as the Green's function: 
\begin{equation}
	\boldsymbol{\Sigma}_{\sigma} = 
 \left(
		\begin{array}{cc}
			\Sigma^{\R}_{\sigma} & \Sigma^{\K}_{\sigma}  \\
			0 & \Sigma^{\A}_{\sigma}
		\end{array}
	\right)
\end{equation}
The lowest order contribution to the electron-hydron self-energy corresponds to the "Fock" diagram (Fig.~\ref{fig:electron self energy}). 
Applying the Feynman rules in Eqs.~\eqref{eq:Ge0}-\eqref{eq:chiw0} yields 
\begin{equation}
\begin{split}
\Sigma^{\R, \A}_{\sigma}(x,x') = \frac{1}{2}& \left[ D_\w^{\R,\A}(x,x') G_{\sigma}^{\K}(x,x')  \right. \\ & \left. + D_\w^{\K}(x,x')G_{\sigma}^{\R,\A}(x,x')\right] 	
\end{split} 
\end{equation}
\begin{equation}
	\begin{split}
\Sigma_{\sigma}^{\K}(x,x') = \frac{1}{2}[&D_\w^{\R}(x,x') G_{\sigma}^{\R}(x,x') + D_\w^{\A}(x,x')G_{\sigma}^{\A}(x,x') \\ &+ D_\w^{\K}(x,x')G_{\sigma}^{\K}(x,x')],  
	\end{split}
\end{equation}
with
\begin{equation}\label{eq:def Dw}
	\begin{split}
	\left(
\begin{array}{cc}
D_\w^{\K} & D_\w^{\R} \\
 D_\w^{\A} & 0
\end{array}
\right)_{(x_1,x_2)} 
= 
&\int \d \r \d \r' V(\r - \r_1) V(\r' - \r_2) \dots \\
&\dots	\left(
\begin{array}{cc}
-2iS_\w/\hbar & \chi_\w^{\R} \\
\chi_\w^{\A} & 0
\end{array}
\right)_{(\r, \r', t_1 - t_2)}
\end{split}
\end{equation}
The further evaluation of the self-energy is cumbersome in the general case of a three-dimensional electronic system, but tractable expressions have been obtained in \cite{coquinotquantumfeedback2023} for a two-dimensional solid in contact with a semi-infinite liquid. In that case, the self-energy is expressed in terms of the liquid's surface response function $g_\w^{\R}(\q, \omega)$. Note that, to account for self-consistent screening between the liquid and the solid at the same level as in the evaluation of the friction force, Coquinot \emph{\emph{et al.}} use a renormalized surface response function 
\begin{equation}
\tilde g_\w^{\R}(\q, \omega) = \frac{g_\w^{\R}(\q, \omega)[ 1+ V(q) \chi_{\e, \rm RPA}^{\R}(\q, \omega)]^2 }{1 + V(q) g_\w^{\R}(\q, \omega) \chi_{\e, \rm RPA}^{\R} (\q, \omega)},
\label{eq:g_tilde}
\end{equation}
which contains the contribution of all the diagrams in Fig. \ref{fig:sigma_diag}. 

In the case where the typical energies of the hydron modes are smaller than the characteristic energies of the electron dynamics (Fermi energy, plasmon energy), electron-hydron scattering behaves essentially like impurity scattering, so that one may assume a simplified form for the self-energy: 
\begin{align}
& \Sigma^{\R,\A}(\q, \omega) \approx \Sigma^{\R,\A} (\q, \xi_{\q}) \approx \mp i \gamma^{\w}_{\q}, \\
& \Sigma^{\K}(\q, \omega) = 2i \, \tanh\left(\frac{\hbar(\omega - \q \v) }{2 \kB T}\right)\im{\Sigma^{\R}}(\q, \omega). 
\label{eq:sigmaK}
\end{align}
Here, $\gamma^{\w}_{\q}$ represents the hydron scattering rate for electrons at wavevector $\q$ (and energy $\xi_{\q}$), and $\v$ is the flow velocity. Eq.~\eqref{eq:sigmaK} represents a Doppler-shifted or \emph{quasi-equilibrium} FDT. Under these conditions, the Green's function itself also satisifies a quasi-equilibrium FDT 
\begin{equation}
G^{\K}(\q, \omega) = 2i \, \tanh\left(\frac{\hbar(\omega - \q \v_\e(\q)) }{2 \kB T}\right)\im{G^{\R}}(\q, \omega), 
\label{eq:quasiFDT}
\end{equation}
with the electronic drift velocity $\v_\e(\q)$ given by 
\begin{equation}
	\label{eq:ve_gamma}
\v_\e(\q) = \frac{\gamma^{\w}_{\q}}{\gamma_{\q}^{\w} + \gamma_{\q}^0} \v,
\end{equation}
$\gamma_{\q}^0$ being the intrinsic electron scattering rate (with impurities, phonons, other electrons) at wavevector $\q$. Then, Eq.~\eqref{current0} yields for the electric current 
\begin{equation}\label{currentexact}
\langle \mathbf{j} \rangle
=
\frac{e }{\kB T }
\int \frac{\d\omega \d \q}{(2\pi)^3}
\frac{[ \q \cdot \v_\e(\q)] \nabla_{\q} \xi_{\q} }
{\cosh^{2}\!\left[\dfrac{\hbar \omega+\xi_{\q}}{2\kB T}\right]}
\,
\frac{\gamma_{\q}}{\gamma_{\q}^{2} + \omega^{2}}. 
\end{equation}
Assuming an isotropic Fermi surface and small broadening of the electronic states ($\hbar \gamma_{\q}, \kB T \ll \xi_{k_F}$), one obtains the simplified expression 
for the  current density \cite{coquinotquantumfeedback2023}
 \begin{equation} \label{currentapprox}
\langle\textbf{j}\rangle\approx e  v_{\rm F} \times  N_F \times [ \hbar k_{\rm F} \v_{\e}(k_{\rm F}) / 2],  
\end{equation}
where $v_{\rm F}$ is the Fermi velocity and $N_{\rm F}$ is the density of states at the Fermi level. This expression highlights the microscopic meaning of the electronic drift velocity: $\Delta \epsilon = \hbar \kF \v_\e / 2$ represents the angle-averaged width of the energy window in which the non-equilibrium electron distribution is skewed. The number of electrons in this window is $N_F \Delta \epsilon$, and they each contribute $e v_{\rm F}$ to the electric current. For a 2D electron gas, $v_F = \hbar \kF /m $, $N_F = m / (\pi \hbar)^2$ and the total electron density is $n_\e = \kF^2 /(2 \pi)$, so that Eq.~\eqref{currentapprox} reduces to the Drude model result $\langle \mathbf{j} \rangle = n_\e e \v_\e$; one can check that this identification holds in the case of graphene as well. 

It is worth comparing the expression for the electronic drift velocity Eq.~\eqref{eq:ve_gamma} to the one obtained in Sec. V from momentum balance arguments:
\begin{equation}
\v_\e = \frac{\lambda_{\rm Q}}{\lambda_{\rm Q} + \lambda_\e^0}.
\end{equation} 
The two expressions identify if the friction coefficients $\lambda$ are proportional to the scattering rates $\gamma$. This is true by definition for $\lambda_\e^0$, but for $\lambda_{\rm Q}$ one can show that it is only the case if the electrons are non-interacting \cite{yuelectroncooling2023}. Indeed, the QF coefficient quantifies the rate of momentum transfer from the flowing liquid to the solid's electrons. If the electrons are interacting, then part of this momentum is transferred into collective (plasmon) modes, that do not carry electric current. The results of Sec. V should thus be used with caution in systems with a strong plasmonic contribution to quantum friction. Nevertheless, QF preferentially transfers momentum to short-wavelength plasmons that are often Landau-damped, so that a large part of this momentum may still contribute to HE drag. 

\subsection{Non-equilibrium quantum friction as Landauer transport of hydrons}\label{section quasiparticle picture}

In the presence of a flow-induced electronic current, or if the solid is set out of equilibrium through any other external influence, the quantum friction coefficient can no longer be evaluated according to Eq.~\eqref{eq:QF_result}, since the electronic surface response function no longer satisfies the FDT. The friction force then takes the general form \cite{coquinotmomentumtunnelling2025}:
\begin{equation}
\frac{\langle\textbf{F}_\parallel\rangle}{\mathcal{A}}=-\int\frac{\d\q}{(2\pi)^2} (\hbar\q)  (\Gamma_{\w \rightarrow  \e}(\q) - \Gamma_{\e \rightarrow \w}(\q))
\label{eq:noneqQF}
\end{equation}
with
\begin{equation}
 \Gamma_{\rm a\rightarrow b}(\q)=\int\frac{\d\omega}{2\pi\hbar} \frac{\im{g_{\rm b}^{\R}(\q,\omega)}g^{\S}_{\rm a}(\q,\omega)}{|1-g_{\rm a}^{\R}(\q,\omega)g_{\rm b}^{\R}(\q,\omega)|^2}.
\end{equation}
The result in Eq.~\eqref{eq:noneqQF} is, in essence, a generalized Landauer formula for the transport of bosonic charge excitations between the liquid and the solid. To make this analogy more explicit, it is useful to specialize to the case of a solid that supports a flow-induced electric current, such that the electron Green's functions may be evaluated in the quasi-equilibrium approximation (Eq.~\eqref{eq:quasiFDT}). Then, it can be shown that the electronic density response functions also satisfy a quasi-equilibrium FDT \cite{coquinotquantumfeedback2023}: 
\begin{equation}
g_\e^{\S}(\q, \omega) = -\hbar \, \coth \left(\frac{\hbar[ \omega - \q \v_\e(\q)]}{2 \kB T}\right) \im{g^{\R}_\w}(\q, \omega).
\end{equation}
This allows recasting Eq.~\eqref{eq:noneqQF} as
\begin{equation}
\frac{\langle\textbf{F}_\parallel\rangle}{\mathcal{A}} = - \frac{2}{\hbar} \int \frac{\d \q \d \omega}{(2\pi)^3} (\hbar \q) \,  \mathcal{T}_{\q, \omega} \, \left[ n^{\rm B}_{\omega - \q \v_\w} - n^{\rm B}_{\omega - \q \v_\e(\q)} \right], 
\label{eq:QF_Landauer}
\end{equation}
with the dimensionless transmission coefficient 
\begin{equation}
\mathcal{T}_{\q, \omega} = \left( \frac{\im{g_\w^{\R}} \im{ g_\e^{\R}}}{| 1 - g_\e^{\R} g_\w^{\R}|^2} \right)_{\q, \omega}, 
\end{equation}
the response functions being evaluated at equilibrium; $n^{\rm B}_{\omega} = 1/(e^{\hbar \omega / \kB T} - 1)$ is the Bose ditribution. 

An elementary charge excitation (hydron for the liquid, plasmon or electron-hole pair for the solid) at wavevector $\q$ carries a momentum $\hbar \q$. Eq.~\eqref{eq:QF_Landauer} is indeed a sum over all the modes $\q$ of the Landauer transport rates of these elementary excitations between the solid and the liquid. It is worth noting that such an interpretation of fluctuation-induced friction in terms of quantum transport still holds for a purely classical system. Then, the value of $\hbar$ has no physical significance: it only serves to formally discretize momentum. But formal quantization does allow for new physical insight: for instance, the flow tunneling phenomenon (see Sec. V) can be directly interpreted as tunneling of hydron quasiparticles between the two liquid slabs \cite{coquinotmomentumtunnelling2025}, and the scaling of the tunneling efficiency with the separation $d$ between the slabs can be obtained without calculation by directly applying the Landauer formula. Indeed, the analogue of the wavefunction for a hydron with wavevector $\q$ and frequency $\omega_{\q}$ is the evanescent plane wave $\phi_\q (\r, t) = e^{i(\q \boldsymbol{\rho} - \omega t)} e^{-qz}$, and the transmission coefficient is proportional the squared overlap of the wavefuctions between the slabs, yielding $\mathcal{T}_{\q} \propto e^{-2qd}$ and a scaling as $1/d^4$ of the tunneling efficiency. 

To linear order in the velocities (and assuming $\v_\e(\q) \approx \v(k_{\rm F})$), Eq.~\eqref{eq:QF_Landauer} becomes 
\begin{equation}
\frac{\langle\textbf{F}_{\parallel}\rangle}{\mathcal{A}}=-\lambda_{\rm Q} (\v_\w-\v_\e(k_{\rm F})), 
\label{eq:boundary_current}
\end{equation}
with $\lambda_{\rm Q}$ evaluated for an equilibrium solid as per Eq.~\eqref{eq:QF_result}. Despite its simplicity, Eq.~\eqref{eq:boundary_current} represents a challenge to conventional hydrodynamics: the boundary condition for the liquid flow depends on the non-equilibrium state of the solid boundary, induced by the flow itself.

\section{Molecular simulations at the water-electron nexus}

Molecular dynamics (MD) simulations have been widely used in nanofluidics to gain microscopic insight into the physical mechanisms that underlie system-scale transport properties. While classical MD, based on pairwise interatomic potentials, has allowed for the study of complex systems of up to tens of thousands of atoms, \emph{ab initio} MD, based on density functional theory (DFT) for electronic degrees of freedom, has allowed for the first quantum-level simulation insights into the dynamics of liquid-solid interfaces. Most \emph{ab initio} MD simulations used in nanofluidics rely on converging equilibrium DFT for the electronic degrees of freedom at every timestep of the nuclear dynamics. In this way, the simulation intrinsically enforces the Born-Oppenheimer approximation, and thus cannot by itself capture liquid-solid quantum friction, which relies on mutual retardation effects between nuclear dynamics in the liquid and electron dynamics in the solid. QF is also not captured by the framework of classical MD, where electronic degrees of freedom are represented by interatomic potentials that act instantaneously at every timestep. 

Nevertheless, over the last twenty years, MD simulations have provided key insights into the complex physics of nanofluidic systems. Many simulations have highlighted the importance of the channel wall and its internal degrees of freedom, beyond a simple boundary condition, for the structure and dynamics of the confined liquid -- including fluctuation effects that do not necessarily rely on electron-nuclear retardation. Indeed, fluctuation-induced friction is not a quantum effect \emph{per se}, as it can arise from purely classical thermal fluctuations. This idea has been exploited in recent years to develop simulation frameworks that reproduce QF-like physics in a classical MD setting. 

In this Section, we give a bird's-eye overview of molecular simulations of nanofluidic systems, which somewhat parallels the experimental overview in Sec. II. We first highlight the most important effects of channel wall structure and dynamics, as observed in classical and \emph{ab initio} MD.
Then, we discuss recent semi-classical approaches to electron dynamics at liquid-solid interfaces and perspectives on tackling QF physics at the level of molecular simulations.

\subsection{Classical molecular dynamics}
Carbon surfaces are generally non-polar and therefore hydrophobic. However, \textcite{hummerwaterconductionhydrophobic2001} demonstrated that, in the framework of force-field MD, water spontaneously enters an unpolarizable, hydrophobic CNT as a single file. This surprising observation was explained by strong hydrogen bonding along the 1D water chain, as well as entropic stabilization of the confined water \cite{pascal2011entropy}. Subsequently, a wide range of confined water and ion properties have been investigated in force-field MD studies of both 1D nanotubes and 2D nanoslits. These include the structure and interfacial layering of water and ions, water flow enhancement in nanotubes, ion dehydration and selectivity, water polarizability, water phase transitions and the influence of surface chemistry. For detailed discussions of the individual phenomena, we refer to recent comprehensive reviews \cite{licarbonnanotubenanofluidics2025, lynchwaternanoporesbiological2020, chatzichristoscurrentunderstandingwater2022} and the references therein.

Notably, most quantitative results and several qualitative findings are sensitive to the choice of force field and simulation setup. For example, the predicted water slip lengths in single-digit CNTs span about 5 orders of magnitude across different studies \cite{kannamhowfastdoes2013}. 
Despite these uncertainties, some remarkable qualitative features of water confined in CNTs are primarily determined by geometry and thus remain robust across different force fields and water models. In particular, water is structured in layers close to the wall \cite{alexiadismolecularsimulationwater2008}, the water dielectric response becomes strongly anisotropic in angstrom-scale channels (strongly reduced in the direction of confinement, but strongly enhanced in the perpendicular direction) \cite{lochegiantaxialdielectric2019}, a significant dehydration energy barrier hinders ion entry \cite{corrydesigningcarbonnanotube2008}, and curvature affects the commensurability between water and CNT molecular structures, and therefore the roughness-induced friction \cite{falk2010}. 

Most force-field MD studies implement the confining walls as frozen and unpolarizable. However, due to the high surface-to-volume ratio in nanofluidic systems, both mechanical fluctuations of the wall (i.e., phonons) and the electronic polarization of the solid can be expected to play crucial roles. We shall discuss classical descriptions of polarizability in Sec. VII.C; here, we focus on phonon effects, which are naturally included in classical MD by "unfreezing" the channel wall. 
The impact of phonons on water transport through double-walled CNTs has been investigated by \textcite{mawatertransport2015} using non-equilibrium MD simulations. 
MD simulations of liquid-solid friction are typically carried out in an equilibrium setting, and the friction coefficient is then computed in the Green-Kubo formalism from the equilibrium fluctuations of the total lateral force $F$ exerted by the liquid on the solid:
\begin{equation}
\label{eq:friction-coeff-GK}
	\lambda = \frac{1}{\kB T \mathcal{A}} \int_0^{\infty} \dd{t} \expval{ F(t) F(0) },
\end{equation}
where $\mathcal{A}$ is the area of the interface. \textcite{mawatertransport2015}, however, computed the time-dependent water-nanotube friction force directly, and observed periodic oscillations that correlated with longitudinal phonon vibrations in the nanotube. From equilibrium simulations, they report a 300\% increase in the water center of mass diffusion coefficient in the presence of phonons \cite{mawatertransport2015} compared to a rigid CNT.
While the precise numbers are under debate \cite{cruz-chuphononswaterflow2017}, other simulations have found a reduction of interfacial friction for water in CNTs and boron nitride nanotubes due to phonons \cite{thiemannwaterflow2022}, as well as on flat graphene \cite{tocciinitionanofluidics2020}. 

Another important insight from classical MD is the role of surface charge in liquid-solid friction \cite{xieliquidsolidslip2020}. The authors implemented two models for graphene-like surfaces: one in which the surface charge is homogeneously distributed across all carbon atoms, and another where full unit charges are placed on randomly selected atoms. The simulations revealed that the heterogeneous charge distribution hinders interfacial water flow more strongly because counterions in solution are tightly bound to the charged sites, thereby contributing to additional surface roughness. 
The authors also demonstrate, based on analytical modeling, that a smaller interatomic distance in the solid, e.g., the carbon-carbon distance in graphene, reduces the dependence of slippage on surface charge.

\subsection{\emph{Ab initio} and machine learning approaches}
\emph{Ab initio} MD simulations use density functional theory (DFT) to describe the electron density in the liquid-solid system explicitly, accounting for electron dynamics within the Born-Oppenheimer approximation. However, computational costs limit these simulations to smaller system sizes than in force-field MD.
As with the many possible force field parameter sets for classical MD, there is a vast collection of density functionals. 
\textcite{maAdsorptionDiffusion2011} benchmarked several DFT functionals against diffusion Monte Carlo (and later benchmarks by \textcite{brandenburgInteractionWaterCarbon2019} included coupled cluster methods) based on the interaction of a single water molecule with a graphene sheet. The interaction energies were found to vary between functionals within roughly $\pm 200~\rm meV$, which is sufficient to change the macroscopic behavior of the graphene surface from weakly hydrophobic to fully hydrophilic. Thus, ab initio MD has been a reliable tool for the study of surface chemistry processes such as ion adsorption (with energy scales in the range $100~\rm meV - 1~\rm eV$), while the predictions of liquid-solid friction coefficients -- that involve corrugations of the interfacial energy landscape on the order of a few meV \cite{toccifrictionwater2014} -- need to be handled with care. 

The Born-Oppenheimer approximation underlying ab initio MD can be partially relaxed through molecular dynamics with electronic friction (MDEF), a method developed in the surface science community \cite{douPerspectiveHow2018}. It replaces Newton's equations of motion for the nuclei with generalized Langevin equations, in which a friction tensor and a corresponding random force account for the excitation of electron-hole pairs by nuclear motion \cite{Head-Gordon1995}. The friction tensor is evaluated either in the local-density friction approximation, which embeds each atom in a homogeneous electron gas at the local density \cite{juaristiRoleElectronHole2008}, or, more accurately, using first-order perturbation theory on the Kohn-Sham DFT states computed at a given timestep \cite{askerkaRoleTensorial2016,maurerInitioTensorial2016}. The costly evaluation of the friction tensor has thus far restricted MDEF to simulating single molecules or clusters on metal surfaces and has precluded its application to liquid-solid interfaces. Compared to the analytical QF theory, MDEF represents an approximation where the solid's surface response function is linearized around $\omega = 0$ and electron-electron interactions are neglected.

In its usual Born-Oppenheimer version, ab initio MD has provided important insight into the nature of surface charge on carbon and boron nitride surfaces in water. Ion conductivity measurements in CNTs and BNNTs indicated large surface charges on the tube walls, up to \SIrange[range-phrase=-, range-units=single]{0.01}{0.1}{\coulomb\per\square\meter} for CNTs and up to \SI{1}{\coulomb\per\square\meter} for BNNTs \cite{secchiscalingbehaviorionic2016, siriagiantosmoticenergy2013}. \textcite{grosjean2019versatile} demonstrated that this surface charge could arise from hydroxide ion adsorption on the nanotube surfaces. They showed that hydroxide chemisorbs on hBN, forming an immobile surface charge, whereas it physisorbs on graphene, where it retains finite surface mobility. \textcite{advinculaProtonsAccumulateGraphene2025a} later demonstrated that hydronium ions have an even stronger affinity for the water-graphene interface than hydroxide ions.

\emph{Ab initio} MD has also been a guiding line for understanding, at least qualitatively, the roughness-induced friction of water on 2D material surfaces and nanotubes -- subject, however, to the above-mentioned limitations in terms of quantitative accuracy.  
\textcite{toccifrictionwater2014, tocciinitionanofluidics2020} investigated water friction on planar graphene, hBN, and MoS$_2$ surfaces. Graphene and hBN have structurally almost identical lattices, and the water density profile was almost identical at all three interfaces. In contrast, the simulations found water friction on graphene $(\lambda \approx 5 \times 10^4~\rm N \cdot s \cdot m^{-3})$ to be about 3-5 times weaker than on hBN and about 11 times weaker than on MoS$_2$. The enhanced friction on hBN can be attributed to the more pronounced corrugation of the surface energy landscape. The authors analyze the obtained friction coefficients by decomposing the Green-Kubo formula as the product of a static contribution, $\expval{F^2}$, and a dynamic contribution, $\tau_F$:
\begin{equation}
\label{eq:friction-coeff-GK-disentangled}
	\lambda = \frac{1}{\kB T \mathcal{A}} \expval{F^2} \cdot \tau_F \qq{with} \tau_F = \int_0^{+\infty} \dd{t} \frac{\expval{ F(t) F(0) }}{\expval{F^2}}.
\end{equation}
They find that the static component, $\expval{F^2}$, increases from graphene to hBN to MoS$_2$, whereas the relaxation time $\tau_F$ is similar for water on graphene or hBN, but about twice as long on MoS$_2$. 

Recently, the boundaries of time and length scales accessible to ab initio MD accuracy have been extended with the advent of machine-learning-based interaction potentials \cite{schranMachineLearningPotentials2021}. The starting point is a representative \emph{ab initio} MD simulation which provides the reference data. Then a collection of neural network potentials, the \emph{committee}, is trained to reproduce the structure-energy relation of a randomly selected subset of structures from the reference trajectory. The committee is evaluated over the entire reference trajectory, and the variation in results among the committee members serves as an uncertainty estimate. Iteratively, a small subset of structures with the highest committee uncertainty is added to the training set until the model converges. The resulting neural network potentials were shown to yield accurate results relative to the reference data at orders-of-magnitude lower computational cost \cite{schranMachineLearningPotentials2021, thiemannwaterflow2022, kapilFirstprinciplesPhaseDiagram2022}.
For example, \cite{kapilFirstprinciplesPhaseDiagram2022} used this approach to map the temperature-pressure phase diagram of 2D water confined between graphene sheets. Among other phases, they observed a hexatic phase -- an intermediate state between solid and liquid in 2D phase-transition theory -- as well as a superionic phase with rapid proton conduction.

Another example of machine learning-accelerated MD with DFT accuracy is the investigation of the radius dependence of water friction in CNTs and BNNTs \cite{thiemannwaterflow2022}. As for the planar interfaces, the water friction in hBN systems is about 4-5 times higher than in the corresponding carbon systems due to a more pronounced surface corrugation. Microscopically, a strong attraction force between the nitrogen atoms of hBN and the hydrogen atoms of water hinders water flow in BNNTs.
We note that hydroxide adsorption to the hBN surface, which is not considered in those simulations, is expected to further increase the hydrodynamic resistance and contribute to the small permeability of BNNTs observed experimentally \cite{grosjean2016, secchimassiveradiusdependent2016, xieliquidsolidslip2020}.
The simulations show that decreasing the tube radius reduces surface corrugation and, in turn, water friction in both tube types. This radius dependence agrees qualitatively with the classical MD results for CNTs reported by \textcite{falk2010}. For tubes with radii larger than 2-\SI{3}{\nm}, the friction coefficient reaches the values of the planar graphene or hBN interfaces, with slip lengths of about \SI{22}{\nm} or \SI{6}{\nm}, respectively \cite{thiemannwaterflow2022}. These results are approximately consistent with the data of \textcite{tocciinitionanofluidics2020}, who reported \SI{20}{\nm} on graphene and \SI{4}{\nm} on hBN.

Despite significant progress in recent years, hydrodynamic slip lengths remain very difficult to predict quantitatively using ab initio or machine-learning-based MD simulations. For example, experimental water slip lengths in CNTs \cite{secchimassiveradiusdependent2016,yangfastwatertransport2023} remain significantly higher than predicted in simulations, for reasons that remain unclear.

\begin{figure*}
	\centering
	\includegraphics[width=\textwidth]{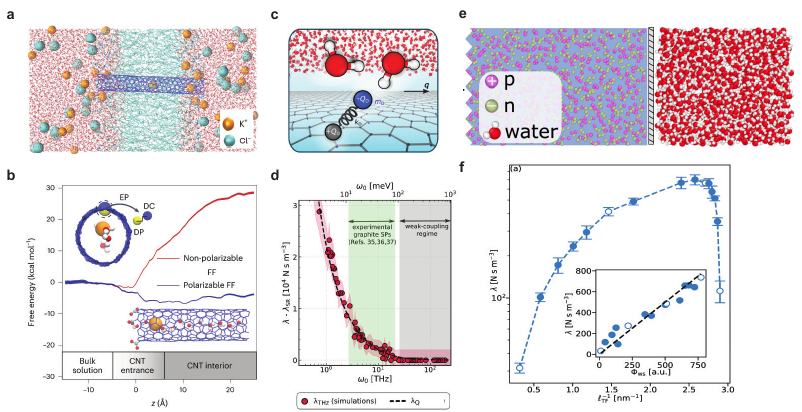}
	\caption{\textbf{Simulations with polarizable nanochannels.}
	(a) Simulation snapshot of trans-membrane potassium diffusion through a metallic (8,4)-chirality CNT with \SI{0.83}{\nm} diameter. (b) The polarizability of the CNT is implemented based on Drude oscillators. While potassium ions have a substantial free energy barrier to enter the CNT based on non-polarizable force fields, the polarizable force-field simulations yield a small free energy gain upon K$^+$ ion entry into the CNT. (a,b) Adapted from \textcite{liBreakdownNernstEinstein2023}. (c) Schematic of Drude oscillators with finite mass attached to the carbon atoms. (d) Simulated FI friction coefficient $\lambda$ for water on polarizable graphene-like lattice as a function of the Drude oscillator frequency $\omega_0$. The simulation result is in excellent agreement with the theoretical prediction (Eq.~\eqref{lambdaQ_cl}). (c,d) Adapted from \textcite{buiclassicalquantumfriction2023}. (e) Simulation of water in contact with a metallic Thomas-Fermi fluid. (f) The water FI friction coefficient $\lambda$ depends on the Thomas-Fermi screening length. It scales linearly with the overlap $\Phi_{\w\s}$ of the liquid and solid fluctuation spectra (inset). (e,f) Adapted from \textcite{herrerofluidselectrostaticallyactive2026}.
	}
	\label{fig:simulations-polarizable}
\end{figure*}

\subsection{Semi-classical approaches to electron dynamics}
Substantial progress in aligning force-field MD simulations with experiments stems from the treatment of induced charges and polarizability. Conductive wall materials can be modeled as perfect metals by enforcing a constant potential at all atoms on the electrode surfaces. In the Born-Oppenheimer approximation, in which the electrode charges adapt instantaneously to the dynamics of the liquid, the constant potential can be implemented exactly, either via a self-consistent model \cite{reedelectrochemicalinterfaceionic2007} or via an extended-Lagrangian approach \cite{coretticommunicationconstrainedmolecular2018}. Both methods account for point charges $\rho^l(\vb{r}) = \sum_i Q^\l_i \,\delta(\vb{r}-\vb{r}_i)$ in the liquid and they introduce time-dependent charges $Q^s_i(t)$ with a Gaussian charge distribution $\rho^\s(\vb{r}) = \sum_i Q^\s_i(t) A \, \exp(-\abs{\vb{r}-\vb{r}_i}^2 \eta^2)$ at each electrode atom $r_i$, with $A = \eta^3 \pi^{-3/2}$. The total Coulomb energy of the system is
\begin{equation}
	U(\vec{Q}) = \frac{1}{2} \int \dd[3]{r} \int \dd[3]{r'} \frac{\rho(\vb{r}) \rho(\vb{r'})}{4\pi \epsilon \, \abs{\vb{r}-\vb{r'}}},
\end{equation} 
where $\vec{Q}$ is a vector of all charges $Q^\l_i$ and $Q^\s_i$, $\rho(\vb{r}) = \rho^l(\vb{r}) + \rho^s(\vb{r})$ is the resulting total charge density, and $\epsilon$ is the dielectric constant of a homogeneous background, say vacuum. In systems with periodic boundary conditions, $U(\vec{Q})$ is typically computed using the Ewald summation method.
The Coulomb potential acting on any charge $Q_j$ is $V_j(\vec{Q}) = \pdv{U(\vec{Q})}{Q^\s_j}$, where all charges $Q^{\s/\l}_i$, with $i \neq j$, are kept constant. Enforcing a constant potential $\psi_j$ (which is equal for all atoms $j$ on the same electrode) amounts to
\begin{equation}
\label{eq:constant-potential-MD}
	V_j(\vec{Q}) - \psi_j = \pdv{U(\vec{Q})}{Q^\s_j} - \psi_j = 0 .
\end{equation}
Since $U(\vec{Q})$ is quadratic in the charges $\vec{Q}$, Eq.~\eqref{eq:constant-potential-MD} defines a system of linear equations, which can be inverted to obtain the electrode charges. In fact, the matrix inversion problem depends only on the electrode geometry, so that, for a static electrode, it can be precomputed. The electrostatic force on the mobile point charges in the liquid due to the constant-potential electrode follows as $\vb{F}_i = -\pdv{U(\vec{Q})}{\vb{r}_i}$.

An equivalent approach is to treat the charges $Q^\s_j$ of the electrode atoms as additional, massless dynamical variables in the system's Lagrangian and to enforce Eq.~\eqref{eq:constant-potential-MD} as a holonomic constraint. The resulting equations of motion for the additional dynamical variables can be solved using symplectic and time-reversible algorithms, which render the implementation more stable \cite{coretticommunicationconstrainedmolecular2018, bonellaadiabaticmotionstatistical2020, corettifluctuationrelationssystems2020}. Both methods, the one with an external solver for the electrode charges and the one with additional dynamical variables, are implemented in the MetalWalls simulation package \cite{marin-laflechemetalwallsclassical2020}.

Semiconducting and insulating materials cannot redistribute charge freely. Instead, they can be assumed to polarize locally in response to an electric field $\vb{E}$ with polarization $\vb{P} = (\vb*{\epsilon} - \epsilon_0 \mathds{1}) \vb{E}$, where $\vb*{\epsilon}$ is the dielectric tensor and $\epsilon_0$ is the vacuum dielectric constant. The macroscopic polarization is the volume density of the microscopic dipole moments $\vb*{\mu} = \vb*{\alpha} \vb{E}$ of each atom, determined through the polarizability tensor $\vb*{\alpha}$. For a homogeneous and isotropic material, the polarizability $\vb*{\alpha} = \alpha \mathds{1}$ can be modeled by Drude oscillators \cite{lamoureuxmodelinginducedpolarization2003}. In this model, each atom is dressed with a mobile point charge $Q_D$, bound to the atom through a force constant $k_D$, and the atom charge is replaced by $Q - Q_D$ in order to preserve the net charge. The Drude particle oscillates around an equilibrium position, which, in the presence of a uniform electric field $\vb{E}$, is displaced from the atom position by the vector $\vb{d}$, leading to the dipole $\vb*{\mu} = \alpha \cdot \vb{E}$ with $\alpha = Q_D^2 / k_D$. The force constant $k_D$ can be chosen to be large, such that the effective displacement $\vb{d}$ is small compared to atomic distances and the atom-Drude pair can be treated as a point-like dipole. The Drude particles contribute to the system's internal energy $U$ through interactions with one another and with other charges of the system. In the Born-Oppenheimer approximation, the positions of the Drude particles need to be determined self-consistently at each time step in the simulation from $\pdv{U}{\vb{d_i}} = \vb{0}$ for each Drude particle $i$. After relaxing the Drude particles, the atomic coordinates are updated accordingly.
This self-consistent relaxation of the Drude particles is, however, computationally expensive and prone to systematic errors from a non-converged energy minimization \cite{lamoureuxmodelinginducedpolarization2003}.
An approximate alternative implementation is to assign a mass $m_D$ to the Drude particles and to treat them as additional, independent dynamical degrees of freedom \cite{lamoureuxmodelinginducedpolarization2003}. The motion of the Drude particles can be effectively decoupled from the atomic motions by a sufficiently small mass $m_D$ and a separate thermostat at a much lower temperature than for the atoms \cite{lamoureuxmodelinginducedpolarization2003}.

\textcite{misrainsightsrole2017} used this Drude oscillator model, in conjunction with the Thole model for dipole interactions \cite{tholemolecularpolarizabilitiescalculated1981}, to describe the atomic polarizability of graphene. Due to interactions between the Drude particles, they effectively describe an anisotropic polarizability, reproducing the out-of-plane polarizability of graphene from \emph{ab initio} calculations \cite{yuinitiostudypolarizability2008} and a ratio of 1/3.5 between the out-of-plane and the in-plane polarizability. \textcite{misrainsightsrole2017} showed that this polarizability plays an important role in determining the contact angle of water on graphene. It is important to note, however, that the Drude model describes specifically atomic polarizability, and does not incorporate the non-local in-plane polarizability resulting from the presence of free charge carriers \cite{hwangDielectricFunction2007}. 

The same method has been applied to the study of water, ion and proton transport through CNT porins \cite{limoleculartransport2024}. The authors report that potassium ion entry into a \SI{0.8}{\nm} diameter CNT porin is unfavorable with an unpolarizable CNT model due to a large dehydration energy barrier, whereas a polarizable CNT porin is spontaneously filled with potassium ions \cite{liBreakdownNernstEinstein2023}, see figure~\ref{fig:simulations-polarizable}(a,b). Furthermore, they model the difference between a metallic and a semiconducting CNT porin by introducing a large (but local) axial polarizability in the metallic case. They find that the transport of water and protons is enhanced upon increasing the axial polarizability, whereas ion transport is almost unaffected \cite{limoleculartransport2024} -- in qualitative agreement with experimental observations.

\textcite{buiclassicalquantumfriction2023} modified the Drude oscillator model to serve as a classical proxy for non-adiabatic electron dynamics. 
They modeled a graphene-water interface in a manner similar to \textcite{misrainsightsrole2017}, but with an increased mass $m_D$ of the Drude particles. In this way, the Drude frequency $\omega_D = \sqrt{k_D/m_D}$ was reduced, and the polarization of the graphene lattice -- then described by the surface response function introduced in Sec. V -- acquired retardation with respect to the liquid. The study demonstrates that increasing frequency overlap between the surface response functions of water and Drude particles increases the water-solid friction coefficient. The increase in friction is shown to have a purely dynamical origin and is in excellent quantitative agreement with the fluctuation-induced friction formula in Eq.~\eqref{eq:QF_result} of \cite{kavokinefluctuationinducedquantum2022}, see figure~\ref{fig:simulations-polarizable}(c,d). 

While the Drude oscillator model can capture polarizability, it cannot describe long-range charge transport by mobile conduction electrons. However, \textcite{schlaichelectronicscreeningusing2022} developed an effective model that can capture both effects. This approach describes the solid using the Thomas-Fermi model -- a semi-classical model of a free electron gas characterized by its screening length. The authors map the Thomas-Fermi model onto a virtual Thomas-Fermi liquid, consisting of light charged particles confined inside the solid. The virtual particles are chosen to be light and strongly thermalized, ensuring rapid charge relaxation. By varying the charge and density of the virtual Thomas-Fermi liquid, the system's metallicity can be tuned from an insulator to a perfect metal. Based on this method, \textcite{schlaichelectronicscreeningusing2022} studied the confinement-induced freezing of an ionic liquid and its dependence on the metallicity of the confining solid. Their results agree with the experimental data of \textcite{comtetnanoscalecapillary2017}.

Going further, \textcite{herrerofluidselectrostaticallyactive2026} used the virtual Thomas-Fermi liquid to study fluctuation-induced friction at the interface between water and a conductive solid. To avoid non-physical temperature transfer from the solid to the liquid, they simulated both media at the same temperature and implemented fast electronic relaxation via a small mass for the virtual Thomas-Fermi liquid particles, confined behind a perfectly reflective wall. However, the relaxation was still slow enough to observe fluctuation-induced friction between the water and the solid, in the range $\lambda_{\rm SI} \sim \SIrange[range-phrase=-, range-units=single]{10}{1000}{\N\s\per\m^3}$. The friction showed a maximum as a function of the Thomas-Fermi length, corresponding to maximal overlap between the water and virtual fluid surface response functions, see figure~\ref{fig:simulations-polarizable}(e,f). Finally, \textcite{herrerofluidselectrostaticallyactive2026} used the electrical conduction property of the virtual Thomas-Fermi fluid to directly demonstrate electro-hydrodynamic drag in a non-equilibrium simulation. For these simulations, they fixed the positive charges in the virtual fluid while leaving the negative charges mobile. Upon applying a voltage to the solid, they observed an induced water flow.

Classical proxies for quantum electron dynamics thus represent a rich playground for studying nanoscale liquid-electron couplings at a qualitative level. 
\textcite{succikeldyshlatticeboltzmann2025} proposed to take this idea even further by implementing water, electrons and phonons as classical fluids within a lattice Boltzmann framework, with the quantum liquid-electron couplings taken into account as friction coefficients between the fluids -- echoing the momentum balance approach to the liquid-electron transport matrix described in Sec. V. Given accurate friction coefficients, this method enables, in fact, a quantitative description of coupled liquid-electron transport in complex geometries. Going further, \textcite{succinonequilibriumgreensfunctions2025} proposed to describe the electron fluid at the level of the Wigner equation, which incorporates quantum corrections to the Boltzmann equation in a systematic way -- the first of which could be tackled within the lattice Boltzmann numerical scheme.

\section{Perspective: nanofluidics meets condensed-matter physics} 

Throughout this Article, we have shown how framing nanofluidics as a condensed-matter physics problem yields new physical insight. As fluid transport in extreme confinement is exquisitely sensitive to the material properties of the confining walls, it is natural to view it within a framework that places the fluid and the wall on equal footing. Ultimately, a nanofluidic channel is a hybrid quantum-classical system in which the largely classical dynamics of the liquid intertwine with the quantum dynamics of the solid's internal degrees of freedom. 

\textbf{\emph{The physics of quantum friction.}} We focused in this Article on the role of electronic degrees of freedom, which were the subject of most studies so far. Their key role is to provide a momentum-dissipation channel for the flowing liquid. The resulting quantum friction effect is the liquid-solid analogue of van der Waals/Casimir friction between two solids or electronic friction of gas phase adsorbates on metal surfaces (Sec. III). The term "quantum friction" reflects the role of quantum electron dynamics -- beyond semi-classical models -- in mediating the effect (see Sec. IV.E). Quantum friction is thus a particular case of fluctuation-induced friction -- friction that results from coupled fluctuations of two media, which can be of thermal or quantum nature. The idea of quantum friction challenges the paradigm of a featureless "hydrodynamic wall". At the nanoscale, the channel wall is no longer a passive boundary condition: it is a piece of quantum matter, whose dynamical properties shape interfacial transport. 

The quantitative determinants of quantum friction are the surface response functions $g_\l(q, \omega)$ and $g_\e(q, \omega)$, that encode the charge fluctuation spectra of the liquid and electron subsystems. Efficient momentum transfer requires a matching of the surface response functions in both frequency and wavevector, with thermally excited modes at frequencies $\omega \lesssim \kB T / \hbar \approx 6 ~\rm THz$ making the largest contribution. This singles out collective excitations of polar liquids, along with particle-hole excitations and low-energy plasmons in metallic systems, as the key players in interfacial liquid-electron couplings. 

\textbf{\emph{Hydro-electronic couplings.}} The main qualitative consequence of quantum friction is to couple liquid flows and electronic currents. When a flowing liquid transfers momentum to a solid's electrons, the electrons are set in motion: this is hydro-electronic drag. The reciprocal effect is electro-hydrodynamic drag, in which a solid-state electronic current induces a liquid flow. These are not small effects: electric currents induced by the flow of salt-free water over graphene, with values quantitatively compatible with quantum friction theory (see Sec. V), have been measured in microfluidic geometries \cite{takedaenhancingelectricitygeneration2025,leeflowinducedvoltagegeneration2013}. The corresponding electro-hydrodynamic drag is expected to be on the order of typical electro-osmosis. These effects extend the usual nanofluidic transport matrix to a \emph{hydro-electronic transport matrix}, in which the electronic current and the applied voltage \emph{inside} the channel wall enter as a new force-flux pair. 

It is worth noting that, at the water-graphene interface -- despite sizeable hydro-electronic couplings -- quantum friction provides only a small fraction of the total hydrodynamic friction, and a similar situation is expected for typical water-metal interfaces. At the water-graphite interface, however, due to the extreme smoothness of the carbon lattice and the emergence of an interlayer plasmon mode, quantum friction has been predicted to contribute a significant fraction of the total friction -- which may account for the observed radius dependence of water slippage in multiwall carbon nanotubes \cite{secchimassiveradiusdependent2016}. Overall, water-carbon interfaces have so far been the most promising systems for observing interfacial liquid-electron couplings. 

\textbf{\emph{A blueprint for new experiments.}} The new nanofluidic phenomena unlocked by quantum friction physics call for careful experimental characterization. Nanofluidic platforms with electrically connected channel walls would be of particular interest for probing coupled transport phenomena unambiguously. For instance, a simultaneous pressure-driven flow and hydro-electronic drag current disentangles the roughness-induced and quantum contributions to hydrodynamic friction. An extensive parameter space would be opened by the ability to tune the electronic properties of the channel wall material: either extrinsically (e.g., through a gating voltage or optical excitation) or by changing the material in a given device geometry (e.g., different 2D material heterostructures). The perspective lies in transposing the rich phenomenology of quantum materials (e.g., Mott transition, slow plasmons, flat band and Moir\'e physics) to nanofluidic transport. Furthermore, it remains an open question to what extent the presence of the liquid can affect the confining solid's electronic properties -- especially if the electrons are strongly correlated. Here, spectroscopic probes could be particularly powerful. For instance, optical-pump-terahertz-probe spectroscopy has been used to measure direct electron-liquid energy transfer at the water-graphene interface and benchmark quantum friction theory \cite{yuelectroncooling2023}. Sum frequency generation spectroscopy, thanks to its surface specificity, could provide indirect information on electron-liquid couplings by probing the orientation of interfacial water \cite{wangspontaneoussurfacecharging2025}. For these techniques, the challenge lies in reducing the optical spot size to probe interfaces at the same spatial scale as relevant for nanofluidic transport. 
While a detailed characterization of the underlying physics is ongoing, it is already clear that hydro-electronic couplings can be harnessed for functional applications. From nanoscale hydroelectricity without electrochemistry to miniature pumps based on electro-hydrodynamic drag -- we should not wait to put the physics of quantum friction to practical use.

\textbf{\emph{Perspectives for theory and simulations.}} The concept of liquid-solid quantum friction developed throughout this Article originates from extending the ideas of van der Waals/Casimir friction to the case of liquid-solid interfaces. However, liquid-solid interfaces differ qualitatively from their solid-solid counterparts and thus require a different theoretical approach. First, liquid dynamics are usually described classically, while electron dynamics are intrinsically quantum: a theory that includes both needs to seamlessly bridge classical and quantum descriptions. Second, if the theory is to rigorously account for coupled liquid-electron transport, then it cannot describe the solid's electrons at the coarse-grained level of their dielectric function, as is usually done in the theory of non-contact friction between solids. In Sec. VI, we outlined how the Keldysh formalism of quantum field theory satisfies both of these criteria. The theory rests on the assumption of Gaussian charge fluctuations in the liquid: these fluctuations can then be represented as harmonic oscillators and formally quantized. Once quantized, the liquid can be coupled to the electrons within a single Keldysh action. The resulting field theory benefits from the well-rounded machinery of Feynman diagram expansions, which yield systematic results for electron-liquid couplings. 

The benefit of using Keldysh machinery to describe nanoscale liquid-solid interfaces is not only in capturing the quantum nature of electron dynamics. In fact, the Keldysh formalism has proven to be powerful in describing even purely classical systems, as in the case of flow tunneling through a system of Drude oscillators \cite{coquinotmomentumtunnelling2025}. The reason is that the Keldysh formalism (or its classical counterpart, the Martin-Siggia-Rose field theory) is primarily a tool for describing fluctuations in a many-body system, regardless of their thermal or quantum nature. 
Since many-body fluctuations have been mostly studied in condensed-matter systems, it is the quantum version of the formalism that is more well-rounded today. But a nanoconfined fluid is a condensed-matter system, together with its confining wall: with increasing confinement, the dynamics are increasingly governed by fluctuations -- not only electronic, but also phononic, ionic, hydrodynamic. An exciting perspective is now to include all these fluctuations in a single Keldysh action -- thus providing a general framework that one can turn to when the mean-field PNP-Stokes description breaks down. 

New developments in analytical theory will need to go hand in hand with simulations methods. Currently, the theory is lacking molecular-level benchmarks, since state-of-the-art ab initio simulations cannot capture quantum friction nor hydro-electronic drag (see Sec. VII). Classical proxies for electron dynamics help build qualitative intuition, but can hardly provide quantitative accuracy. Molecular dynamics with electronic friction could provide an approximate treatment of non-adiabatic liquid-electron couplings, but have so far been restricted to single adsorbate molecules or clusters on metal surfaces \cite{askerkaRoleTensorial2016}. An extension of these methods to true liquid-solid systems would mark a major stride in our ability to accurately simulate nanofluidic transport. Most importantly, such simulations would illuminate the interplay of interfacial chemistry (ion adsorption, redox processes) and quantum-friction-like couplings. 

\textbf{\emph{A bridge between fields.}} Nanofluidics has long been fueled by condensed matter physics analogies -- {p-n} junctions, transistors, Coulomb blockade. What the physics of quantum friction makes clear is that the two fields are connected by more than just analogy. A nanofluidic channel is a setting where classical liquid dynamics and quantum electron dynamics are forced into intimate contact. The flowing liquid then offers condensed matter physics a new probe and a new control knob: a tunable source of charge fluctuations, capable of driving and reading out electronic currents through hydro-electronic coupling. The other way around, solid-state electron dynamics represent a new degree of freedom for nanofluidics, with the potential to support new concepts for blue energy harvesting or ionic computing. 
Bridging nanofluidics and condensed-matter physics thus promises more than a transfer of tools -- it opens the prospect of liquids and solids studied as a single, mutually responsive quantum-classical system, with each field enriching the questions the other can ask. The challenge ahead, shared equally among theory and experiment, is to build the common language in which these questions can be posed.

\appendix 
\section{Fourier transform and sign conventions}

For a function $f(\r, t)$, where $\r$ is a vector in $d$-dimensional space, we define de Fourier transform 
\begin{equation}
f(\q, \omega) = \int \d \r \int_{-\infty}^{+\infty} \d t \, e^{i(\omega t - \q \cdot \r)} f(\r, t). 
\end{equation}
The inverse Fourier transform is 
\begin{equation}
f(\r, t) = \int \frac{\d \q}{(2\pi)^d} \int_{-\infty}^{+\infty} \frac{\d \omega}{2 \pi} \, e^{-i(\omega t - \q \cdot \r)} f(\q, \omega).
\end{equation}
Note that the Fourier transform of $\d f(t) / \d t $ is $-i\omega f(\omega)$. For a function $\chi(\r,\r',t,t')$ that depends only on the differences $\r - \r'$, $t - t'$, we define the Fourier transform as 
\begin{equation}
f(\q, \omega) = \int \d \r \int_{-\infty}^{+\infty} \d t \, e^{i(\omega t - \q \cdot \r)} f(\r, 0, t, 0). 
\end{equation}
With this convention, the Fourier transform of a retarded response function (that is non-zero only for positive time arguments) is analytical in the upper half of the complex frequency plane. Then, it satisfies the Karmers-Kronig relation
\begin{equation}
\re{\chi^{\R}(\omega)} = \frac{1}{\pi} \int_{-\infty}^{+\infty} \d \omega' \, \frac{\im{\chi^{\R}(\omega')- \chi^{\R}(\omega)}}{\omega' - \omega}. 
\end{equation}
Note that, in this form, the integral is non-singular and there is no need to introduce a Cauchy principal value. A response function being real in the time domain, its imaginary part is odd: $\chi^{\R}(\omega)^* = \chi^{\R}(-\omega)$. Then, at zero frequency, the Kramers-Kronig relation becomes 
\begin{equation}
\re{\chi^{\R}(0)} = \frac{2}{\pi} \int_0^{\infty} \d \omega \frac{\im{\chi^{\R}(\omega)}}{\omega}.
\end{equation}
Thus, for a retarded response function, the imaginary part at positive frequencies has the same sign as the real part at zero frequency. In particular, a density response function is (in general) negative at zero frequency, because the the charge density decreases in response to a positive potential: then, a retarded density response function has a negative imaginary part at positive frequencies. Conversely, a surface response function is positive at zero frequency, beacause it represents a reflection coefficient for the potential. A retarded surface response function has a positive imaginary part at positive frequencies. Note that, for advanced functions (analytical in the lower half-plane) the Kramers-Kronig relation picks up an overall minus sign, and the opposite conclusions hold. \\

\section{Detailed computation of fluctuation-induced friction}

In this appendix, we propose to write out in detail the computational steps that leads from Eq.~\eqref{Fqchi} to Eq.~\eqref{lambdaQ_cl}, which consist essentially in carrying Fourier transforms parallel to the interface. Let us introduce the water or solid structure factor, defined as:
\begin{equation}
	S_{\s, \w} (\r, \r', t-t') = \langle n_{\s,\l, \v=0 }^0 (\r,t) n_{\s,\l, \v=0}^0 (\r',t') \rangle,
\end{equation}
and assume translational invariance parallel to the interface, so that $S_{\s, \l} (\r, \r' , t-t') = S_{\s, \l}( \r_\parallel - \r'_\parallel, z,z',t-t')$. Eq.~\eqref{Fqchi} then becomes, after going to the reference frame of the liquid's center of mass,
\begin{equation}
	\begin{split} \label{eq: starting point Fq}
		&\langle \F(t) \rangle = - \int \d \r_\s \d \r_\s' \d \r_\l \d \r_\l' \d t' \, \nabla_\l V(\r_\l - \r_\s) V(\r_\l'- \r_\s') \cdots \\ &  \quad \cdots \chi_\s(\r_\s + \v t, \r_\s' + \v t', t-t')
		   S_\l(\r_\l, \r_\l', t-t') \\ 
		   &  - \int \d \r_\s \d \r_\s' \d \r_\l \d \r_\l' \d t' \, \nabla_\l V(\r_\l - \r_\s) V(\r_\l'- \r_\s') \cdots \\ & \quad \cdots \chi_\l(\r_\l, \r_\l', t-t')
		   S_\s(\r_\s + \v t, \r_\s' + \v t', t-t'). 
	\end{split}
\end{equation}
Let us first focus on the first term of this equation, and write out explicitly the 2D spatial Fourier decompositions along the interface as:
\begin{equation}
\begin{split}
	- & \int \d \r_\s \d z_\s \d \r_\s' \d z_\s' \d \r_\l \d z_\l  \d\r_\l' \d z_\l' \d t' \cdots \\
	& \cdots \int \frac{\d \q_1 }{(2 \pi)^2} e^{i \q_1 \cdot (\r_\l - \r_\s)} i\q_1 V(\q_1, z_\l, z_\s) \cdots \\
		& \cdots \int \frac{\d \q_2 }{(2 \pi)^2} e^{i \q_2 \cdot (\r_\l' - \r_\s')}  V(\q_2, z_\l', z_\s') \cdots \\
		& \cdots \int \frac{\d \q_3 }{(2 \pi)^2} e^{i \q_3 \cdot (\r_\s - \r_\s')} e^{i \q_3\cdot \v(t-t')} \chi_\s(\q_3, z_\s, z_\s', t-t') \cdots \\
		& \cdots \int \frac{\d \q_4 }{(2 \pi)^2} e^{i \q_4 \cdot (\r_\l - \r_\l')} S_\l(\q_4, z_\l, z_\l', t-t').
\end{split}
\end{equation}
Carrying out for example the integration over $\r_\s$ yields $(2 \pi)^2 \delta^{(2)}(\q_1 - \q_3)$, and similarly for $\r_\s', \r_\l$ and the last integration over $\r_\l'$ introduces an overall factor $\mathcal{A}$, so that Eq.~\eqref{eq: starting point Fq} becomes:
\begin{equation}
\begin{split}
	\langle \F(t) \rangle & = - \mathcal{A} \int \frac{\d \q}{(2 \pi)^2} \int \d z_\s \d z_\s' \d z_\l \d z_\l' \int \d t' e^{i \q \cdot \v (t-t')} \cdots \\ 
	& \quad \cdots i \q V(\q, z_\l, z_\s) V(-\q , z_\l', z_\s')\cdots \\ & \quad \cdots \Big[ \chi_\s(\q , z_\s, z_\s', t-t') S_\l( - \q, z_\l, z_\l', t-t') \\
	& \quad \cdots + S_\s(\q , z_\s, z_\s', t-t') \chi_\l( - \q, z_\l, z_\l', t-t') \Big]. 
\end{split}
\end{equation}
We identify in this equation the temporal Fourier transform evaluated at $\omega = \q \cdot \v$ of a product, which takes the form of a convolution in Fourier space. Using the fact that $\langle \F (t) \rangle$ is real, we further simplify:
\begin{equation}
	\begin{split}
		\langle \F(t) \rangle & = \frac{\mathcal{A}}{(2 \pi)^3} \int \d \q \d \omega \int \d z_\s \d z_\s' \d z_\l \d z_\l' \cdots \\ &  \cdots \q V(\q, z_\l, z_\s) V(-\q , z_\l', z_\s')\cdots \\ &  \cdots \Big[ \im{\chi_\s(\q , z_\s, z_\s',\omega)} S_\l(  \q, z_\l, z_\l', \q \cdot \v - \omega) \\
	&  \cdots + S_\s(\q , z_\s, z_\s', \omega) \im{\chi_\l(  \q, z_\l, z_\l', \q \cdot \v - \omega)} \Big],
	\end{split}
\end{equation}
where we also used the spatial symmetry $S(-\q, z, z', \omega) = S(\q, z,z',\omega)$. Finally, according to the fluctuation-dissipation theorem:
\begin{equation}
	S_{\s, \l} (\q, z, z', \omega) = - \frac{2 \kB T}{\omega} \im{\chi_{\s, \l} (\q, z, z', \omega)}, 
\end{equation}
such that we can proceed, using $\im{\chi_\l(\q, - \omega)} = - \im{\chi_\l(\q, \omega)}$, with:
\begin{equation}
\begin{split}
	&\langle \F(t) \rangle = - \frac{\kB T \mathcal{A}}{4 \pi^3} \int \d \q \d \omega \int \d z_\s \d z_\s' \d z_\l \d z_\l' \cdots \\ & \cdots  \q \Big[ \frac{1}{\omega - \q \cdot \v}-\frac{1}{\omega} \Big] V(\q, z_\l, z_\s) V(-\q , z_\l', z_\s')\cdots \\ & \cdots \im{ \chi_\s(\q, z_\s, z_\s', \omega)} \im{\chi_\l(\q, z_\l, z_\l', \omega - \q \cdot \v)}. 
\end{split}
\end{equation}
To linear order in the flow velocity $\v$, the latter equation reduces to $\langle \F(t) \rangle = - \lambda_{\rm FI } \mathcal{A} \v$ with:
\begin{equation}\label{eq: dem almost there}
	\begin{split}
		\lambda_{\rm FI} &=  \frac{\kB T}{4 \pi^3} \int \d \q \frac{(\q \cdot \v)^2}{v^2} \int_{-\infty}^{+\infty} \frac{\d \omega}{\omega^2}  \int \d z_\s \d z_\s' \d z_\l \d z_\l' \cdots \\ & \cdots   V(\q, z_\l, z_\s) V(-\q , z_\l', z_\s') \cdots \\ & \cdots \im{ \chi_\s(\q, z_\s, z_\s', \omega)} \im{\chi_\l(\q, z_\l, z_\l', \omega)}. 
	\end{split}
\end{equation}
The spatial 2D Fourier transform of the Coulomb potential is given by: 
\begin{equation}
V(\q, z_\l, z_\s) = -\frac{e^2}{2 \epsilon_0 q} e^{-q \vert z_\l - z_\s \vert} = -\frac{e^2}{2 \epsilon_0 q} e^{-q(z_\s+z_\l)}
\end{equation}
as the geometry of the problem imposes $z_\l<0$ and $z_\s>0$. Recombining the exponential factors and identifying in Eq.~\eqref{eq: dem almost there} the surface response functions as defined by Eqs.~\eqref{surface response function water} and \eqref{surface response function solid}, we are left with:
\begin{equation}
\begin{split}
		\lambda_{\rm FI} &=  \frac{\kB T}{4 \pi^3} \int \d \q \frac{(\q \cdot \v)^2}{v^2} \int_{-\infty}^{+\infty} \frac{\d \omega}{\omega^2} \cdots \\ & \qquad \cdots \im{g_\l(\q, \omega)} \im{g_\s(\q, \omega)}. 
\end{split}
\end{equation}
Assuming that the surface reponse functions are isotropic and carrying out the integration over the polar angle, we recover Eq.~\eqref{lambdaQ_cl} from the main text.

\section{Fluctuation-dissipation theorem for response functions}\label{appendix: FDT}

In this Appendix, we derive the fluctuation-dissipation theorem (FDT) for response functions that is used throughout the Article. It can be stated most generally in the following way. Consider a quantum system described by a time-independent Hamiltonian $\hat H_0$ and an observable $\hat n$ of that system. The system is perturbed by a small external potential $\phi(\r,t)$ according to the interaction Hamiltonian
\begin{equation}
	\hat H_{\rm int}(t) = \int \d\r \, \hat n(\r,t)\, \phi(\r, t).
\end{equation}
We denote as $\hat n_0(\r, t)$ the observable in the absence of perturbation and $\delta n(\r,t) = \langle \hat n(\r, t) \rangle - \langle \hat n_0(\r, t) \rangle$. 
Define the response function $\chi^{\R}$ such that, to first order in $\phi$, 
\begin{equation}
	\delta n(\r,t) = \int \d\r' \int \d t' \, \chi^{\R}(\r, \r', t-t')\, \phi(\r',t').
	\label{App_defchi}
\end{equation}
Define the structure factor $S$ as the equilibrium expectation value
\begin{equation}
S(\r, \r', t - t') = \frac{1}{2} \langle \{ \hat n_0 (\r, t) , \hat n_0(\r', t') \} \rangle.
\label{App_defS}
\end{equation}
Then, the structure factor and the response function are related according to
\begin{equation}\label{eq: FDT quantum statement}
	S(\r, \r', \omega) = - \hbar\, \coth\!\left(\frac{\hbar \omega}{2 \kB T}\right) \im{\chi^{\R}(\r, \r', \omega)}.
\end{equation}
In the classical limit, the operator $\hat n$ becomes a scalar dynamical variable $n$, the structure factor reduces to $S(\r, \r', t - t') = \langle n_0(\r, t) n_0(\r', t') \rangle$, and the FDT becomes
\begin{equation}\label{eq: FDT classical}
	S(\r, \r', \omega) = - \frac{2 \kB T}{\omega} \im{\chi^{\R}(\r, \r', \omega)}.
\end{equation}
Note that, in the Keldysh/MSRJD formalism representing either the quantum or the classical system, the response function and structure factor defined in Eqs.~\eqref{App_defchi} and \eqref{App_defS} correspond to the field correlators $\chi^{\R}(x, x') = -i \langle n(x)\, \tilde n(x') \rangle_{\rm eq}$ and $S(x, x') = \langle n(x)\, n(x') \rangle_{\rm eq}$, with the notations of Sec. VI.

\paragraph{Kubo formula.}
We first express the response function as an equilibrium expectation value in the operator formalism. In the interaction picture with respect to $\hat H_0$, the observable evolves as $\hat n(\r,t) = \hat U^\dag(t)\, \hat n_0(\r,t)\, \hat U(t)$, with
\begin{equation}
	\hat U(t) = \mathcal{T} \exp\!\left( - \frac{i}{\hbar} \int_{-\infty}^t \d t' \int \d\r' \, \hat n_0(\r',t')\, \phi(\r',t') \right),
\end{equation}
where $\mathcal{T}$ is the time-ordering operator. Expanding $\hat U(t)$ to first order in $\phi$,
\begin{equation}
\begin{split}
	\hat n(\r,t) =& \hat n_0(\r,t) - \frac{i}{\hbar} \int \d\r' \int_{-\infty}^{t} \d t'\, \dots \\
	&[\hat n_0(\r,t), \hat n_0(\r',t')]\, \phi(\r',t') + \mathcal{O}(\phi^2).
\end{split}
\end{equation}
Taking the expectation value and comparing with Eq.~\eqref{App_defchi}, the response function is the retarded commutator
\begin{equation}\label{eq: kubo}
	\chi^{\R}(\r, \r', t-t') = - \frac{i}{\hbar}\, \theta(t-t')\, \langle [\hat n_0(\r,t), \hat n_0(\r',t')] \rangle.
\end{equation}

\paragraph{Spectral representation.}
Let $\{\vert \nu \rangle\}$ be a complete set of eigenstates of $\hat H_0$ with energies $\{\hbar \omega_\nu\}$, and let $p_\nu = e^{-\beta \hbar \omega_\nu}/Z$ be the thermal weight of $\vert \nu \rangle$, with $\beta = 1/(\kB T)$. Using time-translation invariance to set the time arguments to $t$ and $0$, and inserting the identity $\sum_{\nu'} \vert \nu' \rangle \langle \nu' \vert$ between the operators, the commutator in Eq.~\eqref{eq: kubo} becomes
\begin{equation}
\begin{split}
	\langle [\hat n_0(\r,t), &\hat n_0(\r',0)] \rangle = \sum_{\nu, \nu'} p_\nu \Big[ n_{\nu \nu'}(\r) n_{\nu' \nu}(\r')\, e^{i \omega_{\nu \nu'} t} \dots \\
	&\dots - n_{\nu \nu'}(\r') n_{\nu' \nu}(\r)\, e^{-i \omega_{\nu \nu'} t} \Big],
\end{split}
\end{equation}
where $n_{\nu \nu'}(\r) = \langle \nu \vert \hat n_0(\r,0) \vert \nu' \rangle$ and $\omega_{\nu \nu'} = \omega_\nu - \omega_{\nu'}$. Multiplying by $-\tfrac{i}{\hbar}\theta(t)$ and taking the temporal Fourier transform $\int \d t\, e^{i \omega t}$, the step function turns each oscillating exponential into $(\omega \mp \omega_{\nu \nu'} + i0^+)^{-1}$, whose imaginary part is $-\pi \delta(\omega \mp \omega_{\nu \nu'})$. Relabeling $\nu \leftrightarrow \nu'$ in the second term, we obtain
\begin{equation}\label{eq: FDT part chi}
\begin{split}
	&\im{\chi^{\R}(\r, \r', \omega)} = - \frac{\pi}{\hbar}\,(1 - e^{-\beta \hbar \omega}) \sum_{\nu, \nu'} p_\nu\, \dots \\
	&\quad \dots\, n_{\nu \nu'}(\r) n_{\nu' \nu}(\r')\, \delta (\omega + \omega_{\nu} - \omega_{\nu'}),
\end{split}
\end{equation}
where the factor $(1 - e^{-\beta \hbar \omega})$ arises because the relabeling sends $p_{\nu'} \to p_\nu e^{-\beta \hbar \omega}$ on the support of the delta function (since $\hbar \omega_{\nu' \nu} = \hbar \omega$ there).

\paragraph{Structure factor.}
We treat the structure factor Eq.~\eqref{App_defS} in the same way. Inserting eigenstates,
\begin{equation}
\begin{split}
	S(\r, \r', t) = &\frac{1}{2}\sum_{\nu, \nu'} p_\nu \Big[ n_{\nu \nu'}(\r) n_{\nu' \nu}(\r')\, e^{i \omega_{\nu \nu'} t} \dots \\
	&\dots + n_{\nu \nu'}(\r') n_{\nu' \nu}(\r)\, e^{-i \omega_{\nu \nu'} t} \Big],
\end{split}
\end{equation}
the anticommutator giving a relative $+$ sign between the two orderings instead of the $-$ of the commutator. Taking the temporal Fourier transform and relabeling $\nu \leftrightarrow \nu'$ in the second term as before,
\begin{equation}\label{eq: FDT part S}
\begin{split}
	S(\r, \r', \omega)& = \pi\,(1 + e^{-\beta \hbar \omega}) \sum_{\nu, \nu'} p_\nu\, \dots \\
	&\dots\, n_{\nu \nu'}(\r) n_{\nu' \nu}(\r')\, \delta (\omega + \omega_{\nu} - \omega_{\nu'}).
\end{split}
\end{equation}
The only difference with Eq.~\eqref{eq: FDT part chi} is that the anticommutator carries the factor $(1 + e^{-\beta \hbar \omega})$ in place of $(1 - e^{-\beta \hbar \omega})$; the spectral sums are otherwise identical.

\paragraph{Result.}
Dividing Eq.~\eqref{eq: FDT part S} by Eq.~\eqref{eq: FDT part chi}, the spectral sums cancel and we obtain the quantum FDT,
\begin{equation}
\begin{split}
	S(\r, \r', &\omega) = - \hbar\, \frac{1 + e^{-\beta \hbar \omega}}{1 - e^{-\beta \hbar \omega}}\, \im{\chi^{\R}(\r, \r', \omega)} \\
	&= - \hbar\, \coth\!\left(\frac{\hbar \omega}{2 \kB T}\right) \im{\chi^{\R}(\r, \r', \omega)}.
\end{split}
\end{equation}
In the limit $\hbar \rightarrow 0$, $\coth(\hbar\omega/2\kB T) \to 2\kB T/\hbar\omega$ and we recover the classical formulation Eq.~\eqref{eq: FDT classical}.

\bibliography{cleaned}

@article{schmittMesoscopicKleinSchwingerEffect2023,
  title = {Mesoscopic {{Klein-Schwinger}} Effect in Graphene},
  author = {Schmitt, A. and Vallet, P. and Mele, D. and Rosticher, M. and Taniguchi, T. and Watanabe, K. and Bocquillon, E. and F{\`e}ve, G. and Berroir, J. M. and Voisin, C. and Cayssol, J. and Goerbig, M. O. and Troost, J. and Baudin, E. and Pla{\c c}ais, B.},
  year = 2023,
  month = jun,
  journal = {Nature Physics},
  volume = {19},
  number = {6},
  pages = {830--835},
  issn = {1745-2473, 1745-2481},
  doi = {10.1038/s41567-023-01978-9},
  urldate = {2023-10-27},
  langid = {english}
}

@article{demeryDragForces2010,
  title = {Drag {{Forces}} in {{Classical Fields}}},
  author = {D{\'e}mery, Vincent and Dean, David S.},
  year = 2010,
  month = feb,
  journal = {Physical Review Letters},
  volume = {104},
  number = {8},
  pages = {080601},
  issn = {0031-9007, 1079-7114},
  doi = {10.1103/PhysRevLett.104.080601},
  urldate = {2026-06-04},
  copyright = {http://link.aps.org/licenses/aps-default-license},
  langid = {english}
}

@article{deanNonequilibriumTuning2016,
  title = {Nonequilibrium {{Tuning}} of the {{Thermal Casimir Effect}}},
  author = {Dean, David S. and Lu, Bing-Sui and Maggs, A. C. and Podgornik, Rudolf},
  year = 2016,
  month = jun,
  journal = {Physical Review Letters},
  volume = {116},
  number = {24},
  pages = {240602},
  issn = {0031-9007, 1079-7114},
  doi = {10.1103/PhysRevLett.116.240602},
  urldate = {2026-06-04},
  copyright = {http://link.aps.org/licenses/aps-default-license},
  langid = {english}
}

@article{coquinotElectronElectrolyte2026,
  title = {Electron--Electrolyte Coupling in {{AC}} Transport through Nanofluidic Channels},
  author = {Coquinot, Baptiste and Liz{\'e}e, Mathieu and Bocquet, Lyd{\'e}ric and Kavokine, Nikita},
  year = 2026,
  month = apr,
  journal = {The Journal of Chemical Physics},
  volume = {164},
  number = {13},
  pages = {134704},
  issn = {0021-9606, 1089-7690},
  doi = {10.1063/5.0313352},
  urldate = {2026-05-26},
  langid = {english}
}

@article{narozhnyCoulombDrag2016,
  title = {Coulomb Drag},
  author = {Narozhny, B. N. and Levchenko, A.},
  year = 2016,
  journal = {Reviews of Modern Physics},
  volume = {88},
  number = {2},
  eprint = {1505.07468},
  pages = {1--55},
  issn = {15390756},
  doi = {10.1103/RevModPhys.88.025003},
  archiveprefix = {arXiv}
}

@article{maAdsorptionDiffusion2011,
  title = {Adsorption and Diffusion of Water on Graphene from First Principles},
  author = {Ma, Jie and Michaelides, Angelos and Alf{\`e}, Dario and Schimka, Laurids and Kresse, Georg and Wang, Enge},
  year = 2011,
  month = jul,
  journal = {Physical Review B},
  volume = {84},
  number = {3},
  pages = {033402},
  issn = {1098-0121, 1550-235X},
  doi = {10.1103/PhysRevB.84.033402},
  urldate = {2026-05-24},
  copyright = {http://link.aps.org/licenses/aps-default-license},
  langid = {english}
}

@article{douPerspectiveHow2018,
  title = {Perspective: {{How}} to Understand Electronic Friction},
  author = {Dou, Wenjie and Subotnik, Joseph E.},
  year = 2018,
  journal = {Journal of Chemical Physics},
  volume = {148},
  number = {23},
  issn = {00219606},
  doi = {10.1063/1.5035412}
}

@article{Head-Gordon1995,
  title = {Molecular Dynamics with Electronic Frictions},
  author = {{Head-Gordon}, Martin and Tully, John C.},
  year = 1995,
  journal = {The Journal of Chemical Physics},
  volume = {103},
  number = {23},
  pages = {10137--10145},
  issn = {00219606},
  doi = {10.1063/1.469915}
}

@article{juaristiRoleElectronHole2008,
  title = {Role of {{Electron-Hole Pair Excitations}} in the {{Dissociative Adsorption}} of {{Diatomic Molecules}} on {{Metal Surfaces}}},
  author = {Juaristi, J. I. and Alducin, M. and Mui{\~n}o, R. D{\'i}ez and Busnengo, H. F. and Salin, A.},
  year = 2008,
  month = mar,
  journal = {Physical Review Letters},
  volume = {100},
  number = {11},
  pages = {116102},
  issn = {0031-9007, 1079-7114},
  doi = {10.1103/PhysRevLett.100.116102},
  urldate = {2026-05-25},
  copyright = {http://link.aps.org/licenses/aps-default-license},
  langid = {english}
}

@article{askerkaRoleTensorial2016,
  title = {Role of {{Tensorial Electronic Friction}} in {{Energy Transfer}} at {{Metal Surfaces}}},
  author = {Askerka, Mikhail and Maurer, Reinhard J. and Batista, Victor S. and Tully, John C.},
  year = 2016,
  month = may,
  journal = {Physical Review Letters},
  volume = {116},
  number = {21},
  pages = {217601},
  issn = {0031-9007, 1079-7114},
  doi = {10.1103/PhysRevLett.116.217601},
  urldate = {2026-05-25},
  copyright = {http://link.aps.org/licenses/aps-default-license},
  langid = {english}
}

@article{maurerInitioTensorial2016,
  title = {{\emph{Ab Initio}} Tensorial Electronic Friction for Molecules on Metal Surfaces: {{Nonadiabatic}} Vibrational Relaxation},
  shorttitle = {{\emph{Ab Initio}} Tensorial Electronic Friction for Molecules on Metal Surfaces},
  author = {Maurer, Reinhard J. and Askerka, Mikhail and Batista, Victor S. and Tully, John C.},
  year = 2016,
  month = sep,
  journal = {Physical Review B},
  volume = {94},
  number = {11},
  pages = {115432},
  issn = {2469-9950, 2469-9969},
  doi = {10.1103/PhysRevB.94.115432},
  urldate = {2026-05-25},
  copyright = {http://link.aps.org/licenses/aps-default-license},
  langid = {english}
}

@article{belotcerkovtcevaExtremeCurrent2025,
  title = {Extreme Current Density and Breakdown Mechanism in Graphene on Diamond Substrate},
  author = {Belotcerkovtceva, Daria and Datt, Gopal and Nameirakpam, Henry and Aitkulova, Aisuluu and Suntornwipat, Nattakarn and Majdi, Saman and Isberg, Jan and Kamalakar, M. Venkata},
  year = 2025,
  month = apr,
  journal = {Carbon},
  volume = {237},
  pages = {120108},
  issn = {00086223},
  doi = {10.1016/j.carbon.2025.120108},
  urldate = {2026-05-14},
  langid = {english}
}

@article{songIonConveying2026,
  title = {Ion {{Conveying Electron Enabling Electrodeless Osmotic Energy Harvesting}}},
  author = {Song, Fei and Fu, Jialin and Tian, Wenbo and Xue, Shuangchen and Zhang, Qingliang and Wu, Tianxu and Song, Dongxing and Wang, Ke},
  year = 2026,
  month = apr,
  journal = {Advanced Functional Materials},
  volume = {36},
  number = {33},
  pages = {e28563},
  issn = {1616-301X, 1616-3028},
  doi = {10.1002/adfm.202528563},
  urldate = {2026-05-14},
  langid = {english}
}

@article{kuriyaOutputDensity2020,
  title = {Output Density Quantification of Electricity Generation by Flowing Deionized Water on Graphene},
  author = {Kuriya, Kei and Ochiai, Kotaro and Kalita, Golap and Tanemura, Masaki and Komiya, Atsuki and Kikugawa, Gota and Ohara, Taku and Yamashita, Ichiro and Ohuchi, Fumio S. and Meyyappan, M. and Samukawa, Seiji and Washio, Katsuyoshi and Okada, Takeru},
  year = 2020,
  month = sep,
  journal = {Applied Physics Letters},
  volume = {117},
  number = {12},
  pages = {123905},
  issn = {0003-6951, 1077-3118},
  doi = {10.1063/5.0018862},
  urldate = {2026-05-14},
  langid = {english}
}

@article{takedaInvestigatingCorrelation2024,
  title = {Investigating the Correlation between Flow Dynamics and Flow-Induced Voltage Generation},
  author = {Takeda, Hikaru and Iwamoto, Naoya and Honda, Mitsuhiro and Tanemura, Masaki and Yamashita, Ichiro and Komiya, Atsuki and Okada, Takeru},
  year = 2024,
  month = oct,
  journal = {Applied Physics Letters},
  volume = {125},
  number = {18},
  pages = {184101},
  issn = {0003-6951, 1077-3118},
  doi = {10.1063/5.0230115},
  urldate = {2026-05-14},
  langid = {english}
}

@article{kavokineInteractionConfinement2022,
  title = {Interaction Confinement and Electronic Screening in Two-Dimensional Nanofluidic Channels},
  author = {Kavokine, Nikita and Robin, Paul and Bocquet, Lyd{\'e}ric},
  year = 2022,
  month = sep,
  journal = {The Journal of Chemical Physics},
  volume = {157},
  number = {11},
  pages = {114703},
  issn = {0021-9606},
  doi = {10.1063/5.0102002}
}

@article{esfandiarSizeEffect2017,
  title = {Size Effect in Ion Transport through Angstrom-Scale Slits},
  author = {Esfandiar, A. and Radha, B. and Wang, F. C. and Yang, Q. and Hu, S. and Garaj, S. and Nair, R. R. and Geim, A. K. and Gopinadhan, K.},
  year = 2017,
  journal = {Science},
  volume = {358},
  number = {6362},
  pages = {511--513},
  issn = {10959203},
  doi = {10.1126/science.aan5275}
}

@article{parkRedoxgovernedCharge2019,
  title = {Redox-Governed Charge Doping Dictated by Interfacial Diffusion in Two-Dimensional Materials},
  author = {Park, Kwanghee and Kang, Haneul and Koo, Seonghyun and Lee, DaeEung and Ryu, Sunmin},
  year = 2019,
  month = oct,
  journal = {Nature Communications},
  volume = {10},
  number = {1},
  pages = {4931},
  issn = {2041-1723},
  doi = {10.1038/s41467-019-12819-w},
  urldate = {2025-09-25},
  langid = {english}
}

@article{gravelleAnomalousCapillary2016,
  title = {Anomalous Capillary Filling and Wettability Reversal in Nanochannels},
  author = {Gravelle, Simon and Ybert, Christophe and Bocquet, Lyd{\'e}ric and Joly, Laurent},
  year = 2016,
  journal = {Physical Review E},
  volume = {93},
  number = {3},
  pages = {1--7},
  issn = {24700053},
  doi = {10.1103/PhysRevE.93.033123}
}

@article{thomasWaterFlow2009,
  title = {Water Flow in Carbon Nanotubes: {{Transition}} to Subcontinuum Transport},
  author = {Thomas, John A. and McGaughey, Alan J.H.},
  year = 2009,
  journal = {Physical Review Letters},
  volume = {102},
  number = {18},
  pages = {1--4},
  issn = {00319007},
  doi = {10.1103/PhysRevLett.102.184502}
}

@article{agrawalObservationExtreme2017,
  title = {Observation of Extreme Phase Transition Temperatures of Water Confined inside Isolated Carbon Nanotubes},
  author = {Agrawal, Kumar Varoon and Shimizu, Steven and Drahushuk, Lee W. and Kilcoyne, Daniel and Strano, Michael S.},
  year = 2017,
  month = mar,
  journal = {Nature Nanotechnology},
  volume = {12},
  number = {3},
  pages = {267--273},
  issn = {1748-3387, 1748-3395},
  doi = {10.1038/nnano.2016.254},
  urldate = {2025-10-12},
  langid = {english}
}

@article{parkIdentificationDropletFlowInduced2017,
  title = {Identification of {{Droplet-Flow-Induced Electric Energy}} on {{Electrolyte}}--{{Insulator}}--{{Semiconductor Structure}}},
  author = {Park, Junwoo and Song, Suhwan and Yang, YoungJun and Kwon, Soon-Hyung and Sim, Eunji and Kim, Youn Sang},
  year = 2017,
  month = aug,
  journal = {Journal of the American Chemical Society},
  volume = {139},
  number = {32},
  pages = {10968--10971},
  issn = {0002-7863, 1520-5126},
  doi = {10.1021/jacs.7b05030},
  urldate = {2026-04-04},
  langid = {english}
}

@article{shinSaltWater2026,
  title = {Salt {{Water Drops Slide Faster}}: {{Ionic Modulation}} of {{Drop Friction}}},
  shorttitle = {Salt {{Water Drops Slide Faster}}},
  author = {Shin, Dongho and Lathia, Rutvik and Hinduja, Chirag and Cheon, Hyunbae and Park, Seongmin and Butt, Hans-J{\"u}rgen and Park, Junwoo},
  year = 2026,
  month = mar,
  journal = {Advanced Science},
  volume = {13},
  number = {17},
  pages = {e21659},
  issn = {2198-3844, 2198-3844},
  doi = {10.1002/advs.202521659},
  urldate = {2026-04-04},
  langid = {english}
}

@article{perssonElectronicFriction2004,
  title = {Electronic Friction and Liquid-Flow-Induced Voltage in Nanotubes},
  author = {Persson, B. N J and Tartaglino, U. and Tosatti, E. and Ueba, H.},
  year = 2004,
  journal = {Physical Review B - Condensed Matter and Materials Physics},
  volume = {69},
  number = {23},
  pages = {1--5},
  issn = {01631829},
  doi = {10.1103/PhysRevB.69.235410}
}

@article{biehsNearfieldRadiativeHeat2021,
  title = {Near-Field Radiative Heat Transfer in Many-Body Systems},
  author = {Biehs, S.-A. and Messina, R. and Venkataram, P. S. and Rodriguez, A. W. and Cuevas, J. C. and {Ben-Abdallah}, P.},
  year = 2021,
  month = jun,
  journal = {Reviews of Modern Physics},
  volume = {93},
  number = {2},
  pages = {025009},
  publisher = {American Physical Society},
  issn = {0034-6861},
  doi = {10.1103/RevModPhys.93.025009}
}

@article{volokitinVanWaals2008,
  title = {Van Der {{Waals}} Frictional Drag Induced by Liquid Flow in Low-Dimensional Systems},
  author = {Volokitin, A. I. and Persson, B. N.J.},
  year = 2008,
  journal = {Physical Review B},
  volume = {77},
  number = {3},
  pages = {2--5},
  issn = {10980121},
  doi = {10.1103/PhysRevB.77.033413}
}

@article{wuGraphiticCarbonwater2013,
  title = {Graphitic Carbon-Water Nonbonded Interaction Parameters},
  author = {Wu, Yanbin and Aluru, N. R.},
  year = 2013,
  journal = {Journal of Physical Chemistry B},
  volume = {117},
  number = {29},
  pages = {8802--8813},
  issn = {15206106},
  doi = {10.1021/jp402051t},
  pmid = {23802763}
}

@article{rysselbergheRemarksConcerning1932,
  title = {Remarks Concerning the {{Clausius-Mossotti Law}}},
  author = {Rysselberghe, Pierre Van},
  year = 1932,
  month = apr,
  journal = {The Journal of Physical Chemistry},
  volume = {36},
  number = {4},
  pages = {1152--1155},
  issn = {0092-7325, 1541-5740},
  doi = {10.1021/j150334a007},
  urldate = {2026-04-10},
  langid = {english}
}

@article{Tudorovskiy2010,
  title = {Intervalley Plasmons in Graphene},
  author = {Tudorovskiy, T. and Mikhailov, S. A.},
  year = 2010,
  journal = {Physical Review B - Condensed Matter and Materials Physics},
  volume = {82},
  number = {7},
  pages = {4--7},
  issn = {10980121},
  doi = {10.1103/PhysRevB.82.073411}
}

@article{hwangDielectricFunction2007,
  title = {Dielectric Function, Screening, and Plasmons in Two-Dimensional Graphene},
  author = {Hwang, E. H. and Das Sarma, S.},
  year = 2007,
  journal = {Physical Review B - Condensed Matter and Materials Physics},
  volume = {75},
  number = {20},
  eprint = {cond-mat/0610561},
  pages = {1--6},
  issn = {10980121},
  doi = {10.1103/PhysRevB.75.205418},
  archiveprefix = {arXiv}
}

@article{wunschDynamicalPolarization2006,
  title = {Dynamical Polarization of Graphene at Finite Doping},
  author = {Wunsch, B and Stauber, T and Sols, F and Guinea, F},
  year = 2006,
  journal = {New Journal of Physics},
  volume = {8},
  issn = {13672630},
  doi = {10.1088/1367-2630/8/12/318}
}

@article{maldagueManybodyCorrections1978,
  title = {Many-Body Corrections to the Polarizability of the Two-Dimensional Electron Gas},
  author = {Maldague, Pierre F.},
  year = 1978,
  month = may,
  journal = {Surface Science},
  volume = {73},
  pages = {296--302},
  issn = {00396028},
  doi = {10.1016/0039-6028(78)90507-1},
  urldate = {2026-04-26},
  copyright = {https://www.elsevier.com/tdm/userlicense/1.0/},
  langid = {english}
}

@article{friedelXIVDistribution1952,
  title = {{{XIV}}. {{The}} Distribution of Electrons Round Impurities in Monovalent Metals},
  author = {Friedel, J.},
  year = 1952,
  month = feb,
  journal = {The London, Edinburgh, and Dublin Philosophical Magazine and Journal of Science},
  volume = {43},
  number = {337},
  pages = {153--189},
  issn = {1941-5982, 1941-5990},
  doi = {10.1080/14786440208561086},
  urldate = {2026-04-26},
  langid = {english}
}

@Article{	  abdelghani-idrissiresonantosmotic2025,
  title		= {Resonant Osmotic Diodes for Voltage-Induced Water
		  Filtration across Composite Membranes},
  author	= {{Abdelghani-Idrissi}, Soufiane and Ries, Lucie and Monet,
		  Geoffrey and {Perez-Carvajal}, Javier and Pilo, Zacharie
		  and Sarnikowski, Paulina and Siria, Alessandro and Bocquet,
		  Lyd{\'e}ric},
  year		= 2025,
  month		= jul,
  journal	= {Nature Materials},
  volume	= {24},
  number	= {7},
  pages		= {1109--1115},
  issn		= {1476-1122, 1476-4660},
  doi		= {10.1038/s41563-025-02257-z},
  urldate	= {2025-08-31},
  langid	= {english}
}

@Article{	  alexiadismolecularsimulationwater2008,
  title		= {Molecular {{Simulation}} of {{Water}} in {{Carbon
		  Nanotubes}}},
  author	= {Alexiadis, Alessio and Kassinos, Stavros},
  year		= 2008,
  month		= dec,
  journal	= {Chemical Reviews},
  volume	= {108},
  number	= {12},
  pages		= {5014--5034},
  publisher	= {American Chemical Society},
  issn		= {0009-2665},
  doi		= {10.1021/cr078140f},
  urldate	= {2026-02-28}
}

@Book{		  altlandcondensedmatter2023,
  title		= {Condensed {{Matter Field Theory}}},
  author	= {Altland, Alexander and Simons, Ben},
  year		= 2023,
  month		= sep,
  edition	= {3},
  publisher	= {Cambridge University Press},
  doi		= {10.1017/9781108781244},
  urldate	= {2026-01-04},
  copyright	= {https://www.cambridge.org/core/terms},
  isbn		= {978-1-108-78124-4 978-1-108-49460-1}
}

@Article{	  alurufluidselectrolytesconfinement2023,
  title		= {Fluids and {{Electrolytes}} under {{Confinement}} in
		  {{Single-Digit Nanopores}}},
  author	= {Aluru, Narayana R. and Aydin, Fikret and Bazant, Martin Z.
		  and Blankschtein, Daniel and Brozena, Alexandra H. and {de
		  Souza}, J. Pedro and Elimelech, Menachem and Faucher,
		  Samuel and Fourkas, John T. and Koman, Volodymyr B. and
		  Kuehne, Matthias and Kulik, Heather J. and Li, Hao-Kun and
		  Li, Yuhao and Li, Zhongwu and Majumdar, Arun and Martis,
		  Joel and Misra, Rahul Prasanna and Noy, Aleksandr and Pham,
		  Tuan Anh and Qu, Haoran and Rayabharam, Archith and Reed,
		  Mark A. and Ritt, Cody L. and Schwegler, Eric and Siwy,
		  Zuzanna and Strano, Michael S. and Wang, YuHuang and Yao,
		  Yun-Chiao and Zhan, Cheng and Zhang, Ze},
  year		= 2023,
  month		= mar,
  journal	= {Chemical Reviews},
  volume	= {123},
  number	= {6},
  pages		= {2737--2831},
  publisher	= {American Chemical Society},
  issn		= {0009-2665},
  doi		= {10.1021/acs.chemrev.2c00155},
  urldate	= {2025-07-29}
}

@Article{	  barrat1999,
  title		= {Influence of Wetting Properties on Hydrodynamic Boundary
		  Conditions at a Fluid/Solid Interface},
  author	= {Barrat, Jean-Louis and Bocquet, Lyd{\'e}ric},
  year		= {1999},
  journal	= {Faraday Discussions},
  volume	= {112},
  number	= {0},
  pages		= {119--128},
  publisher	= {Royal Society of Chemistry},
  doi		= {10.1039/A809733J},
  urldate	= {2024-02-08},
  langid	= {english}
}

@Article{	  bocquet1994,
  title		= {Hydrodynamic Boundary Conditions, Correlation Functions,
		  and {{Kubo}} Relations for Confined Fluids},
  author	= {Bocquet, Lyd{\'e}ric and Barrat, Jean-Louis},
  year		= {1994},
  month		= apr,
  journal	= {Physical Review E},
  volume	= {49},
  number	= {4},
  pages		= {3079--3092},
  publisher	= {American Physical Society},
  doi		= {10.1103/PhysRevE.49.3079},
  urldate	= {2024-02-08}
}

@Article{	  bocquet2007,
  title		= {Flow Boundary Conditions from Nano- to Micro-Scales},
  author	= {Bocquet, Lyd{\'e}ric and Barrat, Jean-Louis},
  year		= {2007},
  journal	= {Soft Matter},
  volume	= {3},
  number	= {6},
  pages		= {685--693},
  publisher	= {Royal Society of Chemistry},
  doi		= {10.1039/B616490K},
  urldate	= {2022-05-04},
  langid	= {english}
}

@Article{	  bocquetgreenkuborelationship2013,
  title		= {On the {{Green-Kubo}} Relationship for the Liquid-Solid
		  Friction Coefficient},
  author	= {Bocquet, Lyd{\'e}ric and Barrat, Jean Louis},
  year		= 2013,
  journal	= {Journal of Chemical Physics},
  volume	= {139},
  number	= {4},
  issn		= {00219606},
  doi		= {10.1063/1.4816006}
}

@Article{	  bocquetnanofluidicsbulkinterfaces2010,
  title		= {Nanofluidics, from Bulk to Interfaces},
  author	= {Bocquet, Lyd{\'e}ric and Charlaix, Elisabeth},
  year		= 2010,
  journal	= {Chem. Soc. Rev.},
  volume	= {39},
  number	= {3},
  pages		= {1073--1095},
  issn		= {0306-0012, 1460-4744},
  doi		= {10.1039/B909366B},
  urldate	= {2024-11-05},
  langid	= {english}
}

@Article{	  bocquetphysicstechnological2014,
  title		= {Physics and Technological Aspects of Nanofluidics},
  author	= {Bocquet, Lyderic and Tabeling, Patrick},
  year		= 2014,
  journal	= {Lab on a Chip},
  volume	= {14},
  number	= {17},
  pages		= {3143--3158},
  issn		= {14730189},
  doi		= {10.1039/c4lc00325j}
}

@Article{	  bonellaadiabaticmotionstatistical2020,
  title		= {Adiabatic Motion and Statistical Mechanics via Mass-Zero
		  Constrained Dynamics},
  author	= {Bonella, Sara and Coretti, Alessandro and Vuilleumier,
		  Rodolphe and Ciccotti, Giovanni},
  year		= 2020,
  month		= may,
  journal	= {Physical Chemistry Chemical Physics},
  volume	= {22},
  number	= {19},
  pages		= {10775--10785},
  publisher	= {The Royal Society of Chemistry},
  issn		= {1463-9084},
  doi		= {10.1039/D0CP00163E},
  urldate	= {2025-12-14},
  langid	= {english}
}

@Article{	  bonthuis2012,
  title		= {Profile of the {{Static Permittivity Tensor}} of {{Water}}
		  at {{Interfaces}}: {{Consequences}} for {{Capacitance}},
		  {{Hydration Interaction}} and {{Ion Adsorption}}},
  shorttitle	= {Profile of the {{Static Permittivity Tensor}} of {{Water}}
		  at {{Interfaces}}},
  author	= {Bonthuis, Douwe Jan and Gekle, Stephan and Netz, Roland
		  R.},
  year		= {2012},
  month		= may,
  journal	= {Langmuir},
  volume	= {28},
  number	= {20},
  pages		= {7679--7694},
  publisher	= {American Chemical Society},
  issn		= {0743-7463},
  doi		= {10.1021/la2051564},
  urldate	= {2022-02-04}
}

@Article{	  buiclassicalquantumfriction2023,
  title		= {Classical {{Quantum Friction}} at {{Water}}--{{Carbon
		  Interfaces}}},
  author	= {Bui, Anna T. and Thiemann, Fabian L. and Michaelides,
		  Angelos and Cox, Stephen J.},
  year		= {2023},
  month		= jan,
  journal	= {Nano Lett.},
  volume	= {23},
  number	= {2},
  pages		= {580--587},
  issn		= {1530-6984, 1530-6992},
  doi		= {10.1021/acs.nanolett.2c04187},
  urldate	= {2023-06-19},
  langid	= {english}
}

@Article{	  carlson2020,
  title		= {Exploring the {{Absorption Spectrum}} of {{Simulated
		  Water}} from {{MHz}} to {{Infrared}}},
  author	= {Carlson, Shane and Br{\"u}nig, Florian N. and Loche,
		  Philip and Bonthuis, Douwe Jan and Netz, Roland R.},
  year		= {2020},
  month		= jul,
  journal	= {The Journal of Physical Chemistry A},
  volume	= {124},
  number	= {27},
  pages		= {5599--5605},
  publisher	= {American Chemical Society},
  issn		= {1089-5639},
  doi		= {10.1021/acs.jpca.0c04063},
  urldate	= {2024-02-03}
}

@Article{	  casimir1948a,
  title		= {The {{Influence}} of {{Retardation}} on the {{London-van}}
		  Der {{Waals Forces}}},
  author	= {Casimir, H. B. G. and Polder, D.},
  year		= {1948},
  month		= feb,
  journal	= {Physical Review},
  volume	= {73},
  number	= {4},
  pages		= {360--372},
  issn		= {0031-899X},
  doi		= {10.1103/PhysRev.73.360},
  urldate	= {2025-08-27},
  copyright	= {http://link.aps.org/licenses/aps-default-license},
  langid	= {english}
}

@Article{	  casimir:1948dh,
  author	= "Casimir, H. B. G.",
  title		= "{On the attraction between two perfectly conducting
		  plates}",
  journal	= "Indag. Math.",
  volume	= "10",
  number	= "4",
  pages		= "261--263",
  year		= "1948"
}

@Article{	  chatzichristoscurrentunderstandingwater2022,
  title		= {Current {{Understanding}} of {{Water Properties}} inside
		  {{Carbon Nanotubes}}},
  author	= {Chatzichristos, Aris and Hassan, Jamal and Chatzichristos,
		  Aris and Hassan, Jamal},
  year		= 2022,
  month		= jan,
  journal	= {Nanomaterials},
  volume	= {12},
  number	= {1},
  publisher	= {publisher},
  issn		= {2079-4991},
  doi		= {10.3390/nano12010174},
  urldate	= {2025-12-11},
  copyright	= {http://creativecommons.org/licenses/by/3.0/},
  langid	= {english}
}

@Article{	  cheninducingelectriccurrent2023,
  title		= {Inducing {{Electric Current}} in {{Graphene Using Ionic
		  Flow}}},
  author	= {Chen, Fanfan and Zhao, Yunhong and Saxena, Anshul and
		  Zhao, Chunxiao and Niu, Mengdi and Aluru, Narayana R and
		  Feng, Jiandong},
  year		= {2023},
  month		= may,
  journal	= {Nano Lett.},
  volume	= {23},
  number	= {10},
  pages		= {4464--4470},
  issn		= {1530-6984, 1530-6992},
  doi		= {10.1021/acs.nanolett.3c00821},
  urldate	= {2025-03-03},
  copyright	= {https://doi.org/10.15223/policy-029},
  langid	= {english}
}

@Article{	  chensilverelectrodepositionag2025,
  title		= {Silver {{Electrodeposition}} from {{Ag}}/{{AgCl
		  Electrodes}}: {{Implications}} for {{Nanoscience}}},
  shorttitle	= {Silver {{Electrodeposition}} from {{Ag}}/{{AgCl
		  Electrodes}}},
  author	= {Chen, Chuhongxu and Wang, Ziwei and Chen, Guilin and
		  Zhang, Zhijia and Bedran, Zakhar and Tipper, Stephen and
		  {Diaz-N{\'u}{\~n}ez}, Pablo and Timokhin, Ivan and
		  Mishchenko, Artem and Yang, Qian},
  year		= 2025,
  month		= jun,
  journal	= {Nano Letters},
  volume	= {25},
  number	= {23},
  pages		= {9427--9432},
  issn		= {1530-6984, 1530-6992},
  doi		= {10.1021/acs.nanolett.5c01929},
  urldate	= {2025-06-12},
  copyright	= {https://creativecommons.org/licenses/by-nc-nd/4.0/},
  langid	= {english}
}

@Article{	  chmiolaanomalousincreasecarbon2006,
  title		= {Anomalous {{Increase}} in {{Carbon Capacitance}} at {{Pore
		  Sizes Less Than}} 1 {{Nanometer}}},
  author	= {Chmiola, J. and Yushin, G. and Gogotsi, Y. and Portet, C.
		  and Simon, P. and Taberna, P. L.},
  year		= 2006,
  month		= sep,
  journal	= {Science},
  volume	= {313},
  number	= {5794},
  pages		= {1760--1763},
  publisher	= {American Association for the Advancement of Science},
  doi		= {10.1126/science.1132195},
  urldate	= {2025-07-29}
}

@Article{	  collismeasurementnavierslip2021,
  title		= {Measurement of {{Navier Slip}} on {{Individual
		  Nanoparticles}} in {{Liquid}}},
  author	= {Collis, Jesse F. and Olcum, Selim and Chakraborty, Debadi
		  and Manalis, Scott R. and Sader, John E.},
  year		= 2021,
  month		= jun,
  journal	= {Nano Letters},
  volume	= {21},
  number	= {12},
  pages		= {4959--4965},
  issn		= {1530-6984, 1530-6992},
  doi		= {10.1021/acs.nanolett.1c00603},
  urldate	= {2025-03-04},
  copyright	= {https://doi.org/10.15223/policy-029},
  langid	= {english}
}

@Article{	  comtetnanoscalecapillary2017,
  title		= {Nanoscale Capillary Freezing of Ionic Liquids Confined
		  between Metallic Interfaces and the Role of Electronic
		  Screening},
  author	= {Comtet, Jean and Nigu{\`e}s, Antoine and Kaiser, Vojtech
		  and Coasne, Benoit and Bocquet, Lyd{\'e}ric and Siria,
		  Alessandro},
  year		= {2017},
  month		= jun,
  journal	= {Nature Mater},
  volume	= {16},
  number	= {6},
  pages		= {634--639},
  issn		= {1476-1122, 1476-4660},
  doi		= {10.1038/nmat4880},
  urldate	= {2025-04-20},
  langid	= {english}
}

@Article{	  coquinotcollectivemodes,
  author	= "Coquinot, Baptiste and Becker, Maximilian and Netz, Roland
		  R. and Bocquet, LydÃ©ric and Kavokine, Nikita",
  title		= "Collective modes and quantum effects in two-dimensional
		  nanofluidic channels",
  journal	= "Faraday Discuss.",
  year		= "2024",
  volume	= "249",
  issue		= "0",
  pages		= "162-180",
  publisher	= "The Royal Society of Chemistry",
  doi		= "10.1039/D3FD00115F",
  url		= "http://dx.doi.org/10.1039/D3FD00115F"
}

@Article{	  coquinothydroelectricenergy2024,
  title		= {Hydroelectric Energy Conversion of Waste Flows through
		  Hydroelectronic Drag},
  author	= {Coquinot, Baptiste and Bocquet, Lyd{\'e}ric and Kavokine,
		  Nikita},
  year		= {2024},
  month		= oct,
  journal	= {Proc. Natl. Acad. Sci. U.S.A.},
  volume	= {121},
  number	= {43},
  pages		= {e2411613121},
  issn		= {0027-8424, 1091-6490},
  doi		= {10.1073/pnas.2411613121},
  urldate	= {2025-01-24},
  langid	= {english}
}

@Article{	  coquinotmomentumtunnelling2025,
  title		= {Momentum Tunnelling between Nanoscale Liquid Flows},
  author	= {Coquinot, Baptiste and Bui, Anna T. and Toquer, Damien and
		  Michaelides, Angelos and Kavokine, Nikita and Cox, Stephen
		  J. and Bocquet, Lyd{\'e}ric},
  year		= {2025},
  month		= mar,
  journal	= {Nat. Nanotechnol.},
  volume	= {20},
  number	= {3},
  pages		= {397--403},
  issn		= {1748-3387, 1748-3395},
  doi		= {10.1038/s41565-024-01842-8},
  urldate	= {2025-04-20},
  langid	= {english}
}

@Article{	  coquinotquantumfeedback2023,
  title		= {Quantum {{Feedback}} at the {{Solid-Liquid Interface}}:
		  {{Flow-Induced Electronic Current}} and {{Its Negative
		  Contribution}} to {{Friction}}},
  shorttitle	= {Quantum {{Feedback}} at the {{Solid-Liquid Interface}}},
  author	= {Coquinot, Baptiste and Bocquet, Lyd{\'e}ric and Kavokine,
		  Nikita},
  year		= {2023},
  month		= feb,
  journal	= {Phys. Rev. X},
  volume	= {13},
  number	= {1},
  pages		= {011019},
  issn		= {2160-3308},
  doi		= {10.1103/PhysRevX.13.011019},
  urldate	= {2023-10-27},
  langid	= {english}
}

@Article{	  coretticommunicationconstrainedmolecular2018,
  title		= {Communication: {{Constrained}} Molecular Dynamics for
		  Polarizable Models},
  shorttitle	= {Communication},
  author	= {Coretti, Alessandro and Bonella, Sara and Ciccotti,
		  Giovanni},
  year		= 2018,
  month		= nov,
  journal	= {The Journal of Chemical Physics},
  volume	= {149},
  number	= {19},
  pages		= {191102},
  issn		= {0021-9606},
  doi		= {10.1063/1.5055704},
  urldate	= {2025-12-11}
}

@Article{	  corettifluctuationrelationssystems2020,
  title		= {Fluctuation Relations for Systems in a Constant Magnetic
		  Field},
  author	= {Coretti, Alessandro and Rondoni, Lamberto and Bonella,
		  Sara},
  year		= 2020,
  month		= sep,
  journal	= {Physical Review E},
  volume	= {102},
  number	= {3},
  pages		= {030101},
  publisher	= {American Physical Society},
  doi		= {10.1103/PhysRevE.102.030101},
  urldate	= {2026-02-28}
}

@Article{	  corrydesigningcarbonnanotube2008,
  title		= {Designing {{Carbon Nanotube Membranes}} for {{Efficient
		  Water Desalination}}},
  author	= {Corry, Ben},
  year		= 2008,
  month		= feb,
  journal	= {The Journal of Physical Chemistry B},
  volume	= {112},
  number	= {5},
  pages		= {1427--1434},
  publisher	= {American Chemical Society},
  issn		= {1520-6106},
  doi		= {10.1021/jp709845u},
  urldate	= {2026-02-28}
}

@Article{	  cruz-chuphononswaterflow2017,
  title		= {On Phonons and Water Flow Enhancement in Carbon
		  Nanotubes},
  author	= {{Cruz-Ch{\'u}}, Eduardo R. and Papadopoulou, Ermioni and
		  Walther, Jens H. and Popadi{\'c}, Aleksandar and Li,
		  Gengyun and Praprotnik, Matej and Koumoutsakos, Petros},
  year		= 2017,
  month		= dec,
  journal	= {Nature Nanotechnology},
  volume	= {12},
  number	= {12},
  pages		= {1106--1108},
  issn		= {1748-3387, 1748-3395},
  doi		= {10.1038/nnano.2017.234},
  urldate	= {2025-12-13},
  langid	= {english}
}

@article{cuicouplingiontransport2025,
  title = {Coupling between Ion Transport and Electronic Properties in Individual Carbon Nanotubes},
  author = {Cui, Guandong and Xu, Zhi and Zhang, Shuchen and Siria, Alessandro and Ma, Ming},
  year = 2025,
  month = aug,
  journal = {Science Advances},
  volume = {11},
  number = {34},
  pages = {eadu7410},
  issn = {2375-2548},
  doi = {10.1126/sciadv.adu7410},
  urldate = {2025-08-31},
  langid = {english}
}

@Article{	  dajornadauniversalslow2020,
  title		= {Universal Slow Plasmons and Giant Field Enhancement in
		  Atomically Thin Quasi-Two-Dimensional Metals},
  author	= {{da Jornada}, Felipe H. and Xian, Lede and Rubio, Angel
		  and Louie, Steven G.},
  year		= 2020,
  journal	= {Nature Communications},
  volume	= {11},
  number	= {1},
  pages		= {1--10},
  issn		= {20411723},
  doi		= {10.1038/s41467-020-14826-8}
}

@Article{	  dedkovelectromagneticfluctuationelectromagneticforces2002,
  title		= {Electromagnetic and Fluctuation-Electromagnetic Forces of
		  Interaction of Moving Particles and Nanoprobes with
		  Surfaces: {{A}} Nonrelativistic Consideration},
  shorttitle	= {Electromagnetic and Fluctuation-Electromagnetic Forces of
		  Interaction of Moving Particles and Nanoprobes with
		  Surfaces},
  author	= {Dedkov, G. V. and Kyasov, A. A.},
  year		= {2002},
  month		= oct,
  journal	= {Physics of the Solid State},
  volume	= {44},
  number	= {10},
  pages		= {1809--1832},
  issn		= {1090-6460},
  doi		= {10.1134/1.1514767},
  urldate	= {2025-05-15},
  langid	= {english}
}

@Article{	  dhimanharvestingenergywater2011,
  title		= {Harvesting {{Energy}} from {{Water Flow}} over
		  {{Graphene}}},
  author	= {Dhiman, Prashant and Yavari, Fazel and Mi, Xi and
		  Gullapalli, Hemtej and Shi, Yunfeng and Ajayan, Pulickel M.
		  and Koratkar, Nikhil},
  year		= {2011},
  month		= aug,
  journal	= {Nano Lett.},
  volume	= {11},
  number	= {8},
  pages		= {3123--3127},
  issn		= {1530-6984, 1530-6992},
  doi		= {10.1021/nl2011559},
  urldate	= {2025-03-03},
  langid	= {english}
}

@Article{	  dudkasuperionicliquidsconducting2019,
  title		= {Superionic Liquids in Conducting Nanoslits: {{A}} Variety
		  of Phase Transitions and Ensuing Charging Behavior},
  shorttitle	= {Superionic Liquids in Conducting Nanoslits},
  author	= {Dudka, Maxym and Kondrat, Svyatoslav and B{\'e}nichou,
		  Olivier and Kornyshev, Alexei A. and Oshanin, Gleb},
  year		= 2019,
  month		= nov,
  journal	= {The Journal of Chemical Physics},
  volume	= {151},
  number	= {18},
  pages		= {184105},
  issn		= {0021-9606},
  doi		= {10.1063/1.5127851},
  urldate	= {2025-08-06}
}

@Article{	  dzyaloshinksiigeneraltheory1961,
  title		= {The General Theory of van Der {{Waals}} Forces},
  author	= {Dzyaloshinksii, I.E. and Lifshitz, E.M. and Pitaevskii,
		  L.P. and Priestley, M.G.},
  year		= {1961},
  journal	= {Advances in Physics},
  pages		= {165--209},
  doi		= {10.1016/b978-0-08-036364-6.50039-9}
}

@Article{	  emmerichenhancednanofluidictransport2022,
  title		= {Enhanced Nanofluidic Transport in Activated Carbon
		  Nanoconduits},
  author	= {Emmerich, Theo and Vasu, Kalangi S. and Nigu{\`e}s,
		  Antoine and Keerthi, Ashok and Radha, Boya and Siria,
		  Alessandro and Bocquet, Lyd{\'e}ric},
  year		= 2022,
  month		= jun,
  journal	= {Nature Materials},
  volume	= {21},
  number	= {6},
  pages		= {696--702},
  publisher	= {Nature Publishing Group},
  issn		= {1476-4660},
  doi		= {10.1038/s41563-022-01229-x},
  urldate	= {2025-09-23},
  copyright	= {2022 The Author(s), under exclusive licence to Springer
		  Nature Limited},
  langid	= {english}
}

@Article{	  emmerichnanofluidics2024,
  title		= {Nanofluidics},
  author	= {Emmerich, Theo and Ronceray, Nathan and Agrawal, Kumar
		  Varoon and Garaj, Slaven and Kumar, Manish and Noy,
		  Aleksandr and Radenovic, Aleksandra},
  year		= 2024,
  month		= sep,
  journal	= {Nature Reviews Methods Primers},
  volume	= {4},
  number	= {1},
  pages		= {69},
  issn		= {2662-8449},
  doi		= {10.1038/s43586-024-00344-0},
  urldate	= {2025-10-15},
  langid	= {english}
}

@Article{	  falk2010,
  title		= {Molecular {{Origin}} of {{Fast Water Transport}} in
		  {{Carbon Nanotube Membranes}}: {{Superlubricity}} versus
		  {{Curvature Dependent Friction}}},
  shorttitle	= {Molecular {{Origin}} of {{Fast Water Transport}} in
		  {{Carbon Nanotube Membranes}}},
  author	= {Falk, Kerstin and Sedlmeier, Felix and Joly, Laurent and
		  Netz, Roland R. and Bocquet, Lyd{\'e}ric},
  year		= {2010},
  month		= oct,
  journal	= {Nano Letters},
  volume	= {10},
  number	= {10},
  pages		= {4067--4073},
  publisher	= {American Chemical Society},
  issn		= {1530-6984},
  doi		= {10.1021/nl1021046},
  urldate	= {2024-02-15}
}

@Article{	  fauchercriticalknowledge2019,
  title		= {Critical {{Knowledge Gaps}} in {{Mass Transport}} through
		  {{Single-Digit Nanopores}}: {{A Review}} and
		  {{Perspective}}},
  author	= {Faucher, Samuel and Aluru, Narayana and Bazant, Martin Z.
		  and Blankschtein, Daniel and Brozena, Alexandra H. and
		  Cumings, John and {Pedro de Souza}, J. and Elimelech,
		  Menachem and Epsztein, Razi and Fourkas, John T. and Rajan,
		  Ananth Govind and Kulik, Heather J. and Levy, Amir and
		  Majumdar, Arun and Martin, Charles and McEldrew, Michael
		  and Misra, Rahul Prasanna and Noy, Aleksandr and Pham, Tuan
		  Anh and Reed, Mark and Schwegler, Eric and Siwy, Zuzanna
		  and Wang, Yuhuang and Strano, Michael},
  year		= 2019,
  month		= sep,
  journal	= {The Journal of Physical Chemistry C},
  volume	= {123},
  number	= {35},
  pages		= {21309--21326},
  issn		= {1932-7447},
  doi		= {10.1021/acs.jpcc.9b02178}
}

@Article{	  ferrell1979419,
  title		= {Friction parameter of an ion near a metal surface},
  journal	= {Solid State Communications},
  volume	= {32},
  number	= {5},
  pages		= {419-422},
  year		= {1979},
  issn		= {0038-1098},
  doi		= {https://doi.org/10.1016/0038-1098(79)90479-4},
  url		= {https://www.sciencedirect.com/science/article/pii/0038109879904794},
  author	= {T.L. Ferrell and P.M. Echenique and R.H. Ritchie}
}

@Article{	  ghoshcarbonnanotubeflow2003,
  title		= {Carbon {{Nanotube Flow Sensors}}},
  author	= {Ghosh, Shankar and Sood, A. K. and Kumar, N.},
  year		= 2003,
  month		= feb,
  journal	= {Science},
  volume	= {299},
  number	= {5609},
  pages		= {1042--1044},
  publisher	= {American Association for the Advancement of Science},
  doi		= {10.1126/science.1079080},
  urldate	= {2025-09-06}
}

@Article{	  ghoshflowinducedvoltagecurrent2004,
  title		= {Flow-Induced Voltage and Current Generation in Carbon
		  Nanotubes},
  author	= {Ghosh, S. and Sood, A. K. and Ramaswamy, S. and Kumar,
		  N.},
  year		= 2004,
  month		= nov,
  journal	= {Physical Review B},
  volume	= {70},
  number	= {20},
  pages		= {205423},
  publisher	= {American Physical Society},
  doi		= {10.1103/PhysRevB.70.205423},
  urldate	= {2025-07-30}
}

@Misc{		  gispertelectrostaticscreeningnanotubes2025,
  title		= {Electrostatic {{Screening}} in {{Nanotubes}}: {{A Tubular
		  Response Function Framework}}},
  shorttitle	= {Electrostatic {{Screening}} in {{Nanotubes}}},
  author	= {Gispert, Peter and Kavokine, Nikita},
  year		= 2025,
  month		= dec,
  number	= {arXiv:2512.07036},
  eprint	= {2512.07036},
  primaryclass	= {cond-mat},
  publisher	= {arXiv},
  doi		= {10.48550/arXiv.2512.07036},
  urldate	= {2025-12-20},
  archiveprefix	= {arXiv}
}

@Article{	  griffinharris,
  author	= {Griffin, A. and Harris, J.},
  title		= {Sum rules for a bounded electron gas},
  journal	= {Canadian Journal of Physics},
  volume	= {54},
  number	= {13},
  pages		= {1396-1408},
  year		= {1976},
  doi		= {10.1139/p76-164},
  url		= {
		  
		  https://doi.org/10.1139/p76-164
		  
		  },
  eprint	= {
		  
		  https://doi.org/10.1139/p76-164
		  
		  }
}

@Article{	  grosjean2016,
  title		= {Chemisorption of {{Hydroxide}} on {{2D Materials}} from
		  {{DFT Calculations}}: {{Graphene}} versus {{Hexagonal Boron
		  Nitride}}},
  shorttitle	= {Chemisorption of {{Hydroxide}} on {{2D Materials}} from
		  {{DFT Calculations}}},
  author	= {Grosjean, Benoit and Pean, Clarisse and Siria, Alessandro
		  and Bocquet, Lyd{\'e}ric and Vuilleumier, Rodolphe and
		  Bocquet, Marie-Laure},
  year		= {2016},
  month		= nov,
  journal	= {The Journal of Physical Chemistry Letters},
  volume	= {7},
  number	= {22},
  pages		= {4695--4700},
  publisher	= {American Chemical Society},
  doi		= {10.1021/acs.jpclett.6b02248},
  urldate	= {2024-02-15}
}

@Article{	  grosjean2019versatile,
  title		= {Versatile electrification of two-dimensional nanomaterials
		  in water},
  author	= {Grosjean, Beno{\^\i}t and Bocquet, Marie-Laure and
		  Vuilleumier, Rodolphe},
  journal	= {Nature communications},
  volume	= {10},
  number	= {1},
  pages		= {1656},
  year		= {2019},
  publisher	= {Nature Publishing Group UK London}
}

@Article{	  guo2015giant,
  title		= {Giant conductance and anomalous concentration dependence
		  in sub-5 nm carbon nanotube nanochannels},
  author	= {Guo, Shirui and Buchsbaum, Steven F and Meshot, Eric R and
		  Davenport, Matthew W and Siwy, Zuzanna and Fornasiero,
		  Francesco},
  journal	= {Biophysical Journal},
  volume	= {108},
  number	= {2},
  pages		= {175a},
  year		= {2015},
  publisher	= {Elsevier}
}

@Misc{		  herrerofluidselectrostaticallyactive2026,
  title		= {Fluids at an Electrostatically Active Surface: {{Optimum}}
		  in Interfacial Friction and Electrohydrodynamic Drag},
  shorttitle	= {Fluids at an Electrostatically Active Surface},
  author	= {Herrero, Cecilia and Bocquet, Lyderic and Coasne, Benoit},
  year		= 2026,
  month		= jan,
  number	= {arXiv:2601.02539},
  eprint	= {2601.02539},
  primaryclass	= {cond-mat},
  publisher	= {arXiv},
  doi		= {10.48550/arXiv.2601.02539},
  urldate	= {2026-02-19},
  archiveprefix	= {arXiv},
  langid	= {english}
}

@Article{	  herreropoissonboltzmann2024,
  title		= {The {{Poisson}}--{{Boltzmann}} Equation in Micro- and
		  Nanofluidics: {{A}} Formulary},
  shorttitle	= {The {{Poisson}}--{{Boltzmann}} Equation in Micro- and
		  Nanofluidics},
  author	= {Herrero, Cecilia and Joly, Laurent},
  year		= 2024,
  month		= oct,
  journal	= {Physics of Fluids},
  volume	= {36},
  number	= {10},
  pages		= {101801},
  issn		= {1070-6631, 1089-7666},
  doi		= {10.1063/5.0238173},
  urldate	= {2025-12-07},
  langid	= {english}
}

@Article{	  holeeflowinducedvoltagegeneration2013a,
  title		= {Flow-Induced Voltage Generation in Non-Ionic Liquids over
		  Monolayer Graphene},
  author	= {Ho Lee, Seung and Jung, Yousung and Kim, Soohyun and Han,
		  Chang-Soo},
  year		= 2013,
  month		= feb,
  journal	= {Applied Physics Letters},
  volume	= {102},
  number	= {6},
  pages		= {063116},
  issn		= {0003-6951, 1077-3118},
  doi		= {10.1063/1.4792702},
  urldate	= {2025-10-14},
  langid	= {english}
}

@Article{	  holtfastmasstransport2006,
  title		= {Fast {{Mass Transport Through Sub-2-Nanometer Carbon
		  Nanotubes}}},
  author	= {Holt, Jason K. and Park, Hyung Gyu and Wang, Yinmin and
		  Stadermann, Michael and Artyukhin, Alexander B. and
		  Grigoropoulos, Costas P. and Noy, Aleksandr and Bakajin,
		  Olgica},
  year		= 2006,
  month		= may,
  journal	= {Science},
  volume	= {312},
  number	= {5776},
  pages		= {1034--1037},
  issn		= {0036-8075, 1095-9203},
  doi		= {10.1126/science.1126298},
  urldate	= {2025-02-26},
  langid	= {english}
}

@Article{	  huang2008,
  title		= {Water {{Slippage}} versus {{Contact Angle}}: {{A
		  Quasiuniversal Relationship}}},
  shorttitle	= {Water {{Slippage}} versus {{Contact Angle}}},
  author	= {Huang, David M. and Sendner, Christian and Horinek,
		  Dominik and Netz, Roland R. and Bocquet, Lyd{\'e}ric},
  year		= {2008},
  month		= nov,
  journal	= {Physical Review Letters},
  volume	= {101},
  number	= {22},
  pages		= {226101},
  publisher	= {American Physical Society},
  doi		= {10.1103/PhysRevLett.101.226101},
  urldate	= {2024-02-08}
}

@Article{	  hummerwaterconductionhydrophobic2001,
  title		= {Water Conduction through the Hydrophobic Channel of a
		  Carbon Nanotube},
  author	= {Hummer, G. and Rasaiah, J. C. and Noworyta, J. P.},
  year		= 2001,
  month		= nov,
  journal	= {Nature},
  volume	= {414},
  number	= {6860},
  pages		= {188--190},
  publisher	= {Nature Publishing Group},
  issn		= {1476-4687},
  doi		= {10.1038/35102535},
  urldate	= {2025-07-30},
  copyright	= {2001 Macmillan Magazines Ltd.},
  langid	= {english}
}

@Article{	  kaiser2017electrostatic,
  title		= {Electrostatic interactions between ions near Thomas--Fermi
		  substrates and the surface energy of ionic crystals at
		  imperfect metals},
  author	= {Kaiser, V and Comtet, J and Nigu{\`e}s, A and Siria, A and
		  Coasne, Benoit and Bocquet, L},
  journal	= {Faraday discussions},
  volume	= {199},
  pages		= {129--158},
  year		= {2017},
  publisher	= {Royal Society of Chemistry}
}

@Book{		  kamenev2011,
  title		= {Field {{Theory}} of {{Non-Equilibrium Systems}}},
  author	= {Kamenev, Alex},
  year		= {2011},
  publisher	= {Cambridge University Press},
  address	= {Cambridge},
  doi		= {10.1017/CBO9781139003667},
  urldate	= {2022-05-09},
  isbn		= {978-0-521-76082-9}
}

@Article{	  kannamhowfastdoes2013,
  title		= {How Fast Does Water Flow in Carbon Nanotubes?},
  author	= {Kannam, Sridhar Kumar and Todd, B. D. and Hansen, J. S.
		  and Daivis, Peter J.},
  year		= 2013,
  month		= mar,
  journal	= {The Journal of Chemical Physics},
  volume	= {138},
  number	= {9},
  pages		= {094701},
  issn		= {0021-9606, 1089-7690},
  doi		= {10.1063/1.4793396},
  urldate	= {2025-03-16},
  langid	= {english}
}

@Article{	  kavokinefluctuationinducedquantum2022,
  title		= {Fluctuation-Induced Quantum Friction in Nanoscale Water
		  Flows},
  author	= {Kavokine, Nikita and Bocquet, Marie-Laure and Bocquet,
		  Lyd{\'e}ric},
  year		= {2022},
  month		= feb,
  journal	= {Nature},
  volume	= {602},
  number	= {7895},
  eprint	= {2105.03413},
  pages		= {84--90},
  publisher	= {Springer US},
  issn		= {0028-0836},
  doi		= {10.1038/s41586-021-04284-7},
  archiveprefix	= {arXiv}
}

@Article{	  kavokinefluidsnanoscalecontinuum2021,
  title		= {Fluids at the {{Nanoscale}}: {{From Continuum}} to
		  {{Subcontinuum Transport}}},
  shorttitle	= {Fluids at the {{Nanoscale}}},
  author	= {Kavokine, Nikita and Netz, Roland R. and Bocquet,
		  Lyd{\'e}ric},
  year		= 2021,
  month		= jan,
  journal	= {Annual Review of Fluid Mechanics},
  volume	= {53},
  number	= {1},
  pages		= {377--410},
  issn		= {0066-4189, 1545-4479},
  doi		= {10.1146/annurev-fluid-071320-095958},
  urldate	= {2024-06-03},
  langid	= {english}
}

@Article{	  keerthiwaterfrictionnanofluidic2021,
  title		= {Water Friction in Nanofluidic Channels Made from
		  Two-Dimensional Crystals},
  author	= {Keerthi, Ashok and Goutham, Solleti and You, Yi and
		  Iamprasertkun, Pawin and Dryfe, Robert A. W. and Geim,
		  Andre K. and Radha, Boya},
  year		= 2021,
  month		= may,
  journal	= {Nature Communications},
  volume	= {12},
  number	= {1},
  pages		= {3092},
  publisher	= {Nature Publishing Group},
  issn		= {2041-1723},
  doi		= {10.1038/s41467-021-23325-3},
  urldate	= {2025-09-14},
  copyright	= {2021 The Author(s)},
  langid	= {english}
}

@Article{	  keldysh1965,
  title		= {Diagram {{Technique}} for {{Nonequilibrium Processes}}},
  author	= {Keldysh, L. V.},
  year		= 1965,
  month		= apr,
  journal	= {Sov. Phys. JETP},
  volume	= {20},
  pages		= {1018}
}

@Article{	  khalijiplasmonsscreening2020,
  title		= {Plasmons and Screening in Finite-Bandwidth Two-Dimensional
		  Electron Gas},
  author	= {Khaliji, Kaveh and Stauber, Tobias and Low, Tony},
  year		= 2020,
  journal	= {Physical Review B},
  volume	= {102},
  number	= {12},
  pages		= {1--6},
  issn		= {2469-9950},
  doi		= {10.1103/physrevb.102.125408}
}

@Article{	  kingquantumfrictionwater2023,
  title		= {Quantum Friction with Water Effectively Cools Graphene
		  Electrons},
  author	= {King, Sarah B.},
  year		= 2023,
  month		= aug,
  journal	= {Nature Nanotechnology},
  volume	= {18},
  number	= {8},
  pages		= {842--843},
  issn		= {1748-3387, 1748-3395},
  doi		= {10.1038/s41565-023-01424-0},
  urldate	= {2023-10-27},
  langid	= {english}
}

@Article{	  kistwallightinducedquantum2025,
  title		= {Light-Induced Quantum Friction of Carbon Nanotubes in
		  Water},
  author	= {Kistwal, Tanuja and Kanhaiya, Krishan and Buchmann, Adrian
		  and Ma, Chen and Nikoli{\'c}, Jana and Ackermann, Julia and
		  Galonska, Phillip and Nalige, Sanjana S and Havenith,
		  Martina and Sulpizi, Marialore and Kruss, Sebastian},
  year		= {2025},
  journal	= {arXiv},
  doi		= {10.48550/arXiv.2503.12580},
  langid	= {english}
}

@Book{		  kittel2018introduction,
  title		= {Introduction to solid state physics},
  author	= {Kittel, Charles and McEuen, Paul},
  year		= {2018},
  publisher	= {John Wiley \& Sons}
}

@Article{	  kondratsuperionicstatedoublelayer2011,
  title		= {Superionic State in Double-Layer Capacitors with
		  Nanoporous Electrodes},
  author	= {Kondrat, S and Kornyshev, A},
  year		= 2011,
  month		= jan,
  journal	= {Journal of Physics: Condensed Matter},
  volume	= {23},
  number	= {2},
  pages		= {022201},
  publisher	= {IOP Publishing},
  issn		= {0953-8984, 1361-648X},
  doi		= {10.1088/0953-8984/23/2/022201},
  urldate	= {2025-07-23},
  langid	= {english}
}

@Article{	  kondrattheorysimulationsionic2023,
  title		= {Theory and {{Simulations}} of {{Ionic Liquids}} in
		  {{Nanoconfinement}}},
  author	= {Kondrat, Svyatoslav and Feng, Guang and Bresme, Fernando
		  and Urbakh, Michael and Kornyshev, Alexei A.},
  year		= 2023,
  month		= may,
  journal	= {Chemical Reviews},
  volume	= {123},
  number	= {10},
  pages		= {6668--6715},
  issn		= {0009-2665, 1520-6890},
  doi		= {10.1021/acs.chemrev.2c00728},
  urldate	= {2025-08-04},
  copyright	= {https://creativecommons.org/licenses/by/4.0/},
  langid	= {english}
}

@Article{	  kornyshevdoublelayerionicliquids2007,
  title		= {Double-{{Layer}} in {{Ionic Liquids}}:\, {{Paradigm
		  Change}}?},
  shorttitle	= {Double-{{Layer}} in {{Ionic Liquids}}},
  author	= {Kornyshev, Alexei A.},
  year		= 2007,
  month		= may,
  journal	= {The Journal of Physical Chemistry B},
  volume	= {111},
  number	= {20},
  pages		= {5545--5557},
  publisher	= {American Chemical Society},
  issn		= {1520-6106},
  doi		= {10.1021/jp067857o},
  urldate	= {2025-07-29}
}

@Article{	  kralnanotubeelectron2001,
  title		= {Nanotube Electron Drag in Flowing Liquids},
  author	= {Kr{\'a}l, Petr and Shapiro, Moshe},
  year		= {2001},
  journal	= {Physical Review Letters},
  volume	= {86},
  number	= {1},
  pages		= {131--134},
  issn		= {00319007},
  doi		= {10.1103/PhysRevLett.86.131}
}

@Article{	  laine2020nanotribology,
  title		= {Nanotribology of ionic liquids: transition to yielding
		  response in nanometric confinement with metallic surfaces},
  author	= {Lain{\'e}, Antoine and Nigu{\`e}s, Antoine and Bocquet,
		  Lyd{\'e}ric and Siria, Alessandro},
  journal	= {Physical Review X},
  volume	= {10},
  number	= {1},
  pages		= {011068},
  year		= {2020},
  publisher	= {APS}
}

@Article{	  laitenbergerplasmon,
  title		= {Plasmon Dispersion and Damping at the Surface of a
		  Semimetal},
  author	= {Laitenberger, P. and Palmer, R. E.},
  journal	= {Phys. Rev. Lett.},
  volume	= {76},
  issue		= {11},
  pages		= {1952--1955},
  numpages	= {0},
  year		= {1996},
  month		= {Mar},
  publisher	= {American Physical Society},
  doi		= {10.1103/PhysRevLett.76.1952},
  url		= {https://link.aps.org/doi/10.1103/PhysRevLett.76.1952}
}

@Article{	  lamoureuxmodelinginducedpolarization2003,
  title		= {Modeling Induced Polarization with Classical {{Drude}}
		  Oscillators: {{Theory}} and Molecular Dynamics Simulation
		  Algorithm},
  shorttitle	= {Modeling Induced Polarization with Classical {{Drude}}
		  Oscillators},
  author	= {Lamoureux, Guillaume and Roux, Beno{\^\i}t},
  year		= 2003,
  month		= aug,
  journal	= {The Journal of Chemical Physics},
  volume	= {119},
  number	= {6},
  pages		= {3025--3039},
  issn		= {0021-9606},
  doi		= {10.1063/1.1589749},
  urldate	= {2025-07-29}
}

@Article{	  largeotrelationionsize2008,
  title		= {Relation between the {{Ion Size}} and {{Pore Size}} for an
		  {{Electric Double-Layer Capacitor}}},
  author	= {Largeot, Celine and Portet, Cristelle and Chmiola, John
		  and Taberna, Pierre-Louis and Gogotsi, Yury and Simon,
		  Patrice},
  year		= 2008,
  month		= mar,
  journal	= {Journal of the American Chemical Society},
  volume	= {130},
  number	= {9},
  pages		= {2730--2731},
  publisher	= {American Chemical Society},
  issn		= {0002-7863},
  doi		= {10.1021/ja7106178},
  urldate	= {2025-07-31}
}

@Article{	  larkinnonlinearconductivity1975,
  title		= {Nonlinear Conductivity of Superconductors in the Mixed
		  State},
  author	= {Larkin, A. and Ovchinnikov, {\relax Yu}.},
  year		= 1975,
  journal	= {Sov. Phys. JETP},
  volume	= {41},
  pages		= {960},
  issn		= {1063-7761}
}

@Article{	  lavorprobingstructure2020,
  title		= {Probing the Structure and Composition of van Der {{Waals}}
		  Heterostructures Using the Nonlocality of {{Dirac}}
		  Plasmons in the Terahertz Regime},
  author	= {Lavor, Icaro Rodrigues and Cavalcante, L S R and Chaves,
		  Andrey and Peeters, F. M. and Van Duppen, B},
  year		= 2020,
  month		= oct,
  journal	= {2D Materials},
  volume	= {8},
  number	= {1},
  pages		= {015014},
  issn		= {2053-1583},
  doi		= {10.1088/2053-1583/abbecc}
}

@Article{	  leeflowinducedvoltagegeneration2013,
  title		= {Flow-Induced Voltage Generation over Monolayer Graphene in
		  the Presence of Herringbone Grooves},
  author	= {Lee, Seung Ho and Kang, Young Bok and Jung, Wonsuk and
		  Jung, Yousung and Kim, Soohyun and Noh, Hongseok},
  year		= 2013,
  month		= dec,
  journal	= {Nanoscale Research Letters},
  volume	= {8},
  number	= {1},
  pages		= {487},
  issn		= {1556-276X},
  doi		= {10.1186/1556-276X-8-487},
  urldate	= {2025-09-05},
  langid	= {english}
}

@Article{	  leetunablesubnanoporesgraphene2016,
  title		= {Tunable {{Sub-nanopores}} of {{Graphene Flake
		  Interlayers}} with {{Conductive Molecular Linkers}} for
		  {{Supercapacitors}}},
  author	= {Lee, Keunsik and Yoon, Yeoheung and Cho, Yunhee and Lee,
		  Sae Mi and Shin, Yonghun and Lee, Hanleem and Lee,
		  Hyoyoung},
  year		= 2016,
  month		= jul,
  journal	= {ACS Nano},
  volume	= {10},
  number	= {7},
  pages		= {6799--6807},
  publisher	= {American Chemical Society},
  issn		= {1936-0851},
  doi		= {10.1021/acsnano.6b02415},
  urldate	= {2025-08-31}
}

@Article{	  lewandowskiintrinsicallyundamped2019,
  title		= {Intrinsically Undamped Plasmon Modes in Narrow Electron
		  Bands},
  author	= {Lewandowski, Cyprian and Levitov, Leonid},
  year		= 2019,
  journal	= {Proceedings of the National Academy of Sciences of the
		  United States of America},
  volume	= {116},
  number	= {42},
  pages		= {20869--20874},
  issn		= {10916490},
  doi		= {10.1073/pnas.1909069116}
}

@Article{	  li2022,
  title		= {Translucency and Negative Temperature-Dependence for the
		  Slip Length of Water on Graphene},
  author	= {Li, Han and Xu, Zhi and Ma, Chen and Ma, Ming},
  year		= {2022},
  month		= oct,
  journal	= {Nanoscale},
  volume	= {14},
  number	= {39},
  pages		= {14636--14644},
  publisher	= {The Royal Society of Chemistry},
  issn		= {2040-3372},
  doi		= {10.1039/D2NR01481E},
  urldate	= {2024-02-08},
  langid	= {english}
}

@Article{	  licarbonnanotubenanofluidics2025,
  title		= {Carbon Nanotube Nanofluidics},
  author	= {Li, Zhongwu and Noy, Aleksandr},
  year		= 2025,
  month		= aug,
  journal	= {Chemical Society Reviews},
  publisher	= {The Royal Society of Chemistry},
  issn		= {1460-4744},
  doi		= {10.1039/D5CS00233H},
  urldate	= {2025-08-20},
  langid	= {english}
}

@Article{	  liebschdynamicalscreening1987,
  title		= {Dynamical Screening at Simple-Metal Surfaces},
  author	= {Liebsch, A.},
  year		= 1987,
  journal	= {Physical Review B},
  volume	= {36},
  number	= {14},
  pages		= {7378--7388},
  issn		= {01631829},
  doi		= {10.1103/PhysRevB.36.7378}
}

@Article{	  limoleculartransport2024,
  title		= {Molecular Transport Enhancement in Pure Metallic Carbon
		  Nanotube Porins},
  author	= {Li, Yuhao and Li, Zhongwu and Misra, Rahul Prasanna and
		  Liang, Chenxing and Gillen, Alice J. and Zhao, Sidi and
		  Abdullah, Jobaer and Laurence, Ted and Fagan, Jeffrey A.
		  and Aluru, Narayana and Blankschtein, Daniel and Noy,
		  Aleksandr},
  year		= {2024},
  month		= aug,
  journal	= {Nat. Mater.},
  volume	= {23},
  number	= {8},
  pages		= {1123--1130},
  issn		= {1476-1122, 1476-4660},
  doi		= {10.1038/s41563-024-01925-w},
  urldate	= {2024-09-08},
  langid	= {english}
}

@Article{	  liu2010translocation,
  title		= {Translocation of single-stranded DNA through single-walled
		  carbon nanotubes},
  author	= {Liu, Haitao and He, Jin and Tang, Jinyao and Liu, Hao and
		  Pang, Pei and Cao, Di and Krstic, Predrag and Joseph, Sony
		  and Lindsay, Stuart and Nuckolls, Colin},
  journal	= {Science},
  volume	= {327},
  number	= {5961},
  pages		= {64--67},
  year		= {2010},
  publisher	= {American Association for the Advancement of Science}
}

@Article{	  lizee2023,
  title		= {Strong {{Electronic Winds Blowing}} under {{Liquid Flows}}
		  on {{Carbon Surfaces}}},
  author	= {Liz{\'e}e, Mathieu and Marcotte, Alice and Coquinot,
		  Baptiste and Kavokine, Nikita and Sobnath, Karen and
		  Barraud, Cl{\'e}ment and Bhardwaj, Ankit and Radha, Boya
		  and Nigu{\`e}s, Antoine and Bocquet, Lyd{\'e}ric and Siria,
		  Alessandro},
  year		= {2023},
  month		= feb,
  journal	= {Physical Review X},
  volume	= {13},
  number	= {1},
  pages		= {011020},
  publisher	= {American Physical Society},
  doi		= {10.1103/PhysRevX.13.011020},
  urldate	= {2023-05-09}
}

@Article{	  lizeeanomalousfriction2024,
  title		= {Anomalous Friction of Supercooled Glycerol on Mica},
  author	= {Liz{\'e}e, Mathieu and Coquinot, Baptiste and Mariette,
		  Guilhem and Siria, Alessandro and Bocquet, Lyd{\'e}ric},
  year		= {2024},
  month		= jul,
  journal	= {Nat Commun},
  volume	= {15},
  number	= {1},
  pages		= {6129},
  issn		= {2041-1723},
  doi		= {10.1038/s41467-024-50232-0},
  urldate	= {2025-03-18},
  langid	= {english}
}

@Article{	  lochegiantaxialdielectric2019,
  title		= {Giant {{Axial Dielectric Response}} in {{Water-Filled
		  Nanotubes}} and {{Effective Electrostatic Ion}}--{{Ion
		  Interactions}} from a {{Tensorial Dielectric Model}}},
  author	= {Loche, Philip and Ayaz, Cihan and Schlaich, Alexander and
		  Uematsu, Yuki and Netz, Roland R.},
  year		= 2019,
  month		= dec,
  journal	= {The Journal of Physical Chemistry B},
  volume	= {123},
  number	= {50},
  pages		= {10850--10857},
  issn		= {1520-6106, 1520-5207},
  doi		= {10.1021/acs.jpcb.9b09269},
  urldate	= {2025-03-16},
  copyright	= {https://doi.org/10.15223/policy-029},
  langid	= {english}
}

@Article{	  london1930,
  title		= {{{\"U}ber das Verh{\"a}ltnis der van der Waalsschen
		  Kr{\"a}fte zu den hom{\"o}opolaren Bindungskr{\"a}ften}},
  author	= {London, F. and Eisenschitz, R. },
  year		= {1930},
  month		= jul,
  journal	= {Zeitschrift f{\"u}r Physik},
  volume	= {60},
  number	= {7},
  pages		= {491--527},
  issn		= {0044-3328},
  doi		= {10.1007/BF01341258},
  urldate	= {2025-08-27},
  langid	= {ngerman}
}

@Article{	  lynchwaternanoporesbiological2020,
  title		= {Water in {{Nanopores}} and {{Biological Channels}}: {{A
		  Molecular Simulation Perspective}}},
  shorttitle	= {Water in {{Nanopores}} and {{Biological Channels}}},
  author	= {Lynch, Charlotte I. and Rao, Shanlin and Sansom, Mark S.
		  P.},
  year		= 2020,
  month		= sep,
  journal	= {Chemical Reviews},
  volume	= {120},
  number	= {18},
  pages		= {10298--10335},
  publisher	= {American Chemical Society},
  issn		= {0009-2665},
  doi		= {10.1021/acs.chemrev.9b00830},
  urldate	= {2025-12-10}
}

@Article{	  maalimeasurementsliplength2008a,
  title		= {Measurement of the Slip Length of Water Flow on Graphite
		  Surface},
  author	= {Maali, Abdelhamid and {Cohen-Bouhacina}, Touria and
		  Kellay, Hamid},
  year		= 2008,
  month		= feb,
  journal	= {Applied Physics Letters},
  volume	= {92},
  number	= {5},
  publisher	= {AIP Publishing},
  issn		= {0003-6951, 1077-3118},
  doi		= {10.1063/1.2840717},
  urldate	= {2025-07-23},
  langid	= {english}
}

@Article{	  majumderenhancedflowcarbon2005,
  title		= {Enhanced Flow in Carbon Nanotubes},
  author	= {Majumder, Mainak and Chopra, Nitin and Andrews, Rodney and
		  Hinds, Bruce J.},
  year		= 2005,
  month		= nov,
  journal	= {Nature},
  volume	= {438},
  number	= {7064},
  pages		= {44--44},
  issn		= {0028-0836, 1476-4687},
  doi		= {10.1038/438044a},
  urldate	= {2025-03-17},
  langid	= {english}
}

@Article{	  marbachosmosismolecular2019,
  title		= {Osmosis, from Molecular Insights to Large-Scale
		  Applications},
  author	= {Marbach, Sophie and Bocquet, Lyd{\'e}ric},
  year		= 2019,
  journal	= {Chemical Society Reviews},
  volume	= {48},
  number	= {11},
  pages		= {3102--3144},
  issn		= {0306-0012, 1460-4744},
  doi		= {10.1039/C8CS00420J},
  urldate	= {2025-08-31},
  langid	= {english}
}

@Article{	  marin-laflechemetalwallsclassical2020,
  title		= {{{MetalWalls}}: {{A}} Classical Molecular Dynamics
		  Software Dedicated to the Simulation of Electrochemical
		  Systems},
  shorttitle	= {{{MetalWalls}}},
  author	= {{Marin-Lafl{\`e}che}, Abel and Haefele, Matthieu and
		  Scalfi, Laura and Coretti, Alessandro and Dufils, Thomas
		  and Jeanmairet, Guillaume and Reed, Stewart and Serva,
		  Alessandra and Berthin, Roxanne and Bacon, Camille and
		  Bonella, Sara and Rotenberg, Benjamin and Madden, Paul and
		  Salanne, Mathieu},
  year		= {2020},
  month		= sep,
  journal	= {JOSS},
  volume	= {5},
  number	= {53},
  pages		= {2373},
  issn		= {2475-9066},
  doi		= {10.21105/joss.02373},
  urldate	= {2023-10-30},
  langid	= {english}
}

@Article{	  martinstatisticaldynamics1973,
  title		= {Statistical {{Dynamics}} of {{Classical Systems}}},
  author	= {Martin, P. C. and Siggia, E. D. and Rose, H. A.},
  year		= 1973,
  month		= jul,
  journal	= {Physical Review A},
  volume	= {8},
  number	= {1},
  pages		= {423--437},
  issn		= {0556-2791},
  doi		= {10.1103/PhysRevA.8.423},
  urldate	= {2026-01-06},
  copyright	= {http://link.aps.org/licenses/aps-default-license},
  langid	= {english}
}

@Article{	  mawatertransport2015,
  title		= {Water Transport inside Carbon Nanotubes Mediated by
		  Phonon-Induced Oscillating Friction},
  author	= {Ma, Ming and Grey, Fran{\c c}ois and Shen, Luming and
		  Urbakh, Michael and Wu, Shuai and Liu, Jefferson Zhe and
		  Liu, Yilun and Zheng, Quanshui},
  year		= {2015},
  journal	= {Nature Nanotechnology},
  volume	= {10},
  number	= {8},
  pages		= {692--695},
  issn		= {17483395},
  doi		= {10.1038/nnano.2015.134}
}

@Article{	  merletmolecularorigin2012,
  title		= {On the Molecular Origin of Supercapacitance in Nanoporous
		  Carbon Electrodes},
  author	= {Merlet, C{\'e}line and Rotenberg, Benjamin and Madden,
		  Paul A. and Taberna, Pierre Louis and Simon, Patrice and
		  Gogotsi, Yury and Salanne, Mathieu},
  year		= {2012},
  journal	= {Nature Materials},
  volume	= {11},
  number	= {4},
  pages		= {306--310},
  issn		= {14764660},
  doi		= {10.1038/nmat3260}
}

@Article{	  misrainsightsrole2017,
  title		= {Insights on the {{Role}} of {{Many-Body Polarization
		  Effects}} in the {{Wetting}} of {{Graphitic Surfaces}} by
		  {{Water}}},
  author	= {Misra, Rahul Prasanna and Blankschtein, Daniel},
  year		= {2017},
  journal	= {Journal of Physical Chemistry C},
  volume	= {121},
  number	= {50},
  pages		= {28166--28179},
  issn		= {19327455},
  doi		= {10.1021/acs.jpcc.7b08891}
}

@Article{	  mouterde2018,
  title		= {Interfacial Transport with Mobile Surface Charges and
		  Consequences for Ionic Transport in Carbon Nanotubes},
  author	= {Mouterde, Timoth{\'e}e and Bocquet, Lyd{\'e}ric},
  year		= {2018},
  month		= dec,
  journal	= {The European Physical Journal E},
  volume	= {41},
  number	= {12},
  pages		= {148},
  issn		= {1292-895X},
  doi		= {10.1140/epje/i2018-11760-2},
  urldate	= {2022-02-04},
  langid	= {english}
}

@Article{	  mouterdemolecularstreamingits2019,
  title		= {Molecular Streaming and Its Voltage Control in \AA
		  ngstr\"om-Scale Channels},
  author	= {Mouterde, T. and Keerthi, A. and Poggioli, A. R. and Dar,
		  S. A. and Siria, A. and Geim, A. K. and Bocquet, L. and
		  Radha, B.},
  year		= 2019,
  month		= mar,
  journal	= {Nature},
  volume	= {567},
  number	= {7746},
  pages		= {87--90},
  publisher	= {Nature Publishing Group},
  issn		= {1476-4687},
  doi		= {10.1038/s41586-019-0961-5},
  urldate	= {2025-09-28},
  copyright	= {2019 The Author(s), under exclusive licence to Springer
		  Nature Limited},
  langid	= {english}
}

@Article{	  nunez1980,
  title		= {The Energy Loss of Energetic Ions Moving near a Solid
		  Surface},
  author	= {Nunez, R. and Echenique, P. M. and Ritchie, R. H.},
  year		= {1980},
  month		= aug,
  journal	= {Journal of Physics C: Solid State Physics},
  volume	= {13},
  number	= {22},
  pages		= {4229},
  issn		= {0022-3719},
  doi		= {10.1088/0022-3719/13/22/017},
  urldate	= {2025-09-02},
  langid	= {english}
}

@Article{	  pangorigingiantionic2011,
  title		= {Origin of {{Giant Ionic Currents}} in {{Carbon Nanotube
		  Channels}}},
  author	= {Pang, Pei and He, Jin and Park, Jae Hyun and Krsti{\'c},
		  Predrag S. and Lindsay, Stuart},
  year		= 2011,
  month		= sep,
  journal	= {ACS Nano},
  volume	= {5},
  number	= {9},
  pages		= {7277--7283},
  publisher	= {American Chemical Society},
  issn		= {1936-0851},
  doi		= {10.1021/nn202115s},
  urldate	= {2025-09-23}
}

@Article{	  parkmembranestrategies2022,
  title		= {Membrane {{Strategies}} for {{Water Electrolysis}}},
  author	= {Park, Eun Joo and Arges, Christopher G. and Xu, Hui and
		  Kim, Yu Seung},
  year		= 2022,
  month		= oct,
  journal	= {ACS Energy Letters},
  volume	= {7},
  number	= {10},
  pages		= {3447--3457},
  issn		= {2380-8195, 2380-8195},
  doi		= {10.1021/acsenergylett.2c01609},
  urldate	= {2025-12-01},
  copyright	= {https://doi.org/10.15223/policy-029},
  langid	= {english}
}

@Article{	  pascal2011entropy,
  title		= {Entropy and the driving force for the filling of carbon
		  nanotubes with water},
  author	= {Pascal, Tod A and Goddard, William A and Jung, Yousung},
  journal	= {Proceedings of the National Academy of Sciences},
  volume	= {108},
  number	= {29},
  pages		= {11794--11798},
  year		= {2011},
  publisher	= {National Academy of Sciences}
}

@Article{	  pashayevelectronicfingerprintsconfined2025,
  title		= {Electronic Fingerprints of Confined and Adsorbed Water in
		  {{SWCNTs}}},
  author	= {Pashayev, Said and Lhermerout, Romain and Roblin,
		  Christophe and Alibert, Eric and Jelinek, Remi and Izard,
		  Nicolas and Jabbarov, Rasim and Henn, Francois and Noury,
		  Adrien},
  year		= 2025,
  month		= sep,
  journal	= {Nanoscale},
  publisher	= {The Royal Society of Chemistry},
  issn		= {2040-3372},
  doi		= {10.1039/D5NR01939G},
  urldate	= {2025-10-14},
  langid	= {english}
}

@Article{	  pendry1997,
  title		= {Shearing the Vacuum - Quantum Friction},
  author	= {Pendry, J B},
  year		= {1997},
  month		= nov,
  journal	= {Journal of Physics: Condensed Matter},
  volume	= {9},
  number	= {47},
  pages		= {10301--10320},
  issn		= {0953-8984, 1361-648X},
  doi		= {10.1088/0953-8984/9/47/001},
  urldate	= {2025-09-26}
}

@Article{	  perssonelectronicfriction1995,
  title		= {Electronic Friction of Physisorbed Molecules},
  author	= {Persson, B. N.J. and Volokitin, A. I.},
  year		= {1995},
  journal	= {The Journal of Chemical Physics},
  volume	= {103},
  number	= {19},
  pages		= {8679--8683},
  issn		= {00219606},
  doi		= {10.1063/1.470125}
}

@Article{	  phillipfutureseawater2011,
  title		= {The {{Future}} of {{Seawater Desalination}}: {{Energy}},
		  {{Technology}}, and the {{Environment}}},
  author	= {Phillip, William A. and Elimelech, Menachem},
  year		= 2011,
  journal	= {Science},
  volume	= {333},
  number	= {6043},
  pages		= {712--717}
}

@Article{	  picallonanofluidicosmotic2013,
  title		= {Nanofluidic Osmotic Diodes: {{Theory}} and Molecular
		  Dynamics Simulations},
  author	= {Picallo, Clara B. and Gravelle, Simon and Joly, Laurent
		  and Charlaix, Elisabeth and Bocquet, Lyd{\'e}ric},
  year		= 2013,
  journal	= {Physical Review Letters},
  volume	= {111},
  number	= {24},
  pages		= {1--5},
  issn		= {00319007},
  doi		= {10.1103/PhysRevLett.111.244501}
}

@Article{	  pitarketheorysurface2007,
  title		= {Theory of Surface Plasmons and Surface-Plasmon
		  Polaritons},
  author	= {Pitarke, J. M. and Silkin, V. M. and Chulkov, E. V. and
		  Echenique, P. M.},
  year		= 2007,
  journal	= {Reports on Progress in Physics},
  volume	= {70},
  number	= {1},
  pages		= {1--87},
  issn		= {00344885},
  doi		= {10.1088/0034-4885/70/1/R01}
}

@Article{	  portail1999dynamical,
  title		= {Dynamical properties of graphite and peculiar behaviour of
		  the low-energy plasmon},
  author	= {Portail, Marc and Carrere, Marcel and Layet, Jean Marc},
  journal	= {Surface science},
  volume	= {433},
  pages		= {863--867},
  year		= {1999},
  publisher	= {Elsevier}
}

@Article{	  rabinowitzelectricallyactuatedcarbonnanotubebased2020,
  title		= {An {{Electrically Actuated}}, {{Carbon-Nanotube-Based
		  Biomimetic Ion Pump}}},
  author	= {Rabinowitz, Jake and Cohen, Charishma and Shepard, Kenneth
		  L.},
  year		= 2020,
  month		= feb,
  journal	= {Nano Letters},
  volume	= {20},
  number	= {2},
  pages		= {1148--1153},
  issn		= {1530-6984, 1530-6992},
  doi		= {10.1021/acs.nanolett.9b04552},
  urldate	= {2025-05-06},
  copyright	= {https://doi.org/10.15223/policy-029},
  langid	= {english}
}

@Article{	  radhamoleculartransport2016,
  title		= {Molecular Transport through Capillaries Made with
		  Atomic-Scale Precision},
  author	= {Radha, B. and Esfandiar, A. and Wang, F. C. and Rooney, A.
		  P. and Gopinadhan, K. and Keerthi, A. and Mishchenko, A.
		  and Janardanan, A. and Blake, P. and Fumagalli, L. and
		  {Lozada-Hidalgo}, M. and Garaj, S. and Haigh, S. J. and
		  Grigorieva, I. V. and Wu, H. A. and Geim, A. K.},
  year		= {2016},
  journal	= {Nature},
  volume	= {538},
  number	= {7624},
  pages		= {222--225},
  issn		= {14764687},
  doi		= {10.1038/nature19363}
}

@Book{		  rammer2007,
  title		= {Quantum {{Field Theory}} of {{Non-equilibirum States}}},
  author	= {Rammer, Jorgen},
  year		= {2007},
  publisher	= {Cambridge University Press}
}

@Article{	  reedelectrochemicalinterfaceionic2007,
  title		= {Electrochemical Interface between an Ionic Liquid and a
		  Model Metallic Electrode},
  author	= {Reed, Stewart K. and Lanning, Oliver J. and Madden, Paul
		  A.},
  year		= 2007,
  month		= feb,
  journal	= {The Journal of Chemical Physics},
  volume	= {126},
  number	= {8},
  pages		= {084704},
  issn		= {0021-9606},
  doi		= {10.1063/1.2464084},
  urldate	= {2025-12-11}
}

@Article{	  ritchie1966234,
  title		= {The surface plasmon dispersion relation for an electron
		  gas},
  journal	= {Surface Science},
  volume	= {4},
  number	= {3},
  pages		= {234-240},
  year		= {1966},
  issn		= {0039-6028},
  doi		= {https://doi.org/10.1016/0039-6028(66)90003-3},
  url		= {https://www.sciencedirect.com/science/article/pii/0039602866900033},
  author	= {R.H. Ritchie and A.L. Marusak}
}

@Article{	  robinnanofluidicscrossroads2023,
  title		= {Nanofluidics at the Crossroads},
  author	= {Robin, Paul and Bocquet, Lyd{\'e}ric},
  year		= 2023,
  month		= apr,
  journal	= {The Journal of Chemical Physics},
  volume	= {158},
  number	= {16},
  pages		= {160901},
  issn		= {0021-9606, 1089-7690},
  doi		= {10.1063/5.0143222},
  urldate	= {2025-12-01},
  langid	= {english}
}

@Article{	  rochesterinterionicinteractionsconducting2013,
  title		= {Interionic {{Interactions}} in {{Conducting
		  Nanoconfinement}}},
  author	= {Rochester, Christopher C. and Lee, Alpha A. and Pruessner,
		  Gunnar and Kornyshev, Alexei A.},
  year		= 2013,
  journal	= {ChemPhysChem},
  volume	= {14},
  number	= {18},
  pages		= {4121--4125},
  issn		= {1439-7641},
  doi		= {10.1002/cphc.201300834},
  urldate	= {2025-07-31},
  langid	= {english}
}

@Article{	  schlaichelectronicscreeningusing2022,
  title		= {Electronic Screening Using a Virtual {{Thomas}}--{{Fermi}}
		  Fluid for Predicting Wetting and Phase Transitions of Ionic
		  Liquids at Metal Surfaces},
  author	= {Schlaich, Alexander and Jin, Dongliang and Bocquet,
		  Lyderic and Coasne, Benoit},
  year		= {2022},
  month		= feb,
  journal	= {Nat. Mater.},
  volume	= {21},
  number	= {2},
  pages		= {237--245},
  issn		= {1476-1122, 1476-4660},
  doi		= {10.1038/s41563-021-01121-0},
  urldate	= {2023-05-16},
  langid	= {english}
}

@Article{	  schochtransportphenomena2008,
  title		= {Transport Phenomena in Nanofluidics},
  author	= {Schoch, Reto B. and Han, Jongyoon and Renaud, Philippe},
  year		= 2008,
  journal	= {Reviews of Modern Physics},
  volume	= {80},
  number	= {3},
  pages		= {839--883},
  issn		= {00346861},
  doi		= {10.1103/RevModPhys.80.839}
}

@Article{	  secchimassiveradiusdependent2016,
  title		= {Massive Radius-Dependent Flow Slippage in Carbon
		  Nanotubes},
  author	= {Secchi, Eleonora and Marbach, Sophie and Nigu{\`e}s,
		  Antoine and Stein, Derek and Siria, Alessandro and Bocquet,
		  Lyd{\'e}ric},
  year		= {2016},
  journal	= {Nature},
  volume	= {537},
  number	= {7619},
  pages		= {210--213},
  publisher	= {Nature Publishing Group},
  issn		= {14764687},
  doi		= {10.1038/nature19315}
}

@Article{	  secchiscalingbehaviorionic2016,
  title		= {Scaling {{Behavior}} for {{Ionic Transport}} and Its
		  {{Fluctuations}} in {{Individual Carbon Nanotubes}}},
  author	= {Secchi, Eleonora and Nigu{\`e}s, Antoine and Jubin,
		  Laetitia and Siria, Alessandro and Bocquet, Lyd{\'e}ric},
  year		= 2016,
  month		= apr,
  journal	= {Physical Review Letters},
  volume	= {116},
  number	= {15},
  publisher	= {American Physical Society (APS)},
  issn		= {0031-9007, 1079-7114},
  doi		= {10.1103/physrevlett.116.154501},
  urldate	= {2025-07-24},
  copyright	= {http://link.aps.org/licenses/aps-default-license},
  langid	= {english}
}

@Article{	  siriagiantosmoticenergy2013,
  title		= {Giant Osmotic Energy Conversion Measured in a Single
		  Transmembrane Boron Nitride Nanotube},
  author	= {Siria, Alessandro and Poncharal, Philippe and Biance,
		  Anne-Laure and Fulcrand, R{\'e}my and Blase, Xavier and
		  Purcell, Stephen T. and Bocquet, Lyd{\'e}ric},
  year		= 2013,
  month		= feb,
  journal	= {Nature},
  volume	= {494},
  number	= {7438},
  pages		= {455--458},
  publisher	= {Nature Publishing Group},
  issn		= {1476-4687},
  doi		= {10.1038/nature11876},
  urldate	= {2025-07-30},
  copyright	= {2013 Springer Nature Limited},
  langid	= {english}
}

@Article{	  sirianewavenues2017,
  title		= {New Avenues for the Large-Scale Harvesting of Blue
		  Energy},
  author	= {Siria, Alessandro and Bocquet, Marie Laure and Bocquet,
		  Lyd{\'e}ric},
  year		= 2017,
  journal	= {Nature Reviews Chemistry},
  volume	= {1},
  number	= {11},
  pages		= {0091},
  issn		= {23973358},
  doi		= {10.1038/s41570-017-0091}
}

@Article{	  stauberquasiflatplasmonic2016,
  title		= {Quasi-{{Flat Plasmonic Bands}} in {{Twisted Bilayer
		  Graphene}}},
  author	= {Stauber, Tobias and Kohler, Heinerich},
  year		= 2016,
  journal	= {Nano Letters},
  volume	= {16},
  number	= {11},
  pages		= {6844--6849},
  issn		= {15306992},
  doi		= {10.1021/acs.nanolett.6b02587}
}

@Article{	  stein2004surface,
  title		= {Surface-charge-governed ion transport in nanofluidic
		  channels},
  author	= {Stein, Derek and Kruithof, Maarten and Dekker, Cees},
  journal	= {Physical Review Letters},
  volume	= {93},
  number	= {3},
  pages		= {035901},
  year		= {2004},
  publisher	= {APS}
}

@Article{	  succikeldyshlatticeboltzmann2025,
  title		= {Keldysh {{Lattice Boltzmann Approach}} to {{Quantum
		  Nanofluidics}}},
  author	= {Succi, Sauro and Lauricella, Marco and Montessori,
		  Andrea},
  year		= 2025,
  month		= apr,
  journal	= {AIAA Journal},
  volume	= {63},
  number	= {4},
  pages		= {1330--1337},
  issn		= {0001-1452, 1533-385X},
  doi		= {10.2514/1.J064211},
  urldate	= {2025-06-03},
  langid	= {english}
}

@Article{	  succinonequilibriumgreensfunctions2025,
  title		= {From Non-Equilibrium {{Green}}'s Functions to {{Lattice
		  Wigner}}: {{A}} Toy Model for Quantum Nanofluidics
		  Simulations},
  shorttitle	= {From Non-Equilibrium {{Green}}'s Functions to {{Lattice
		  Wigner}}},
  author	= {Succi, S. and Lauricella, M. and Tiribocchi, A.},
  year		= 2025,
  month		= apr,
  journal	= {Physics of Fluids},
  volume	= {37},
  number	= {4},
  pages		= {047122},
  issn		= {1070-6631},
  doi		= {10.1063/5.0260872},
  urldate	= {2025-12-08}
}

@Article{	  takedaenhancingelectricitygeneration2025,
  title		= {Enhancing Electricity Generation at Graphene--Fluid
		  Interface under Disturbance Flow Conditions},
  author	= {Takeda, Hikaru and Honda, Mitsuhiro and Tanemura, Masaki
		  and Yamashita, Ichiro and Komiya, Atsuki and Okada,
		  Takeru},
  year		= 2025,
  month		= dec,
  journal	= {AIP Advances},
  volume	= {15},
  number	= {12},
  pages		= {125305},
  issn		= {2158-3226},
  doi		= {10.1063/5.0294593},
  urldate	= {2025-12-07}
}

@Article{	  thiemannwaterflow2022,
  title		= {Water {{Flow}} in {{Single-Wall Nanotubes}}: {{Oxygen
		  Makes It Slip}}, {{Hydrogen Makes It Stick}}},
  shorttitle	= {Water {{Flow}} in {{Single-Wall Nanotubes}}},
  author	= {Thiemann, Fabian L. and Schran, Christoph and Rowe,
		  Patrick and M{\"u}ller, Erich A. and Michaelides, Angelos},
  year		= {2022},
  month		= jul,
  journal	= {ACS Nano},
  volume	= {16},
  number	= {7},
  pages		= {10775--10782},
  issn		= {1936-0851, 1936-086X},
  doi		= {10.1021/acsnano.2c02784},
  urldate	= {2024-09-10},
  copyright	= {https://creativecommons.org/licenses/by/4.0/},
  langid	= {english}
}

@Article{	  tholemolecularpolarizabilitiescalculated1981,
  title		= {Molecular Polarizabilities Calculated with a Modified
		  Dipole Interaction},
  author	= {Thole, B. T.},
  year		= 1981,
  month		= aug,
  journal	= {Chemical Physics},
  volume	= {59},
  number	= {3},
  pages		= {341--350},
  issn		= {0301-0104},
  doi		= {10.1016/0301-0104(81)85176-2},
  urldate	= {2026-02-28}
}

@Article{	  toccifrictionwater2014,
  title		= {Friction of Water on Graphene and Hexagonal Boron Nitride
		  from {{Ab}} Initio Methods: {{Very}} Different Slippage
		  despite Very Similar Interface Structures},
  author	= {Tocci, Gabriele and Joly, Laurent and Michaelides,
		  Angelos},
  year		= {2014},
  journal	= {Nano Letters},
  volume	= {14},
  number	= {12},
  pages		= {6872--6877},
  issn		= {15306992},
  doi		= {10.1021/nl502837d}
}

@Article{	  tocciinitionanofluidics2020,
  title		= {Ab Initio Nanofluidics: Disentangling the Role of the
		  Energy Landscape and of Density Correlations on
		  Liquid/Solid Friction},
  author	= {Tocci, Gabriele and Bilichenko, Maria and Joly, Laurent
		  and Iannuzzi, Marcella},
  year		= {2020},
  journal	= {Nanoscale},
  volume	= {12},
  number	= {20},
  pages		= {10994--11000},
  issn		= {2040-3364},
  doi		= {10.1039/D0NR02511A}
}

@Article{	  tomassone1997,
  title		= {Electronic Friction Forces on Molecules Moving near
		  Metals},
  author	= {Tomassone, M. and Widom, A.},
  year		= {1997},
  journal	= {Physical Review B - Condensed Matter and Materials
		  Physics},
  volume	= {56},
  number	= {8},
  pages		= {4938--4943},
  issn		= {1550235X},
  doi		= {10.1103/PhysRevB.56.4938}
}

@Article{	  tunuguntlaenhancedwaterpermeability2017,
  title		= {Enhanced Water Permeability and Tunable Ion Selectivity in
		  Subnanometer Carbon Nanotube Porins},
  author	= {Tunuguntla, Ramya H. and Henley, Robert Y. and Yao,
		  Yun-Chiao and Pham, Tuan Anh and Wanunu, Meni and Noy,
		  Aleksandr},
  year		= 2017,
  month		= aug,
  journal	= {Science},
  volume	= {357},
  number	= {6353},
  pages		= {792--796},
  publisher	= {American Association for the Advancement of Science},
  doi		= {10.1126/science.aan2438},
  urldate	= {2025-07-29}
}

@Article{	  uematsu2018,
  title		= {Crossover of the {{Power-Law Exponent}} for {{Carbon
		  Nanotube Conductivity}} as a {{Function}} of {{Salinity}}},
  author	= {Uematsu, Yuki and Netz, Roland R. and Bocquet, Lyd{\'e}ric
		  and Bonthuis, Douwe Jan},
  year		= {2018},
  month		= mar,
  journal	= {The Journal of Physical Chemistry B},
  volume	= {122},
  number	= {11},
  pages		= {2992--2997},
  issn		= {1520-6106, 1520-5207},
  doi		= {10.1021/acs.jpcb.8b01975},
  urldate	= {2023-05-24},
  langid	= {english}
}

@Article{	  vlasov1968vibrational,
  title		= {The vibrational properties of an electron gas},
  author	= {Vlasov, Anatoli Aleksandrovich},
  journal	= {Soviet Physics Uspekhi},
  volume	= {10},
  number	= {6},
  pages		= {721--733},
  year		= {1968}
}

@Article{	  vlassiouknanofluidicdiode2007,
  title		= {Nanofluidic Diode},
  author	= {Vlassiouk, Ivan and Siwy, Zuzanna S.},
  year		= 2007,
  journal	= {Nano Letters},
  volume	= {7},
  number	= {3},
  pages		= {552--556},
  issn		= {15306984},
  doi		= {10.1021/nl062924b}
}

@Article{	  volokitin1999,
  title		= {Theory of Friction: The Contribution from a Fluctuating
		  Electromagnetic Field},
  shorttitle	= {Theory of Friction},
  author	= {Volokitin, A I and Persson, B N J},
  year		= {1999},
  month		= jan,
  journal	= {Journal of Physics: Condensed Matter},
  volume	= {11},
  number	= {2},
  pages		= {345--359},
  issn		= {0953-8984, 1361-648X},
  doi		= {10.1088/0953-8984/11/2/003},
  urldate	= {2025-09-26}
}

@Article{	  volokitin2006,
  title		= {Quantum Field Theory of van Der {{Waals}} Friction},
  author	= {Volokitin, A. I. and Persson, B. N.J.},
  year		= {2006},
  journal	= {Physical Review B - Condensed Matter and Materials
		  Physics},
  volume	= {74},
  number	= {20},
  pages		= {1--11},
  issn		= {10980121},
  doi		= {10.1103/PhysRevB.74.205413}
}

@Article{	  volokitin2007,
  title		= {Near-Field Radiative Heat Transfer and Noncontact
		  Friction},
  author	= {Volokitin, A. I. and Persson, B. N. J.},
  year		= {2007},
  month		= oct,
  journal	= {Reviews of Modern Physics},
  volume	= {79},
  number	= {4},
  pages		= {1291--1329},
  issn		= {0034-6861, 1539-0756},
  doi		= {10.1103/RevModPhys.79.1291},
  urldate	= {2025-08-29},
  copyright	= {http://link.aps.org/licenses/aps-default-license},
  langid	= {english}
}

@Article{	  volokitin2008,
  title		= {Theory of the Interaction Forces and the Radiative Heat
		  Transfer between Moving Bodies},
  author	= {Volokitin, A. I. and Persson, B. N. J.},
  year		= {2008},
  month		= oct,
  journal	= {Physical Review B},
  volume	= {78},
  number	= {15},
  pages		= {155437},
  publisher	= {American Physical Society},
  doi		= {10.1103/PhysRevB.78.155437},
  urldate	= {2025-10-08}
}

@Article{	  wangspontaneoussurfacecharging2025,
  title		= {Spontaneous {{Surface Charging}} and {{Janus Nature}} of
		  the {{Hexagonal Boron Nitride}}--{{Water Interface}}},
  author	= {Wang, Yongkang and Luo, Haojian and Advincula, Xavier R.
		  and Zhao, Zhengpu and Esfandiar, Ali and Wu, Da and Fong,
		  Kara D. and Gao, Lei and Hazrah, Arsh S. and Taniguchi,
		  Takashi and Schran, Christoph and Nagata, Yuki and Bocquet,
		  Lyd{\'e}ric and Bocquet, Marie-Laure and Jiang, Ying and
		  Michaelides, Angelos and Bonn, Mischa},
  year		= 2025,
  month		= aug,
  journal	= {Journal of the American Chemical Society},
  volume	= {147},
  number	= {33},
  pages		= {30107--30116},
  publisher	= {American Chemical Society},
  issn		= {0002-7863},
  doi		= {10.1021/jacs.5c07827},
  urldate	= {2025-10-14}
}

@Article{	  xiefastwatertransport2018,
  title		= {Fast Water Transport in Graphene Nanofluidic Channels},
  author	= {Xie, Quan and Alibakhshi, Mohammad Amin and Jiao, Shuping
		  and Xu, Zhiping and Hempel, Marek and Kong, Jing and Park,
		  Hyung Gyu and Duan, Chuanhua},
  year		= 2018,
  month		= mar,
  journal	= {Nature Nanotechnology},
  volume	= {13},
  number	= {3},
  pages		= {238--245},
  publisher	= {{Springer Science and Business Media LLC}},
  issn		= {1748-3387, 1748-3395},
  doi		= {10.1038/s41565-017-0031-9},
  urldate	= {2025-07-22},
  copyright	= {http://www.springer.com/tdm},
  langid	= {english}
}

@Article{	  xieliquidsolidslip2020,
  title		= {Liquid-{{Solid Slip}} on {{Charged Walls}}: {{The Dramatic
		  Impact}} of {{Charge Distribution}}},
  author	= {Xie, Yanbo and Fu, Li and Niehaus, Thomas and Joly,
		  Laurent},
  year		= {2020},
  journal	= {Physical Review Letters},
  volume	= {125},
  number	= {1},
  eprint	= {2002.02444},
  pages		= {1--7},
  issn		= {10797114},
  doi		= {10.1103/PhysRevLett.125.014501},
  archiveprefix	= {arXiv},
  pmid		= {32678629}
}

@Article{	  yangfastwatertransport2023,
  title		= {Fast {{Water Transport}} through {{Subnanometer Diameter
		  Vertically Aligned Carbon Nanotube Membranes}}},
  author	= {Yang, Da-Chi and Castellano, Richard J. and Silvy, Ricardo
		  Prada and Lageshetty, Sathish K. and Praino, Robert F. and
		  Fornasiero, Francesco and Shan, Jerry W.},
  year		= 2023,
  month		= jun,
  journal	= {Nano Letters},
  volume	= {23},
  number	= {11},
  pages		= {4956--4964},
  issn		= {1530-6984, 1530-6992},
  doi		= {10.1021/acs.nanolett.3c00797},
  urldate	= {2025-02-26},
  copyright	= {https://doi.org/10.15223/policy-029},
  langid	= {english}
}

@Article{	  yingeneratingelectricitymoving2014,
  title		= {Generating Electricity by Moving a Droplet of Ionic Liquid
		  along Graphene},
  author	= {Yin, Jun and Li, Xuemei and Yu, Jin and Zhang, Zhuhua and
		  Zhou, Jianxin and Guo, Wanlin},
  year		= {2014},
  month		= may,
  journal	= {Nature Nanotech},
  volume	= {9},
  number	= {5},
  pages		= {378--383},
  issn		= {1748-3387, 1748-3395},
  doi		= {10.1038/nnano.2014.56},
  urldate	= {2025-03-03},
  copyright	= {http://www.springer.com/tdm},
  langid	= {english}
}

@Article{	  yinharvestingenergywater2012,
  title		= {Harvesting {{Energy}} from {{Water Flow}} over
		  {{Graphene}}?},
  author	= {Yin, Jun and Zhang, Zhuhua and Li, Xuemei and Zhou,
		  Jianxin and Guo, Wanlin},
  year		= 2012,
  month		= mar,
  journal	= {Nano Letters},
  volume	= {12},
  number	= {3},
  pages		= {1736--1741},
  issn		= {1530-6984, 1530-6992},
  doi		= {10.1021/nl300636g},
  urldate	= {2025-10-07},
  langid	= {english}
}

@Article{	  yuelectroncooling2023,
  title		= {Electron Cooling in Graphene Enhanced by Plasmon--Hydron
		  Resonance},
  author	= {Yu, Xiaoqing and Principi, Alessandro and Tielrooij,
		  Klaas-Jan and Bonn, Mischa and Kavokine, Nikita},
  year		= {2023},
  month		= aug,
  journal	= {Nat. Nanotechnol.},
  volume	= {18},
  number	= {8},
  pages		= {898--904},
  issn		= {1748-3387, 1748-3395},
  doi		= {10.1038/s41565-023-01421-3},
  urldate	= {2023-10-27},
  langid	= {english}
}

@Article{	  yuinitiostudypolarizability2008,
  title		= {Ab Initio Study of Polarizability and Induced Charge
		  Densities in Multilayer Graphene Films},
  author	= {Yu, E. K. and Stewart, D. A. and Tiwari, S.},
  year		= 2008,
  month		= may,
  journal	= {Physical Review B},
  volume	= {77},
  number	= {19},
  pages		= {195406},
  publisher	= {American Physical Society},
  doi		= {10.1103/PhysRevB.77.195406},
  urldate	= {2026-02-28}
}

@Article{	  zwanzigdielectricfrictionmoving1963,
  title		= {Dielectric {{Friction}} on a {{Moving Ion}}},
  author	= {Zwanzig, Robert},
  year		= {1963},
  month		= apr,
  journal	= {The Journal of Chemical Physics},
  volume	= {38},
  number	= {7},
  pages		= {1603--1605},
  issn		= {0021-9606, 1089-7690},
  doi		= {10.1063/1.1776929},
  urldate	= {2025-04-14},
  langid	= {english}
}

@article{cetindagIonHydrodynamicTranslucency2024,
	title = {Ion and {{Hydrodynamic Translucency}} in {{1D}} van Der {{Waals Heterostructured Boron-Nitride Single-Walled Carbon Nanotubes}}},
	author = {Cetindag, Semih and Park, Sei Jin and Buchsbaum, Steven F. and Zheng, Yongjia and Liu, Ming and Wang, Shuhui and Xiang, Rong and Maruyama, Shigeo and Fornasiero, Francesco and Shan, Jerry W.},
	year = 2024,
	month = jan,
	journal = {ACS Nano},
	volume = {18},
	number = {1},
	pages = {355--363},
	publisher = {American Chemical Society},
	issn = {1936-0851},
	doi = {10.1021/acsnano.3c07282},
	urldate = {2025-08-21}
}

@article{baekHighPerformanceAntifouling2014,
	title = {High Performance and Antifouling Vertically Aligned Carbon Nanotube Membrane for Water Purification},
	author = {Baek, Youngbin and Kim, Cholin and Seo, Dong Kyun and Kim, Taewoo and Lee, Jeong Seok and Kim, Yong Hyup and Ahn, Kyung Hyun and Bae, Sang Seek and Lee, Sang Cheol and Lim, Jaelim and Lee, Kyunghyuk and Yoon, Jeyong},
	year = 2014,
	month = jun,
	journal = {Journal of Membrane Science},
	volume = {460},
	pages = {171--177},
	issn = {03767388},
	doi = {10.1016/j.memsci.2014.02.042},
	urldate = {2025-03-17},
	langid = {english}
}

@article{buiUltrabreathableProtectiveMembranes2016,
	title = {Ultrabreathable and {{Protective Membranes}} with {{Sub}}-5 Nm {{Carbon Nanotube Pores}}},
	author = {Bui, Ngoc and Meshot, Eric R. and Kim, Sangil and Pe{\~n}a, Jos{\'e} and Gibson, Phillip W. and Wu, Kuang Jen and Fornasiero, Francesco},
	year = 2016,
	month = jul,
	journal = {Advanced Materials},
	volume = {28},
	number = {28},
	pages = {5871--5877},
	issn = {0935-9648, 1521-4095},
	doi = {10.1002/adma.201600740},
	urldate = {2025-03-17},
	copyright = {http://onlinelibrary.wiley.com/termsAndConditions\#vor},
	langid = {english}
}

@article{duMembranesVerticallyAligned2011,
	title = {Membranes of {{Vertically Aligned Superlong Carbon Nanotubes}}},
	author = {Du, Feng and Qu, Liangti and Xia, Zhenhai and Feng, Lianfang and Dai, Liming},
	year = 2011,
	month = jul,
	journal = {Langmuir},
	volume = {27},
	number = {13},
	pages = {8437--8443},
	issn = {0743-7463, 1520-5827},
	doi = {10.1021/la200995r},
	urldate = {2025-03-17},
	langid = {english}
}

@article{jueUltraPermeableSingleWalledCarbon2020,
	title = {Ultra-{{Permeable Single}}-{{Walled Carbon Nanotube Membranes}} with {{Exceptional Performance}} at {{Scale}}},
	author = {Jue, Melinda L. and Buchsbaum, Steven F. and Chen, Chiatai and Park, Sei Jin and Meshot, Eric R. and Wu, Kuang Jen J. and Fornasiero, Francesco},
	year = 2020,
	month = dec,
	journal = {Advanced Science},
	volume = {7},
	number = {24},
	pages = {2001670},
	issn = {2198-3844, 2198-3844},
	doi = {10.1002/advs.202001670},
	urldate = {2025-03-17},
	langid = {english}
}

@article{kimFabricationFlexibleAligned2014,
	title = {Fabrication of Flexible, Aligned Carbon Nanotube/Polymer Composite Membranes by in-Situ Polymerization},
	author = {Kim, Sangil and Fornasiero, Francesco and Park, Hyung Gyu and In, Jung Bin and Meshot, Eric and Giraldo, Gabriel and Stadermann, Michael and Fireman, Micha and Shan, Jerry and Grigoropoulos, Costas P. and Bakajin, Olgica},
	year = 2014,
	month = jun,
	journal = {Journal of Membrane Science},
	volume = {460},
	pages = {91--98},
	issn = {03767388},
	doi = {10.1016/j.memsci.2014.02.016},
	urldate = {2025-03-17},
	langid = {english}
}

@article{leeMostDensifiedVerticallyaligned2016,
	title = {The Most Densified Vertically-Aligned Carbon Nanotube Membranes and Their Normalized Water Permeability and High Pressure Durability},
	author = {Lee, Kwang-Jin and Park, Hee-Deung},
	year = 2016,
	month = mar,
	journal = {Journal of Membrane Science},
	volume = {501},
	pages = {144--151},
	issn = {03767388},
	doi = {10.1016/j.memsci.2015.12.009},
	urldate = {2025-03-17},
	langid = {english}
}

@article{majumderMassTransportCarbon2011,
	title = {Mass {{Transport}} through {{Carbon Nanotube Membranes}} in {{Three Different Regimes}}: {{Ionic Diffusion}} and {{Gas}} and {{Liquid Flow}}},
	shorttitle = {Mass {{Transport}} through {{Carbon Nanotube Membranes}} in {{Three Different Regimes}}},
	author = {Majumder, Mainak and Chopra, Nitin and Hinds, Bruce J.},
	year = 2011,
	month = may,
	journal = {ACS Nano},
	volume = {5},
	number = {5},
	pages = {3867--3877},
	issn = {1936-0851, 1936-086X},
	doi = {10.1021/nn200222g},
	urldate = {2025-03-17},
	langid = {english}
}

@article{mcginnisLargescalePolymericCarbon2018,
	title = {Large-Scale Polymeric Carbon Nanotube Membranes with Sub--1.27-Nm Pores},
	author = {McGinnis, Robert L. and Reimund, Kevin and Ren, Jian and Xia, Lingling and Chowdhury, Maqsud R. and Sun, Xuanhao and Abril, Maritza and Moon, Joshua D. and Merrick, Melanie M. and Park, Jaesung and Stevens, Kevin A. and McCutcheon, Jeffrey R. and Freeman, Benny D.},
	year = 2018,
	month = mar,
	journal = {Science Advances},
	volume = {4},
	number = {3},
	pages = {e1700938},
	issn = {2375-2548},
	doi = {10.1126/sciadv.1700938},
	urldate = {2025-03-17},
	langid = {english}
}

@article{trivediEffectVerticallyAligned2016,
	title = {Effect of Vertically Aligned Carbon Nanotube Density on the Water Flux and Salt Rejection in Desalination Membranes},
	author = {Trivedi, Samarth and Alameh, Kamal},
	year = 2016,
	month = dec,
	journal = {SpringerPlus},
	volume = {5},
	number = {1},
	pages = {1158},
	issn = {2193-1801},
	doi = {10.1186/s40064-016-2783-3},
	urldate = {2025-03-17},
	langid = {english}
}

@article{zhangGasTransportVerticallyaligned2014,
	title = {Gas Transport in Vertically-Aligned Carbon Nanotube/Parylene Composite Membranes},
	author = {Zhang, Lei and Zhao, Bin and Wang, Xianying and Liang, Youxuan and Qiu, Hanxun and Zheng, Guangping and Yang, Junhe},
	year = 2014,
	month = jan,
	journal = {Carbon},
	volume = {66},
	pages = {11--17},
	issn = {00086223},
	doi = {10.1016/j.carbon.2013.08.007},
	urldate = {2025-03-17},
	langid = {english}
}

@article{heiranianRevisitingSampsonsTheory2020,
	title = {Revisiting {{Sampson}}'s Theory for Hydrodynamic Transport in Ultrathin Nanopores},
	author = {Heiranian, Mohammad and Taqieddin, Amir and Aluru, Narayana R.},
	year = 2020,
	month = oct,
	journal = {Physical Review Research},
	volume = {2},
	number = {4},
	pages = {043153},
	issn = {2643-1564},
	doi = {10.1103/PhysRevResearch.2.043153},
	urldate = {2025-03-06},
	langid = {english}
}

@article{liBreakdownNernstEinstein2023,
	title = {Breakdown of the {{Nernst}}--{{Einstein}} Relation in Carbon Nanotube Porins},
	author = {Li, Zhongwu and Misra, Rahul Prasanna and Li, Yuhao and Yao, Yun-Chiao and Zhao, Sidi and Zhang, Yuliang and Chen, Yunfei and Blankschtein, Daniel and Noy, Aleksandr},
	year = 2023,
	month = feb,
	journal = {Nature Nanotechnology},
	volume = {18},
	number = {2},
	pages = {177--183},
	publisher = {Nature Publishing Group},
	issn = {1748-3395},
	doi = {10.1038/s41565-022-01276-0},
	urldate = {2025-07-29},
	copyright = {2022 The Author(s), under exclusive licence to Springer Nature Limited},
	langid = {english}
}

@article{schranMachineLearningPotentials2021,
	title = {Machine Learning Potentials for Complex Aqueous Systems Made Simple},
	author = {Schran, Christoph and Thiemann, Fabian L. and Rowe, Patrick and M{\"u}ller, Erich A. and Marsalek, Ondrej and Michaelides, Angelos},
	year = 2021,
	month = sep,
	journal = {Proceedings of the National Academy of Sciences},
	volume = {118},
	number = {38},
	pages = {e2110077118},
	issn = {0027-8424, 1091-6490},
	doi = {10.1073/pnas.2110077118},
	urldate = {2024-10-21},
	langid = {english}
}

@article{kapilFirstprinciplesPhaseDiagram2022,
	title = {The First-Principles Phase Diagram of Monolayer Nanoconfined Water},
	author = {Kapil, Venkat and Schran, Christoph and Zen, Andrea and Chen, Ji and Pickard, Chris J. and Michaelides, Angelos},
	year = 2022,
	month = sep,
	journal = {Nature},
	volume = {609},
	number = {7927},
	pages = {512--516},
	issn = {0028-0836, 1476-4687},
	doi = {10.1038/s41586-022-05036-x},
	urldate = {2026-05-18},
	langid = {english}
}

@article{advinculaProtonsAccumulateGraphene2025a,
	title = {Protons {{Accumulate}} at the {{Graphene}}--{{Water Interface}}},
	author = {Advincula, Xavier R. and Fong, Kara D. and Michaelides, Angelos and Schran, Christoph},
	year = 2025,
	month = may,
	journal = {ACS Nano},
	volume = {19},
	number = {18},
	pages = {17728--17737},
	publisher = {American Chemical Society},
	issn = {1936-0851},
	doi = {10.1021/acsnano.5c02053},
	urldate = {2026-05-18}
}

@article{brandenburgInteractionWaterCarbon2019,
	title = {Interaction between Water and Carbon Nanostructures: {{How}} Good Are Current Density Functional Approximations?},
	shorttitle = {Interaction between Water and Carbon Nanostructures},
	author = {Brandenburg, Jan Gerit and Zen, Andrea and Alf{\`e}, Dario and Michaelides, Angelos},
	year = 2019,
	month = oct,
	journal = {The Journal of Chemical Physics},
	volume = {151},
	number = {16},
	pages = {164702},
	issn = {0021-9606},
	doi = {10.1063/1.5121370},
	urldate = {2026-05-18}
}

@article{mihaila2011lindhard,
  title={Lindhard function of a d-dimensional Fermi gas},
  author={Mihaila, Bogdan},
  journal={arXiv preprint arXiv:1111.5337},
  year={2011}
}

\end{document}